\documentstyle[aps,preprint,tighten,rmp,eqsecnum,amssymb,harvard,psfig]{revtex}

\def\({\left(} \def\){\right)}

\def\br{{\bf r}} 
\def\bx{{\bf \xi}} 
\def \bs{{\bf s}} 
\def\bk{{\bf k}}
\newcommand{\bd}{{\bf d}}
\newcommand{\bj}{{\bf j}}

\newcommand{\bmu}{{\bf \mu}}
\newcommand{\bsigma}{\bbox{\sigma}}

\newcommand{\cD}{{\cal D}}
\newcommand{\cE}{{\cal E}}
\newcommand{\cF}{{\cal F}}
\newcommand{\cS}{{\cal S}}
\newcommand{\cV}{{\cal V}}

\newcommand{\ue}{\underline{e}}
\newcommand{\umu}{\underline{\mu}}

\newcommand{\dist}{{\rm dist}}
\newcommand{\1}{\openone}
\newcommand{\DDelta}{\Delta \!\!\!\!\Delta} 
\newcommand{\lat}{({\cal Z})^{d}\cap \Lambda } 
\newcommand{\av}[1]{\langle #1 \rangle } 

\newtheorem{theorem}{Theorem}[section]

\newcommand{\nn}{\nonumber}
\newcommand{\proof}{\noindent {\bf Proof. }\hspace{2mm}}
\newcommand{\qed}{\hspace*{\fill} QED \medskip}

\renewcommand{\thepage}{\thesection--\arabic{page}}

\begin{document}

\title{Coulomb systems at low density}

\author{David C. Brydges}
\address{Department of Mathematics\\
University of Virginia\\
Charlottesville, VA 22903}

\author{Ph. A. Martin}
\address{Institut de Physique Th\'eorique\\
Ecole Polytechnique F\'ed\'erale de Lausanne\\
CH-1015 Lausanne\\
Switzerland}

\date{February 16 1998}
\maketitle

\begin{abstract}
Results on the correlations of low density classical and quantum
Coulomb systems at equilibrium in three dimensions are reviewed.  The
exponential decay of particle correlations in the classical Coulomb
system -- Debye-H\"uckel screening -- is compared and contrasted with
the quantum case where strong arguments are presented for the absence
of exponential screening. Results and techniques for detailed
calculations that determine the asymptotic decay of correlations for
quantum systems are discussed. Theorems on the existence of molecules
in the Saha regime are reviewed. Finally, new combinatoric formulas
for the coefficients of Mayer expansions are presented and their role
in proofs of results on Debye-H\"uckel screening is discussed.
\end{abstract}

\newpage
\tableofcontents

\newpage \section{Coulomb systems} \setcounter{page}{1} 
\label{chapter- Coulomb systems}

\subsection{Introduction}
\label{sec- introduction}

The emergence of thermodynamics and of various physical laws from the
more fundamental levels of atomic theory and statistical mechanics
were high points in our education in physics. But by the time we
reached textbook treatments of the complex world of the Coulomb
interaction, for example, formation of atoms and molecules and
equilibria between them, we learned to demand less from theory and to
be more content with reasoning by analogy consistent with
thermodynamics.  We know experimentally that atoms and molecules form,
that they can behave like mixtures of ideal gases, so
phenomenologically they have chemical potentials and we no longer
insist that this emerge in some limit from $N$-body quantum Coulomb
systems.  We also know that Coulomb systems can have screening and
dielectric phases. But when screening is expected to occur, it is
quite a common practice to rather bluntly replace the Coulomb
potential by a screened potential inherited from mean-field theories
without inquiring too deeply into the legitimacy of this
change. Likewise, there does not exist a first principle theory of the
dielectric constant that does not presuppose the existence of atoms
and molecules.  At best, assuming that atoms form, one uses several
laws, such as the Clausius-Mosotti formula, whose microscopic
foundations have to be elucidated. In all these cases, this is a
sensible attitude since most of non-relativistic physics is reputed to
be hidden within the $N$-particle Coulomb Hamiltonian. There are
however basic facts that can be cleanly formulated as limiting
theories and shown to persist near the limit as well.

This review is devoted to these types of results on Coulomb systems at
low density.  At low density the most famous properties of Coulomb
systems are related to screening.  The classical Debye-H\"uckel theory
and its quantum analogue the random phase approximation have been
supplemented in recent years by additional information established
both by rigorous proofs and by closer inspection of resummed
perturbation theory consistent with aspects of the rigorous
proofs. There is also a precise understanding how a dilute assembly of
quantum nuclei and electrons can form gases of atoms and molecules in
the Saha regime. These results are physically well understood but the
challenge to derive them has shown that standard perturbation theories
miss some effects; indeed it turns out that quantum Coulomb systems do
not screen, in the strong sense that the classical system does. This
is a principal focus of this review.

The first rigorous results were at the level of thermodynamics, the
celebrated stability of matter theorem of Dyson and Lenard and the
existence of thermodynamic behavior established by Lieb and
Lebowitz. They hold for all states of nonrelativistic matter. 
The importance of the stability result is that it shows that quantum
mechanics and electrostatics is a consistent simplified statistical
mechanics of nature that needs no intervention from more fundamental
levels of description.

The next level, which is the focus of this review, is to ask for
information on correlations: here the variety of physical phenomena in
question is so vast that we cannot hope for too much generality:
thorough consideration of the low density behavior is the first and
the most natural step in this direction.

There are three parts. The first is devoted to classical systems.
Here a significant achievement is the proof that the Debye-H\"uckel
theory is a limiting theory and Debye screening holds close to this
limit, that is, at sufficiently low density (or high
temperature). This is the subject of chapter~\ref{chapter-
debye-screening}.

Frequently proofs of such results are very long with tedious parts
devoid of physical insight, but there are also components where a
physical insight becomes a precise argument or where a useful new
identity such as a resummation of perturbation theory appears.  The
philosophy here and throughout the review is to state the theorems and
to give only those ideas in the proofs that conform to this
specification.  Many of these arguments are founded on the Sine-Gordon
transformation, described in chapter~\ref{chapter- debye-screening}.
We hope that our applications will raise appreciation for this
transformation.  We hope to display enough arguments without
overwhelming technicalities to satisfy the physicist curious to see
the way between physical concepts and their full mathematical
realization.

In the second part we consider quantum mechanics.  In between there is
a natural transition chapter (chapter~\ref{chapter- dipole}) which
revolves around the absence of screening in classical dipole
systems. We present the simple proof of this lack of screening.  This
chapter prepares for the study of the quantum case because it turns
out that the Feynman-Kac formula reduces quantum equilibrium
statistical mechanics to a classical formalism in which both monopoles
and higher order multipoles appear.

The proper quantum mechanical effects on screening are exhibited in
chapter~\ref{chapter- Semi-classical} in the simplified framework of
the semi-classical approximation. The well known Feynman-Kac
representation of quantum statistical mechanics is recalled and
quantum statistics are neglected. Perturbing around the classical gas
makes immediately clear why intrinsic quantum fluctuations always
destroy Debye screening.  This appears to have been first suggested by
Federbush, see \cite[page 428, section 5]{BrFe81}. One also discovers
that the remaining screening mechanisms are more subtle than in the
classical case, as revealed by the different long distance behaviors
of various types of correlations. This chapter is intended to be a
gentle baptism before the confrontation with the full quantum
mechanical setting in chapter~\ref{chapter-loops}.

Here the mathematics is much harder and less has been proved: in
chapters ~\ref{chapter-loops} and ~\ref{chapter-6} the style is that
of formal perturbation theory. In these chapters, following mainly the
work of Alastuey, Cornu and collaborators, one exploits the full power
of the Feynman-Kac representation including the quantum statistics. In
this language, the quantum Coulomb gas is similar to a certain
classical gas of fluctuating multipoles and all the efficient
techniques of classical statistical mechanics are at hand. In
particular, one can perform the usual partial resummations needed to
cure the Coulomb divergences and end up with well defined Mayer-like
diagrammatic rules.  This offers a new perturbative scheme for the
many-body problem which is particularly suited for low density
expansions, in contrast to the standard Feynman perturbation theory
with respect to the coupling constant.

The formalism is applied in chapter~\ref{chapter-6} to the
determination of the long distance behavior of the correlations and to
the equation of state at low density. The (formally) exact results
presented here provide the most detailed information available now on
these questions. Some of them are new, in particular the explicit
formulae for the tails of the correlations at low density.  Also one
recovers and completes the equation of state already obtained by
Ebeling and coworkers by different methods.

In chapter~\ref{chapter-7} we report on a nice development, initiated
by Fefferman, on atomic and molecular phases at low density (the Saha
regime). One considers a joint limit, the atomic or molecular limit,
of vanishing density and vanishing temperature by fixing the chemical
potentials (negatives) and letting the temperature tend to
zero. Lowering the temperature enhances the probabilities for quantum
mechanical binding whereas lowering the density favors ionization.  It
turns out that in this limit each value of the chemical potentials
determines in principle certain chemical species and the system
behaves as a mixture of these ideal substances. Ionization equilibrium
phases are obtained in this way.  In this asymptotic sense, it gives a
precise meaning to the notion of atoms and molecules in the many-body
problem.

The third part consists of a single
chapter~\ref{chapter-convergent-expansions} that is more technical:
rigorous results at low density require control of the convergence of
various types of cluster expansions.  The proof of Debye screening has
stimulated the development of resummation techniques which may be
useful in other situations. For example there are new combinatoric
formulae for Mayer coefficients. Without aiming at completeness, this
last chapter provides the dedicated reader with some arguments left
out in the main course of the text and gives an updated summary of
some of the available tools in this field.

The considerations in this review are essentially concerned with our
three dimensional world. There has of course been dramatic progress in
the one and two dimensional solvable models of Coulomb systems, which
we do not include in the review, apart from isolated comments included
to demonstrate some contrast with three dimensions. Let us also
mention that basic new results on the stability of Coulombic matter in
presence of magnetic fields and quantized radiation field have been
obtained recently (references in \ref{sec- quantum}).

The history of proofs of fundamental results, for example that
statistical mechanics predicts phase transitions, suggests that
difficult results can simplify over time and serve as consistency
checks on what comes next. Some but not all of the derivations in this
review are far from this stage: nevertheless, the existence of the
subtle tunneling corrections in the Debye asymptotic regime revealed
by the Sine-Gordon representation is not seen by conventional Mayer
graph summations. The suggestive view of the quantum gas as an
assembly of random multipoles displayed by the Feynman-Kac
representation and the related lack of screening are much easier to
miss in the conventional many-body perturbation theory. The physically
very natural molecular limit considered in the main theorem of
chapter~\ref{chapter-7} should be introduced and discussed in any text
book on quantum statistical mechanics.

We conclude this introduction with a list of some questions that seem
to us of interest, to be pursued in the spirit of this review and
perhaps by similar techniques.\\[2mm]

\noindent CLASSICAL: Debye regime\\[2mm]
\hspace*{2mm}
\begin{minipage}{15.5cm}
A more detailed understanding of boundary conditions and
charge expulsion.\\ 
Results on correlations along walls, surface charges.\\
Understanding the improved Debye-H\"uckel theories of Fisher et
al. as limiting theories.\\[2mm]
\end{minipage}
\newpage

\noindent Dipole systems\\[2mm]
\hspace*{2mm}
\begin{minipage}{15.5cm}
More definitive results on the dielectric constant and clarification
of the literature in the light of such results.\\[.2mm]
\end{minipage}

\noindent QUANTUM: Low density\\[2mm]
\hspace*{2mm}
\begin{minipage}{15.5cm}
Prove absence of exponential screening or at least finiteness of
each individual Mayer-like diagram of chapter~\ref{chapter-loops}.\\
Correlations in non-uniform systems, e.g. along walls.\\
Coherent treatment of para and diamagnetism, boundary
currents.\\[.2mm]
\end{minipage}

\noindent Slightly imperfect molecular gases\\[2mm]
\hspace*{2mm}
\begin{minipage}{15.5cm}
Lowest order corrections to the molecular limit, atomic
correlations and van der Waals forces.\\
Possible existence of dielectric phases with true atomic
dipoles.\\[.2mm]
\end{minipage}

\noindent Compatibility of macroscopic electrostatics and statistical
mechanics of quantum charges.\\

\subsection{The classical Coulomb gas}
\label{sec- classical}

We consider several species, $\alpha = 1,\dots, \cS$, of charged
particles with charges $e_{\alpha }$ of both signs, confined to a box
$\Lambda$ in ${\Bbb R}^3$.  If there are $N$ particles in total sitting in
an external electrostatic potential $\phi$ then the potential energy is
given by
\begin{equation}\label{1.1.1}
  U (\br _{1}, e_{\alpha _{1}},\dots ,\br _{N},e_{\alpha _{N}})+ \dots
  + \sum_{i=1}^N e_{\alpha_i}\phi(\br _{i}) \end{equation} where
\begin{equation}\label{1.1.1a}
  U (\br _{1}, e_{\alpha _{1}},\dots ,\br _{N},e_{\alpha _{N}} )=
  \sum_{i<j=1}^N e_{\alpha_i} e_{\alpha_j} V(\br _i - \br _j )
\end{equation}
$V$ is the two body Coulomb potential energy
\begin{equation}\label{1.1.2}
  V(\br) = \frac{1}{ \mid\br\mid}.
\end{equation}
and the ellipsis indicates potentials due to other forces, that are
discussed below.

Suppose that $N_{\alpha }$ denotes the number of particles of species
$\alpha $. Then the equilibrium statistical mechanics is determined by
the partition function
\begin{eqnarray}\label{1.1.3}
&&
  \Xi_\Lambda = \Xi_\Lambda(\beta \phi) = \sum_{\{N_\alpha\}}^\infty
  \prod_{\alpha = 1}^{\cS } \frac{(z_\alpha)^{N_\alpha}}{N_\alpha !}
  \int_{\Lambda} d{\bf r}_{1} \ldots \int_{\Lambda} d{\bf r}_{N}
  \exp\(-\beta U - \beta \sum_{i=1}^N e_{\alpha_i}\phi(\br _{i})\)
\end{eqnarray} and $z_{\alpha }$
is called the activity of species $\alpha$. In particular the pressure
$P$ is given by
\begin{equation}\label{1.1.3c} \beta P
  = \lim_{|\Lambda |\rightarrow \infty } \frac{1}{|\Lambda |}\ln
  \Xi_\Lambda .
\end{equation}
The activities in the partition function are related to the chemical
potential $\mu _{\alpha }$ of species $\alpha $ by the standard
formula
\begin{equation}\label{1.1.3cc}
  z_{\alpha } = \frac{e^{\beta \mu_{\alpha }}} {(2\pi \lambda_{\alpha
      }^{2})^{3/2}}
\end{equation}
where $\lambda_{\alpha }$ is the de Broglie length given below in
(\ref{1.1.3ccc}).  This formula results from the classical limit of
quantum mechanics.

In three dimensions a purely Coulomb Boltzmann factor $\exp \( - \beta
U \)$ is not integrable because of the Coulomb singularities at
$\br_{i} = \br_{j}$: additional forces are necessary for stability. We
suppose there is a length scale $\lambda $ where the Coulomb potential
gives way to these other forces. We used $\lambda $ because it
suggests a de~Broglie length
\begin{equation}\label{1.1.3ccc} \lambda_{\alpha } = 
  \hbar(\beta /m_{\alpha })^{1/2} 
\end{equation} 
  where $m_{\alpha }$ is the mass of species $\alpha $.  For example,
for individual point particles, classical mechanics gives way to
quantum mechanics when their wave functions overlap, which occurs at
separation $O (\lambda _{\alpha })$. For ions, $\lambda $ would instead
be the radius of an outer orbital. For Fermions, one attempts to model
the Pauli principle together with electrostatic repulsion by classical
effective short range forces. We will either represent these short
range forces   by a hard core of radius $\lambda $ or we will smooth
the Coulombic   singularity by replacing $|\br|^{-1}$ by
\begin{equation}\label{1.1.3bc}
        V_{\infty ,\lambda }(\br ) = 
        |\br|^{-1}(1-\exp(-|\br|/\lambda ))
\end{equation}
Neither choice is likely to be a realistic description of the physics
at this length scale, so these are only claimed to be useful when one
can show that there are predictions which are not very dependent on
how the short range force is chosen.

The notation $V_{\infty ,\lambda }$ conveys the equation
\begin{equation}\label{1.1.3be}
V_{\infty ,\lambda } = V_{\infty } - V_{\lambda }
\end{equation}
where we set
\begin{equation}\label{1.1.3bd}
        V_{\lambda }(\br ) = 
        |\br|^{-1}\exp(-|\br|/\lambda ); \ \ \ 
        V_{\infty }(\br ) = 1/|\br |
\end{equation}
The potential $V_{\infty ,\lambda }$ is the Green's function for
$(-\Delta + \lambda ^{2}\Delta ^{2})/(4\pi ) $ with the boundary condition
appropriate for a system confined to a box $\Lambda $ with insulating
walls.  We shall also consider systems confined in a box with
conducting grounded walls. In these cases $V_{\lambda }(\br - \br ')$
is replaced by $V_{\lambda }(\br ,\br ')$ which then denotes the
Greens' function with zero boundary conditions.

Consider the case where the ellipsis in (\ref{1.1.1}) represent a hard
core of radius $\lambda $. By Newton's theorem each point charge can
then be spread out into an equivalent charge on a sphere of radius
$\lambda $ without changing the Coulomb interaction energy. If
self-energies of the spheres are added into the Coulomb energy then it
is positive because it can be written as the integral of the square of
the electric field. Since the self-energy of $N$ spheres is
$\frac{1}{2}\sum _{\alpha } e_{\alpha}^{2}/\lambda$ this argument
shows that
\begin{equation}\label{1.1.3b} 
  \mbox{Stability:} \ \ \ \ \ U \geq - B N; \ \ \ B =
  O(\frac{e_{\alpha }^{2}}{2 \lambda })
\end{equation} 
Thus the introduction of the hard cores has made $\Xi_\Lambda$ finite
\cite{Ons39}.

The same stability bound holds in the other case where the additional
terms in the ellipsis in (\ref{1.1.1}) smooth out the Coulombic
singularity. The total potential energy with self-energies added is
\begin{equation}
\label{1.1.3bb} \frac{1}{2}\sum _{i,j}
e_{\alpha _{i}}e_{\alpha _{j}} \frac{1 - e^{-\mid\br _{i}-\br _{j}\mid
    /\lambda }}{\mid\br _{i}-\br _{j}\mid}
\end{equation} which is positive because it equals, in Fourier space,
\begin{equation}\label{1.1.3bbb} \frac{1}{2} \frac{4\pi }{(2\pi )^{3}}
\int d \bk \, \left( \frac{1}{|\bk|^{2} } - \frac{1}{\lambda ^{-2} +
|\bk|^{2} } \right) |\sum _{i} e_{\alpha _{i}}e^{i\bk\cdot \br
_{i}}|^{2} \geq 0
\end{equation}
The self energies are again $\frac{1}{2}\sum _{\alpha
  }e_{\alpha}^{2}/\lambda$ so we obtain (\ref{1.1.3b}).
Analogous arguments using the appropriate eigenfunctions in place of
$\exp (i\bk \cdot \br )$ show that stability bounds hold also for
particles in an grounded container.

We use the summation identity
\begin{equation}\label{1.1.4}
  \sum_{\{N_\alpha\}}^\infty\prod_{\alpha}\frac{1}{N_{\alpha}\,!}\cdots
  = \sum_{N=0}^\infty\frac{1}{N!}
  \sum_{\alpha_1,\ldots,\alpha_N}^{\cS}\cdots
\end{equation}
to write the partition function (\ref{1.1.3}) in the form
\begin{equation} \label{1.1.5}
  \Xi_\Lambda = \sum_{N=0}^\infty\frac{1}{N!} \int \prod_{k=1}^N
  d{\cal E}_k z({\cal E}_k) \exp\(-\beta U({\cal E}_1, \ldots, {\cal
    E}_k) -\beta \sum_{i=1}^N \phi(\cE _{i})\)
\end{equation}
where $\cE = (\br, \alpha )$ unites the spatial and species coordinate
so that $\phi(\cE ) = e_{\alpha }\phi(\br )$ and the $d{\cal E}$
integration means
\begin{equation}\label{1.1.6}
  \int d{\cal E}\cdots= \sum_{\alpha =1}^{\cS}\int_{\Lambda}d{\bf r}
  \cdots
\end{equation}
and $z(\cE) = z_{\alpha }$.

We denote by $\omega=(\cE _{1},\ldots,\cE _N)$ a configuration of the
particles,  $U(\omega ) = U(\cE _{1},\ldots,\cE _N)$ and $z(\omega ) =
z(\cE _{1})\cdots z(\cE _{N})$. The grand canonical summation over
these configurations is \begin{equation}\label{1.1.7} \int _{\Lambda }
d\omega \cdots = \sum_{N=0}^\infty\frac{1}{N!} \int \prod_{k=1}^{N}
d{\cal E}_k \cdots .
\end{equation}
and the grand canonical average is, for zero external potential,
\begin{equation}\label{1.1.7a} \langle\cdots \rangle =
  \frac{1}{\Xi_\Lambda} \int _{\Lambda } d\omega z(\omega ) e^{-\beta
    U(\omega )} \cdots
\end{equation}

A system in which the species occur in pairs of equal but opposite
charge with equal activities is said to be charge-symmetric.

The observable that measures density of particles of species $\alpha $
at $\br $ is given by
\begin{equation}\label{1.1.8a}
  \hat{\rho }(\cE ,\omega) = \sum_{j=1}^N \delta(\cE ,\cE _j) =
  \sum_{j=1}^N \delta_{\alpha ,\alpha _{j}}\delta({\bf r} -{\bf r}_j).
\end{equation} and $n$ - particle distributions for points $\br _{a}$,
$a = 1,\dots , n$ are defined by
\begin{eqnarray}\label{1.1.9a}
  \rho (\cE _{1},\dots ,\cE _{n}) = \langle \prod _{a=1}^{n} \hat{\rho
}(\cE _{a})\rangle 
\end{eqnarray}
These definitions coincide with the standard definitions of
distribution functions in statistical mechanics when $\cE_{1},\dots
,\cE _{n}$ are distinct. If they are not all distinct (coincident
points) then our definition includes squares or higher powers of delta
functions.  At various points in the text we will use the phrase
``does not include contributions from coincident points'' to indicate
that such powers of delta functions are omitted. To illustrate what
``including contributions from coincident points'' means, consider
$\int d\cE \, f (\cE) \rho (\cE ,\cE _{1})$.  The integral smooths the
delta function in (\ref{1.1.8a}) so that the diagonal terms $\sum
_{j}\delta (\cE ,\cE _{j})\delta (\cE _{1},\cE _{j})$ in $\hat{\rho
}(\cE ,\omega) \hat{\rho }(\cE_{1} ,\omega) $ give the ``coincident
contribution'' $\sum _{j} f (\cE _{j}) \delta (\cE_{1} ,\cE _{j})$.

We also introduce the charge density observable
\begin{equation}\label{1.1.8} \hat{c}({\bf r},\omega) = \sum_{\alpha }
e_{\alpha }\hat{\rho }_{\alpha } (\br ,\alpha,\omega ) = \sum_{j=1}^N
e_{\alpha_j} \delta({\bf r} -{\bf r}_j). \end{equation} and the charge
distributions
\begin{eqnarray}\label{1.1.9}
  c(\br _{1},\dots ,\br _{n}) = \langle \prod _{a} \hat{c}({\bf
r_{a}})\rangle \end{eqnarray} More generally one obtains mixed charge
- particle distributions by taking the average of products of density
observables and charge observables (\ref{1.1.8a}) and (\ref{1.1.8}).

We shall also consider the corresponding truncated distributions
(correlations) $\rho _{T}$ and $c_{T}$ defined in the usual way
\begin{equation}\label{1.1.9b}
  \rho _{T}(\cE _{1},\cE _{2}) = \rho (\cE _{1},\cE _{2}) - \rho (\cE
_{1}) \rho (\cE _{2}) \end{equation} and so on.

Correlation functions of charge densities can also be obtained as the
response to an infinitesimal external potential by variational
differentiation of the logarithm of the partition function
$\Xi_{\Lambda }(\beta \phi)$ with respect to $-\beta \phi$. In
particular, the charge density induced in the system by a potential
perturbation at $\br_{1}$ is \begin{equation} c(\br_{1})=\frac{\delta
}{\delta (-\beta \phi(\br _{1}))}\ln \Xi_{\Lambda }(\beta \phi)
\label{1.1.9o}
\end{equation} and the truncated distribution of two charge densities is
\begin{equation}\label{1.1.10b} c_{T} (\br _{1},\br _{2}) = \frac{\delta }
  {\delta (-\beta \phi(\br _{1}))} \frac{\delta } {\delta (-\beta
    \phi(\br _{2}))} \ln \Xi_{\Lambda }(\beta \phi)
\end{equation}
By this definition, we find that \begin{equation} c_{T} (\br _{1},\br _{2}) =
\langle \hat{c}(\br _{1}) \hat{c}(\br _{2}) \rangle - \langle
\hat{c}(\br _{1}) \rangle \langle \hat{c}(\br _{2}) \rangle .
\label{1.1.10c}
\end{equation}

The physical properties of an homogeneous infinitely extended charged
fluid are conveniently described by its linear response to an external
classical charge density $c^{{\rm ext}}(\br)$.  One defines the static
susceptibility $\chi(\br_{2}-\br_{1})$ as the linear response at
$\br_{2}$ of the charge density of the fluid to the external density
at $\br_{1}$ \begin{equation} \chi(\br_{2}-\br_{1})=\left[\frac{\delta
} {\delta c^{{\rm ext}}(\br_{1})}c(\br_{2})\right]_{ c^{{\rm ext}}=0}
\label{1.1.10d} \end{equation} The function $\chi(\br)$ describes the
shape of the screening cloud around an infinitesimal point test charge
at the origin. In Fourier space,
\begin{equation}\label{1.1.fourier}
\tilde{\chi}(\bk) = \int \chi(\br )e^{-ik\cdot \br }d\br 
\end{equation} 
is related to the Fourier transform $S(\bk)$ of the charge-charge
correlations $c_{T}(\br,{\bf 0})$ (\ref{1.1.10c}) at zero potential
(the structure factor of the fluid) by \begin{equation}
\tilde{\chi}(\bk)=-\frac{4\pi\beta}{|\bk|^{2}}S(\bk) \label{1.1.10e}
\end{equation} 
Finally, the total effective potential due to the test charge plus its
screening cloud is 
\begin{equation} \tilde{V}^{{\rm
eff}}(\bk)=(1+\tilde{\chi}(\bk))\tilde{V}(\bk)=\varepsilon^{-1}(\bk)\tilde{V}(\bk)
\label{1.1.10f} \end{equation} 
where we have defined the static
dielectric function $\varepsilon(\bk)$ by $\varepsilon^{-1}(\bk)
=\tilde{\chi}(\bk)+1$.  Equivalently we immerse two test charges
$(\br _{a},e_{a})_{a=1,2}$ in the ensemble. Their (nonlinear)
effective potential energy is the excess free energy due to the
external charge distribution given by
\begin{eqnarray}\label{1.1.eff1}
&& 
\exp \left(- \beta V^{{\bf eff}}((\br _{a},e_{a})_{a=1,2}  \right) 
\nn \\
&& \hspace{.2in}= 
\frac{\langle \exp \left(-\beta e_{1}V\ast \hat{c }(\br _{1})-
\beta e_{2}V\ast
\hat{c }(\br _{2})  - \beta e_{1}e_{2}V(\br _{1} - \br _{2})\right) \rangle}
{\langle\exp(-\beta e_{1}V\ast \hat{c }(\br _{1}))\rangle
\langle\exp(-\beta e_{2}V\ast \hat{c }(\br _{2}))\rangle}
\end{eqnarray}
But then we take the test charges infinitesimal and find the leading
term
\begin{eqnarray}\label{1.1.eff}
&&
	V^{{\bf eff}}((\br _{a},e_{a})_{a=1,2} ) 
\sim e_{1}e_{2}
	V^{{\bf eff}}(\br _{1} - \br _{2} )
\nn \\
&&
	V^{{\bf eff}}(\br _{1} - \br _{2} ) 
= 
	V(\br _{1}-\br _{2}) - \beta V\ast c_{T} \ast V(\br _{1} - \br _{2}) 
\end{eqnarray}
which is equivalent to our previous formula (\ref{1.1.10f}).

A well known criterion for the plasma phase of the Coulomb system is the
vanishing of $\varepsilon^{-1}(\bk)$ as $\bk\to 0$. This implies with (\ref{1.1.10e}) that
$S(\bk)\simeq (4\pi\beta)^{-1}|\bk|^{2},\; \bk\to 0$, which is equivalent
with the second moment rule (Stillinger-Lovett rule) for the charge-charge correlation 
\begin{equation}
\int d\br |\br|^{2} c_{T}(\br,{\bf 0})=-\frac{3}{2\pi\beta}
\label{1.1.1sti}
\end{equation}

If the Coulomb interaction $U$ is set to zero the partition function
is
\begin{eqnarray}\label{1.1.11}
  \Xi_{{\rm ideal},\Lambda }(\beta \phi) &=&
  \sum_{N=0}^\infty\frac{1}{N!} \int \prod_{k=1}^{N} d{\cal E}_k
  z({\cal E}_k) \exp\(-\beta \sum _{j=1}^{N}e_{\alpha _{j}}\phi(\br
  _{j})\) \nn \\ &=& \exp\( \int d{\cal E}\, z({\cal E}) e^{-\beta
    e_{\alpha }\phi(\br )} \).
\end{eqnarray}
We will call the resulting system an ideal gas, because the external
potential is just a multiplication of the activities by $\exp (-\beta
e_{\alpha_{1} }\phi(\br_{1}))$. In this case all distributions are
computable, e.g., 
\begin{eqnarray}\label{1.1.12} \langle \hat{c}({\bf
  r_{1}}) \rangle_{{\rm ideal},\beta \phi} &=& \frac{\delta } {\delta
  (-\beta \phi(\br _{1}))} \ln \Xi_{{\rm ideal},\Lambda }(\beta
\phi) = \sum _{e_{\alpha _{1}}} e_{\alpha_{1} } z_{\alpha
  _{1}} e^{-\beta e_{\alpha_{1} }\phi(\br_{1})} 
\end{eqnarray}
where we used (\ref{1.1.9o}).

\subsection{The quantum Coulomb gas}
\label{sec- quantum}

We consider $\cS $ species of quantum point charges (electrons and
nuclei) in a box $\Lambda $ in ${\Bbb R}^3$, with masses $m_{\alpha }$,
charges $e_{a}$ and spins $s_{\alpha }$, $\alpha = 1,\dots, \cS$. Each
species obeys Fermi or Bose statistics and at least one species is
Fermionic. The system is governed by the non-relativistic N-particle
Hamiltonian
\begin{equation}\label{1.2.1} H_{\Lambda, N} =
  - \sum _{i=1}^{N}\frac{\hbar^{2}}{2m_{\alpha _{i}}} \Delta _{i} +
  \sum _{i<j}e_{\alpha _{i}}e_{\alpha _{j}} V(\br _{i} - \br _{j}) +
  \sum _{i=1}^{N} e_{\alpha _{i}}\phi(\br_{i}) \end{equation} acting
  on the $N$-particle Hilbert space appropriately symmetrized
  according to the statistics of each species. The Laplacians $\Delta
  _{i}$ in (\ref{1.2.1}) have Dirichlet boundary conditions on the
  boundary of $\Lambda $. The external electrostatic potential is set to
  zero unless we warn the reader otherwise.

  We denote by $H_{N}$ the Hamiltonian for particles in infinite
  space.  The stability of the Hamiltonian is the statement that if at
  least one of the particle species obeys Fermi statistics, then
  \begin{equation}\label{1.2.2} H_{N} \geq -B N
\end{equation}
with $B$ a positive constant independent of $N$
\cite{DyLe67,DyLe68}. The proof has been greatly simplified by Lieb
and Thirring with the use of the Thomas-Fermi theory.  This is
reviewed in \cite{Lie76}. A new proof \cite{Gra97} is based on a
remarkable electrostatic inequality for the classical Coulomb
interaction (\ref{1.1.1}).  For a review on recent developments on
these questions, see \cite{Lie90}. We mention here that the problem of
the stability of matter in presence of magnetic fields
\cite{LiLoSo95,Fef95} and the quantized radiation field\footnote{The
matter is non-relativistic and an ultraviolet cutoff is imposed.}
\cite{BuFrGr96,FeFrGr98} has received much attention lately.

Let \begin{equation} P_\alpha = \frac{1}{N_\alpha !} \sum_{p_\alpha}
(\eta_\alpha)^{p_\alpha}{p_\alpha}, \;\;\;\; \eta_\alpha=\pm 1
\label{1.2.3}
\end{equation} be the projection onto the symmetric ($\eta_\alpha=+1$) or
antisymmetric ($\eta_\alpha=-1$) states of particles of species
$\alpha$.The sum runs on all permutations $p_\alpha$ of the $N_\alpha$
particles and $(\eta_\alpha)^{p_\alpha}$ is the signature of
$p_\alpha$; $p_\alpha$ acts both on the position $\br$ and the spin
variable $\sigma$ of the particle. The projection onto the subspace of
states of a many-body system with $N=\sum_{\alpha } N_\alpha$
particles having the appropriate statistics is given by \begin{equation} P =
\prod_\alpha P_\alpha = \sum_p \( \prod_\alpha
\frac{(\eta_\alpha)^{p_\alpha}}{N_\alpha !} \) p \label{1.2.4} \end{equation}
where the sum runs now on all permutations of the $N$ particles that
are compositions $p=p_{\alpha_{1}}\ldots p_{\alpha_{\cal S}}$ of
permutations of particles of each species.

Associating to each species a chemical potential $\mu _{a}$, the grand
canonical partition function of the quantum gas is given by the sum
\begin{eqnarray} \Xi_{\Lambda} &=& \sum_{\{N_\alpha\}}\mbox{Tr}
P\exp\( -\beta\(
  H_{\Lambda,N} - \sum_{\alpha} \mu_\alpha N_\alpha \) \) \nonumber\\ 
  &=& \sum_{\{N_\alpha\}} \sum_p \prod_\alpha
  \frac{(\eta_\alpha)^{p_\alpha} \exp(\beta \mu_\alpha N_\alpha)} {N_\alpha
    !} \int_\Lambda d\br_1\ldots\br_N \nonumber\\ &\times &
  \sum_{\{\sigma_{\alpha_{i}}\}}
  \langle\{\br_{p(i)},\sigma_{\alpha_{p(i)}}\} | \exp\(-\beta
  H_{\Lambda,N}\) | \{\br_i,\sigma_{\alpha_{i}}\}\rangle
\label{1.2.5}
\end{eqnarray}
In (\ref{1.2.5}), $|\{\br_i,\sigma_i\}\rangle =\prod
|\br_i,\sigma_i\rangle$ is a product of states of individual particles
that also diagonalize the spin component along a fixed direction with
quantum numbers $\sigma_\alpha$ taking the values
$-s_\alpha,\;-s_\alpha+1\;,\ldots,\;s_\alpha$.

The pressure is defined as in (\ref{1.1.3c}).  Particle and charge
distributions are defined as in the classical case,
replacing the average (\ref{1.1.7a}) by
\begin{equation}\label{1.2.6}
  \langle \cdots \rangle = \frac{1}{\Xi_\Lambda} {\rm Tr}P (e^{-\beta
    H_{\Lambda N}}\cdots ) 
\end{equation} 
Here the particle and charge distributions and correlations are
defined as in (\ref{1.1.9a}) and (\ref{1.1.9}) where now the particle
coordinates $\br _{j}$ in (\ref{1.1.8a}) are the quantum mechanical
position operators\footnote{We use the same symbol $\br$ for the
quantum mechanical position operator and for the argument of a
correlation function. We do not consider spin correlations. All the
quantum correlations in the review are purely positional with spin
variables averaged out.}.

In contrast to the classical situation, the response to an external
electrostatic potential yields a new type of correlation functions,
the imaginary time Green's functions. They are obtained by variational
differentiation with respect to the external potential.  Of special
interest is the charge distribution at $\br_{2}$ induced by an
external potential perturbation at $\br_{1}$ in the linear response
regime
\begin{eqnarray}
\label{1.2.7b}
c^{{\rm ind}}(\br_{1}|\br_{2}) &=&\left[\frac{\delta } {\delta (-
  \beta\phi(\br _{1}))}\langle \hat{c}(\br_{2})
\rangle\right]_{\phi=0}\nonumber\\ &=& \left[\frac{\delta } {\delta (-
  \beta\phi(\br _{1}))} \(\Xi_{\Lambda}^{-1}{\rm Tr} P\exp\( -\beta
H_{\Lambda,N} \) \hat{c}(\br_{2}) \)\right]_{\phi=0}\nn\\ &=&
\frac{1}{\beta}\int _{0}^{\beta } d\tau \, 
c_{T}(\br_{1},\tau,\br_{2})
\end{eqnarray}
In (\ref{1.2.7b}) \begin{equation} c_{T}(\br_{1},\tau,\br_{2})=\left\langle e^{\tau
  H_{\Lambda,N}}\hat{c}(\br_{1})e^{-\tau
  H_{\Lambda,N}}\hat{c}(\br_{2})\right\rangle-\langle
\hat{c}(\br_{1})\rangle \langle \hat{c}(\br_{2})\rangle
\label{1.2.7c}
\end{equation} 
is the imaginary time displaced charge-charge correlation (or Duhamel
function): it reduces to the static charge-charge truncated distribution at
$\tau=0$.

For an infinitely extended quantum charged fluid, we introduce the
spatial Fourier transform $S(\bk,\tau)$ of $c_{T}(\br,\tau,{\bf 0})$.
The static structure factor of the fluid is then
\begin{equation}
S(\bk)=S(\bk,\tau=0)
\label{1.2.7cc}
\end{equation}
and the generalization of the classical
relations (\ref{1.1.10e}) and (\ref{1.1.10f}) to the quantum
mechanical situation is \begin{equation} 
\tilde{\chi} (\bk)
=\varepsilon^{-1}(\bk)-1=-\frac{4\pi}{|\bk|^{2}}\int_{0}^{\beta}d\tau
S(\bk,\tau) 
\label{1.2.7d}
\end{equation} Notice that in the latter case, the susceptibility $\tilde{\chi}
  (\bk)$ (defined as in (\ref{1.1.10d})) is no longer proportional to
the static structure factor.

Equivalence of ensembles and existence of thermodynamic functions was
established in \cite{Gri69,LeLi69,LiLe72,LiNa75}. The
analogous results for classical systems are corollaries.

There are no results on existence of the infinite volume limit of
distribution functions except for charge symmetric Bose systems with
positive-definite interactions.\footnote{(\ref{1.1.3bc}) is a
positive-definite interaction but hard cores plus Coulomb are not.} In
this case at any density and temperature for which the system is
stable\footnote{For sufficiently large activity the grand canonical
ensemble diverges. For the ideal gas this happens at the Bose-Einstein
condensation.} all particle distribution functions have unique
infinite volume limits \cite{FrPa78,FrPa80}\footnote{Their proof only
works for some choices of boundary conditions but this is probably not
a physical limitation.}. These results include charge-symmetric
classical systems with positive-definite potentials. They rely on the
Sine-Gordon transformation described later.

The physical content of these results on the existence of the infinite
volume limit is that the limit exists independently of the way the
container is enlarged.  For example, we could consider two sequences
of increasingly large containers related by a fixed translation.  Both
sequences would have the same infinite volume limit, so the limit must
be translation invariant.  In particular the limiting distributions
could not describe a crystal with a low density of defects. On the
other hand one would expect to be able to select such a state by
choosing a sequence of containers whose boundaries match the natural
planes of the crystal, unless the system never forms crystals. Therefore
these results are evidence that stable charge symmetric systems with
positive-definite potentials are always in a fluid phase.

Further results on existence of the infinite volume limit of classical
distribution functions at low density are included in the work on
Debye screening \cite{BrFe78,Imb83} reviewed in the next chapter, but
these are low density results.

\subsection{Neutrality}
\label{sec- neutrality}

\def\ue{\underline{\bf e}}
\def\umu{\underline{\mbox{\boldmath $\mu$\unboldmath}}}
\def\bmu{\mbox{\boldmath $\mu$\unboldmath}}

In systems with short range interactions the densities of different
species can be varied independently, but Coulomb systems maintain
charge neutrality
\begin{equation}\label{2.1.1bb}
  \sum _{\alpha } \rho _{\alpha }e_{\alpha } = 0 \end{equation} by
expelling excess charge to the boundary \cite{LiLe72,GrSc95b}.  We
introduce a vector notation
\begin{eqnarray}
\ue &=& (e_{1},\ldots,e_{{\cal S}}), \ \ \
\umu = (\mu_{1},\ldots,\mu_{{\cal S}})
\label{2.1.1cc}
\end{eqnarray}
for the set of charges and chemical potentials, and decompose $\umu$
\begin{equation}
\umu=\bmu+\nu\ue,\;\;\;\;\;\;\bmu\cdot\ue =0, \;\;\;\;\;\;\;\nu=\frac{\umu\cdot\ue}{|\ue|^{2}}
\label{2.1.dd}
\end{equation}
into its components perpendicular and parallel to the charge vector.
Then, the charge neutrality is equivalent with the fact that the infinite volume limit of grand
canonical pressure does not depend on the component $\nu$ of $\umu$ along the charge vector
\begin{equation}
P(\beta,\umu)=P(\beta,\bmu+\nu\ue) =P(\beta, \bmu)
\label{2.1.1ee}
\end{equation}
This means that different
  choices of chemical potentials $\mu_{\alpha}$ or activities $z_{\alpha }$ do not necessarily
lead to
  different densities. It is common to break this redundancy by
  imposing \begin{equation}\label{2.1.1b} \mbox{pseudo-neutrality:} \ 
  \ \ \sum _{\alpha } e_{\alpha }z_{\alpha } = 0.
\end{equation}
The system will be neutral regardless of whether pseudo-neutrality is
imposed or not, but in section~\ref{sec-charge-expulsion} we will
argue that charges that are expelled to the boundary create a constant
external electrostatic potential $\psi$ which renormalizes the
activities $z_{\alpha } \rightarrow z_{\alpha } \exp [-\beta
e_{\alpha} f]$ in such a way as to restore the condition for the
renormalized activities. Thus imposing the condition amounts to
working with the renormalized activities.

\newpage \section{Debye screening}\label{chapter- debye-screening} \setcounter{page}{1}

This chapter is devoted to screening in classical Coulomb systems. We
begin with a review of the original theory of Debye and H\"uckel
followed by an account of some rigorous theorems. Then there follow
several sections that illustrate the main ideas used in the proofs
with discussions concerning open problems.

\subsection{Debye - H\"uckel Theory}
\label{sec-debye-huckel}

The mean field theory approximation for the dilute plasma phase
\cite{DeHu23} is at the center of our discussion. We will briefly
review their argument to prepare the way for further developments. For
further background see \cite{McQ76,HaMc76}.

The activity $z_{\alpha }$ has dimension ${\rm length}^{-3}$ and
$\beta e_{\alpha }^{2}$ and $\lambda $ have dimensions ${\rm length}.$
Thus if $l_{D}$ is defined by \begin{equation}\label{2.1.1} l_{D} =
(4\pi \sum_{\alpha } e_{\alpha }^{2} z_{\alpha } \beta
)^{-\frac{1}{2}}
\end{equation}
then $l_{D}$ is a length, called the Debye length. The activities in
this definition of $l_{D}$ are required to satisfy the
pseudo-neutrality condition (\ref{2.1.1b}). It is standard to define
the Debye length in terms of densities by the formula
\begin{equation}\label{2.1.1a} {\kappa ^{-1}} = (4\pi \sum_{\alpha }
e_{\alpha }^{2} \rho _{\alpha } \beta )^{-\frac{1}{2}}
\end{equation}
and we have used the notation $l_{D}$ instead of the standard
$\kappa^{-1}$ because the two definitions are not equal, although
their ratio tends to one in the Debye-H\"uckel limit described below.

The conclusions of Debye-H\"uckel theory include that the pressure $P$
and charge-charge correlation obey
\begin{eqnarray}\label{2.1.2}
&&
\beta (P - P_{{\rm ideal}}) \sim
\frac{1}{12 \pi l_{D} ^{3}}\nn \\
&& c_{T}(\br _{1},\br _{2}) \sim -\left( (4\pi )^{2} \beta l_{D}^{4}
|\br _{1} - \br _{2}| \right)^{-1} \exp( - \frac{|\br _{1} - \br
_{2}|}{l_{D}}) + (4\pi \beta l_{D}^{2})^{-1}\delta (\br _{1} - \br _{2})
\end{eqnarray}
 The Debye-H\"uckel argument suggests that these are really limiting
laws, holding in a limit in which the number of particles in a volume
$l_{D}^{3}$ tends to infinity, while the number in a volume $\lambda
^{3}$ tends to zero. Perturbation expansions
\cite{May50,Mee61,LeSt68,HaMc76,MaMa77} are also likely to be
asymptotic as opposed to convergent, but it is still natural to
believe that the second result holds in a stronger sense: that without
taking the limit there can be exponential decay of correlations,
particularly if one is reasonably close to the limit. We shall refer
to this stronger statement as Debye screening.

The first law in (\ref{2.1.2}) can be expressed in terms of densities
by using $\rho _{\alpha } = z_{\alpha } \partial \beta P/\partial
z_{\alpha} $ and $\beta P_{{\rm ideal}} = \sum z_{\alpha }$. One finds
that
\begin{equation}\label{2.1.2b} \beta (P - \sum _{\alpha }\rho _{\alpha
}) \sim
-\frac{\kappa ^{3}}{24 \pi}\nn \\
\end{equation}

The second result (\ref{2.1.2}) was based on an argument that can be
paraphrased as follows. Consider a grand canonical ensemble of
particles interacting with each other and also with a charge $e_{0}$
held fixed at the origin. This is the same as adding to $\beta U$ the
external potential term $\beta\int e_{0}\hat{c}(\br ,\omega ) V(\br )
d\br$. Then the potential at another point $\br $ will be the sum of
$V(\br) $ and another potential $\phi$ created by the charges in the
gas. Debye and H\"uckel approximate this $\phi$ by assuming that the
gas reacts to the combined potential $f = e_{0}V + \phi$ as an ideal
gas, which leads to the self-consistent scheme
\begin{equation}\label{2.1.4}
\frac{1}{4\pi }\Delta f(\br,\omega  )
=
- \hat{c}(\br ,\omega ) - e_{0}\delta (\br )
\approx
- \langle\hat{c}(\br)\rangle_{{\rm ideal},\beta f} - e_{0}\delta (\br
) 
\end{equation} 
We wrote $f(\br ,\omega )$ to emphasize that the exact field is
configuration dependent but having made this approximation we write $f
(\br )$ from now on. From (\ref{1.1.12}),
\begin{equation}\label{2.1.7}
\frac{1}{4\pi }\Delta f(\br )
=
- \sum _{\alpha }z_{\alpha } e_{\alpha } \exp \left [ - \beta
e_{\alpha }f(\br) \right]  - e_{0}\delta (\br )
\end{equation}
If $\beta f(\br)$ is a weak potential
then $\exp(x) \approx 1 + x$ and pseudo-neutrality (\ref{2.1.1b}) leads to
\begin{equation}\label{2.1.10}
[\Delta - l_{D} ^{-2}] f(\br )
=
- 4\pi e_{0} \delta (\br )
\end{equation}
which has the solution
\begin{equation}\label{2.1.9}
f(\br ) = e_{0} |\br | ^{-1} \exp \(-|\br|/l_{D} \) \end{equation} The
exponential decay of $f$ is the origin of exponential decay of charge
density correlations within this mean field approximation. Indeed
noting that $f$ is related to the effective potential introduced in
section~\ref{sec- classical} by $f = e_{0}V^{{\rm eff}}$ we can obtain
the second Debye-H\"uckel law from (\ref{1.1.eff}) using the Fourier
transform.

In (\ref{2.1.4}) the $\langle\hat{c}(\br)\rangle_{{\rm ideal},\beta f}$
is the mean density of a spherically symmetric opposite charge
``screening'' cloud that forms around the positive charge at the
origin. Note that within the linear approximation $\exp \left(x
\right) = 1 +x$ our calculations imply
\begin{equation}\label{2.1.13}
\int d \br \,
\langle \hat{c}(\br)\rangle _{{\rm ideal},\beta f} \sim
- e_{0}
\end{equation}
By Newton's theorem, such a cloud is equivalent to an additional
negative charge at the origin that neutralizes the fixed charge up to
a remainder exponentially small in the radius to the point $\br $. Of
course no single charge configuration of point particles can be
spherically symmetric. The cloud is an ensemble average over
configurations. This cancellation of charge is captured in ``sum
rules'' which will be discussed in
section~\ref{sec-Debye-screening}. Related to this point is the fact
that the integral over $\br _{2}$ of the right hand side of
(\ref{2.1.2}) vanishes.

\subsection{Theorems on Debye screening} \label{sec-Debye-screening}

In this section we will survey the rigorous results on Debye screening
and the Debye-H\"uckel approximation concentrating on three
dimensional systems but  with some remarks about other dimensions.

The one dimensional Coulomb system without hard core has been solved
exactly \cite{Len61,EdLe62}. Lenard's solution has Debye screening for
charge density observables, but one dimensional systems have
exceptional screening properties, because the linear potential
$-|x-y|$ has no multipoles. For example, the electric field is
piecewise-constant with jumps at the positions of particles so that
outside an interval containing any neutral configuration, one can
achieve a zero electric field. On the other hand if a single
fractional charge is placed in a system of integer charges then the
resulting potential can never be screened because the ensemble of
integer charges can do no more than cancel integer parts of the
corresponding electric field. One dimensional Coulomb systems have
been reviewed in \cite{CKMN81}. Since then, highly non-trivial results
have been obtained on the correlations of classical charges confined
to a circle and interacting with a logarithmic potential
\cite{For92,For93a,For93b}.  Some comments on two dimensional systems
are given at the end of this section.

For three dimensions the results are complicated to state completely
and the complications are probably artifacts of the proofs. Therefore
let us restrict ourselves to a simple case of a hard core plasma with
two species, whose charges $e_{\alpha }$ have rational ratio. Recall
that the combination $\beta e_{\alpha }^{2}$ has dimensions of
length. By absorbing a fundamental unit of charge into $\beta $ we
assume that $\beta $ carries the dimension of length while $e_{\alpha
}$ is dimensionless. To put it another way, we state results regarding
$\beta , z_{\alpha }, \lambda $ as parameters, and keeping $e_{\alpha
}$ fixed. The results below are valid in a thermodynamic limit in
which there are boundary conditions on the Coulomb potential. Details
may be found on page 199 of \cite{BrFe80}; the boundary conditions are
a little artificial but their essence is that the walls of the
container are grounded, so that all electrostatic potentials vanish at
the boundary $\partial \Lambda $.

\begin{theorem}\label{th-1.3.1}
Suppose that
\begin{eqnarray}
&&
\mbox{Pseudo-Neutrality: }
\sum _{\alpha }z_{\alpha } e_{\alpha } =0 \label{2.2.1}\\
&&
\mbox{Debye Sphere Assumptions: }
\left\{\begin{array}{l}
z_{\alpha } \l_{D} ^{3} e^{-\frac{\beta}{2\lambda}} \gg 1\\ 
z_{\alpha } \lambda ^{3} \ll 1
\end{array}
\right.
\label{2.2.2}
\end{eqnarray}
then all (truncated) $n$-point particle and charge density
correlations decay exponentially on length scale $l\approx
l_{D}$. Furthermore, if $\Xi_{\Lambda }(\br _{1},\br _{2})$ is the
grand canonical charge symmetric ensemble of integer charges with two
fractional charges held fixed at $\br _{1},\br _{2}$, then the two
fractional charges are screened in the sense that
\begin{equation}\label{2.2.3} \frac{\Xi_{\Lambda }(\br _{1},\br
_{2})}{\Xi_{\Lambda }} = 
\frac{\Xi_{\Lambda }(\br _{1})}{\Xi_{\Lambda }} 
\frac{\Xi_{\Lambda }(\br _{2})}{\Xi_{\Lambda }}
 + \ O \( \exp \(-|\br_{1} - \br _{2}|/l \)
\). \end{equation}
\end{theorem}

The Debye sphere assumptions say that the number of particles in a
volume of size $\lambda ^{3}$ is small, so that the hard core is out
of play, and the number of particles inside a sphere of size $l_{D}$
is large. For comparing these statements with the ones in the
references it is useful to realize that
\begin{equation}\label{2.2.3b}
4\pi \sum _{\alpha }e_{\alpha }^{2} z_{\alpha } l_{D} ^{3} =
\frac{l_{D}}{\beta }
\end{equation}
so the first Debye sphere condition is equivalent to $\beta/l_{D}$
being small. Theorem~\ref{th-1.3.1} is a combination of results from
\cite{Bry78,BrFe80,Imb83a,BrKe87}.

It is shown in \cite{Imb83} that there is Debye screening in Jellium
and in systems with irrational charges. Jellium is a limit in which
all species with positive charges are smeared out into a background
charge density by letting their charge tend to zero as their density
increases. The results of Brydges and Federbush were not proved with
enough uniformity to persist as this limit is taken.

The Debye-H\"uckel limiting laws should hold in the 
\begin{equation}\label{2.2.4} 
\mbox{Debye-H\"uckel Limit: }
z_{\alpha}\lambda ^{3}
\rightarrow 0; \ \ \
z_{\alpha}l_{D} ^{3} e^{-\frac{\beta}{2\lambda}} \rightarrow
\infty
\end{equation}
but published proofs \cite{Ken83,Ken84} have the much more stringent
conditions that $\beta /\lambda $ is kept bounded and the system be
charge symmetric. Similar results for quantum systems have been
established by \cite{Fon86}. The original Debye-H\"uckel results
included correction terms containing a hard
core\footnote{\label{footnote-dh} $\beta (P - \sum _{\alpha }\rho
_{\alpha }) \sim -\frac{\kappa ^{3}}{24 \pi}\sigma (\kappa a)$ where
$\sigma (\kappa a) = \frac{3}{(\kappa a)^{3}} \bigg ( 1 + \kappa a -
\frac{1}{1+\kappa a} - 2\ln (1+\kappa a)\bigg)$.  It may be better to
consider $a$ to be a length scale that characterizes ``effective''
short range forces created by cooperation between Coulomb forces and
hard cores length as opposed to a genuine hard core.} length scale
$a$.  More systematically, hard core corrections have been obtained by
resummations of Mayer expansions \cite{Mee58,Mee61}\footnote{This is
compared with extended Debye-H\"uckel theories in \cite{BeFi98}.} and
by the Kac limit developed in \cite{LeSt68} which develop corrections
in $\kappa a$.  There is also a long history of intuitive extensions
of the Debye-H\"uckel theory to include corrections from hard cores
and dipolar effects.  These are reviewed and advanced in
\cite{FiLe93,Fis94,LeFi95,Fis96,LeFi96b,LeFi97,ZuFiLe97}\footnote{There
is a competing program \cite{Ste95,YeZhSt96} based in part on the mean
spherical approximation instead of Debye-H\"uckel theory.}.  However
consistency of these approximations is guesswork: is there an
asymptotic expansion in $\kappa a$? What physical effects provide the
next largest corrections to Debye-H\"uckel theory?

In the proofs of screening complications at the boundary have been
avoided by artificial boundary conditions, but only laziness stands in
the way of proofs for the case where the particles are in a conducting
grounded container so that the Coulomb potential vanishes at the
boundary. Particles with $1/r$ forces, i.e., particles in an
insulating container, have not been completely analyzed, but Debye
screening for a simplified model is established \cite{FeKe85}. This
case is discussed some more in section~\ref{sec-tunneling}. It is
difficult because screening fails at the boundary; there are power law
forces between charges on the boundary.

It should be possible to omit the pseudo-neutrality condition
(\ref{2.2.1}), but to do it rigorously may require insight into the
boundary condition problem. In section \ref{sec-charge-expulsion} we
will discuss these issues along with charge expulsion.

We return to (\ref{2.1.13}). The $\sim$ involves approximations but
the theorems known as sum rules say that if truncated distributions
decay integrably or better then this is exact in the limit $e_{0}
\rightarrow 0$: the screening cloud around a fixed infinitesimal
charge neutralizes it. More generally any fixed set of charges
surrounds itself with a screening cloud in such a way that all
multipoles are canceled. In particular, the function of $\br $ defined
by
\begin{eqnarray}\label{2.2.5}
&&
c_{T}(\br|\br_{1},e_{\alpha_{1}},\ldots,\br _{n},e_{\alpha _{n}})=
\nn \\
&& \hspace{-1in}
 \langle \hat{c}(\br ) \hat{\rho} (\br _{1},e_{\alpha _{1}}) \cdots
\hat{\rho} (\br _{n},e_{\alpha _{n}})
\rangle -
\langle \hat{c}(\br ) \rangle
\langle
\hat{\rho} (\br _{1},e_{\alpha _{1}}) \cdots
\hat{\rho} (\br _{n},e_{\alpha _{n}})
\rangle
\end{eqnarray}
which represents the excess charge density at $\br$ when particles in
the system are fixed at $\br_{1},\ldots,\br_{n}$ has no multipole
moments in the Debye regime \begin{equation}\label{2.2.5a} \int d\br
g_{l}(\br)c_{T}(\br|\br_{1},e_{\alpha_{1}},\ldots,\br _{n},e_{\alpha
_{n}})=0 \end{equation}for all harmonic polynomials $g_{l}(\br)$ of order
$l=0,1,\ldots $. In (\ref{2.2.5a}) the contribution of coincident
points $\br =\br_{i}$ is included.  The case $l=0,\; g_{0}(\br)=1$,
called the charge sum rule, is expected to hold very generally in
homogeneous phases of Coulomb systems (classical and quantum).  These
results are reviewed in \cite{Mar88}. Results on the effect of slow
decay at boundaries are also covered in this review.  In the next
section, we present a derivation of the sum rules with the help of the
Sine-Gordon formalism, extending the arguments of \cite{FoMa84} to
charges with hard cores.

There are also results \cite{AlMa85} that say that if truncated
distributions are integrable and decay monotonically at infinity then
they must decay faster than any inverse power. A different result of
the same genre was obtained by \cite{Fed79}.

\emph{Two dimensions, no short range forces}: in units where the
Coulomb interaction between two unit charges is $-\ln r$ let $\beta
_{1}= 2, \beta _{\infty } = 4$ and define the sequence of intermediate
thresholds $\beta _{n} = \beta _{\infty }(2n-1)/(2n)$ with $n = 1, 2,
\dots $.  The Coulomb potential is not stable in the sense of
(\ref{1.1.3b}) but the instability is weak enough to permit
thermodynamic behavior anyway \cite{HaHe71,DeLa74,Fro76}, at least for
$\beta < \beta _{1}$. At $\beta _{1}$ the Gibbs factor $\exp { -\beta
_{1}V_{2}}$ for two oppositely charge particles is no longer
integrable and the partition function diverges. The partition function
for the Yukawa gas also diverges and, in the Yukawa gas Mayer
expansion, the diagram with two vertices becomes infinite at $\beta =
\beta _{1}=2$.  However, all diagrams with more vertices remain
finite. Similarly, at $\beta _{2}$, neutral diagrams with 2, 4
vertices diverge, all diagrams with more vertices remain finite, and
so on.  It has been proved \cite{BrKe87} that for $\beta < 4/3 \beta
_{1}$ the Mayer expansion without the infinite two vertex term is
convergent for small activity. The natural extension of this
result to the higher thresholds almost certainly holds for $\beta <
\beta _{\infty}$ \footnote{An almost equivalent result appeared in
\cite{DiHu93}, but the authors have reported a serious error
invalidating their proof \cite{Dim98}}.  These results imply that
there are natural infinite renormalizations of the partition
function. Presumably, this enables some version of the Yukawa and
Coulomb gases to be defined in the range $\beta \in [\beta _{1}, \beta
_{\infty })$ as a renormalized limit of systems with regularized
interaction, because dropping a divergent term in the Mayer expansion
is the same as multiplying the partition function by an infinite
factor.  This has not been discussed clearly in the literature, but
see \cite{GaNi85b,Spe86}.

It has been proven \cite{Yan87} that the two species charge symmetric
Coulomb gas of point particles in a grounded container has exponential
screening if $\beta \ll \beta _{1}$ and the activity are small.  The
grand canonical ensemble with insulating boundary conditions is harder
to analyze because it has to be restricted to neutral configurations,
whereas image charges provide neutrality for free in the grounded
container case.

As the last paragraph suggests, boundary conditions are more important
in two dimensions than in three.  For example it is known
\cite[Theorem 4.1]{FrSp81b} that a pair of oppositely charged
fractional charges immersed in a two dimensional Coulomb gas are not
screened from each other when the thermodynamic limit is taken with
insulating container boundary conditions. On the other hand, although
the details are not published anywhere, it ought to follow from the
results of Yang that they are screened when the thermodynamic limit is
taken with grounded containers. This influence of boundary conditions
does not happen in three dimensions.  We will discuss this further in
section~\ref{sec-tunneling}.

\emph{Two dimensions, with stabilizing short range forces}: The
significance of $\beta _{\infty }$ is that the Kosterlitz-Thouless
transition takes place at a critical $\beta_{c} \approx \beta _{\infty
}$\footnote{$ \beta_{c} \rightarrow \beta _{\infty } $ as the density
tends to zero}.  By stabilizing the interaction with short range
forces such as hard cores one can reach $\beta _{c}$ without infinite
renormalizations and see the $\beta > \beta _{c}$ dipole phase.  The
original argument \cite{KoTh73} was given a complete proof
\cite{FrSp81} for the Coulomb system on a lattice. A more detailed
analysis of the Kosterlitz-Thouless phase, based on term by term
analysis of the Mayer expansion, is given in
\cite{AlCo92,AlCo97a,AlCo97b}. There is confusion over whether the
thresholds $\beta _{n}$ should be interpreted as the successive
collapse into neutral clusters of 2, 4,\dots particles.
\cite{GaNi85b,Spe86} proposed such an interpretation based on
successive divergences in the Mayer expansion.  This is analogous to
the Yukawa gas alluded to above, except that now infra-red divergences
cause the thresholds.  To complete their argument one must show that
there are bulk observables whose correlations have associated
singularities.  On the other hand \cite{FLL95} argue that no
singularities show up in thermodynamic functions and consequently the
thresholds are not physical.

The heuristic renormalization group argument \cite{KoTh73} predicts
that exponential screening of particle distributions holds for all
$\beta < \beta _{c}$ if the density is sufficiently small. A
complete proof of screening at low activity for $\beta $ in this range
would be a wonderful achievement.

\emph{Exact formulas in two dimensions} for the excess free energy of
a single charge fixed at the origin are conjectured in \cite{LuZa97}.
See also \cite{Smi92} for other interesting formulas. These formulas
in principle could be used to express the excess free energy of a
fractional charge in terms of the density ( or the activity) and
temperature.  The authors study the Sine-Gordon functional integral
with boundary conditions that correspond to a grounded conducting
container so that violating neutrality by fixing a fractional charge
makes sense.  Their conjecture translates, via the Sine-Gordon
transformation to the two component charge symmetric gas with a purely
Coulomb potential at least for $\beta < \beta _{1}$. With limitations
on the size of the fractional charge their formulas apply for $\beta
\in [\beta _{1}, \beta _{\infty })$. 

In two dimensions, there are a number of exactly solvable models for
the special value of the temperature that corresponds to $\beta_{1}$,
starting with the two dimensional one component plasma
\cite{Jan81}. These models \cite{Ala87,Jan90,Jan92} exhibit
"super-Debye screening", namely Gaussian decay of the
correlations. They have played an important role in testing various
refined screening properties in presence of walls and of different
types of inhomogeneities.

Although a characteristic of Debye phases is the rapid decay of the
particle correlations, one should be aware that it is never the case
for potential and electric field fluctuations \cite{LeMa84,Mar88}.
For instance, correlations of the components of the electric field
$E_{\mu}$ always have an asymptotic dipolar behavior
\begin{equation}\langle E_{\mu}(\br) E_{\nu}({\bf 0})\rangle \simeq
-\frac{1}{\beta}
\nabla_{\mu}\nabla_{\nu}\left(\frac{1}{\mid\br\mid}\right)
\label{2.2.2field}
\end{equation}This observation has recently led to the development of
the interesting view point that there are some universal properties of
Debye phases that closely resemble those of a system with short range
forces at its critical temperature
\cite{JMP94,Jan96,FoJaT96}. Actually, potential and field fluctuations
are critical in Debye phases, and this leads to universal finite size
corrections to the free energy as in critical systems.

\subsection{The Sine-Gordon
transformation}\label{sec-sinegordon}

\subsubsection{The Fourier transform of a Gaussian is a Gaussian}
In chapter~\ref{chapter- Coulomb systems} we have defined a potential $$
V_{\infty,\lambda }(\br )
=
|\br|^{-1}(1-\exp(-|\br|/\lambda ))
$$
which is a Coulomb potential smoothed at the origin. We saw that
$V_{\infty,\lambda }$ has a positive Fourier transform. For such
potentials there is an identity that expresses the interaction in
terms of external potentials --- an exact version of the mean field
idea of Debye-H\"uckel. This is the Sine-Gordon transformation
\cite{Kac59,Sie60}. Kac and Siegert noticed that if self-energies are
included, then the Gibbs factor
\begin{equation}\label{2.3.3a}
e^{ -\frac{\beta }{2} \sum_{i, j=1}^N e_{\alpha_i} e_{\alpha_j}
V_{\infty,\lambda }(\br _i ,\br _j ) } = e^{ -\frac{\beta }{2} \int d
\br \, \int d\br ' \, \hat{c}(\br ) V_{\infty ,\lambda }(\br ,\br')
\hat{c}(\br ') }
\end{equation}
is Gaussian in $\hat{c}(\br)$. By a functional integral generalization
\cite{Sim79} of ``the Fourier transform of a Gaussian is a Gaussian''
there exists a Gaussian probability measure, intuitively described by
\begin{equation}\label{2.3.3} d\mu_{\lambda } (\phi) = \cD \phi \, e^{
- \frac{1}{2} \int \phi(\br ) V_{\infty,\lambda }^{-1} \phi(\br ) d\br
}
\end{equation}
such that
\begin{equation}\label{2.3.1}
e^{
-\frac{\beta }{2} \sum_{i, j=1}^N e_{\alpha_i} e_{\alpha_j} V_{\infty,\lambda }(\br _i ,\br _j )
}
=
\int d\mu _{\lambda }(\phi)
e^{
-i\beta ^{1/2} \int d \br \,
\hat{c}(\br ) \phi(\br )
}.
\end{equation}
$V_{\infty,\lambda }^{-1}$ is the operator inverse of the operator
whose kernel is $V_{\infty,\lambda }$. In our case we find from
(\ref{1.1.3bd}) that it is the partial differential operator
\begin{equation}\label{2.3.3b} 
V_{\infty,\lambda }^{-1} = \frac{1}{4\pi }\( -\Delta + \lambda ^{2}
\Delta ^{2} \) \end{equation} possibly with boundary conditions on
$\Delta $ .  Equation (\ref{2.3.1}) says that the Gibbs factor is a
superposition of Gibbs factors for external potentials $i\phi$. It
follows that if an ideal gas of particles in an external potential is
integrated over the external potentials then the result is a partition
function for a gas with two-body interaction $V_{\infty,\lambda }$,
including self-energies. Self-energies are equivalent to a shift in
the activities. We define \begin{equation}\label{2.3.4a}
z^{(\lambda)}(\cE ) = z(\cE) e^{ \frac{\beta }{2}e_{\alpha}^{2}
V_{\infty,\lambda }(\br ,\br ) }
\end{equation}
Then the partition function with potential energy $U(\omega )$ built
out of $V_{\infty,\lambda }$ and an external potential $\psi $,
\begin{equation}\label{2.3.4aa} \Xi_\Lambda (\beta ^{1/2}\psi ) = \int
_{\Lambda } d\omega \, z(\omega ) e^{ -\beta U(\omega ) - \beta ^{1/2}
\int \psi \hat{c}(\omega )d\br }
\end{equation}
is given by
\begin{equation}\label{2.3.4c}
\Xi_\Lambda (\beta ^{1/2}\psi )
=
\int d\mu _{\lambda }(\phi) \,
\Xi_{{\rm ideal},\Lambda,z^{(\lambda)}}
(\beta ^{1/2}[i\phi + \psi]).
\end{equation}
where
\begin{equation}\label{2.3.4b}
\Xi_{{\rm ideal},\Lambda,z^{(\lambda)}}(f) =
\int _{\Lambda } d\omega \, z^{(\lambda)}(\omega ) e^{-\int f\hat{c}d\br }
\end{equation}
is the ideal gas partition function (\ref{1.1.11}) with the
renormalized activities (\ref{2.3.4a}). (\ref{2.3.4c}) is the
Sine-Gordon representation. It says that the partition function for
the interacting system exactly equals a (functional) integral over
\emph{imaginary} external fields of the ideal gas partition
function.

Gaussian measures have been studied in the mathematical literature and
it is known that for this one the typical potential $\phi$ is a
continuous function \cite[Theorem~A.4.4]{GlJa87}. This result holds
provided $\lambda \not =0$; when $\lambda =0$ the instability of the
Coulomb potential returns to haunt us in the shape of the typical
$\phi$ becoming a distribution instead of a function.

The right hand side of formula (\ref{2.3.3}) is best understood as a
mnemonic for the translation formula, also known as the Cameron-Martin
formula, 
\begin{equation}\label{2.3.5}
d\mu _{\lambda }(\phi+g) = d\mu _{\lambda }(\phi) \,
e^{-\frac{1}{2}\int g V_{\infty,\lambda }^{-1} g d\br - \int \phi
V_{\infty,\lambda }^{-1} g d\br } \end{equation} which is valid for
any function $g$ for which $\int g V_{\infty,\lambda }^{-1}g$
exists\footnote{In the sense $\int (\mid \nabla g\mid ^{2} + \mid
\lambda \Delta g\mid ^{2})\, d\br < \infty $.}. Translations $\phi
\rightarrow - ig$ are legitimate as well whenever the integrand
permits analytic continuation to $\alpha = i$ after the change of
variable $\phi \rightarrow \phi + \alpha g$.  The translation formula
is the reason why computations with the intuitive formula
(\ref{2.3.3}) are usually correct.

It is not physically very reasonable to try to capture the very local
fluctuations on length scales much smaller than $l_{D}$ in a mean
field picture. For example, the activity $z^{(\lambda )}$ is very
different from $z$ when $\lambda $ is small, but we shall see in
section~\ref{sec-debye-huckel2} that $z$ is the correct effective
activity, in the sense that an ideal gas with this activity is the
best approximation in the Debye-H\"uckel limit. Also hard cores do not
have positive Fourier transforms and cannot be represented by a
Sine-Gordon potential. For these reasons we now consider a better
class of representations in which the particle picture is retained on
scales less than $O(l_{D})$ and the mean field or Sine-Gordon is used
to represent interactions on larger scales.  Define the Yukawa
interaction with a decay length $L \geq \lambda $,
\begin{equation}\label{2.3.6} V_{L}(\br ) = |\br |^{-1}e^{-|\br |/L};
\ \ \ V_{L,\lambda }(\br ) = |\br |^{-1} \left( e^{-|\br |/L} -
e^{-|\br |/\lambda } \right) \end{equation} We split the interaction
$V_{\infty,\lambda }$ according to \begin{equation} V_{\infty,\lambda
} = V_{\infty,L} + V_{L,\lambda } \end{equation} and use the
Sine-Gordon transformation to represent just the part
$V_{\infty,L}$. The result is
\begin{equation}\label{2.3.4}
\Xi_\Lambda (\beta ^{1/2}\psi ) =\int d\mu _{L} (\phi) \,
\Xi_{L,\Lambda,z^{(L)}} (\beta ^{1/2} [i\phi + \psi])
\end{equation}
where
\begin{equation}\label{2.3.7}
\Xi_{L,\Lambda,z^{(L)}}(f)
=
\int _{\Lambda } d\omega \, z^{(L)}(\omega ) e^{-\beta U_{L}(\omega ) -
\int f \hat{c}(\omega )d\br }
\end{equation}
is no longer ideal; $U_{L}$ is built out of two-body interactions
$V_{L,\lambda }$. If hard cores are used in place of the cutoff
$\lambda $ in $V_{L,\lambda }$ then $U_{L}$ is built out of $V_{L}$
and hard cores.

The replacement of $V$ by $V_{\infty,L}$ is the same as
smoothing the Coulomb interaction by introducing form factors for the
charges so they are smeared out into spherical distributions of
characteristic radius $L$. Note that 
\begin{equation}\label{2.3.formfactor}
\tilde{V}_{\infty,L}(\bk) = \tilde{F}(L^{2}k^{2}) \frac{4\pi}{k^{2}}
\tilde{F}(L^{2}k^{2}), \ \ \ \tilde{F}(L^{2} k^{2} ) = \left(
\frac{1}{1+L^{2} k^{2}}\right)^{1/2}
\end{equation}
which shows that $\tilde{F}$ is the Fourier transform of the form
factor. 

\subsubsection{Sum rules}

\newcommand{\EE}[1]{\langle #1 \rangle }
\newcommand{\Ephi}[1]{\langle #1 | \phi \rangle } 
\newcommand{\Ephibar}[1]{\langle #1 | \psi \rangle } 
\newcommand{\E}[1]{E\!\!\!\!E [ #1 ] }

Sum rules provide a good illustration of aspects of the Sine-Gordon
transformation that will reappear in the sequel.  Here we sketch how
the argument in \cite{FoMa84}\footnote{This paper uses the smooth
short range regularization (\ref{1.1.3bc}).} applies to a Coulomb
system with hard cores. We will show that sum rules result from an
invariance of the Sine-Gordon measure under infinitesimal translation
by harmonic functions.  Formulas arising from infinitesimal symmetries
are called Ward identities in quantum field theory. While it is
interesting that sum rules are Ward identities, it is also easy to
derive them without the Sine-Gordon transformation
\cite{Mar88,BrMa99}.

We consider charged particles with hard core interactions of radius
$\lambda $ in a box $\Lambda $ whose boundary is grounded and
impenetrable to the hard cores.  We spread each charge in the ensemble
uniformly onto the surface of its hard core sphere.  By Newton's
theorem the interaction energy $U$ is unchanged by the spreading
out. However the self-energy of each particle becomes finite and of
the order of $\frac{1}{\lambda }$. It can be added into $U$ and the
change in $U$ is compensated by replacing $z_{\alpha }$ by $ z_{\alpha
}^{(\lambda )} = z_{\alpha } \exp ( \beta e_{\alpha }^{2}/ (2\lambda)
) $.  By the Sine-Gordon transformation
\begin{eqnarray}\label{sum-rules.2}
&&
	\Xi_\Lambda 
=
	\int \Xi_\Lambda (i \beta^{1/2} \bar{\phi}_{\lambda} ) 
        \, d\mu (\phi ) \nn \\
&&
        d\mu (\phi ) = e^{-1/(8\pi ) \int_\Lambda 
        (\partial \phi)^{2} } \cD \phi 
\end{eqnarray}
where $\bar{\phi}_{\lambda}(\br)$ is the average of $\phi $ over a
sphere of radius $\lambda $ and center $\br$ and $\Xi_\Lambda (\phi )$
is the hard core gas in external field $\phi $:
\begin{eqnarray}\label{sum-rules.3}
&&
	\Xi_\Lambda (\phi )
=
	\sum_{\{N_\alpha\}}^\infty \prod_{\alpha = 1}^{\cS } 
        \frac{(z_\alpha^{(\lambda )} )^{N_\alpha}}{N_\alpha !}
        \int_{\Lambda} d{\bf r}_{1} \ldots \int_{\Lambda} d{\bf r}_{N}
        \exp\(
        - \sum _{j} e_{\alpha_j} \phi (\br _{j})
        - {\rm hard \ core}
  \)
\end{eqnarray}
$d\mu (\phi) $ is characterized by
\begin{equation}\label{sum-rules.4}
	d\mu (\phi + g) = d\mu (\phi ) 
	e^{1/(4\pi ) \int_\Lambda \phi \Delta g 
	- 1/(8\pi ) \int_\Lambda (\partial g)^{2} };
\end{equation}
where $g$ is any function which vanishes on the boundary with
$\int_\Lambda (\partial g)^{2} < \infty $.

The complete expectation 
\begin{eqnarray}\label{sum-rules.5}
	\langle F \rangle 
&=&
        \frac{1}{\Xi_\Lambda }
	\sum_{\{N_\alpha\}}^\infty \prod_{\alpha = 1}^{\cS } 
        \frac{(z_\alpha^{(\lambda )} )^{N_\alpha}}{N_\alpha !}
        \int_{\Lambda} d{\bf r}_{1} \ldots \int_{\Lambda} d{\bf r}_{N}
        \nn\\
&\times&
        \exp\(
        - \beta U 
        - {\rm hard \ core} 
        \) F
\end{eqnarray}
can be taken in two stages; the first stage is an expectation
conditioned on $\phi $
\begin{eqnarray}\label{sum-rules.6}
	\Ephi{F}
&=&
        \frac{1}{\Xi_\Lambda ( \phi ) }
	\sum_{\{N_\alpha\}}^\infty \prod_{\alpha = 1}^{\cS } 
        \frac{(z_\alpha^{(\lambda )} )^{N_\alpha}}{N_\alpha !}
        \int_{\Lambda} d{\bf r}_{1} \ldots \int_{\Lambda} d{\bf r}_{N}
        \nn\\
&\times&
        \exp\(
        - \sum _{j} e_{\alpha_j} \phi (\br _{j})
        - {\rm hard \ core} 
        \) F\nn 
\end{eqnarray}
and the second is the remaining integration over $\phi $
\begin{eqnarray}\label{sum-rules.7}
        \E{F}
&=&
        \int \ e^{\ln \Xi_\Lambda (i \beta^{1/2} \phi )}  \
        F  \ 
        d\mu (\phi ) \ 
        / \int \ e^{\ln \Xi_\Lambda (i \beta^{1/2} \phi )}  \
        d\mu (\phi )
\end{eqnarray}
and these fit together by 
\begin{equation}\label{sum-rules.8}
        \langle F \rangle
=
        \E{ \Ephibar{F} } \mbox{ where } 
	\psi = i \beta^{1/2} \bar{\phi}_{\lambda} 
\end{equation}
 
To obtain a sum rule we make the change of variables $\phi \rightarrow
\phi + t g$ in the $d\mu (\phi )$ integral on the right hand side and
differentiate with respect to $t$ at $t = 0$. The derivative must
vanish since the left hand side is independent of $t$.  The derivative
of $\ln \Xi_\Lambda (\phi )$ with respect to $\phi (\br )$ is
$-\Ephi{\hat{c}(\br )}$, which accounts for the first line in the next
equation.  The second line arises because the derivative of $\Ephi{F}$
with respect to $\phi (\br)$ is the truncated expectation $ - \Ephi{F
\hat{c}(\br )} + \Ephi{F} \Ephi{\hat{c}(\br )} $. 
\begin{eqnarray}\label{sum-rules.9}
	0 
&=&
        \E{
        \Ephibar{F } \
        \Ephibar{\hat{c}(\bar{g}_{\lambda })}}
        - \E{ \Ephibar{F } } \
          \E{ \Ephibar{\hat{c}(\bar{g}_{\lambda } ) } }
\nn \\
&+&
        \E{
        \Ephibar{F \hat{c}(\bar{g}_{\lambda })}
        - \Ephibar{F } \
        \Ephibar{\hat{c}(\bar{g}_{\lambda })}
        } + 
        \mbox{contributions from } (\ref{sum-rules.4})
\end{eqnarray}
The contributions from (\ref{sum-rules.4}) are proportional to
\begin{equation}\label{sum-rules.9b}
	\E{\Ephibar{F} \int \phi \Delta g } - 
	\E{\Ephibar{F}} \E{ \int \phi \Delta g }
\end{equation} 
We would like to choose $g$ to be a harmonic polynomial because then
these contributions vanish. However this is not a legal choice of $g$
because no harmonic polynomial can vanish on the boundary of $\Lambda
$. Instead choose $g$ to be a harmonic polynomial multiplied by
another function $h$ that is one in the interior of $\Lambda $ and
which goes to zero near the boundary. Now there are non-vanishing
contributions from $\int \phi g \Delta h$ and $\int \phi \partial g
\partial h$ localized near the boundary.  Suppose that the container
$\Lambda $ is enlarged: there is competition between the growth of the
harmonic polynomial $g (\br )$ for $\br $ near the boundary and the
decay of the truncated expectation (\ref{sum-rules.9b}).  If the
system is in a screening phase where truncated expectations of local
functions of $\phi $ decay exponentially then the polynomial growth of
$g$ is crushed as the container is enlarged and the contributions in
(\ref{sum-rules.9b}) vanish in the thermodynamic limit. It is
true that the conditional expectations $\Ephibar{\hat{c}(\br ) }$ are
not exactly local in $\phi $ but if the Mayer expansion of
$\Ephibar{\hat{c}(\br ) }$ is convergent then $\Ephi{\hat{c}(\br )}$
is almost (exponentially) local in $\phi $.  This captures a
physically natural condition that the system should not be so dense that
there are long range correlations caused by the hard core. If the
Mayer expansion converges at $\psi =0$ then it converges uniformly for
all $\psi $ because $\psi \not = 0 $ just means that activities are
multiplied by modulus one phase factors.

Thus in the infinite volume limit (\ref{sum-rules.9}) simplifies to $
0 = \E{ \Ephibar{F \hat{c}(\bar{g}_{\lambda })} } - \E{ \Ephibar{F } }
\ \E{ \Ephibar{\hat{c}(\bar{g}_{\lambda }) } } $ where $g$ is a
harmonic polynomial. Now we simplify further using (\ref{sum-rules.8}
) and Newton's theorem (spherical average of harmonic = value at
center) $\bar{g}_{\lambda }(\br) = g(\br)$ to reach
\begin{eqnarray}\label{sum-rules.11}
	0 
&=&
        \EE { F \hat{c}(g) } - \EE{F} \EE{\hat{c}(g)}
	\ \ \ g \mbox{ any harmonic polynomial}
\end{eqnarray}
For example by choosing $F = \hat{\rho}(\br_{1},\alpha_{1})\ldots
\hat{\rho}(\br_{n},\alpha_{n}) $ (\ref{sum-rules.11}) becomes the same
as (\ref{2.2.5a}). In conclusion, the sum rules (\ref{2.2.5a}) for a
system of classical charges with hard cores hold if the Mayer
expansion for the hard core system without Coulomb interaction is
convergent and if the system is in a screening phase.

\subsection{Debye spheres}\label{sec-debyesphere}

In the previous section we divided the interaction into the Yukawa
part $V_{L,\lambda }$ whose decay length is $L$ and a remaining part,
the slow part $V_{\infty,L}$, which carries the slowly varying long
range piece of the Coulomb interaction. In this section, and in detail
in section~\ref{sec-mayer2}, the Yukawa part is shown to leave the gas
essentially ideal, provided its range $L$ is of order the Debye length
$l_{D}$ and the Debye sphere assumptions (\ref{2.2.2}) are
imposed. The underlying idea is that the effect of the $V_{L,\lambda
}$ part of the interaction is to associate some of the monopoles into
aggregates which can be regarded as new species so that the resulting
multi-species gas is ideal. The Mayer expansion is used to express
this. Instead of leaving out the long range part of the interaction,
one can instead say that the gas inside a sphere of radius less than
the Debye length is close to ideal. This illuminates the
Debye-H\"uckel argument in which the charge density was computed by an
ideal gas formula: the mean field captures the part of the interaction
with range larger than $L$ and the interactions of shorter range are
small enough to use an ideal gas calculation.

Consider the partition function for a Yukawa gas in an external
potential $\phi$,
\begin{equation}\label{2.4.1.1a}
\Xi_{L,\Lambda }(\beta ^{1/2} \phi)
=
\int d\omega \, z(\omega )
e^{-\beta U_{L}(\omega ) - \beta ^{1/2}
\int \phi \hat{c}(\omega )d\br }
\end{equation}
The Mayer expansion is an expansion for the logarithm of the partition
function or the pressure $P$ in powers of the activities $z_{\alpha }$
whose first term $\beta P V = \sum z_{\alpha } + \dots $ gives the
ideal gas law when $\phi=0$: it has the form 
\begin{eqnarray}
\label{2.4.1.2}
&& 
\Xi_{L,\Lambda}(\beta ^{1/2}\phi) = \nn \\
&& \hspace{.2in}
\exp \left(\sum _{m \geq
1} \frac{1}{m!}  \int \prod _{k} d\cE_{k} z(\cE_{k}) e^{-\beta ^{1/2}
e_{\alpha _{k}}\phi(\br _{k})} u_{m}(\cE _{1},\dots ,\cE _{m})  \right)
\end{eqnarray}
where $u_{m}$ are given \cite{Rue69,HaMc76} as sums of contributions
from connected ``Mayer'' graphs. In particular $u_{1}(\cE) = 1$ so
that the $m=1$ contribution to the product is $$ \exp \left[\sum
_{\alpha } \int d\br \, z_{\alpha} e^{-\beta ^{1/2}
e_{\alpha}\phi}\right] $$ which equals the ideal gas partition
function $\Xi_{{\rm ideal},\Lambda }(\beta ^{1/2} \phi) $ that
appeared above in (\ref{1.1.11}).

The higher $m$ terms in (\ref{2.4.1.2}) have the form of additional
ideal gas species, e.g., the $m = 2$ term is an ideal gas partition
function for a gas whose particles are the possible aggregates of two of
original particles, with a cluster activity $\frac{1}{2!}z(\cE_{1})
z(\cE_{2})u_{2}(\cE _{1},\cE _{2})$. Such a cluster describes two
particles bound together because the functions $u_{m}$ decay
exponentially in the separation of any pair of arguments, because the
graphs that contribute to them are connected.  Although these
``Mayer'' aggregates do capture effects of aggregates of particles they
are not directly physical because their activities need not be
positive.

This decomposition of the interacting gas into species of ideal gases
is only useful if the activities of large aggregates are small. In
fact we need them to be summable which is the same as demanding that
the Mayer expansion is convergent. The standard condition \cite{Rue69}
for convergence for the activity expansion for a gas with activity
$z(\cE )$ and two-body potential $v(\cE , \cE ')$ is $$ \sup _{\cE }
\int |e^{-\beta v(\cE ,\cE ')}-1| |z(\cE )e^{-\beta ^{1/2}\phi(\cE )}|
d\cE ' e^{2B\beta } < e^{-1}
$$
where $B$ is the stability constant, i.e., the best constant such that
$$ \sum _{1 \leq i<j \leq N}v(\cE _{i}, \cE _{j}) \geq -B N. $$
Suppose for simplicity that there are two species with equal but
opposite charges $e_{\alpha } = \pm 1$ and equal activities $$ z(\cE )
= z_{\alpha } = z
$$
Using the stability estimate (\ref{1.1.3b}) we find the Mayer
expansion converges if
\begin{equation}\label{2.4.ruelle}
|z| \beta L^{2}
\exp(\beta/\lambda ) \ll 1
\end{equation}
which we calculated using $\exp (\pm V_{L,\lambda }) - 1 \approx \pm
V_{L,\lambda }$. This condition is sufficient for convergence of the
Mayer expansion, but it is not a very accurate condition when $\lambda
< \beta < L$. In section~\ref{sec-mayer2} we will show that there is a
better condition, namely
\begin{equation}\label{2.4.1.5}
z \beta ^{3} e^{\frac{\beta }{ \lambda }} \ll 1; \ \ \ z \beta L ^{2}
e^{\frac{\beta }{ \beta }} \ll 1. \end{equation} The $\beta /\beta $
emphasizes why the estimate is an improvement. For the Yukawa
potential and hard core, one needs
\begin{equation}\label{2.4.1.5b}
z_{\alpha} \lambda ^{3} \ll 1
\end{equation}
as well.  Now observe that when $L = \gamma l _{D}$ the second
condition in (\ref{2.4.1.5}) drops out, provided $\gamma $ is small,
but independent of $z, \beta, \lambda $. The other two conditions are
implied by our Debye sphere hypotheses (\ref{2.2.2}), when combined
with the definition of $l_{D}$. Thus the Debye sphere hypotheses say
that the gas inside a sphere of radius $L = \gamma l _{D}$ is a
multi-species ideal gas whose activities are small according to
\begin{equation}\label{2.4.1.5c}
\frac{1}{m!}
\int \prod _{k} d\cE_{k} |z(\cE_{k})|
|u_{m}(\cE _{1},\dots ,\cE _{m})|
\leq
O\left(\frac{L}{l_{D}}\right)^{2(m-1)} |z| |\Lambda | 
\end{equation}
where $z$ is the largest of the $z_{\alpha }$. For details on the
derivation of this estimate see section ~\ref{sec-mayer2}.

The conditions for convergence of the Mayer expansion when there is an
external potential are the same except that the exponential of the
potential is absorbed into the activity.  Thus an imaginary potential
has no effect in (\ref{2.4.1.5c}) because of the absolute value in
$|z(\cE_{k})|$.

\subsection{The Debye-H\"uckel limit}\label{sec-debye-huckel2}

The standard derivation \cite{May50} of the thermodynamics, the first
part of the Debye-H\"uckel law (\ref{2.1.2}), proceeded by resumming a
class of ring diagrams in the Mayer expansion or low density
expansion. A more systematic basis for this resummation is based on
the Kac limit, see \cite{LeSt68}.  In this section, following
\cite{Ken82,Ken83,Ken84}, we sketch a proof of the Debye-H\"uckel law
based on the fact that the representation (\ref{2.3.4}) becomes an
explicitly integrable Gaussian integral in the Debye-H\"uckel limit
(\ref{2.2.4}). The Debye-H\"uckel mean field emerges as the stationary
point for the Sine-Gordon action.  This line of argument has the
advantage of not using any expansion which might be divergent.

Our definition of the Debye-H\"uckel limit is low density in the
extreme sense that the hard core becomes irrelevant and the effects of
association into dipoles and other aggregates are out of play. The
original Debye-H\"uckel theory (footnote~\ref{footnote-dh}) has terms
in $\kappa a$ that would disappear in this limit if $a$ is taken to be
the hard core radius. Debye-H\"uckel theories (see the discussion
below (\ref{2.2.4})) have been advocated as a good description for
some transitions in ionic fluids at higher densities in which phase
separation occurs. These are extensions of Debye-H\"uckel theory in
which hard core corrections remain and effective dielectric constants,
which represent the effect of charges pairing to form dipoles, are
used. For consistency considerations it is desirable to show that
these theories are also limiting cases of the Coulomb system or
obtainable in some systematic approximation.

Let us work in units where $l_{D} = 1$. This means that in the right
hand side of the representation (\ref{2.3.4}) we rescale lengths by
$\br = l_{D}\br '$ so that the Debye length becomes ${l'}_{D} = 1$ and
$\br '$ becomes dimensionless. Then we have exactly the same formulas
but the parameters are replaced by dimensionless parameters: 
\begin{eqnarray}\label{2.5.1}
\beta'
=
\beta /l_{D}, \ \ \
\lambda'
=
\lambda /l_{D}, \ \ \
{z'}_{\alpha }
=
{z}_{\alpha } l_{D} ^{3}, \ \ \
L'
=
L/l_{D}, \ \ \
\Lambda'
=
l_{D}^{-3}\Lambda
\end{eqnarray}
In terms of the primed parameters the Debye-H\"uckel limit becomes
\begin{eqnarray}\label{2.5.2} 
{z'}_{\alpha }
\rightarrow
\infty , \ \ \
\beta '
\rightarrow
0 \mbox{ with } \left\{
\begin{array}{l}
4\pi \sum_{\alpha } e_{\alpha }^{2}
{z'}_{\alpha } \beta' = 1 \\
{z'}_{\alpha }{\lambda '} ^{3}
\rightarrow
0\\
{z'}_{\alpha } {l'}_{D}^{3} e^{-\beta' /2\lambda' } \rightarrow
\infty
\end{array}\right.
\end{eqnarray}
From now on in this section all formulas involve the primed
quantities, so we will save on notation by dropping the primes.

We write (\ref{2.3.4}) in the form
\begin{eqnarray}\label{2.5.3}
\Xi_{\Lambda }(\beta ^{1/2}\psi )
&=&
\int d\mu_{L} (\phi) \, \exp \left( F(i\phi + \psi ) \right)
\end{eqnarray}
where
\begin{equation}\label{2.5.3a}
F(i\phi + \psi )
=
\ln \Xi_{L , \Lambda, z^{(L)}}
(i\beta ^{1/2} \phi + \beta ^{1/2} \psi )
\end{equation} 
We define the action functional $S(\psi,g)$  by 
\begin{equation}\label{2.5.3b}
-S(\psi,g )
=
\frac{1}{2} \int g V_{\infty,L }^{-1} g d\br + F(g + \psi )
\end{equation}
The partition function $\Xi_{\Lambda }(\beta ^{1/2}\psi )$ does not
depend on $L$ in the left hand side of (\ref{2.5.3}) so we may take
the infinite volume and Debye-H\"uckel limits at fixed $L$ and then
afterwards let $L \rightarrow 0$. The correct order of limits is
infinite volume limit followed by Debye-H\"uckel limit, but we do it
in the opposite order. A serious technical point in the papers we
cited is to show that these limits can be interchanged.

In the Debye-H\"uckel limit the action $S(\psi,i\phi )$ becomes a
quadratic polynomial in $\phi$. We forestall the appearance of some
terms linear in $\phi$ by using the translation formula (\ref{2.3.5})
with $\phi \rightarrow \phi - ig$ applied to (\ref{2.5.3}) 
\begin{eqnarray}\label{2.5.4}
&&
\Xi_{\Lambda }(\beta ^{1/2}\psi )
=
e^{\frac{1}{2}\int g V_{\infty,L }^{-1} g d\br } \int d\mu_{L} (\phi) \, \nn
\\
&&\hspace{1in}
\times \exp \left[
\int i\phi V_{\infty,L}^{-1} g d\br + F(i\phi + \psi + g)
\right]
\end{eqnarray}
$g$ will be chosen later. By (\ref{2.4.1.5c}) the Mayer expansion for
$F(i\phi + g + \psi )$ 
is convergent uniformly in the Debye-H\"uckel limit, provided $L \ll
1$: indeed the $O({z}^{N+1})$ term in the Mayer expansion is
$O({L}^{N}) z |\Lambda |$ uniformly in the Debye-H\"uckel limit. The
$z|\Lambda |$ is a common volume factor. Divide it out and fix $L \ll
1$. Then the first term in the Mayer expansion dominates the sum of
all the others uniformly in the Debye-H\"uckel limit. Therefore the
Debye-H\"uckel limit can be taken term by term under the sum in the
Mayer expansion and under the $d\mu $ integral by standard theorems of
analysis (dominated convergence). Every term becomes a quadratic
polynomial in $\phi$ and only the first term in the Mayer expansion
contributes in the limit $L \rightarrow 0$. To see this in more
detail, note that the first term is 
\begin{equation}\label{2.5.6}
\sum _{\alpha} \int_{\Lambda} d\br \,
z^{(L)}_{\alpha} 
e^{-e_{\alpha}\beta ^{1/2}(i\phi + g + \psi) } \end{equation}
By (\ref{2.5.2}) terms of order $O(z^{(L)}_{\alpha} \beta ^{3/2})$ 
vanish in the Debye-H\"uckel limit. The $O(z^{(L)}_{\alpha}
\beta^{1/2} )$ term vanishes in the limit by pseudo-neutrality
(\ref{2.2.1}). Therefore, in the Debye-H\"uckel limit, the
non-vanishing part of $F(i\phi + g + \psi )$  is
\begin{eqnarray}\label{2.5.7} 
&&
 F_{{\rm DH}}(i\phi + g + \psi ) = 
\int_{\Lambda} d\br \( \sum _{\alpha }
z_{\alpha} + 
\frac{1}{8\pi } (i\phi + g + \psi )^{2} + \frac{1}{8\pi } V_{\infty,L
}(0 )\) \nn \\
&& \hspace{1in}
+ R_{2}(i\phi+g+\psi) + R_{0}
\end{eqnarray} 
where $V_{\infty,L }(0 )$ arose from the difference between $z_{\alpha
}$ and $z^{(L)}_{\alpha}$; see (\ref{2.3.4a}). $R_{2}$ is a quadratic
remainder term  
\begin{equation}\label{2.5.7a} R_{2}(f) = \int \int f(\br _{1}) r(\br _{1},\br _{2})f(\br_{2})d\br
_{1}d\br _{2} 
\end{equation} 
$R_{0}$ is independent of $f = i\phi + g + \psi $.  $R_{0}$ and
$R_{2}$ account for all the remaining terms in the Mayer
expansion. There is no term linear in $f$: inspection of $O(f)$
contributions from the Mayer expansion is required to verify that the
pseudo-neutrality condition makes them all vanish in the
Debye-H\"uckel limit.  Furthermore $R_{0}$ and $R_{2}$ are independent
of $z, \beta $ and vanishing as $L \rightarrow 0$. If we substitute
$F_{{\rm DH}}$ back into (\ref{2.5.4}) and regroup into terms
independent of $\phi$, terms linear in $\phi$ and terms quadratic in
$\phi$, the result is 
\begin{eqnarray}\label{2.5.8} 
\Xi_{\Lambda 
}/\Xi_{{\rm ideal},\Lambda } &\sim & e^{-S_{{\rm DH}}(\psi , g)} \int
d\mu_{L } (\phi) \, e^{ - \frac{1}{8\pi } \int _{\Lambda } \(\phi^{2}
- V_{\infty,L }(0)\) d\br} \
e^{-{S'}_{{\rm DH}}(\psi , g;i\phi)+R_{2}(i\phi)} 
\end{eqnarray}
Note that the activity in the ideal gas $\Xi_{{\rm ideal},\Lambda }$
is $z$, whereas a naive calculation based on (\ref{2.3.4c}) might draw
us into using $z^{(\lambda )}$.  The $\sim $ means that we keep only
the  terms that are non-vanishing in the Debye-H\"uckel  limit.  We
have defined the quadratic functional 
\begin{equation}\label{2.5.13} 
- S_{{\rm DH}}(\psi, g)
=
\frac{1}{2} \int g V_{\infty,L}^{-1} g \, d\br + \frac{1}{8\pi } \int
(g + \psi )^{2} \, d\br + R_{2}(\psi +g) + R_{0}
\end{equation}
whose first variation in direction $\phi$ is  
\begin{equation}\label{2.5.13b}
-{S'}_{{\rm DH}}(\psi , g;\phi)
=
\int \phi V_{\infty,L}^{-1} g \, d\br
+ \frac{1}{4\pi } \int \phi (g + \psi ) \, d\br + R_{2}'(\psi +g;\phi)
\end{equation}
$R_{2}'(\psi +g;\phi)$ is the first variation of $R_{2}(\psi +g)$ in
direction $\phi$.  It is natural to choose $g$ to make $S_{{\rm DH}}'$
vanish, but instead we settle for making it almost vanish by choosing
$g$ to solve \begin{equation}\label{2.5.5} -\Delta g + g + \psi =0
\end{equation}
because then by (\ref{2.3.3b}) 
\begin{equation}\label{2.5.13c}
-{S'}_{{\rm DH}}(\psi , g;\phi)
=
\frac{1}{4\pi } L^{2} \int \phi \Delta ^{2} g \, d\br + R_{2}'(\psi
+g; \phi)
\end{equation}
which means that ${S'}_{{\rm DH}}(\psi , g;\phi)$ disappears at the
end when we take $L \rightarrow 0$. However we will keep these terms
around for a few more lines because we want to derive some formulas
that will be used in later sections.

Define
\begin{equation}\label{2.5.14}
\Xi_{{\rm DH},\Lambda }
=
\int d\mu_{L } (\phi) \,
e^{- \frac{1}{8\pi } \int _{\Lambda }
\( \phi^{2} - V_{\infty,L }(0)\) d\br + R_{2}(i\phi)} \end{equation}
which is the normalization for a new Gaussian probability measure
\begin{equation}\label{2.5.14b} 
d\mu _{{\rm DH}}(\phi)
=
\Xi_{{\rm DH},\Lambda }^{-1} d\mu _{L}(\phi)\, e^{- \frac{1}{8\pi } \int _{\Lambda }
\( \phi^{2} - V_{\infty,L }(0)\) d\br + R_{2}(i\phi)} \end{equation}
Then (\ref{2.5.8}) is the same as
\begin{eqnarray}\label{2.5.12}
\Xi_{\Lambda }/\Xi_{{\rm ideal},\Lambda } &\sim &
e^{-S_{{\rm DH}}(\psi , g)} \
\Xi_{{\rm DH},\Lambda } \
\int d\mu _{{\rm DH}}(\phi) \,
e^{-{S'}_{{\rm DH}}(\psi , g; i\phi)}
\end{eqnarray}
As we have explained above, when $L \rightarrow 0$, we can drop the
${S'}_{{\rm DH}}$ so that the normalized $d\mu _{{\rm DH}}$ also disappears
and we are down to the evaluation of $\Xi_{{\rm DH},\Lambda }$.

The normalization $\Xi_{{\rm DH},\Lambda }$ can be expressed as a
determinant using 
\begin{equation}\label{2.5.9}
\int d\mu _{L}(\phi) \, e^{- \frac{\alpha }{8\pi }
\int _{\Lambda } \phi^{2} d\br }
=
\det\(I + \frac{\alpha }{4\pi } V_{\infty,L }\)^{-\frac{1}{2}}
\end{equation} 
We have dropped the $R_{2}$ term because (\ref{2.4.1.5c}) implies that
\begin{equation}\label{2.5.9a}
|R_{2}(i\phi)| \leq O(L^{2})\int_{\Lambda} \phi^{2}d\br
\end{equation}
and so it gives no contribution in the limit $L \rightarrow 0$. To
obtain this estimate from (\ref{2.4.1.5c}) note that $R_{2}$ involves
two variational derivatives with respect to $\phi $.  Each derivative
brings a factor $\beta ^{1/2}$. These factors combine with $z$ in
(\ref{2.4.1.5c}) and become a Debye length $l_{D} = 1$.  We want
$\alpha = 1$, but we put it there because taking $\alpha =0$ shows
that both sides are correctly normalized. $V_{\infty,L }$ is the
operator whose kernel is the potential $V_{\infty,L }(\br _{1}- \br
_{2})$. The arguments $\br _{1},\br_{2}$ are confined to the finite
volume $\Lambda $ either by boundary conditions on the Coulomb
potential, if the particles are in a grounded conducting container, or
by restriction on the range of integration, if the particles are in an
insulating container. It can be shown that the thermodynamic limit
washes out the difference, but the easier case is the first; then
there are zero boundary conditions on the Laplacian in $V_{\infty,L}$
so that the eigenfunctions are trigonometric functions when $\Lambda $
is a box. Using $\det (I + X) = \exp [\sum _{\xi } \ln(1 + \xi
)]$, where $\xi $ runs over the eigenvalues of the operator $X$,
one can show that the leading term in the infinite volume limit is
\begin{equation}\label{2.5.10} 
\Xi_{{\rm DH},\Lambda }
\approx
\exp \left[
-\frac{1}{2} (2\pi)^{-3} |\Lambda |
\int d \bk \,\(
\ln (1 + \frac{1}{4\pi }\tilde{V}_{\infty,L}) - \frac{1}{4\pi }\tilde{V}_{\infty,L}
\) \right]
\end{equation}
where the Fourier transform $\tilde{V}_{\infty,L}$ is given in
(\ref{1.1.3bbb}). The limit as $L \rightarrow 0$ of this expression is
\begin{equation}\label{2.5.11} \exp \left[ -\frac{1}{2} (2\pi)^{-3}
|\Lambda | \int d \bk \,\( \ln (1 + \frac{1}{\bk^{2}}) -
\frac{1}{\bk^{2}} \) \right] = \exp [| \Lambda | /(12\pi )]
\end{equation}

Returning to (\ref{2.5.12}) we have shown that that the leading terms
in the Debye-H\"uckel limit followed by infinite volume limit followed
by $L \rightarrow 0$ are
\begin{equation} \label{2.5.16}
\ln \left( \Xi_{\Lambda }/\Xi_{{\rm ideal},\Lambda } \right) \sim
\frac{|\Lambda |}{12 \pi} + \frac{1}{8\pi}\int g (-\Delta ) g d\br +
\frac{1}{8\pi} \int (g + \psi )^{2} d\br \end{equation} where $g$
satisfies the stationarity condition (\ref{2.5.5}). When $\psi =0$ we
recover the thermodynamic part of the Debye-H\"uckel law,
(\ref{2.1.2}). Furthermore the stationarity condition coincides with
the linearized Debye-H\"uckel equation (\ref{2.1.10}) in units where
$l_{D} = 1$ when
\begin{equation}\label{2.5.15}
f = \beta ^{-1/2}(g + \psi), \
\psi (\br ) = e_{0}\beta ^{1/2} V(\br )
\end{equation}
The strange factors of $\beta ^{1/2}$ appear because of the way we
normalized the Sine-Gordon transformation. The Debye-H\"uckel result
(\ref{2.1.2}) on correlations, in units where $l_{D}=1$, is obtained
by taking two functional derivatives of the approximation
(\ref{2.5.16}) for $\ln\( \Xi_{\Lambda }/\Xi_{{\rm ideal},\Lambda
}\)$ with respect to $-\beta ^{1/2}\psi $. This interchange of
functional derivatives with Debye-H\"uckel limit might be hard to
justify, but convexity helps in the case of the single charge density
expectation \cite{Ken83}.

There may be inequalities comparing the Debye-H\"uckel limit with the
exact system. Some results follow from Jensen's inequality, but they
are limited to the unrealistic case of charge-symmetric systems
\cite{Ken82}. See also theorem~\ref{theorem-DHupperbound} in the next
section.

Questions: Are extended Debye-H\"uckel theories such as the ones
considered by Fisher et al. limiting theories? One should no longer
let $L \rightarrow 0$ and should choose $g$ to be the exact stationary
configuration for $S_{{\rm DH}}(\psi , g)$, which contains quadratic
remainder terms $R_{2}$ giving corrections to the dielectric
constant. Also the activity may be renormalized, whereas in our case
$L \rightarrow 0$ causes $z^{(L)}$ to coincide with $z$. Is the
expansion \cite{LeSt68} asymptotic in $\kappa a$?

\subsection{Charge expulsion}
\label{sec-charge-expulsion}

We continue to use units in which $l_{D} = 1$ introduced in
section~\ref{sec-debye-huckel2}. In our preliminary discussion of the
Debye-H\"uckel theory in section~\ref{sec-debye-huckel} we claimed
that the activities should be constrained by the pseudo-neutrality
condition (\ref{2.2.1}). If we start with a system that does not
satisfy this condition, then even when $\psi =0$, the action
$S(0,i\phi)$ defined in (\ref{2.5.3b}) is not stationary in $\phi$ at
$\phi = 0$. However, a translation $\phi \rightarrow \phi - ig$ such
that $S(0,g + i\phi)$ is stationary in $\phi$ at $\phi = 0$ reveals
the physical effect of charge expulsion and an attendant
renormalization of activities. This is a nonlinear effect, not small
enough to fit within the approximation of the last section. The
physical interpretation is that excess charge is expelled to the
boundary where it forms a boundary layer of total charge proportional
to the surface area. This layer modifies the electrostatic potential
in the interior in such a way that the activities are renormalized to
satisfy pseudo-neutrality, which is why the excess charge is of order
surface area and not volume in this grand canonical ensemble.

For simplicity we omit the complications that arose in the last
section due to the length scale $\lambda $ by considering a system
with interaction $V_{\infty ,L }$ with insulating boundary conditions
and no other forces.  In this case we can use the Sine-Gordon
representation (\ref{2.3.4c}) to represent the entire interaction. We
can also apply the translation formula (\ref{2.3.5}) for $\phi
\rightarrow \phi - ig$. The result is
\begin{eqnarray}\label{2.6.1}
&&
        \Xi_L (\beta ^{1/2}\psi )
= 
        \int d\mu _{L }(\phi) \, 
        e^{\frac{1}{2}\int g V_{\infty,L }^{-1} g d\br 
        + \int i\phi V_{\infty,L }^{-1} g d\br } \nn \\
&& \hspace{1in} \times 
        \Xi_{{\rm ideal},z^{(L)}}
        (\beta ^{1/2}[i\phi + g + \psi]).
\end{eqnarray}
The translation $g$ is chosen so that the linear term $\int \phi
V_{\infty,L }^{-1} g d\br$ is canceled by a corresponding term
from $\Xi_{{\rm ideal},z^{(L)}}$.  This is the same as
choosing $g$ to be the minimizer $g$ of
\begin{equation}\label{S-infty-lambda}
        - S_{\infty, L }(\psi ,g) 
=
        \frac{1}{2 } \int g
        V_{\infty,L }^{-1} g d\br \, + \ln 
        \Xi_{{\rm ideal},z^{(L)}}
        (\beta ^{1/2}[g + \psi])
\end{equation}
Insulating boundary conditions mean that $V_{\infty,L }^{-1}$ is given
by (\ref{2.3.3b}).  The $\int d\br $ integral extends over infinite
volume but the $\Xi_{{\rm ideal},z^{(L)}}$ is an ideal gas in the
finite volume $\Lambda $.  There exists \cite{Ken84} a unique $g$ that
minimizes this functional.  Uniqueness is an easy consequence of it
being convex.  Let $\psi = 0$. Then in the deep interior of $\Lambda $
$g$ is close to the constant $g_{0}$ that minimizes
\begin{equation}\label{2.6.2a} 
        \ln \Xi_{{\rm ideal},z^{(L)}}
        (\beta ^{1/2} g_{0})
=
        \int_{\Lambda } d\cE \, z^{(L)}(\cE )
        e^{-\beta ^{1/2} e_{\alpha } g_{0}}
\end{equation}
Outside $\Lambda $, $g$ solves $V_{\infty,L }^{-1}g = 0$ and $g
\rightarrow 0$ at infinity. Define renormalized activities
$z^{(L,R)}$ by 
\begin{equation}
        z_{\alpha }^{(L,R)} 
=
        \exp(- e_{\alpha } \beta ^{1/2} g_{0})
        z_{\alpha }^{(L)}
\end{equation}
and note that
\begin{equation}
\Xi_{{\rm ideal},z^{(L)}} (\beta ^{1/2} f) 
=
\Xi_{{\rm ideal},z^{(L,R)}} (\beta ^{1/2} [f-g_{0}])
\end{equation}
Since $g \approx g_{0}$ inside $\Lambda $ the bulk contribution to
$S_{\infty ,L}(g)$ comes from the ideal gas term which has the
renormalized activities $z^{(L,R)}$.  But since $\Delta g$ is non-zero
near the boundary there is a boundary contribution to the pressure in
$\frac{1}{2 } \int g V_{\infty,L }^{-1} g d\br $ and also in the ideal
gas term because $g \not = g_{0}$ at the boundary. These arise from
the surface charge expelled from the interior and account for the
capacity term in the free energy, discussed in the analysis of systems
with net charge \cite[page 55]{LiLe72}, \cite{GrSc95b}\footnote{If
$L=0$ and $g$ is chosen to be constant in the interior of $\Lambda $
and harmonic in the exterior, then the boundary contribution in
(\ref{S-infty-lambda}) exactly equals the electrostatic energy stored
in a perfect conductor with shape $\Lambda$.}.

From (\ref{2.6.1})
\begin{eqnarray}
        \Xi_L (0)
&=&
        e^{-S_{\infty ,L}(g)}
        \int d\mu _{L }(\phi) \, 
        e^{\int i\phi V_{\infty,L }^{-1} g d\br } \  
        \frac{
        \Xi_{{\rm ideal},z^{(L,R)}}(\beta ^{1/2}[i\phi + g - g_{0}])
        }{
        \Xi_{{\rm ideal},z^{(L,R)}}(\beta ^{1/2}[g - g_{0}])
        }\nn 
\end{eqnarray}
By our choice of translation $g$ we have arranged that the $i\phi$
fluctuates around zero so that $\exp \left(-S_{\infty ,L}(g) \right)$
should be the leading term in the Debye-H\"uckel limit.  The next term
in the approximation would be a Debye-H\"uckel term that comes from
integrating over the $\phi$ fluctuations using a quadratic
approximation as in section~\ref{sec-debye-huckel2}.

\subsection{Symmetries and tunneling}\label{sec-tunneling}

We continue to work in units in which $l_{D} = 1$. There is a loss of
intuition entailed by the Sine-Gordon transformation because the
integration extends over imaginary potentials, but there are also
aspects one can see more easily in the Sine-Gordon language. In
particular if all charges $e_{\alpha }$ are integral multiples of a
fundamental unit of charge, which for simplicity we assume to be one,
then the action $S(\psi,g)$ in (\ref{2.5.3b}) is invariant under the
discrete symmetry
\begin{equation}\label{2.7.1} 
S(\psi,g)
=
S(\psi,g + i \delta h), \ \ \
\delta h
=
2\pi \beta^{-1/2}
\end{equation}
This section is concerned with the breaking of this symmetry, which
plays a role in the proofs of screening. We will see also that there
are tunneling corrections associated with the symmetry.

Let us first note that the Coulomb system in an insulating container
in three or more dimensions cannot enjoy this symmetry because the
Gaussian measure $d\mu _{L}(\phi)$ satisfies 
\begin{equation}\label{2.7.1b} 
\int d\mu_{L} (\phi) \, \phi^{2}(0)
=
V_{\infty,L}(0) < \infty
\end{equation}
precluding any symmetry of the form $\phi \rightarrow \phi + c$.

We have seen in section~\ref{sec-debye-huckel2} that the
Debye-H\"uckel approximation is a quadratic approximation around the
stationary configuration $\phi = 0$. The symmetry (\ref{2.7.1}) tells
us that there are infinitely many stationary configurations related by
the symmetry to $\phi = 0$. We claim that even though (\ref{2.7.1b})
implies that the stationary configuration $\phi= 0$ is favored, in the
thermodynamic limit there are arbitrarily large regions where $\phi$
is trapped in the other stationary configurations. This is analogous
to the Ising model at low temperature in which there occur arbitrarily
large islands of $-$ spins even though the boundary is set at $+$ and
the majority of spins are $+$. In the zero temperature limit in the
Ising model, the islands of $-$ spins disappear, but if one wants to
understand low temperature as opposed to zero temperature, then they
are important. In the same way, to obtain screening near, as opposed
to in the Debye H\"uckel limit, the presence of large islands where
the $\phi = 0$ approximation breaks down has to be taken into account.

To take into account the presence of many stationary configurations
related by the symmetry we use a functional version of the
Villain approximation 
\begin{equation}\label{2.7.1a}
e^{\cos (x)}
\approx
\sum _{n} e^{-\frac{1}{2}(x-2\pi n)^{2}+1} 
\end{equation}
on the term $\exp(F)$ in (\ref{2.5.3}). In (\ref{A.villain}) in
section~\ref{sec-polymer-rep1} we explain in terms of this one
dimensional analogue why taking into account the extra wells in this
way improves the convergence of perturbation theory.

$\exp \left(F \right)$ is invariant under a translation $\phi
\rightarrow \phi + h$ where $h = h(\br )$ is any function of $\br $
that takes values in the set $\{n\delta h\}$ of periods, because
$\Xi_{L , \Lambda, z^{(L)}}$ clearly has this property. Also from
(\ref{2.5.7}) it has the quadratic approximation
\begin{eqnarray}\label{2.7.2} 
F_{{\rm DH}}(\psi + i\phi) &=& 
\int_{\Lambda} d\br
\( \sum _{\alpha } z_{\alpha} + \frac{1}{8\pi } (i\phi + \psi )^{2} +
\frac{1}{8\pi } V_{\infty,L }(0 )\) \nn \\
&&\hspace{.5in}
+ R_{2}(i\phi + \psi ) + R_{0}
\end{eqnarray}
Thus an analogue of (\ref{2.7.1a}) is
\begin{equation}\label{2.7.villain}
e^{ F(\psi + i\phi)}
=
e^{R_{3}(\psi + i\phi)}
\sum _{h} e^{F_{{\rm DH}}(\psi - ih + i\phi)} 
\end{equation}
where $R_{3}$ is an $O(\beta ^{1/2})$ error term representing cubic
corrections to each well and put there to obtain exact equality. $h =
h(\br )$ is summed over all functions that vanish outside $\Lambda $,
take values in the set $\{n\delta h\}$ of periods and are piecewise
constant on a lattice of unit cubes filling $\Lambda $.  

If we think of a typical piecewise constant function $h(\br)$ as a
height, then the resulting landscape splits into plateaus where $h(\br )$
is constant but more important are the jumps from one plateau to
another. In analogy with the Ising model we call the connected
surfaces in $\Lambda $ where $h$ jumps ``contours.'' Our objective is
to argue that near the Debye-H\"uckel limit the sum over $h$ is
dominated by terms in which these contours are small and very
dilute. The value of $h$ in a given plateau fixes the well in which
$\phi $ is trapped and the contour is where $\phi $ tunnels from one
well to another.

We substitute (\ref{2.7.villain}) into the Sine-Gordon transformation
(\ref{2.5.3}) and retrace the analysis of
section~\ref{sec-debye-huckel2} leading up to the Debye-H\"uckel
approximation (\ref{2.5.12}) with $\psi $ replaced by $\psi - ih$ and
$g$ replaced by $ig$
\begin{eqnarray}\label{2.7.6}
&&\hspace{-.5in} \Xi_{\Lambda } /\Xi_{{\rm ideal},\Lambda } =
 \Xi_{{\rm DH},\Lambda }\sum _{h} e^{- S_{{\rm DH}}(\psi - ih, ig)}
 \nn \\ && \times \int d\mu _{{\rm DH}}(\phi) \, e^{-{S'}_{{\rm DH}}(\psi
 -ih , ig; i\phi) + R_{3}(\psi+ig+i\phi)}
\end{eqnarray}
For each $h$ in the sum $g(\br)$ can be any function. We choose it to
make $S_{{\rm DH}}(\psi -ih, ig)$ stationary so that ${S'}_{{\rm DH}}$
vanishes.

For the moment we drop the term $R_{3}$, then the $d\mu _{{\rm DH}}$
integral drops out because it is normalized and we are left with an
approximation
\begin{equation}\label{2.7.6a}
\Xi_{\Lambda } /\Xi_{{\rm ideal},\Lambda } \approx 
        \Xi_{{\rm DH},\Lambda }
        \sum _{h} e^{- S_{{\rm DH}}(\psi - ih, ig)} 
\end{equation}
where
\begin{eqnarray} \label{2.7.Sdef}
S_{{\rm DH}}(\psi - ih, ig)
&=&
\frac{1}{2 }\int g V_{\infty,L}^{-1} g d\br +
\frac{1}{8\pi }\int (g - h -i\psi )^{2} d\br \nn \\
&&\hspace{1in} + R_{2} + R_{0}
\end{eqnarray}

When $\psi =0$ $g$ minimizes $S_{{\rm DH}}(-ih, ig)$ by trying to make
$(g-h)^{2}$ vanish but cannot quite succeed because $\int g
V_{\infty,L}^{-1} g d\br$ forces $g$ to be smooth whereas $h$ has
jumps.  Suppose $h(\br ) = h_{\Gamma }(\br )$ has only one contour
$\Gamma $ and $h$ vanishes at the boundary of $\Lambda $.  Then
$\Gamma $ has an exterior ``sea-level'' plateau where $h = 0$ and an
interior plateau where $h \not =0$.  In the interiors of both plateaus
$g(\br ) \approx h(\br ) $.  The error in the $\approx$ is $O(\exp
\left( - \dist(\br ,\Gamma \right))$.

Any $h$ with $h = 0$ at the boundary can be decomposed in a unique way
\begin{equation}\label{2.7.hdecomp}
 h = \sum _{j} h_{j}
\end{equation}
where each $h_{j}$ has only one contour $\Gamma _{j}$ and $h_{j}$
vanishes at the boundary.  Thus $h_{j}$ is defined on the whole box
$\Lambda $ but it assumes just two values: zero outside the contour
$\Gamma _{j}$ and $h_{j}({\rm int})$ at all points in the interior of
the contour. Across the contour $\Gamma _{j}$ $h_{j}$ jumps by
$h_{j}({\rm int})$ which is the same as the jump of $h$ across $\Gamma
_{j}$.

The minimizers are linear functions of $h$
so $g = \sum g_{j}$ where $g_{j}$ is the minimizer for the single
contour $h_{j}$.  If the contours are well separated then by $g(\br ) =
h(\br ) + O(\exp \left(\dist(\br ,\Gamma \right))$.
\begin{equation}\label{2.7.Sdecomp}
S_{{\rm DH}}(\psi - ih, ig) \approx \sum _{j} S_{{\rm DH}}(\psi - ih_{j},
ig_{j}) 
\end{equation}
Thus 
\begin{equation}\label{2.7.ising}
\Xi_{\Lambda } /\Xi_{{\rm ideal},\Lambda } \approx 
        \Xi_{{\rm DH},\Lambda }
        \sum _{h} \prod _{j} e^{- S_{{\rm DH}}(\psi - ih_{j}, ig_{j})} 
\end{equation}
The approximation in these two equations was to leave out
exponentially small cross terms in $S_{{\rm DH}}$ involving $(g_{i}
-h_{i})(g_{j} - h_{j})$ and derivatives of $g_{i}$ times derivatives
of $ g_{j}$.

The sum over $h$ is equivalent to summing over the number $N$ of
contours, the shapes $\Gamma _{j}$ of the contours and the jumps
$h_{j}({\rm int})$ at each contour so that 
\begin{equation}\label{2.7.contour}
\Xi_{\Lambda } /\Xi_{{\rm ideal},\Lambda } \approx 
        \Xi_{{\rm DH},\Lambda }
        \sum _{N} \frac{1}{N!}
        \sum _{\Gamma _{1},\dots ,\Gamma _{N}}
        \prod _{j} 
        \sum _{h_{j}({\rm int})} e^{- S_{{\rm DH}}(\psi - ih_{j}, ig_{j})} 
\end{equation}

The exponential of $S_{{\rm DH}}(\psi - ih_{j}, ig_{j}) $ suppresses
the jump $h_{j}({\rm int})$ at $\Gamma _{j}$ by a factor
\begin{equation}\label{2.7.suppress}
\exp[ - O(|\Gamma_{j} |) h_{j}({\rm int})^{2}]
\end{equation}
where $|\Gamma_{j} |$ is the area of the contour. The $h_{j}({\rm
int})^{2}$ dependence is because $S_{{\rm DH}}$ is quadratic in $g,h$
when $\psi =0$, $g_{j}$ is linear in $h_{j}$ and $h_{j}$ is just
$h_{j}({\rm int}) \times $ a unit jump configuration.

In (\ref{2.7.contour}) the term with no contours, $N = 0$, is our
original Debye-H\"uckel approximation (\ref{2.5.16}). From
(\ref{2.7.contour},\ref{2.7.suppress}) we see that the contours should
behave like an ideal gas with an exponentially tiny density near the
Debye-H\"uckel limit. Nevertheless they represent a phenomenon that
probably makes the standard perturbation series asymptotic but not
convergent and incapable of proving that there is screening near as
opposed to at the limit. In contrast (\ref{2.7.contour}) is the basis
for a convergent expansion which does imply screening near the
Debye-H\"uckel limit. The terms we left out in arriving at this
representation are exponentially small interactions between the dilute
contours.  There are also constraints (hard core interactions) in the
sum that forbid intersections of contours.  Such interactions are
within the purview of conventional convergent Mayer expansions, which
is why it is possible to develop a convergent expansion.  However this
expansion is clumsy in comparison with standard perturbation theory.
One has to show for example that the $R_{3}$ error term does not
destroy the picture we have just developed.  One of the principal
accomplishments \cite{GlJa87} of the constructive quantum field theory
program was to develop a calculus for the analysis of ``almost
Gaussian'' integrals and this is what is used at this point.  The
reader who suffers from residual curiosity can turn to
chapter~\ref{chapter-convergent-expansions} for a compressed and
updated account of this calculus and its applications, or to
\cite{BrFe81} for an older review.

Note the role of grounded container boundary conditions is to anchor
$g$ and therefore $h$ at zero on the boundary. The $1/r$ potential
with no boundary conditions, i.e.  particles in an insulating
container, is more difficult than the case of grounded boundary
because the anchoring is weaker, since it comes from
(\ref{2.7.1b}). This has been partially investigated
\cite{FeKe85}. Their argument is hard because there are long range
forces at the boundary which make it difficult to control the size of
boundary effects.

In the case of two dimensions with insulating boundary conditions,
there is no anchoring at the boundary because the Gaussian
measure\footnote{defined by taking a limit outside the integral as a
small mass tends to zero} $d\mu _{L}$ does not break the symmetry
(\ref{2.7.1}).  We mentioned in section~\ref{sec-Debye-screening} a
result of Fr\"ohlich and Spencer \cite[Theorem 4.1]{FrSp81b} that
fractional charges are not screened when insulating boundary
conditions are imposed.  This is because of this symmetry. Intuitively
there is no preference for which potential well $\phi(\br _{1})$ is
trapped in, but once it has made up its mind, $\phi(\br _{2})$ wants
to choose the same well, which is a long range correlation.  This
argument predicts that the correlation decays to a constant.  On the
other hand, if the boundary is grounded, then the $0$ well is selected
essentially everywhere, except for a very dilute gas of contours that
enclose regions where $\phi$ is in other wells.  In this case all
local observables are expected to have exponential decay, but this is
only proved for integer charge observables \cite{Yan87}.  Recall in
analogy that the Ising model at low temperature with plus boundary
conditions has exponential decay but this decay is lost if the
boundary spins are not tied down because the state is no longer pure.
It is an open problem to prove that two dimensional systems with
insulating boundary conditions have screening of integer charge
observables.

Returning to three dimensions when there are irrational charges the
symmetry is replaced by a quasi-periodicity, nevertheless there is
screening in these systems without exact symmetry \cite{Imb83}. The
argument requires Pirogov-Sinai theory.

There is an intriguing difference between fractional charge
observables and integer charge observables. Suppose the gas consists
of unit positive and negative charges but there is an additional
half-integer charge fixed at the origin $\br = 0$, represented by a
factor $\exp \left(-\beta ^{1/2}i\phi (0)/2 \right)$. The $\phi(0) $
fluctuates around $g(0) \approx h(0)$, so there is a factor $\exp
\left(-\beta ^{1/2}ih(0)/2 \right)$ which is $\pm 1$ in a way that
depends on all contours surrounding the origin, which is a more
non-local (topological) effect than when an integer charge is placed
at the origin.  To put it another way: a phase factor $\exp
\left(-\beta ^{1/2}ih_{j}(0)/2 \right)$ must be included in the sum
over $h_{j}({\rm int})$ in (\ref{2.7.contour}) whenever $\Gamma _{j}$
encloses the origin.  This does not destroy screening in three
dimensions, but it might be worth further study.

We will show, in appendix \ref{sec-polymer-rep1}, that the tunneling
corrections are of size $\exp {( - O (\beta /l_{D}))}$ so they are not of
any consequence in the Debye-H\"uckel asymptotic regime. However, they
should be important if $\beta /l_{D}$ is not particularly small
because then the period is small and the potential wells that trap
$\phi$ are close together and shallow.  There could be a phase
transition in which the imaginary potential $i\phi$ ceases to be well
localized in wells. If as a consequence there are more fluctuations in
the typical $\phi$, then the gas will be organized into more tightly
bound neutral aggregates because when the position $\br $ of a cluster
or particle with charge $e_{\alpha }$ is integrated, the rapidly
varying phase factor $\exp (-ie_{\alpha }\beta^{1/2} \phi(\br ))$ will
give cancellations. \footnote{We thank James Glimm for conversations
on this point.} Note that the Kosterlitz-Thouless phase transition
fits this description.

Tunneling represents a phenomenon which arises because the system
consists of discrete charges as opposed to infinitely divisible charge
distributions, which would give a single well as in the Debye-H\"uckel
approximation.

\newpage \section{Dipoles} \setcounter{page}{1}

\label{chapter- dipole}

The Coulomb plasma is the main focus of this review. However in
chapters~\ref{chapter-4} and \ref{chapter-loops} we will find that a
system of point quantum charges has a close analogy to a certain
system of dipoles and higher order multipoles. The resulting multipole
forces are at the origin of a breakdown of exponential screening in
quantum mechanics. It is therefore of interest to give here a short 
review concerning the lack of screening in classical dipole systems
and estimates on the dielectric constant.

Concerning the analogy an important caveat should be kept in mind. In
a classical system exponential screening is restored as soon as free
charges are added to classical dipoles. At any non-zero temperature,
the quantum gas always has a proportion of free charges due to
ionization processes, but these free \emph{quantum} charges are not
able to restore exponential screening.  The analogies and differences
between the behavior of quantum point charges and classical dipoles
are made more precise in section~\ref{sec-4.6}

\subsection{The Dipole ensemble}
\label{sec- DipoleDescription}

A dipole is specified by the coordinate $\cE = (\br, \bd)$ that unites
the position $\br $ and the dipole moment $\bd$.  We suppose
that in the absence of interaction a single dipole experiences no
preferred directions so that 
\begin{equation}\label{3.1.1}
        \int d\cE \dots  = \int _{\Lambda } d\br \, \int d\Omega \dots
        \mbox{  with  }
        \int d\Omega = 1,
\end{equation}
where $d\Omega$ is surface measure on the sphere of unit vectors 
$\hat{\bd}$ and $\bd = d\hat{\bd}$ with $d$ a fixed dipole moment.

The interaction between a pair of dipoles is
\begin{equation}\label{3.1.2}
        V(\cE _{1}, \cE _{2} ) = (\bd_{1}\cdot\nabla_{1}) \
        (\bd_{2}\cdot\nabla_{2}) \ V(\br _{1} - \br _{2})
\end{equation}
where $\nabla_{i}$ is the gradient operator that acts on $\br _{i}$.
$V(\br _{1} - \br _{2})$  should be the Coulomb potential
between  two unit charges at $\br _{1}, \br _{2}$, but  the
singularity of the Coulomb potential at short distance will lead
to instability for the dipole system.  We choose a length scale $L$
and a form factor $F$ and set 
\begin{equation}\label{3.1.2a}
\tilde{V}(\bk ) = (4\pi )|\bk|^{-2}\tilde{F}^{2}(L^{2}k^{2})
\end{equation}
The form factor satisfies $\int F (\br) d\br = 1$ and means that point
charges are replaced by ``charge clouds.'' This should be regarded as
an effective interaction which arises by integrating out forces on
length scales less than $L$ in a more realistic description of the
short range physics.  An effective action of this form is not
appropriate unless the gas is dilute so that interactions such as hard
cores are not playing a strong role. See for example the use of a
Mayer expansion as in section \ref{sec-debyesphere} which led to such
an effective interaction.

The partition function is   
\begin{equation}\label{3.1.3}
        \Xi_{\Lambda } = \sum _{N} \frac{z^{N}}{N!} 
        \int d\cE _{1} \, \dots \int d\cE _{N} \, 
        e^{-\beta U(\cE _{1},\dots ,\cE _{N})  },
\end{equation}
where the interaction {\it includes} self-energies,
\begin{equation}\label{3.1.4}
        U(\cE _{1},\dots ,\cE _{N}) 
        = \frac{1}{2}\sum _{i,j = 1}^{N} V(\cE _{i}, \cE _{j} ).
\end{equation}

If instead the interaction is $\sum f(\cE _{j})$ where $f$ is an
external field, then we obtain an ideal gas 
partition function 
\begin{eqnarray}\label{3.1.3b}
        \Xi_{{\rm ideal},\Lambda}(f) 
&=& 
        \sum _{N} \frac{z^{N}}{N!} 
        \int d\cE _{1} \, \dots \int d\cE _{N} \, 
        \exp\(
        - \sum _{j=1}^{N}  f(\cE _{j})
        \) \nn \\
&=& 
        \exp \left(z\int d\cE \,e^{-f(\cE )} \right)\nn 
\end{eqnarray}

As in section~\ref{sec-sinegordon} we can reconstruct the interaction
between the dipoles by integrating over an imaginary external field, 
$f(\cE ) = \beta ^{1/2}\bd\cdot\nabla i\phi(\br)$, with respect to the
measure $d\mu _{L}$ discussed in (\ref{2.3.3}):  
\begin{equation}\label{3.1.6}
        \Xi_\Lambda 
=
        \int d\mu _{L} (\phi) \, \Xi_{{\rm ideal},\Lambda}
        (\beta^{1/2}\bd\cdot\nabla i\phi ).
\end{equation}
Formally this is $\int \cD \phi \exp \left(-S(i\phi ) \right)$ where we
define the action $S$ by 
\begin{eqnarray}\label{3.1.7}
        - S(g) 
&=& 
        \frac{1}{2}\int g V^{-1}g
        + z \int d\cE \, \exp (-\beta ^{1/2} \bd\cdot\nabla g)
\nn \\
&=& 
        \frac{1}{2}\int g V^{-1} g
        + z \int d\cE \, \cosh (\beta ^{1/2} \bd\cdot\nabla g)
\end{eqnarray}
where the inverse operator $V^{-1}$ has a kernel whose Fourier transform is
$\tilde{V}^{-1}(\bk )$.

We shall consider the  observable that measures the density of dipoles
with coordinate $\cE $ in a configuration $\omega = (\cE _{1},\dots
,\cE _{N})$, namely
\begin{equation}\label{3.1.5}
        \hat{\rho }(\cE ,\omega ) = \sum _{i=1}^{N} \delta (\cE , \cE _{j}), 
\end{equation}
Distribution functions such as the two point function $\langle
\hat{c}(\cE_{a}) \hat{c}(\cE_{b}) \rangle$ are defined in analogy to
section~\ref{sec- classical}.  Distributions at non-coincident points
of $\hat{c}(\cE ,\omega )$, where $\cE = (\br , \bd )$, become
expectations of $z\exp (-i\beta^{1/2} \bd \cdot\nabla \phi(\br ) )$
in the Sine-Gordon language .

Another useful quantity is
\begin{equation}\label{3.1.5a}
        \langle e^{-\beta ^{1/2}\int i\phi f}\rangle 
=
        \Xi_\Lambda ^{-1}
        \int d\mu _{L} (\phi) \, \Xi_{{\rm ideal},\Lambda}
        (i\beta^{1/2}\bd\cdot\nabla \phi (\br))
        e^{-i\beta ^{1/2}\int \phi f}
\end{equation}
which measures ($\beta \times$)energy of a charge distribution $f$ in
the sea of dipoles. By unraveling the Sine-Gordon transformation one
finds that the charge distribution $f$ has self-energy determined by
the potential $V(\br _{1} - \br _{2})$ corresponding to (\ref{3.1.2a})
and it interacts with system dipoles according to potential energy
$(\bd_{2}\cdot\nabla_{2}) \ V(\br _{1} - \br _{2})$.

We are using the expected value $\langle \ \rangle $ to indicate
either the finite volume or infinite volume limit(s).  This infinite
volume limit is known to exist for several choices of boundary
conditions \cite{FrPa80,FrSp81b,FuSp97}.

We shall also consider expectations of $\phi$.  The two-point function
measures the effective potential between two infinitesimal
charges: as
in (\ref{1.1.eff1}), but in the Sine-Gordon language, $\beta \times
$ the effective potential between two infinitesimal test charges $(\br
_{a}, e_{a})_{a= 1,2}$ is given by
\begin{equation}\label{3.1.8}
e^{- \beta e_{1}e_{2}V^{{\rm eff}}(\br _{1}-\br _{2})}
=
\frac{
\langle 
e^{-ie_{1}\beta^{1/2} \phi(\br _{1})}
e^{-ie_{2}\beta^{1/2} \phi(\br _{2})}
\rangle 
}{
\langle 
e^{-ie_{1}\beta^{1/2}\phi(\br _{1})}
\rangle \langle 
e^{-ie_{2}\beta^{1/2}\phi(\br _{2})}
\rangle 
}
\end{equation}
The numerator removes the self-energies of the test charges.  Since
$e_{1}, e_{2} \rightarrow 0$ we have
\begin{equation}\label{3.1.9}
V^{{\rm eff}}(\br _{1}-\br _{2})
=
\langle\phi(\br _{1})\phi(\br _{2})\rangle
\end{equation}
Furthermore the dielectric constant $\varepsilon$ is obtained from the
large $\br $ asymptotics,
\begin{equation}\label{3.1.10}
V^{{\rm eff}}(\br ) \sim \frac{1}{\varepsilon |\br |}
\end{equation}

\subsection{A no-screening theorem for dipoles}\label{sec- noscreening}

The action $S(g)$ in (\ref{3.1.7}) is invariant under $g \rightarrow
g + c$, where $c$ is any constant.  This is a continuous symmetry
and the action is local. Locality and continuous symmetry generally 
lead to power law correlations because fluctuations, $\phi
\rightarrow \phi + f$, where $f(\br )$ is slowly varying as a function
of $\br $, are very weakly suppressed\footnote{The Coulomb system
possessed a similar symmetry, but with the essential difference that
$c$ could only have discrete values.}.

No-screening theorems for dipoles have been obtained
\cite{Par79}. Subsequently \cite{FrSp81b} gave the proof we
present. This no-screening theorem is a little remarkable because it
is valid for all activities and all temperatures, but its validity at
high density is dependent on the way the short distance singularity in
the dipole force is stabilized. If there are hard cores, then
crystalline phases are likely.  This is proven \cite{FrSp81b} for
dipoles on a lattice. For models where the Coulomb potential is
smoothed out such as our $V_{\infty ,\lambda }$ potential Fr\"ohlich
and Spencer proved further results to the effect that the two point
function has no oscillations, suggesting that this system is always in
a fluid phase.

\begin{theorem}\label{theorem- noscreening} 
For all activities $z \geq 0$ and all inverse temperatures $\beta \geq
0$ there is no screening in the sense that the two point function has
a singularity at $\bk =0$ in its Fourier transform and
\begin{equation}\label{3.2.1}
        \tilde{V}^{{\rm eff}}(\bk ) \geq  \tilde{\cV}(\bk ),
\end{equation}
where
\begin{equation}\label{3.2.2}
        \tilde{\cV}(\bk ) 
=
        \(\tilde{V}^{-1}(k) + \frac{z\beta d^{2}}{3} \bk^{2}\)^{-1}
\end{equation}
is the Fourier transform of the dipole potential with a dielectric
correction. 
\end{theorem}
$\tilde{\cV}(\bk )$ is the Fourier transform of the kernel
$\cV (\br - \br ')$ which is the inverse of the Hessian of $S$ at $\phi =0$.  A
version of this theorem holds in any dimension and also for a large
class of positive-definite interactions in place of dipole-dipole. The
factor 3 is the dimension of space. We will remark in the proof the
main property of the interaction being used. The reader will see that
the argument is a very general mean field theory bound.

From this theorem and discussion of the dielectric constant in
section~\ref{sec- classical} we immediately obtain a mean field theory
bound for the dielectric constant, assuming there is a dielectric
constant: 
\begin{equation}\label{3.2.3}
        \varepsilon 
\leq
        1 + \frac{z\beta d^{2}}{3} 4\pi 
\end{equation}

There is a large literature on the molecular theory of the dielectric
constant, with atoms and molecules modeled by preformed classical
permanent dipoles, using various short range regularizations.  See,
for example, \cite{StPaHo81} and references therein.  In particular, the
Clausius-Mosotti law $(\varepsilon -1) (\varepsilon +2)^{-1} =
4\pi\rho \beta d^{2}/9$ was derived in \cite{HoSt74,HoSt76} by taking
the Kac limit, in which $V(\br)$ is replaced by $\gamma ^{3}V(\gamma
\br)$ and $\gamma \rightarrow \infty$. However, this law is
contradicted by the bound (\ref{3.2.3}) if one just sets $z = \rho$ at
low density.  The reason is that the result of the Kac limit depends
on the treatment of dipoles in the vicinity of the test charge and the
scaling of test charges. The literature would benefit from
clarification of this point. See the further comment at the end of
this section.

Theorem~\ref{theorem- noscreening} is a consequence of
\begin{theorem}\label{theorem-DHupperbound} (The mean field upper
bound) 
\begin{equation}\label{3.2.4}
        \langle e^{-\int i\phi f}\rangle \
\leq 
        \exp \left( \inf_{g} - S(g,f) \right)
\end{equation}
where
\begin{equation}\label{3.2.5}
        - S(g,f)
=
        \frac{1}{2}\int g V^{-1}g - \int gf
        + z \int d\cE \, \( \cosh (\beta ^{1/2} \bd\cdot\nabla g) - 1\) 
\end{equation}
\end{theorem}

\proof We make a complex translation $\phi \rightarrow \phi - ig$.
This is done using the translation formula (\ref{2.3.5}), but we
arrive at the same place if we do it in the intuitive formula $\int
\cD \phi \, \exp(-S(i\phi) -\int i\phi f)$, which is the numerator of
$\langle e^{-\int i\phi f}\rangle$.  On expanding out the exponent we
obtain an upper bound on $|\exp (\dots )|$ by dropping all the
imaginary terms.  Thus
\begin{eqnarray}\label{3.2.6}
&&\hspace{-.3in}
        -{\rm Re } S(i\phi + g) - {\rm Re } \ \int (i\phi+g)f \nn \\
&=&
        {\rm Re } \frac{1}{2}\int (i\phi+g) V^{-1}(i\phi+g)
        - \int gf + z {\rm Re } \int d\cE \, 
        \cosh (\beta ^{1/2} \bd\cdot\nabla (i\phi+g))\nn \\
&=&
        -\frac{1}{2}\int \phi V^{-1}\phi +
        \frac{1}{2}\int g V^{-1}g\nn - \int gf
        + z \int d\cE \, \cos (\beta ^{1/2} \bd\cdot\nabla \phi)
        \cosh (\beta ^{1/2} \bd\cdot\nabla g)\nn \\
&\leq&
        -S(i\phi) - S(g,f) 
\end{eqnarray}
so that in the resulting bound on $\langle \exp(-\int i\phi f)\rangle$
the $\int \cD \phi \exp \left(-S(i\phi ) \right) $ cancels in numerator
and denominator and we obtain the mean field upper bound, since $g$ is
arbitrary.\qed

\textbf{Proof of Theorem ~\ref{theorem- noscreening}.} We rewrite the
mean field upper bound as
\begin{equation}\label{3.2.7}
        1 - \langle e^{-i\alpha \int \phi f}\rangle \ 
\geq 
        1 - \exp \left( - S(\alpha g,\alpha f) \right)
\end{equation}
which holds for any function $g$. Let $\alpha \rightarrow 0$ to
obtain
\begin{equation}\label{3.2.8}
        \frac{1}{2} \langle(\int \phi f)^{2}\rangle \
\geq 
        -\frac{1}{2}\int g V^{-1}g + \int gf
        - \frac{1}{2} z \int d\cE \, \( \beta ^{1/2} \bd\cdot\nabla g\)^{2} 
\end{equation}
This is valid provided $g$ has enough decay so that the terms of
higher order than $\alpha ^{2}$ are convergent integrals.  {\it This
and positive-definiteness are the only properties of the dipole
interaction used in the proof of (\ref{3.2.1})}.

Now choose  $g = \cV f$. The result is 
\begin{equation}\label{3.2.9}
        \langle(\int \phi f)^{2}\rangle \
\geq
         \int f \cV f.
\end{equation}
This is the same as 
\begin{equation}\label{3.2.10}
        \int f V ^{{\rm eff}} f \geq \int f \cV f
\end{equation}
so that 
\begin{equation}\label{3.2.11}
        \int \tilde{V}^{{\rm eff}}(\bk )|\tilde{f}(\bk )|^{2} d\bk 
\geq 
        \int \tilde{\cV}(\bk )|\tilde{f}(\bk )|^{2} d\bk 
\end{equation}
which, being true for a large class of $f$, is the result we claimed. 
\qed 

A Kac limit can be defined by replacing $V(\br)$ by $\gamma
^{-1}V(\gamma \br)$ and letting $\gamma \rightarrow 0$, which is the
same as taking $L \rightarrow \infty $. By scaling one can show that
it is equivalent to replacing $S(i\phi)$ by $\gamma ^{-3} S(i\phi)$ in
the Sine-Gordon transformation (\ref{3.1.6},\ref{3.1.7}).  This
selects the stationary point of the action so that bounds such as
Theorem~\ref{theorem- noscreening} are saturated in this limit,
\emph{provided} the external charge density $f$ is also scaled
suitably.

\subsection{The scaling limit of the lattice dipole gas}\label{subsec-
scalinglimit}

The results of the last section show absence of screening but one
would like to know that the system is in a phase characterized by
correlations that are asymptotic to canonical power laws prefaced by a
dielectric constant or tensor. In this section we will describe
preliminary attempts to delineate when the system is actually in a
dielectric phase in this more detailed sense.

Unfortunately the most appealing result is established only for
lattice systems, but the removal of the restriction to lattice models
is a feasible mathematical problem.

In a lattice model each site $\br$ in a simple cubic lattice in a $\nu
\geq 1$ dimensional box $\Lambda $ can be occupied by zero, one or
more dipoles, each of which can only point in a lattice direction.
The Sine-Gordon transformation of such systems leads to a partition
function of the form
\begin{equation}\label{3.2.12}
        \Xi_{\Lambda } = \int \cD \phi \, e^{-S(\phi)}
\end{equation}
where $\cD \phi = \prod _{\br \in \lat } d\phi(\br )$ is a finite
dimensional integration and the action $S$ has the general form
\begin{equation}\label{3.2.13}
        S(\phi)
=
        \sum _{<\br \br '> }
        W(\phi(\br ) - \phi(\br ') ) .
\end{equation}
where $<\br \br '> = <\br ' \br>$ is a nearest neighbor pair of sites.
For example, a dipole gas with no interactions other than the dipole 
interaction with lattice regularization is given by
\begin{equation}\label{3.2.14}
        W(\phi(\br ) - \phi(\br ') )
=
        \frac{1}{8\pi}(\phi(\br ) - \phi(\br '))^{2} 
        - \frac{z}{\nu} \cos \(d \beta ^{1/2} 
        (\phi(\br') -   \phi(\br ) ) \).
\end{equation}
The first term is the finite difference version of $(\nabla \phi)^{2}$
and the second term corresponds to the $\cosh(i \beta ^{1/2}\bd \cdot
\nabla \phi )$ we have seen before.  If there is a hard core
interaction preventing more than one dipole per lattice site then
$(z/\nu) \cos$ is replaced by $\ln (1 + (z/\nu) \cos )$.
Some of the results we are about to state were derived assuming
periodic boundary conditions on the lattice.

This partition function is also called the {\it anharmonic bedspring},
in which case the variables $\phi$ are considered to be displacements
of the nodes of a bedspring (in a distinguished direction).

General results \cite{NaSp97} have been obtained under the assumption
that $S$ is convex and is a local (or nearly local) function of the
gradient of $\phi$, namely assume there is a positive number
$\delta >0$ independent of $\phi $ such that the matrix
\begin{equation}\label{3.2.15}
        S'' (\phi; \br ,\br ')
=
        \frac{\partial ^{2}S}
        {\partial \phi(\br )\partial \phi(\br ')} 
\end{equation}
of second derivatives obeys
\begin{equation}\label{3.2.16}
        \sum _{<\br \br '>}
        \zeta (\br ) S'' (\phi; \br ,\br ') \zeta (\br ')
        \geq 
        \delta \sum _{<\br \br '> }
        \(\zeta (\br ) - \zeta (\br ')\)^{2} 
\end{equation}
for all $\zeta $. The finite difference Laplacian $\Delta $ associated
to the lattice is a matrix defined by
\begin{equation}\label{3.2.17}
        \sum_{<\br \br '>} \(\zeta (\br ) - \zeta (\br ')\)^{2}
=
        \sum_{<\br \br '>} \zeta (\br ) (-\Delta) (\br ,\br ')
        \zeta (\br '). 
\end{equation}
Thus this convexity assumption is, by definition, the matrix
inequality  
\begin{equation}\label{3.2.convexity}
        {\rm Convexity:} \ \ 
        S''(\phi) \geq  \delta (-\Delta)
\end{equation}
This type of assumption has a natural continuum analogue: a lower
bound on the second variation of $S$ by $\delta \int (\nabla \phi
)^{2}$ or perhaps a quadratic form in $-\Delta + L^{2} \Delta ^{2}$,
but the next assumption brings in the lattice in a strong way.

The assumption is that there is a constant $C$ such that
\begin{equation}\label{3.2.upper}
        \sum _{<\br \br '> }
        \zeta (\br ) S'' (\phi; \br ,\br ') \zeta (\br ')
        \leq 
        C \sum _{<\br \br '> }
        \zeta (\br )^{2}.
\end{equation}
This is not as natural as the convexity assumption and is false for a
continuum model whose action contains $\delta \int (\nabla \phi
)^{2}$ or higher order derivatives.

Now we come to the {\it scaling limit} which concerns expectations of
functions of fields averaged over a length scale $\ell$ that is taken
to infinity.  For any smooth compactly supported function $f$ with
$\sum f(\br ) = 0$ define
\begin{equation}\label{3.2.20}
        \phi(f_{\ell}) = \sum _{\br } \phi(\br )
        \ell^{-\nu/2+1}f(\frac{\br }{\ell}).
\end{equation}
The $\sum f(\br ) = 0$ means that $\phi(f_{\ell})$ is really an
integral of $\nabla \phi $ against a function of compact support. 

The theorem will say that when $\ell$ is large $\phi(f_{\ell})$
becomes Gaussian.  Gaussian random variables are characterized by
their covariance, which, in the theorem, will be the continuum Green's
function $C(\br , \br ')$ solving
\begin{equation}\label{3.2.21}
        - \sum _{i,j=1,\dots ,d} \varepsilon _{ij}
        \partial _{i} \partial _{j}
        C(\br , \br ') = \delta (\br - \br '),
\end{equation}
where $\partial _{i} = \partial /\partial r_{i}$ and $\varepsilon
_{ij}$ will be the dielectric tensor.

For the theorem we need to define the Gaussian measure 
\begin{equation}\label{3.2.22}
        d\mu _{C}(\phi) = \frac{1}{Z}
        \cD \phi\exp \left(-\frac{1}{2} \int 
        \sum_{i,j} \varepsilon_{ij} \partial _{i}\phi(\br ) 
        \partial_{j}\phi(\br ) d\br 
        \right)
\end{equation}
Such a Gaussian measure exists (essentially the massless free field )
and is characterized by Wick's theorem which says

\begin{theorem}\label{thm-wick}
Define the functional Laplacian
\begin{equation}\label{3.2.23}
        \DDelta = \int d\br \, \int d\br ' \,
        \frac{\delta}{\delta \phi(\br )}
        C(\br , \br ')
        \frac{\delta}{\delta \phi(\br ')}
\end{equation} 
Let $P$ be a polynomial in fields $\phi(\br _{1}),\dots ,\phi(\br
_{n})$, then 
\begin{equation}\label{3.2.24}
        \int d\mu _{C}(\phi)\, P(\phi) 
=
        \exp \left( \frac{1}{2} \DDelta 
        \right)
        P(\phi) \arrowvert_{\phi=0}
\end{equation}
where $\exp \left( \frac{1}{2} \DDelta \right)$ is a power series $I +
\frac{1}{2} \DDelta + \cdots $ which terminates after a finite number
of terms when applied to a polynomial $P$.
\end{theorem}
An immediate consequence is $\int d\mu _{C}(\phi) \phi(\br)\phi(\br ')
= C(\br , \br ')$ which is why $C$ is called a covariance. The theorem
\cite{NaSp97} is

\begin{theorem}\label{theorem- naddaf-spencer}
Under the convexity assumption (\ref{3.2.convexity}) and the uniform
upper bound (\ref{3.2.upper}) the continuum scaling limit of the
anharmonic bedspring is a continuum massless free field.  In other
words there exists a constant positive-definite dielectric tensor
$\varepsilon _{ij}$ such that, for any $f^{1},\dots ,f^{n}$, as $\ell
\rightarrow \infty $,
\begin{equation}\label{3.2.25}
        \langle\prod_{i} \phi(f_{\ell}^{i}) \rangle
\rightarrow
        \int d\mu  _{C} \,  \prod_{i} \phi(f^{i})
\end{equation}
\end{theorem}

Thus, in dimension $\nu =3$, the lattice dipole gas (\ref{3.2.14}) is
in a dipolar phase characterized by a dielectric constant at least
when $\frac{1}{3}z\beta d^{2}4\pi <1$ because this is the parameter
range for which (\ref{3.2.14}) is convex.

In mathematical terms this is a new type of central limit
theorem. Central limit theorems are usually a statement about sums of
independent random variables. In the statistical mechanics of lattice
spin systems in the high temperature phase the spins are somewhat
independent and one expects a central limit theorem that the sum of
the spins in a large block of side $\ell$ normalized by $\ell ^{-\nu
/2}$ is almost Gaussian and independent of similar sums over disjoint
blocks. A beautiful theorem of this type has been proved
\cite{New80,NeWr81,NeWr82}. The most gentle way to encode high
temperature ( approximate independence) is to require only a finite
susceptibility which remarkably is the assumption made by Newman,
together with a type of ferromagnetism.  In the case of the anharmonic
bedspring the susceptibility is not finite; correlations decay with
non-integrable power laws and the variables $\phi $ and their
gradients are very far from independent. Now we no longer have a
standard central limit theorem; indeed the right normalization for a
block is now $\ell^{-\nu/2+1}$ but nevertheless these block sums
become Gaussian.

From the point of view of the renormalization group the central limit
theorem is the case of a high temperature fixed point obtained by
scaling keeping the mass fixed so that the renormalization group
dynamics drives the system to a white noise Gaussian fixed
point. Theorem~\ref{theorem- naddaf-spencer} instead describes a
massless Gaussian fixed point.  Notice that the convexity hypothesis
permits interactions that are not small.

The proof sharpens the estimates of \cite{BrLi75,BrLi76}, in
particular the Brascamp-Lieb bounds:

\begin{theorem}\label{theorem- brascamp-lieb}
Let $F = F(\phi)$ be any continuously differentiable function of
fields. Under the convexity assumption (\ref{3.2.convexity}),
\begin{equation}\label{3.2.26}
        \langle F^{2} \rangle - \langle F \rangle^{2}
\leq
        \sum _{\br ,\br '} 
        \langle 
        \frac{\partial F}{\partial \phi(\br )}
        S''^{-1}(\phi; \br ,\br ')
        \frac{\partial F}{\partial \phi(\br' )} 
        \rangle 
\end{equation}
and
\begin{equation}\label{3.2.27}
        \langle e^{\phi(f)} \rangle
\leq
        \exp \left( \frac{1}{2}\sum _{\br ,\br '}
        f(\br ) (-\delta \Delta )^{-1}(\br ,\br ') f(\br ')
        \right)
\end{equation}
\end{theorem}

It is a corollary of theorem~\ref{theorem- naddaf-spencer} that all
reasonable functions of fields $\phi(f_{\ell})$ converge to massless
free field expectations.  Unfortunately the standard observables are
not included in this class, being local functions of the unaveraged
field $\phi (\br )$. For example the dipole density is   
\begin{equation}\label{3.2.28}
        \hat{\rho }(\cE;\phi) 
=
        ze^{ -i\beta^{1/2} \bd \cdot \nabla \phi(\br ) }
\end{equation}
Theorem~\ref{theorem- naddaf-spencer} is more or less equivalent to
\begin{equation}\label{3.2.29}
        \langle\prod \alpha ^{\nu/2 - 1}\phi(\alpha \br _{i}) \rangle
\longrightarrow 
        \int d\mu _{C} (\phi ) \prod \phi(\br _{i})  \ \mbox{as }
\alpha \rightarrow \infty 
\end{equation}
but to get at the density observables we need information on
asymptotics when some but not all of the points $\br _{1},\dots ,\br
_{n}$ are driven apart by a scaling. Theorem~\ref{theorem-
brascamp-lieb} provides an upper bound but lower bounds are not yet
proven. There is no physical reason to doubt that they hold, but we
are interested in a proof because if the Naddaf-Spencer argument could
be extended to prove that more general functionals of $\phi $ have
power law decay then a complete proof that there is {\it no}
exponential screening in quantum Coulomb systems is within reach using
the strategy in \cite{BrKe94}. An alternative route was started by
\cite{BrKe94b} but it is presently unreasonably complicated.

One further remark about dipole systems is that the pressure and
dipole distributions are analytic in the activity in a small
neighborhood of zero \cite{GaKu83,BrYa90}.  Thus the Mayer expansion
is convergent at small activities. Both proofs are complicated and
despite many efforts, no one has yet obtained the result by direct
attack on the size of the Mayer coefficients.

\def\({\left(}
\def\){\right)}
\def\br{{\bf r}}
\def\cf{{\cal F}}
\def\b\xi{\mbox{\boldmath $\xi$\unboldmath}}
\def\umu{\underline{\mbox{\boldmath $\mu$\unboldmath}}}
\def\uN{\underline{\bf N}}
\newcommand{\dM}{d^{[0,1]}\mu}

\newpage \section{Semi-classical Coulomb gas} \setcounter{page}{1}

\label{chapter- Semi-classical}\label{chapter-4}

\subsection{The Feynman-Kac representation}
\label{sec-4.1 }

In this chapter, we give an introduction to some of the main effects
produced by quantum mechanics in the Coulomb gas. These effects are
most easily seen by perturbing around the classical gas,
i.e. neglecting the quantum statistics and keeping the first relevant
contributions to an expansion in powers of the Planck constant $\hbar$
(the Wigner-Kirkwood expansion)\footnote{Note that to have stability
of the classical reference system, the semi-classical analysis
requires a regularization of the Coulomb potential at the origin,
except for the one component plasma which remains well behaved in the
classical limit.}. We have no control on the possible convergence or
asymptotic character of these $\hbar$-series, but the examination of
the lowest order terms gives immediately a prediction to the main
issue: Debye screening cannot survive if the quantum mechanical nature
of the charges is taken into account. This serves as a gentle training
for the reading of chapter~\ref{chapter-loops} where the fully quantum
mechanical gas is considered, since most of the arguments and of the
mathematical structure will be similar.

The formalism that is best adapted for both a semi-classical and a low
density analysis of the quantum Coulomb gas is the Feynman-Kac
functional integral representation.  This representation is of course
not new and goes back to \cite{Gin65,Gin71} with his study of the low
activity expansion of quantum fluids with short range forces.  But,
for quantum charges, the formalism shows its full capability and
becomes particularly operational to deal with the long range of the
Coulomb potential.

In section \ref{sec-4.3} we discuss the simpler situation of only two
quantum charges immersed in a classical plasma. The model can be
studied without recourse to $\hbar$ expansions and illustrates clearly
how quantum fluctuations will irremediably destroy Debye screening:
one finds an $r^{-6}$ decay of the correlation between the two quantum
charges. In section \ref{sec-4.4} we comment on the relation between
this type of correlation between individual charges and the
conventional van der Waals potential. Section \ref{sec-4.5} comes back
to a system of infinitely many quantum charges and summarizes the
results obtained by the semi-classical analysis.  In the last section,
we combine the Sine-Gordon and the Feynman-Kac functional integrals:
this gives an alternative view on the lack of exponential
screening. Again a simplified model that captures the essence of
quantum fluctuations illustrates how the occurrence of dipole (and
higher multipole) fields will spoil the proof of Debye screening. In
contrast, mixtures of classical particles that carry both charges and
dipoles still show strong screening properties.

We start here and in section \ref{sec-4.2} by recalling familiar ideas
about the representation of the quantum mechanical statistical
operator by the Feynman-Kac functional integral.  For the sake of
simplicity, we consider first a single particle of mass $m$ in three
dimensions moving in an external potential $V({\bf r})$. According to
the original path integration \cite{FeHi65}, the configurational
matrix elements of the statistical operator associated with the
one-particle Hamiltonian \begin{equation} H=-\frac{\hbar^2}{2m}\Delta
+V,\;\;\;\;\,\,\Delta=\mbox{Laplacian in}\; {\Bbb R}^3 \label{4.1}
\end{equation}
read
\begin{equation}
\langle{\bf r}_1 |\exp\left(-\beta\left(-\frac{\hbar^2}{2m}\Delta+V\right)\right)
|{\bf r}_2\rangle= 
\sum_{paths}\exp\left(-\frac{1}{\hbar}S({\bf r}(\cdot))\right)
\label{4.2}
\end{equation}
In (\ref{4.2}) $S({\bf r}(\cdot))$ is the classical action corresponding to the potential $-V$
(the Euclidean action)
\begin{equation}
S({\bf r}(\cdot))=\int_0^{\beta \hbar}dt\left(\frac{m}{2}\left |\frac{d{\bf r}(t)}{dt}\right |^2
+V({\bf r}(t))\right)
\label{4.3}
\end{equation}
associated with the path ${\bf r}(t)$ starting from ${\bf r}_1$ at "time" $t=0$ and ending in $
{\bf r}_2$ at "time" $\beta\hbar$. The summation in (\ref{4.2}) runs over all such paths.

It is very useful to parameterize the path ${\bf r}(t)$ by dimensionless variables, making the change
\begin{eqnarray}
s&=&\frac{t}{\beta\hbar}\;,\;\;\;\;\;\;\;\;\;\;\;\;\;\;\;\;\;\;\;\;0\leq\:s\:\leq 1  \nonumber\\
{\bf r}_{1,2}(s)&=&(1-s){\bf r}_1 +s{\bf r}_2 +\lambda
\b\xi(s)\;,\;\;\;\;\;\;\;\b\xi(0)=\b\xi(1)=0    
\label{4.4}
\end{eqnarray}
where
\begin{equation}
\lambda=\hbar\sqrt{\frac{\beta}{m}}
\label{4.5}
\end{equation}
is the de Broglie thermal wave length. In (\ref{4.4}) $\b\xi(s)$
represents now a closed path, called the Brownian bridge, starting and
returning to the origin within the "time" $s=1$. In terms of these
dimensionless variables, the matrix element (\ref{4.2}) can be written
in the precise form \cite{Sim79,Roe94}
\begin{eqnarray} 
\langle{\bf
r}_1\mid\exp\left(-\beta\left(-\frac{\hbar^2}{2m}\Delta+V\right)\right)\mid{\bf
r}_2\rangle &=&
\left(\frac{1}{2\pi\lambda^2}\right)^{3/2}\exp\left(-\frac{|{\bf r}_1
-{\bf
r}_2|^2}{2\lambda^2}\right) \nonumber \\
&& \hspace{-2.25in} \times\int D(\b\xi)\exp\left(-\beta\int_{0}^{1}
ds V((1-s){\bf r}_1 +s{\bf r}_2 +\lambda \b\xi(s))\right) \label{4.6}
\end{eqnarray}
In (\ref{4.6}) $ D(\b\xi)$ is a Gaussian measure (the Brownian bridge measure) with weight
formally represented (up to normalization) by
$\exp\left(-\frac{1}{2}\int_0^1\left|\frac{d\b\xi(s)}{ds}\right|^2\right)$. It has zero mean and
is entirely defined by its covariance 
\begin{equation}
\int D(\b\xi)\xi_{\mu}(s_1)\xi_{\nu}(s_2)=\delta_{\mu,\nu}(\min(s_1,\:s_2)-s_1 s_2)
\label{4.7}
\end{equation}
where $\xi_{\mu}(s)$ are the Cartesian coordinates of $\b\xi(s)$.

The representation (\ref{4.6}) has several advantages: physical parameters
${\bf r}_1,\;{\bf r}_2,\;\beta,\;\hbar,\;\lambda$ appear explicitly in the formula. Moreover the
diagonal matrix elements
\begin{eqnarray} 
&& \hspace{-.5in}
\langle{\bf r}\mid\exp\left(-\beta\left(-\frac{\hbar^2}{2m}\Delta+V\right)\right)\mid{\bf
r}\rangle=\nonumber\\ 
&& \left(\frac{1}{2\pi\lambda^2}\right)^{3/2}\int 
D(\b\xi)\exp\left(-\beta\int_0^1dsV({\bf r} +\lambda \b\xi(s))\right)
\label{4.8}
\end{eqnarray}
have the form of an integral over a certain $\b\xi$-dependent
Boltzmann factor $\exp(-\beta V({\bf r},\b\xi))$ with $V({\bf
r},\b\xi)=\int_0^1dsV({\bf r} +\lambda\b\xi(s))$, as in classical
statistical mechanics. In this view, one can think of a quantum point
particle as a classical extended object, the closed filament
$\b\xi(s)$ located at ${\bf r}$. This closed filament, which plays
formally the same role as a classical internal degree of freedom,
represents the intrinsic quantum fluctuation with size controlled by
the de Broglie wave length $\lambda$. The statistical weight of a
filament having extension $R$ behaves as $\exp(-R^2/\lambda^2)$; thus,
in the semi-classical regime ($\lambda$ small), only short filaments
contribute to the integral (\ref{4.8}).  Obviously one recovers the
genuine classical Boltzmann factor $\exp(-\beta V({\bf r}))$ in
(\ref{4.8}) as $\lambda \rightarrow 0$. For these reasons, the
Feynman-Kac representation is particularly suitable to study the
semi-classical aspects of equilibrium quantities.

\subsection{The gas of charged filaments}
\label{sec-4.2}

The extension of the representation (\ref{4.8}) to the many particle system is straightforward. 
In the sequel, $V({\bf r}_{1}-{\bf r}_{2})$ represents a two-body regularized Coulomb potential,
for instance of the form (\ref{1.1.3bc})  
. To each particle labeled by $i,\; i=1,\ldots, N$,  we associate an independent Brownian bridge
$\b\xi_i(s)$, distributed with the Gaussian measure $D(\b\xi_i)$, and the diagonal elements
of the $N$- particles statistical operator read
\begin{eqnarray} 
\langle\{{\bf r}_i\}\mid\exp\(-\beta
H_{\Lambda,N}\)\mid\{{\bf
r}_i\}\rangle=\prod_{i=1}^{N}\frac{1}{(2\pi\lambda_{\alpha_{i}}^2)^{3/2}}
\int\prod_{i=1}^N D(\b\xi_i) \nonumber \\ \exp\(-\frac{\beta}{2}\sum_{i\neq j}^N
e_{\alpha_i}e_{\alpha_j}\int _0^1 ds V({\bf r}_i + \lambda_{\alpha_i}\b\xi_i(s)-{\bf r}_j
- \lambda_{\alpha_j}\b\xi_j(s))\) \label{4.9}
\end{eqnarray}
The Dirichlet boundary conditions on $H_{N,\Lambda}$ are implemented by the constraint that all paths
in the integrand of (\ref{4.9}) have to remain inside $\Lambda$ for all $s,\; 0\leq\; s \; \leq 1$. We
do not write this constraint explicitly since it will be trivially removed in the infinite volume
limit. The grand-canonical Maxwell-Boltzmann partition function is defined by keeping in
(\ref{1.2.5}) only the term corresponding to the trivial permutation. Denoting it
still by $\Xi_{\Lambda}$ in this chapter, it is given by the sum
\begin{eqnarray} 
\Xi_\Lambda&=&\sum_{\{N_\alpha\}}^\infty \prod
\frac{((2\sigma_\alpha+1)\exp(\beta\mu_{\alpha}))^{N_\alpha}}{N_\alpha !}\nonumber\\&\times&
\int_\Lambda d{\bf r}_1
\ldots d{\bf r}_N \langle\{{\bf r}_i\}\mid\exp\(-\beta
H_{\Lambda,N}\)\mid\{{\bf r}_i\}\rangle
\label{4.11}
\end{eqnarray}
The factor $2\sigma_\alpha +1$ is the spin degeneracy, since the
Hamiltonian (\ref{1.2.1}) is independent of the spin.

 A look at the Boltzmann-like factor in (\ref{4.9}) suggests naturally
the introduction of an enlarged classical phase space whose elements
${\cal F} $ are
\begin{equation}
{\cal F} = (\alpha,\;{\bf r},\;\b\xi)
\label{4.12}
\end{equation}
where $\alpha$ is the species and $\b\xi$ is the filament shape of a
particle at position ${\bf r}$. We shall call a point ${\cal F}$ in
this phase space a "filament". Since the Hamiltonian is invariant
under the permutations of the particles of the same species, we can
use again the summation identity (\ref{1.1.4}) to write the partition
function (\ref{4.11}) in the form
\begin{equation}
\Xi_\Lambda = \sum_{N=0}^\infty\frac{1}{N!}\int \prod_{k=1}^N d{\cal F}_k z({\cal F}_k)\exp\(-\beta
U({\cal F}_1, \ldots, {\cal F}_N)\) 
\label{4.14}
\end{equation}
where the phase space integration means
\begin{equation}
\int d{\cal F}\cdots=\sum_{\alpha=1}^{\cal S}\int_{\Lambda}d{\bf r}\int D(\b\xi)\cdots
\label{4.15}
\end{equation}
In (\ref{4.14}) we have set
\begin{equation}
z({\cal F})=(2\sigma_\alpha+1)z_\alpha,\;\;\;\;z_{\alpha}=\frac{\exp(\beta\mu_{\alpha})}  
{\(2\pi\lambda_{\alpha}^2\)^{3/2}} 
\label{4.16} 
\end{equation}
\begin{equation}
U({\cal F}_1,\ldots, {\cal F}_N)=\sum_{1=i<j}^N e_{\alpha_i} e_{\alpha_j}V({\cal F}_i,{\cal 
 F}_j) \label{4.17}
\end{equation}
with
\begin{equation}
V({\cal F}_1,{\cal F}_2)=\int_0^1dsV({\bf r}_1+\lambda_{\alpha_1}\b\xi_1(s)-{\bf r}_2-\lambda_{\alpha_2}\b\xi_2(s))
\label{4.18}
\end{equation}
It is clear that the system defined by the relations (\ref{4.14})-(\ref{4.18}) has the structure of a
classical statistical mechanical system of objects having internal degrees of freedom $\alpha,\;
\b\xi$ (the species and the shape of a filament), with activity (\ref{4.16}) and two-body interaction 
(\ref{4.18}). We call it "the system of filaments" and can apply to it all the standard methods of
classical statistical mechanics.

In particular, one can define correlations between filaments in the
usual way (see (\ref{1.1.8a}) and (\ref{1.1.9a})). Introducing the
"$\delta$-function" which identifies the points ${\cal F}_1$ and
${\cal F}_2$ in phase space, i.e. \begin{equation} \int d{\cal
F}_2\delta({\cal F}_1,{\cal F}_2)F({\cal F}_2)=F({\cal
F}_1),\;\;\;\;\;\delta({\cal F}_1,{\cal F}_2)=
\delta_{\alpha_1,\alpha_2}\delta(\b\xi_1,\b\xi_2)\delta({\bf r}_1-{\bf
r}_2) \label{4.19}
\end{equation}
the density of filaments and the two filament distribution are 
\begin{equation}
\rho({\cal F})=\langle\hat{\rho}({\cal F})\rangle,\;\;\;\;\rho({\cal F}_a,{\cal
F}_b)=\langle\hat{\rho}({\cal F}_a)\hat{\rho} ({\cal F}_b)\rangle
\label{4.20}
\end{equation}
where 
\begin{equation}
\hat{\rho}({\cal F})=\sum_i\delta({\cal F}_i,{\cal F})
\label{4.21}
\end{equation}
and $\langle\cdots\rangle$ denotes the grand-canonical average in the system of filaments
(the contribution of coincident filaments, included in (\ref{4.20}), will be specified later if
needed). However, one should remember that the shapes of the filaments are not directly observable
quantities: the physical particle distributions are obtained by integrating them out in
(\ref{4.20}) and (\ref{4.21}). Thus the density of particles of type $\alpha$ is
\begin{equation} \rho(\alpha,{\bf r})=\int D(\b\xi)\rho({\cal F})
\label{4.22}
\end{equation}
the two particle distribution is
\begin{equation}
\rho(\alpha_a,{\bf r}_a,\alpha_b,{\bf r}_b)=\int D(\b\xi_a)D(\b\xi_b)\rho({\cal F}_a,{\cal F}_b) 
\label{4.23}
\end{equation}
and in the same way for the truncated distributions,
\begin{eqnarray}
\rho_T({\cal F}_a,{\cal F}_b)&=&\rho({\cal F}_a,{{\cal 
F}}_b)-\rho({\cal F}_a)\rho({\cal F}_b) \nonumber\\
\rho_T(\alpha_a,{\bf r}_a,\alpha_b,{\bf r}_b)&=&
\int D(\b\xi_a)D(\b\xi_b)\rho_T({\cal F}_a,{{\cal 
F}}_b) 
\label{4.24}
\end{eqnarray}

At this point, one has to make a fundamental observation. Although the
system of filaments can be treated with the rules of classical
statistical mechanics, it differs from a genuine assembly of classical
charged random wires (distributed according to the Gaussian measure
$D(\b\xi)$ ) on an important point: {\em the interaction (\ref{4.18})
inherited from the quantum mechanical nature of the charges is not the
standard electrostatic potential between two closed wires of shapes
$\b\xi_1$ and $\b\xi_2$}. The latter potential, denoted $V_{{\rm
cl}}({\cal F}_1,{\cal F}_2)$, is \begin{equation} V_{{\rm cl}}({\cal
F}_1,{\cal F}_2)=\int_0^1 ds_1\int_0^1 ds_2 V({\bf
r}_1+\lambda_{\alpha_1}\b\xi_1(s_1)-{\bf
r}_2-\lambda_{\alpha_2}\b\xi_2(s_2)) \label{4.25}
\end{equation}
since every element of charge $\b\xi_1(s_1)ds_1$ carried by the first
filament has to interact with every element $\b\xi_2(s_2)ds_2$ of the
other by the Coulomb law, and the corresponding total electrostatic
energy is
\begin{equation}
U_{{\rm cl}}({\cal F}_1,\ldots, {\cal F}_k))=\sum_{i<j=1}^k e_{\alpha_i}
e_{\alpha_j}V_{{\rm cl}}({\cal F}_i,{\cal F}_j) \label{4.26}
\end{equation}
\emph{Although the filaments are quantum mechanical in origin, we will
use this subscript $_{{\rm cl}}$ to denote the filament system with
interaction $U_{{\rm cl}}$}.  Replacing the interaction (\ref{4.18})
by (\ref{4.25}) would indeed lead to a genuine system of charged
filaments obeying exponential Debye screening in the appropriate low
density or high temperature regime, according to the theorems quoted
in chapter~\ref{chapter- debye-screening} and the discussion in
section~\ref{sec-4.6}. Hence different behaviors between quantum and
classical charges in thermal equilibrium must be traced back to the
difference between the interactions (\ref{4.18}) and (\ref{4.25})
\begin{eqnarray}
W({\cal F}_1,{\cal F}_2) &=& e_{\alpha_1} e_{\alpha_2}(V({\cal F}_1,{{\cal 
F}}_2)-V_{{\rm cl}}({\cal F}_1,{\cal F}_2)  ) \nonumber\\
&=& e_{\alpha_1} e_{\alpha_2}\int_0^1 ds_1\int_0^1
ds_2(\delta(s_1 - s_2)-1) \nonumber \\
&&\hspace{-.5in}\times V({\bf r}_1+\lambda_{\alpha_1}\b\xi_1(s_1)-{\bf r}_2-\lambda_{\alpha_2} 
\b\xi_2(s_2))
\label{4.27}
\end{eqnarray}
The large distance behavior of $W({\cal F}_1,{\cal F}_2)$ is obtained from the multipolar expansion of the
potential $V$ written in the form
\begin{eqnarray}
V({\bf r}_1+\lambda_{\alpha_1}\b\xi_1(s_1)-{\bf r}_2-\lambda_{\alpha_2}\b\xi_2(s_2))&=&
V({\bf r}_1-{\bf r}_2)+M_1(s_1)V({\bf r}_1-{\bf r}_2)\nonumber\\
&&\hspace{-2in}
+M_2(s_2)V({\bf r}_1-{\bf r}_2) + M_1(s_1)M_2(s_2)V({\bf r}_1-{\bf r}_2) 
\label{4.28}
\end{eqnarray}
where $M_{i}(s)$ are the multipolar differential operators
\begin{equation}
M_{i}(s)=\sum_{k=1}^{\infty}\frac{\(\lambda_{\alpha_{i}}\b\xi_{i}(s)\cdot
\nabla_{{\bf r}_{i}}\)^k}{k!},\;\;\;\;\;i=1,2 
\label{4.28a}
\end{equation}
In the decomposition (\ref{4.28}), the three first terms correspond to the charge-charge and
charge-multipole interactions, whereas the last term represents the rest of the multipole-multipole
interactions. Since
\begin{equation}
\int_0^1 ds_1 (\delta(s_1-s_2)-1)=\int_0^1 ds_2 (\delta(s_1-s_2)-1)=0
\label{4.29}
\end{equation}
we see that only this latter term contributes to $W$
\begin{equation}
W({\cal F}_1,{\cal F}_2)=e_{\alpha_1} e_{\alpha_2}\int_0^1 ds_1\int_0^1 ds_2(\delta(s_1-s_2)-1) 
M_1(s_1)M_2(s_2)V({\bf r}_1-{\bf r}_2)
\label{4.30}
\end{equation}
The dominant large distance behavior of $W$ 
\begin{eqnarray}
W({\cal F}_1,{\cal F}_2) &\sim& e_{\alpha_1} e_{\alpha_2}\int_0^1 ds_1\int_0^1
ds_2(\delta(s_1-s_2)-1)\nonumber\\ 
&&\hspace{-1in}
\times(\lambda_{\alpha_1}{\b\xi}_1(s_1)\cdot\nabla_{{\bf r}_1})
(\lambda_{\alpha_2}{\b\xi}_2(s_2)\cdot\nabla_{{\bf r}_2})
V({\bf r}_1-{\bf r}_2),\;\;\;\;\;\;\;\mid {\bf r}_1-{\bf r}_2\mid
\rightarrow \infty
\nonumber \\
\label{4.31}
\end{eqnarray}
is typically a dipole-dipole potential, decaying as $\mid {\bf r}_1-{\bf r}_2\mid^{-3}$, due to the dipole 
elements $e_{\alpha_{1}}\lambda_{\alpha_1}{\b\xi}_1(s_1)ds_1$ and
$e_{\alpha_{2}}\lambda_{\alpha_2}{\b\xi}_1(s_2)ds_2$ associated with two filaments of arbitrary
shapes. It will be shown (chapter ~\ref{chapter-loops}) that the charge-charge and charge-multipole
interactions have a fast screening in the quantum gas, but not the multipole-multipole  interaction.
This long range term will precisely be the cause for the breakdown of exponential screening in the
quantum system for all values of the density and of the temperature. For the sake of pedagogy, we
illustrate the above assertions in a simplified model involving only two quantum mechanical charges.
This model presents already all the basic mechanisms which will be at work in the general case.

\subsection{Quantum fluctuations destroy exponential screening} 
\label{sec-4.3}

We consider the simplified situation where all particles are classical
but two. For this it suffices to single out two filaments ${\cal
F}_a=(\alpha_a, \,{\bf r}_a, \, \b\xi_a) $, ${\cal F}_b=(\alpha_b,
\,{\bf r}_b, \, \b\xi_b)$ corresponding to two particles with non
vanishing de Broglie lengths $\lambda_a$, $\lambda_b$, and to set the
de Broglie lengths of all the other particles equal to zero. Then the
phase space variable of a classical particle reduces to $(\alpha,\;
{\bf r})$ and its interaction with the quantum particle ${\cal F}_a$
to \begin{equation}V({\cal F}_a, {\bf r})=\int_0^1 dsV({\bf
r}_a+\lambda_a\b\xi_a(s)-{\bf r})= V_{{\rm cl}}({\cal F}_a, {\bf r})
\label{4.32} \end{equation}which is identical to the classical potential
(\ref{4.25}) with one $\lambda$ set equal to $0$, whereas the two
quantum particles still interact with the potential (\ref{4.18}).
Here the quantum mechanical nature of the two specified particles will
be treated non-perturbatively.  The model is semi-classical only in
the sense that their surrounding medium is now a classical plasma at
density $\rho$ and inverse temperature $\beta$.

As in (\ref{1.1.7}) we denote by $\omega=(\alpha_1,{\bf
r}_1,\ldots,\alpha_n,{\bf r}_n)$ a configuration of the classical particles and by
$\hat{c}({\bf r},\omega)=\sum_{j=1}^n e_{\alpha_j}\delta({\bf r} -{\bf r}_j)$
the corresponding microscopic charge density. For this system, the total interaction 
(\ref{4.17}) reads 
\begin{eqnarray}
U({\cal F}_a,{\cal F}_b,\omega)&=& e_{\alpha_a} e_{\alpha_b}V({\cal F}_a,{\cal F}_b)+ 
e_{\alpha_a}\int d{\bf r} V({\cal F}_a, {\bf r})\hat{c}({\bf r},\omega)\nonumber\\
&+&e_{\alpha_b}\int d{\bf r}
V({\cal F}_b, {\bf r})\hat{c}({\bf r},\omega) +U_0(\omega)
\label{4.34}
\end{eqnarray}
where $U_0(\omega)$ is the Coulomb energy of the classical particles.

The distribution of two filaments $\cf_a$ and $\cf_b$ in
thermal equilibrium with the classical Coulomb gas enclosed in a
finite volume $\Lambda$ is defined by 
\begin{equation}
\rho(\cf_a,\cf_b)=\frac{1}{\Xi_0}\int_\Lambda d\omega\exp\(-\beta U(\cf_a,\cf_b,\omega)\)
\label{4.35}
\end{equation}
In (\ref{4.35}), $\int_\Lambda d\omega$ means (canonical or grand
canonical) summation on the configurations $\omega$ of the classical
particles in $\Lambda$ and $\Xi_0=\int_\Lambda d\omega\exp(-\beta
U_0(\omega))$ is the corresponding partition function (in this
definition, there is no contribution of coincident filaments). For a
single filament we define in the same way\footnote{$-\beta ^{-1}\ln
\rho (\cF )$ and $-\beta ^{-1}\ln \rho (\cF_{a},\cF _{b} )$ are the
excess free energies when one or two filaments are embedded in the
classical plasma.}
\begin{equation}
\rho(\cf)=\frac{1}{\Xi_0}\int_{\Lambda } d\omega\exp\(-\beta
U(\cf,\omega)\) 
\label{4.36}
\end{equation}
with
\begin{equation}
U(\cf,\omega)=e_\alpha\int d\br \,
V(\cf,\br)\hat{c}(\br,\omega)+U_0(\omega) \label{4.37}
\end{equation}
Let us also introduce the auxiliary genuine classical distributions of
charged filaments $\rho_{{\rm cl}}(\cf_a,\cf_b), \;\rho_{{\rm
cl}}(\cf)\;$ defined in the same way as (\ref{4.36}, \ref{4.37}), but
with $V$ and $U$ replaced by $V_{{\rm cl}}$ and $U_{{\rm cl}}$.
Comparing now the quantum and this classical system, we observe that
in view of (\ref{4.32})
\begin{equation}
 U(\cf,\omega)= U_{{\rm
cl}}(\cf,\omega)\;,\;\;\;\;\,\rho(\cf)=\rho_{{\rm cl}}(\cf)
\label{4.38}
\end{equation}
but taking into account the definitions (\ref{4.26}) and (\ref{4.27})
\begin{eqnarray}
 U(\cf_a,\cf_b,\omega)&=&W(\cf_a,\cf_b)+U_{{\rm cl}}(\cf_a,\cf_b,\omega)\nonumber\\
\rho(\cf_a,\cf_b)&=&\exp\(-\beta W(\cf_a,\cf_b)\)\rho_{{\rm
cl}}(\cf_a,\cf_b) \label{4.39}
\end{eqnarray}

We consider now these distributions in the thermodynamic limit, keeping
the same notations. Because of translational invariance,
$\rho(\cf)=\rho_{\alpha }(\b\xi)$ is independent of $\br$ and
$\rho(\cf_a,\cf_b)$ depends on $|\br_{a}-\br_{b}|$. In the Debye
regime, we know that $\rho_{{\rm cl}}(\cf_a,\cf_b)$ clusters
exponentially fast as $|\br_a-\br_b| \rightarrow\infty$ (with
$\b\xi_a,\; \b\xi_b$ fixed)
\begin{eqnarray} 
\rho_{{\rm cl}}(\cf_a,\cf_b)&=&\rho_{{\rm cl}}(\cf_a)\rho_{{\rm cl}}(\cf_b)\nonumber\\
&+&O\(\exp\(-C\inf_{0\leq s_1,s_2\leq 1}|{\bf
r}_a+\lambda_{\alpha_a}\b\xi_a(s_1)-
{\bf r}_b-\lambda_{\alpha_b}\b\xi_b(s_2)|\)\)\nonumber\\
\label{4.40}
\end{eqnarray}
Thus the asymptotic behavior of $\rho(\cf_a,\cf_b)$ in (\ref{4.39}) is
governed by that of $W(\cf_a,\cf_b)$, i.e. according to (\ref{4.31}),
$\rho(\cf_a,\cf_b)$ clusters as $|\br_a-\br_b|^{-3}$. The
two-particle distribution is obtained by averaging
$\rho(\cf_a,\cf_b)$ on the shapes of the filaments. Using
(\ref{4.38}, \ref{4.40}) and expanding the factor $\exp\(-\beta
W(\cf_a,\cf_b)\)$ yields
\begin{eqnarray}
\rho(\br_a,\br_b)&=&\int D(\b\xi_a)D(\b\xi_b)
\exp\(-\beta W(\cf_a,\cf_b)\)\rho_{{\rm cl}}(\cf_a,\cf_b)\nonumber\\ 
&=&\rho_a\rho_b-\beta \int D(\b\xi_a)D(\b\xi_b)W(\cf_a,\cf_b)\rho_a(\b\xi_a)\rho_b(\b\xi_b)\nonumber\\
&+&\frac{1}{2}\beta^2\int
D(\b\xi_a)D(\b\xi_b)W^2(\cf_a,\cf_b)\rho_a(\b\xi_a)\rho_b(\b\xi_b) + 
\cdots +O\(\exp(-C|\br_a-\br_b|)\)
\label{4.41} 
\end{eqnarray}
where $\rho _{a} = \rho (\alpha _{a},\br _{a})$. The remainder in
(\ref{4.41}) still gives an exponentially decreasing contribution
since the probability of long filaments is Gaussian small.  Let us
show that the term linear in $W$ in (\ref{4.41}) does not contribute
to the asymptotic behavior of $\rho(\br_a,\br_b)$. Indeed one notes
that the measure $ D(\b\xi)$ is invariant under reflections and
rotations of the filament in three-dimensional space (see
(\ref{4.7})). The same is true for $\rho(\b\xi)$ in an homogeneous
phase of the classical plasma. Hence introducing the multipolar
expansion of $W$ with the help of (\ref{4.28a}), (\ref{4.30}) one sees
the occurrence of the $\b\xi$-averages
\begin{eqnarray}
&&
\int D(\b\xi_a) D(\b\xi_b)\rho(\b\xi_a)\rho(\b\xi_b)\nonumber \\
&&\hspace{-1in} [\lambda_a\b\xi_a(s_1)\cdot\nabla_{{\bf r}_{a}}]^k
[\lambda_b\b\xi_b(s_2)\cdot\nabla_{{\bf
r}_{b}}]^lV(\br_a-\br_b)\;\,\,\, k,\:l\geq 1 \label{4.42}
\end{eqnarray}
In view of these symmetry considerations, the terms with $k$ or $l$
odd vanish, and the terms with both $k$ and $l$ even are necessarily
proportional to $\(\nabla_{\br}^2\)^{\frac{k+l}{2}}V(\br),
\br=\br_a-\br_b$, which is a rapidly decreasing function since
$\nabla_{\br}^2\frac{1}{|\br|}=0, \br\neq 0$. Since all moments of the
Gaussian measure $ D(\b\xi)$ are finite, one concludes that the
$W$-linear term in (\ref{4.41}) decays faster than any power of
$|\br_a-\br_b|$. The leading behavior is therefore determined by the
quadratic term. According to (\ref{4.41}) and (\ref{4.31}) one finds
for the normalized truncated distribution
\begin{eqnarray}
\frac{\rho_{T}(\br_{a},\br_{b})}{\rho_a\rho_b}&=&\frac{\rho(\br_a,\br_b)}{\rho_a\rho_b}-1\nonumber\\
&=&\frac{\beta^{2}}{2\rho_{a}\rho_{b}}\int
D(\b\xi_a)D(\b\xi_b)W^2(\cf_a,\cf_b)\rho_a(\b\xi_a)\rho_b(\b\xi_b)+{\rm O}(W^{3})\nonumber\\
&=&\frac{B(\beta,\rho)}{|\br_a-\br_b|^6}+O\(\frac{1}{|\br_a-\br_b|^8}\) 
\label{4.43}
\end{eqnarray}
where $B(\beta,\rho)>0$ can be expressed, from (\ref{4.31}), as the
Brownian integral of a positive functional, depending on the
temperature and density of the surrounding plasma. The correction
$\mid r_{a} - r_{b}\mid^{-8}$ in (\ref{4.43}) arises from the square
of the dipole-quadrupole interaction.  Note that the result
(\ref{4.43}) holds in any screening phase of the classical plasma and
is non-perturbative in $\hbar$.

In the classical limit or in the high temperature limit, one finds that the dominant contribution
to $B(\beta,\rho)$  is the same \cite{AlMa89}   
\begin{equation}
B(\beta,\rho)\simeq\hbar^4\frac{\beta^4}{240}\frac{e_a^2e_b^2}{m_am_b},\;\;\;\hbar\rightarrow 0
\;\;\;\mbox{or}\;\;\;\beta\rightarrow 0
\label{4.44}
\end{equation}

The role of the quantum fluctuations is clearly demonstrated: the
dipole-like interaction $W$, which originates from these intrinsic
quantum fluctuations, can definitely not be shielded by the classical
charges, even if the latter are in the Debye phase. Nevertheless a
partial screening remains: the monopole is still screened and this,
together with rotation invariance, reduces the bare Coulomb decay
$r^{-1}$ to $r^{-6}$. It will be shown in section \ref{sec-4.5} as
well in chapters~\ref{chapter-loops} and \ref{chapter-6} that these
facts remain true in the full quantum gas.


\subsection{Origin of the van der Waals forces}
\label{sec-4.4}

One can interpret the result (\ref{4.43}) by saying that there exists an effective potential
$\Phi_{1}(r)$ between two individual quantum charges defined by
$\frac{\rho_{T}(\br_{a},\br_{b})}{\rho_a\rho_b}=e^{-\beta\Phi_{1}(|\br_{a}-\br_{b}|)}-1$.
It is attractive at large distance and behaves as
\begin{equation}
\beta\Phi_{1}(r)\simeq -\frac{B(\beta,\rho)}{r^{6}},\;\;\;\; r\rightarrow\infty
\label{4.44a}
\end{equation}
One may wonder if this potential between individual quantum charges is related to the van der Waals
potential, also decaying as $r^{-6}$. The van der Waals forces are usually computed for preformed
atoms and molecules in empty space, and are attributed to the dipole fluctuations in these bound
entities.

To allow for the possibility of quantum mechanical binding in simple
terms, we can generalize the model of section~\ref{sec-4.3} by
specifying four particles, say two electron-proton (e-p) pairs, in
equilibrium with a classical plasma \cite{Mar96}. The information
about the correlations between these four particles is contained in
the fully truncated four point correlation function
$\rho_{T}(\br_{a},\br_{b},{\bf R}_{1}, {\bf R}_{2})$ where
$\br_{a},\br_{b}$ designate the coordinates of the electrons and ${\bf
R}_{1}, {\bf R}_{2}$ those of the protons. This correlation is defined
by the natural generalization of the expressions
(\ref{4.35})-(\ref{4.39}) to four particles immersed in the plasma. We
are interested in the amount of correlation that exists between the
two (e-p) pairs as their centers of mass are taken far apart, but
ignoring the relative positions of the particles. Assuming for
simplicity that the two protons are infinitely heavy (i.e. classical),
the center of mass of the pairs coincide with ${\bf R}_{1}, {\bf
R}_{2}$ and the desired normalized correlation is \begin{equation}
F(R)=\frac{1}{2}\frac{\int d\br_{a}\,\int d\br_{b}
\,\rho_{T}(\br_{a},\br_{b},{\bf R}_{1}, {\bf R}_{2})} {\int
d\br_{a}\,\rho_{T}(\br_{a},{\bf R}_{1})\int d\br_{b}
\,\rho_{T}(\br_{b},{\bf R}_{2})}, \;\;\;\;R=|{\bf R}_{1}- {\bf R}_{2}|
\label{4.44b} \end{equation}The factor $1/2$ takes into account that
the electrons, treated here as identical particles, can be found in
the neighborhood of either one of the protons (the contribution of
coincident particles is omitted in (\ref{4.44b})\footnote{As before,
exchange effects due to the Fermi statistics of the electrons are
neglected. Exchange correlations have a short range.}).  We define the
effective (temperature and density dependent) potential $\Phi_{2}(R)$
between the two pairs by \begin{equation}
F(R)=e^{-\beta\Phi_{2}(R)}-1\simeq -\beta
\Phi_{2}(R),\;\,\;R\rightarrow\infty \label{4.44c} \end{equation}In
contrast to the standard calculation of van der Waals forces,
$\Phi_{2}$ incorporates now the effects of the thermal fluctuations of
the electrons and of the screening provided by the medium.

As in (\ref{4.41}), the asymptotic form of $\Phi_{2}(R)$ is determined by the quadratic term in 
the quantum potential $W$, namely 
$$
-\beta \Phi_{2}(R)\simeq \int \bar{D}(\b\xi_{a})\bar{D}(\b\xi_{b})
W^{2}(\cf_{a},\cf_{b})
$$
where now $ \bar{D}(\b\xi_{a}),\;\bar{D}(\b\xi_{b})$ are dressed Brownian
measures involving the single (e-p) pair correlation. One finds  
\begin{equation}
\beta\Phi_{2}(R)\simeq -\frac{C(\beta,\rho)}{R^{6}},\;\,\,R \rightarrow\infty,\;\,\,C(\beta,\rho)>0
\label{4.44d}
\end{equation}
There are two situations of interest, the high temperature limit ($\beta\rightarrow 0, \;\rho$ fixed)
and the atomic limit. The atomic limit is obtained by simultaneously lowering the temperature
($\beta\rightarrow\infty$) to favor binding and reducing the density at an exponential rate
($\rho\simeq e^{-\delta\beta}\rightarrow 0,\;\delta>0$) to get independent atoms. In this limit (e-p)
pairs form noninteracting hydrogen atoms in their ground states (the atomic limit will be discussed
in chapter~\ref{chapter-7} for the full quantum gas). 

In the high temperature limit, one recovers the result
(\ref{4.44}), i.e. 
\begin{equation}
 C(\beta,\rho)\simeq \hbar^{4}\frac{\beta^{4}e^{4}}{240m^{2}},\;\;\;\beta\rightarrow 0
\label{4.44e}
\end{equation}
As $\beta\rightarrow 0$, the electrons tend to be fully ionized and
the correlation becomes that found between individual charges.

In the atomic limit, for $\delta>0$ not too large, one has
\begin{equation}
\lim_{\beta\rightarrow\infty,\rho\rightarrow
0}\beta^{-1}C(\beta,\rho)=-C_{{\rm van\; der\; Waals}} \label{4.44f}
\end{equation}
where $C_{{\rm van\; der\; Waals}}<0$ is the usual coefficient found
in text books from a second order perturbation calculation of the
residual Coulomb interaction between two hydrogen atoms in their
ground states.  

A similar analysis has now been carried out in the full quantum
mechanical electron-proton gas with the diagrammatic methods of
chapter~\ref{chapter-loops}.  One finds indeed that the coefficient of
the $r^{-6}$ tail of the proton-proton correlation approaches the
value $C_{{\rm van\; der\;Waals}}$ in the scaling limit where hydrogen
atoms form \cite{AlCoMa98b}.  In the medium, density and temperature
effects modify the amplitude of the usual van der Waals potential, but
not its $r^{-6}$ range. In particular, the free charges that are
always present in the medium only partially screen the dipole
interaction between atoms, contrary to a naive mean field description
that would lead to an exponential screening of these interactions. The
partial screening of protons is due in part to collective effects, and
also to individual electrons that will bind with the proton.  As the
temperature goes to zero the former effects vanish, and the screening
is eventually entirely due to the formation of a neutral bound state.
Hence the atomic limit coincides with a zero density calculation with
two preformed neutral atoms in empty space. In fact, the coefficient
$C(\beta,\rho)$ interpolates continuously between the high temperature
and atomic regimes. There is no qualitative difference between the
mechanism producing the van der Waals forces between bound entities
and that causing the $r^{-6}$ tail (\ref{4.43}) between individual
charges.  Both have their common origin in the same basic quantum
fluctuations put into evidence in the preceding section.  One sees
that the difference is only quantitative. In the atomic regime, the
coefficient $C(\beta,\rho)$ has a non-zero value close to $|C_{{\rm
van\; der\; Waals}}|$ whereas it becomes vanishingly small at high
temperature.

\subsection{Semi-classical analysis of Coulombic correlations}
\label{sec-4.5}

We come now to the full quantum
mechanical gas, but treated semi-classically as a whole. We use again
the classical gas as a reference system by splitting the interaction between the
filaments (\ref{4.17}) into
\begin{equation}
U(\cf_1,\ldots,\cf_n)=U_0(\alpha_1,\br_1,\ldots,\alpha_n,\br_n)+
U_m(\cf_1,\ldots,\cf_n)
\label{4.45}
\end{equation}
where
\begin{equation}
U_0(\alpha_1,\br_1,\ldots,\alpha_n,\br_n)=\sum_{i<j}^n
 e_{\alpha_{i}}e_{\alpha_{j}}V(\br_i-\br_j)
\label{4.46}
\end{equation}
is the Coulomb energy of $n$ classical point particles and $U_m$ is the
residual multipolar interaction. It is defined as the sum of two-body
interactions $V_m(\cf_1,\cf_2)=V(\cf_1,\cf_2)-V(\br_{1}-\br_{2})$ obtained from 
(\ref{4.18}) by removing the charge-charge potential:
\begin{eqnarray}
V_m(\cf_1,\cf_2)&=&\int_0^1ds(M_1(s)+M_2(s))V(\br_1-\br_2)\nonumber\\
&+&\int_0^1ds M_1(s) M_2(s)V(\br_1-\br_2)\label{4.47}\\
U_m(\cf_1,\ldots,\cf_n)&=&\sum_{i<j}^n
e_{\alpha_{i}}e_{\alpha_{j}}V_m(\cf_i,\cf_j)\label{4.48}
\end{eqnarray}
Since the Planck constant occurs only through the de Broglie lengths
$\lambda_\alpha$, it is clear that $V_m$ and $U_m$ are power series in
$\hbar$ starting with a linear term (see (\ref{4.5}) and
(\ref{4.28a})). Thus considering $\hbar$ as a small parameter, we are
left with the standard problem of perturbing the classical equilibrium
state (in the framework of the system of filaments) by the small
modification (\ref{4.48}) of the potential.  This will generate
formally power series in $\hbar$ for the distributions of filaments,
and in turn, after averaging on their shapes, for the truncated
distributions of the quantum charges 
\begin{equation}
\rho_T(\alpha_1,\br_1,\ldots,\alpha_n,\br_n)=\sum_{k\geq 0} \hbar^k
\rho_T^{(k)}(\alpha_1,\br_1,\ldots,\alpha_n,\br_n) \label{4.49} 
\end{equation}
We shall not give here the rather complex algorithm for calculating
the coefficients $\rho_T^{(k)}$ of this expansion, but only describe
some qualitative aspects and state results.

From now on we assume that the classical reference system is in the
Debye phase: its correlations $\rho_{{\rm
cl},T}(\alpha_1,\br_1,\ldots,\alpha_n,\br_n)$ cluster exponentially
fast and obey the multipolar sum rule (\ref{2.2.5a}). By expanding
around the classical gas, each $\rho_T^{(k)}$ involves combinations of
convolutions of classical correlations with the multipole potentials
coming from (\ref{4.47}), so that the $\rho_T^{(k)}$ are expressed in
terms of the supposedly known correlations $\rho_{{\rm cl},T}$ of the
reference system.  The main point is that, although the classical
correlations have an exponential fall-off, the law of formation of the
$\rho_T^{(k)}$ is such that some of these convolutions decay only
algebraically (there are two exceptions, namely if the potential is
convoluted with the total charge density, then multipolar sum rules insure a
fast decay, or if derivatives of the Coulomb potential are contracted
to form a Laplacian $\nabla^2V(\br)$).

The results are as follows \cite{AlMa88,AlMa89}. Obviously 
\begin{equation}
\rho_T^{(0)}(\alpha_1,\br_1,\ldots,\alpha_n,\br_n)=
\rho_{{\rm cl},T}(\alpha_1,\br_1,\ldots,\alpha_n,\br_n)
\label{4.50}
\end{equation}
\begin{equation}
\rho_T^{(k)}(\alpha_1,\br_1,\ldots,\alpha_n,\br_n)=0,\;\;\;\;k\; \mbox{odd}
\label{4.51}
\end{equation}
This is because, according to (\ref{4.28a}), odd powers of $\hbar$ in
the expansion (\ref{4.49}) correspond to odd moments of
filaments. Such odd moments have zero Gaussian average.

The coefficients $\rho_T^{(2)}(\alpha_1,\br_1,\ldots,\alpha_n,\br_n)$
are decreasing faster than any inverse power for large spatial
separations of the arguments (for the two point function see
\cite{Jan76}).

At the order $\hbar^4$ the truncated two particle correlation behaves
as
$$
\langle \hat{\rho}(\alpha_{1},\br_{1})\hat{\rho}(\alpha_{2},\br_{2})\rangle^{(4)}_{T}=
\rho_T^{(4)}(\alpha_1,\br_1,\alpha_2,\br_2)\phantom{---------}
$$
\begin{equation}
\sim
\frac{2}{5}\(\frac{\beta^2}{24}\)^2\int d\br \,\int d\br^{\prime}\,
f(\br_1-\br_2+\br-\br^{\prime})G(\br,\alpha_1)G(\br^{\prime},\alpha_2)\phantom{----}
\label{4.51a}
\end{equation}
In (\ref{4.51a}) $f(\br)$ is the square of the dipole-dipole potential
\begin{equation}
f(\br)=\sum_{\mu,\nu=1}^3\(\partial_\mu\partial_\nu V(\br)\)^2
\label{4.52}
\end{equation}
and
\begin{equation}
G(\br,\alpha_i)=\sum_{\alpha}
\frac{e_\alpha^2}{m_\alpha}\langle\hat{\rho}(\alpha,{\bf 0})
\hat{\rho}(\alpha_i,\br)\rangle_{{\rm cl},T},\;\;\;\,i=1,2
\label{4.53}
\end{equation}
where $\langle\cdots\rangle_{{\rm cl}}$ means equilibrium average with respect to the classical
reference system. Since $f(\br)\sim\frac{6}{|\br|^6}$, $|\br|\rightarrow\infty$,
the leading contribution to (\ref{4.51a})
is
\begin{equation}
\langle \hat{\rho}(\alpha_{1},\br_{1})\hat{\rho}(\alpha_{2},\br_{2})\rangle^{(4)}_{T}\sim
\frac{1}{|\br_1-\br_2|^6}\frac{\beta^4}{240} \(\int d\br \, G(\br,\alpha_1)\)\(\int d\br \,
G(\br,\alpha_2)\)  
\label{4.54}
\end{equation}

If one examines the correlation of a particle with the total charge
density $\hat{c}(\br)$, one must replace $\hat{\rho}(\alpha,\br)$ in
one factor $G$ by the charge $\hat{c}(\br)$. Then the coefficient of
$|\br_1-\br_2|^{-6}$ vanishes in (\ref{4.54}) because of the charge
sum rule (\ref{2.2.5a}), and one has to carry the asymptotic expansion
of $f(\br)$ further on. For the particle-charge correlation, one finds
\begin{eqnarray}
\langle \hat{\rho}(\alpha_{1},\br_{1})\hat{c}(\alpha_{2},\br_{2})\rangle^{(4)}_{T}&=&
\sum_{\alpha_{2}}e_{\alpha_{2}}\rho_T^{(4)}(\alpha_1,\br_1,\alpha_2,\br_2)\nonumber\\
&\sim&\frac{1}{|\br_1-\br_2|^8}\frac{\beta^4}{16}\(\int d\br \, G(\br,\alpha_1)\)
\(\frac{1}{3}\int d\br \, |\br|^2K(\br)\)\nonumber\\ 
\label{4.55}
\end{eqnarray}
and for the charge-charge correlation
\begin{eqnarray}
\langle \hat{c}(\alpha_{1},\br_{1})\hat{c}(\alpha_{2},\br_{2})\rangle^{(4)}_{T}&=&
\sum_{\alpha_1,\alpha_2}e_{\alpha_1}e_{\alpha_2}\rho_T^{(4)}
(\alpha_1,\br_1,\alpha_2,\br_2)\nonumber\\&\sim&\frac{1}{|\br_1-\br_2|^{10}}\frac{7\beta^4}
{4}\(\frac{1}{3}\int d\br \, |\br|^2K(\br)\)^2
\label{4.56}
\end{eqnarray}
with
\begin{equation}
K(\br)=\sum_\alpha\frac{e_\alpha^2}{m_\alpha}
\langle\hat{\rho}(\alpha,{\bf 0})\hat{c}(\br)\rangle_{{\rm cl},T}
\label{4.57}
\end{equation}
In fact, as discussed in sections~\ref{sec-6.1.2} and
\ref{subsec-6.1.3}, the formulae (\ref{4.54}), (\ref{4.55}) and
(\ref{4.56}) become asymptotically exact for the quantum Coulomb gas
in the low density limit. In this limit the functions $G(\br)$ and
$K(\br)$ are given by their Debye-H\"{u}ckel
approximation\footnote{Note that the exchange contributions vanish
exponentially fast with $\hbar$, so leading contributions in $\hbar$
can be obtained from Maxwell-Boltzmann statistics.}.

The formulae simplify in the case of the jellium when one has only one
species of charges. Then there is no distinction between particle and
charge correlations up to a factor $e$
$(eG(\br)=K(\br)=\frac{e}{m}\langle\hat{c}({\bf}0)\hat{c}(\br)\rangle_{{\rm
cl},T}$), so (\ref{4.56}) applies in all cases. Furthermore, in view
of the second moment condition (\ref{1.1.1sti}), the correlations at
order $\hbar^4$ in the jellium behaves as \begin{equation}
\rho_T^{(4)}(\br_1,\br_2)\sim\frac{7}{16\pi^2}\(\frac{\beta}{m}\)^2
\frac{1}{|\br_1-\br_2|^{10}},\;\;\;\;\;|\br_1-\br_2|\rightarrow\infty
\label{4.58}
\end{equation}
For the jellium, \cite{GoLe94} have given the complete formula for
$\rho_T^{(4)}(\br_1,\br_2)$ at the order $\hbar^{4}$ including the short range contributions.

Correlations involving more than two particles also have slow decays
at order $\hbar^4$. For instance
$\rho_T^{(4)}(\alpha_1,\br_1,\alpha_2,\br_2,\alpha_3,\br_3)$ decays
also as $|\br|^{-6}$ when the particle 1 is sent to infinity, the
other being kept fixed. If the three particles are taken
simultaneously far apart (say,
$|\br_1-\br_2|=r,\;|\br_1-\br_3|=ar,\;|\br_2-\br_3|=br,\;r
\rightarrow\infty$), this function decays as $r^{-9}$. Moreover, the
four point function
$\rho_T^{(4)}(\alpha_1,\br_1,\alpha_2,\br_2,\alpha_3,\br_3,
\alpha_4,\br_4)$ decays only as $r^{-3}$ when two pairs of particles
are sent far away (say, $|\br_1-\br_2|,|\br_3-\br_4|$ fixed,
$r=|\br_1-\br_3|\rightarrow\infty$). This slow decay is due to the
breaking of rotational invariance around the points $\br_1$ and
$\br_2$ when the directions $\br_1-\br_2$ and $\br_3-\br_4$ are
fixed. It is precisely because of rotation invariance that the pure
dipolar decay $r^{-3}$ is reduced to $r^{-6}$ in the two point
function, as in the model of section \ref{sec-4.3}.

We have reported on the semi-classical tails (\ref{4.54})-(\ref{4.56}) in some
details with the purpose of illustrating the following points. There is
definitely no exponential screening in the quantum gas, even in the high
temperature phase and what ever the density may be. Moreover, in contrast with a
fluid with short range forces, various types of decay laws have to be expected
depending on what type of observables are concerned, and what cluster of
particles are taken far apart. The occurrence of these different types of decay
are due to  the underlying multipolar forces displayed in the filament formalism
in conjunction with the effect of screening sum rules (\ref{2.2.5a}). 

We finally discuss the possible observable implications of these findings. Since the 
calculations pertain to the semi-classical regime, we only consider real systems
under conditions that the quantum effects are small. We determine the crossover distance
$r_{0}$ above which quantum algebraic tails dominate the classical behavior by equating
(\ref{4.58}) to the usual Debye law
\begin{equation}
\frac{\rho e^{- r_{0}/l_{D}}}{4\pi l_{D}^{2}
r_{0}}=\frac{7}{16\pi^{2}}\left(\frac{\beta}{m}\right)^{2}
\frac{\hbar^{4}}{r_{0}^{10}}
\label{4.59}
\end{equation}
The classical exponential is normalized in order to obey the charge
sum rule.  For instance, for sodium chloride at room temperature, one
finds $r_{0}\sim 60\;l_{D} $. For a white dwarf under prevailing
conditions, one has $r_{0}\sim 40\;l_{D} $. In both cases, the quantum
effects on the clustering turn out to be very small and the use of the
Debye potential is still legitimate from a quantitative point of
view. However if we apply crudely (\ref{4.59}) (disregarding its
semi-classical nature) to the electrons in a metal (replacing $l_{D}$
by the Thomas-Fermi screening length $\lambda_{TF}$), one finds
$r_{0}$ of the order of $\lambda_{TF}$. This indicates that the use of
the exponential potential might be less reliable in this case.

\subsection{Breakdown of exponential screening in the Quantum
Sine-Gordon representation}\label{sec-4.6}

In this section we show how the Sine-Gordon transformation of
section~\ref{sec-sinegordon} extends to the quantum mechanical system
\cite{FrPa78}. We saw in (\ref{2.3.4c}) that the classical partition
function is a Gaussian superposition of ideal gas partition functions
in external fields $i\phi (\br )$.  The same holds for quantum
mechanics except that one must use time dependent fields $i\phi
(\br,s)$.  

After the Sine-Gordon transformation is discussed we use it to
illustrate the natural appearance of dipole forces in the quantum system
and make some comparisons with classical systems of particles that
simultaneously have charge and dipole moments.

We consider a quantum Coulomb system as in (\ref{1.2.1}), but 
replace the Coulomb interaction (\ref{1.1.1}) by 
\begin{equation}\label{4.6.0}
	U_{\infty,L} 
= 
	\frac{1}{2}\sum_{i,j}^N e_{\alpha_i} e_{\alpha_j} 
	V_{\infty ,L}(\br _i - \br _j )
\end{equation}
where $V_{\infty ,L}$ is the potential defined in (\ref{1.1.3be})
whose singularity at the origin is smoothed out on length scale $L$.
Note also that self-energies are included in $U_{\infty,L}$.  This
choice of interaction simplifies the Sine-Gordon transformation.  It
should be possible to reduce the $1/r$ to this case by a Mayer
expansion as was done in the classical case in
section~\ref{sec-debyesphere} because the Mayer expansion is known to
converge \cite{BrFe76,BrFe77}.

From (\ref{1.2.5}) with external field $\phi $ set to zero and using
the Trotter product formula
\begin{eqnarray} 
	\Xi_{\Lambda} 
&=&
	\mbox{Tr} P\exp\( -\beta\(
        H_{\Lambda,N} - \umu \cdot \uN \) \) \nonumber\\ 
&=&
        \mbox{Tr} P \lim_{n \rightarrow \infty} \prod_1^n 
        \(
        e^{-\frac{\beta}{n}[H_{\Lambda,N}^{0}- \umu \cdot \uN]} 
        e^{-\frac{\beta}{n} U_{\infty ,L}} 
        \)  \label{4.6.1}
\end{eqnarray}
where the $H_{\Lambda,N}^{0}$ is the free Hamiltonian and $\umu \cdot \uN
=\sum_{\alpha}\mu_{\alpha}N_{\alpha}$.  
We insert
\begin{equation}\label{4.6.2}
	e^{-\frac{\beta}{n} U_{\infty ,L}} 
=
	\int d\mu _{L} (\phi) \, e^{
	-i(\beta/n) ^{1/2} \int d \br \,\hat{c}(\br ) \phi(\br )
	}
\end{equation}
Each factor $e^{-\frac{\beta}{n} U_{\infty ,L}}$ requires its own
auxiliary field $\phi_i(\br)$, $i = 1, \ldots, n$, so that the right
hand side of (\ref{4.6.1}) contains
\begin{equation}
        \int d \mu_{L}(\phi_{1})\cdots d \mu_{L}(\phi_{n}) \,
        \mbox{Tr} P 
        \prod_1^n \(
	e^{-\frac{\beta}{n} H_{\Lambda,N}^{0}} e^{
        -(\beta /n)^{1/2} i \int d \br \,\hat{c}(\br ) \phi_{j}(\br )
        } \)
\end{equation}
By scaling the fields $\phi_{j} \rightarrow n^{-1/2}\phi_{j}$ and
uniting them into one time dependent field $\phi(\br,s) \equiv
\phi_j(\br )$ when $s \in [\frac{(j-1)}{n},\frac{j}{n})$,
we obtain
\begin{equation} \label{4.6.3}
	\Xi_{\Lambda} 
=
        \int \dM_{L}(\phi) \, 
        \Xi_{{\rm ideal},\Lambda}(\beta ^{1/2}i\phi)
\end{equation}
where $\int  \dM_{L}(\phi)$ is the Gaussian defined by
\begin{equation} \label{4.6.4}
        \int  \dM_{L}(\phi) \, \phi(\br, s ) \phi(\br ',s ')
=
        \delta (s - s ') V_{\infty ,L}(\br  - \br ');
\ \ \
0 \leq s \leq 1
\end{equation}
and 
\begin{eqnarray}\label{4.6.5}
&&
        \Xi_{{\rm ideal},\Lambda}(\phi)
=
	\mbox{Tr} P  
        \exp\( - \int_{0}^{1} \(
        \beta H_{\Lambda,N}^{0} - \beta  \umu \cdot \uN 
        + \int d \br \,\hat{c}(\br ) \phi(\br,s  )
        \)  \, ds \)
\end{eqnarray}
is the ideal gas partition function for a time-dependent external
field and the exponential is time-ordered. In the case of Boltzmann
statistics, which means omitting the sum over permutations in the
projection $P$, the trace factors over particles so that
\begin{eqnarray}
\label{4.6.6}
&&
        \Xi_{{\rm ideal, Bolt},\Lambda}(\beta\phi)
=
        \sum_{\{N_\alpha\}} \prod _{\alpha }\frac{1}{N_{\alpha }!}
	\mbox{Tr}  \nn \\
&& \hspace{-1in} \times 
        \exp\( - \beta\int_{0}^{1} \(
         H_{\Lambda,N}^{0} -   \umu \cdot \uN 
        + \int d \br \,\hat{c}(\br ) \phi(\br,s  )
        \)  \, ds \)	 \nn \\
&& =
        \exp \left( \sum _{\alpha } 
        \mbox{Tr} e^{
        - \beta\int_{0}^{1} ds [h_{\alpha }(s)-\mu _{\alpha }]
        }\right)
\end{eqnarray}
where $h_{\alpha }(s) = \frac{\hbar^{2}}{2m_{\alpha }}(-\Delta) + 
e_{\alpha }\phi(\br, s) $ is the one-particle Hamiltonian for species $\alpha
$ in the time-dependent external field and the exponential is still
time-ordered. Without making the simplification of Boltzmann
statistics one has, instead,
\begin{equation}
\label{4.6.6b}
        \Xi_{{\rm ideal},\Lambda}(\beta\phi)
=
        \exp \left( \sum _{\alpha } \sum _{q=1}^{\infty }
       \frac{ \eta_{\alpha }^{q-1}}{q} \mbox{Tr} e^{
        - \beta\int_{0}^{q}ds  [h_{\alpha }(s)-\mu _{\alpha }]
        }\right)
\end{equation}
where $\phi (\br ,s)$ is extended periodically in $s$.  This
formula will be proved in section~\ref{subsec-5.1.4}.  

Notice that this quantum ideal gas is invariant under 
\begin{equation}\label{4.6.6d}
\phi(\br,\tau ) \rightarrow \phi(\br,\tau ) + f(\tau)
\end{equation}
where $f$ is any function independent of $\br $ such that
$\int_0^1 d\tau f(\tau) = 0$.  The Gaussian $\dM$ is also formally
invariant under this transformation.  This ``Goldstone mode'' is a
signal that there are long range correlations in the $\phi$ field, whereas
in the classical system $\phi$ correlations are exponentially decaying.

By the Feynman-Kac formula,
\begin{eqnarray}\label{4.6.7}
&&
	\Xi_{{\rm ideal, Bolt},\Lambda}(\beta\phi)
=
	\exp \left(
        \int d\cf \, z(\cf ) e^{
        - \beta e_{\alpha} \phi(\cf )
        } \right)\nn \\
&&
       \phi(\cf )
= 
       \int_{0}^{1} \phi (\br + \lambda _{\alpha }\xi(s),s) \,ds
\end{eqnarray}
where $\cf, d\cf $ were defined in (\ref{4.12}, \ref{4.15}). There is
a similar formula for the ideal gas (\ref{4.6.6b}) without the
Boltzmann statistics simplification and it is given in
section~\ref{subsec-5.1.4}. It could be used in the following
considerations, but would not change the discussion in any important
way.  Furthermore we set the spins $s_{\alpha } = 0$.

When $\lambda _{\alpha } = 0$ the phase space integration $d\cf $
reduces to $\sum _{\alpha }\int d\br $ and (\ref{4.6.7}) reduces to
the classical ideal gas.  This suggests \cite{BrKe94} the following
model.  We replace the integration $\int d\cf $ over all Wiener paths
by integration concentrated on just one kind of path which oscillates
about the initial point by a distance $O(\lambda _{\alpha })$ (the
size of the wave packet) in a random direction: let $d\sigma(\vec{e})$
be a spherically symmetric integration on vectors $\vec{e}$. Then
$\int d\cf $ is replaced by
\begin{equation}
	\sum _{\alpha }  
	\int \, d\br  \, \int \, d\sigma(\vec{e})
	\label{4.6.8}
\end{equation}
and each $(\br ,\vec{e})$ labels a path: $\br + \lambda _{\alpha } \xi
(s)$ with $\xi(s) = \vec{e} \sin (2 \pi s)$.  We chose $ \sin (2
\pi s)$ because it is orthogonal to $1$.
 
The consequence of this and another minor approximation
\cite{BrKe94} is that the dependence of (\ref{4.6.6}) on $\phi(\br ,s)
$ is only through two modes
\[
	\phi_1(\br ) 
=
	\int_{0}^{1} \phi(\br , s)\, ds;
\ \ \
	\phi_2(\br )
=
	\sqrt{2} \int_{0}^{1} \phi(\br ,s) \sin (2 \pi s) \, ds 
\]
Since $\phi_1$, $\phi_2$ are Gaussian and $\int \dM _{L} \, \phi_i(\br
) \phi_j(\br ') = V_{\infty ,L}(\br -\br ') \delta_{ij}$, the integral
$\int \dM _{L} $ can be replaced by $\int d\mu _{L}(\phi _{1}) \, \int
d\mu _{L} (\phi _{2})$ and the partition function becomes
\begin{eqnarray}
&&
  	\Xi_{\Lambda} 
=
	\int d\mu_{L}(\phi_2) \int d\mu_{L}(\phi_1) \nn \\
&&\hspace{-1in} \times 
        \exp \( \sum _{\alpha } e^{\beta \mu_{\alpha } } 
        \int d\br  \, \int d\sigma(\vec{e}) \,
        e^{i\beta^{1/2} \phi_1  +
        i(\beta/2)^{1/2} \lambda _{\alpha }\vec{e}\cdot\nabla\phi_2}\) 
\label{4.6.9b}
\end{eqnarray}

Notice that if $d\sigma(\vec{e})$ is set to $\delta(\vec{e})$ we
revert to the classical Coulomb gas.  If $\phi_1$ is set to zero then
by reversing the Sine-Gordon transformation we obtain the partition
function of a classical dipole gas with dipole moments $\vec{e}$
distributed according to $d\sigma$.  Furthermore part of the Goldstone
mode (\ref{4.6.6b}) survives as
\begin{equation}\label{4.6.6c}
	\phi _{2}(\br ) \rightarrow \phi _{2}(\br ) 
	+ {\rm const}
\end{equation}
If the fields $\phi = \phi _{1} = \phi _{2}$ are set equal, this
becomes exactly the Sine-Gordon transform of a classical gas of
particles that are simultaneously carrying charge and a dipole
moment. {\it But in this case the symmetry (\ref{4.6.6c}) no longer
holds and the methods of chapter~\ref{chapter- debye-screening} prove
that correlations decay exponentially, near the Debye-H\"uckel limit.}

Indeed consider the Debye-H\"uckel approximation wherein the exponent
in (\ref{4.6.9b}) is replaced by the quadratic approximation
\begin{eqnarray}
&&
        Q = 
	-\frac{1}{2}
        \sum _{\alpha } e^{\beta \mu_{\alpha } } 
        \int d\br  \, \int d\sigma(\vec{e}) \,\(
        \beta^{1/2} \phi  +
        (\beta/2)^{1/2} \lambda _{\alpha }\vec{e}\cdot\nabla\phi
        \) ^{2}\nn \\
=&&
	-\frac{1}{2}
        \sum _{\alpha } \beta e^{\beta \mu_{\alpha } } 
        \int d\br  \, \(
        \phi ^{2}  + \frac{1}{6}
        (\lambda _{\alpha }\nabla\phi)^{2}
        \)
\label{4.6.10}
\end{eqnarray}
The Gaussian $d\mu _{L}\exp \left(-Q \right)$ has a covariance whose
Fourier transform is
\begin{equation}\label{4.6.11}
	4\pi
	\(\bk ^{2} + L^{2} \bk ^{4} + l_{D}^{-2} + a \bk ^{2}\)^{-1}
\end{equation}
where $l_{D}^{-2} = 4\pi \sum _{\alpha } \beta e^{\beta \mu_{\alpha
}}$ and $a = 4\pi \sum _{\alpha } \beta \lambda _{\alpha }^{2} e^{\beta
\mu_{\alpha }}$. There is no pole for any real $\bk $ which indicates
exponential decay of all correlations.
It has been wrongly said \cite[section II G]{Mar88} that classical ions
with structure might have weaker screening properties than pure charges. The
present discussion shows that such systems have Debye screening near the 
Debye-H\"uckel limit. 

The model defined by (\ref{4.6.8}) and (\ref{4.6.9b}) exemplifies
again that the breakdown of exponential screening is really due to
quantum mechanics, manifested here by the existence of the two fields
$\phi _{1}$ and $\phi _{2}$ (and more generally by the time-dependent
field $\phi (\br,s)$ in the Sine-Gordon representation).  Reconsider
indeed that $\phi _{1} \not = \phi _{2}$ are independent Gaussian
fields.  The quadratic approximation is misleading here because the
cross-terms coupling $\phi _{1}, \phi _{2}$ are lost in this
approximation but they are present at cubic and higher order in
(\ref{4.6.9b}).  In \cite{BrKe94} it is argued that these higher order
couplings allow the long range correlations of $\phi _{2}$ to creep
into the correlations of both $\phi _{1}$ and $\phi _{2}$. This gives
an alternative but more qualitative derivation of the no-screening
results.  The massive zero frequency field $\phi_{1}$ mediates (parts
of) the electrostatic potential $V_{{\rm cl}}$ defined in (\ref{4.25})
whereas the non-zero frequency field $\phi_{2}$ mediates (part of) the
non-electrostatic part $W$ defined in (\ref{4.27}) that cannot be
screened.

All these discussions show that screening is destroyed because lack of
commutativity of momentum and position bring time dependent phenomena
in through the back door (e.g., the imaginary time ordering mentioned
below (\ref{4.6.5})).  It is in fact easy to prove that imaginary time
dependent observables are not screened \cite{BrSe86}.

We should also mention that another class of semiclassical models has
been introduced, to deal with polarizability effects in dielectrics
\cite{HoSt81b,BrHo88}.  In these models, internal atomic degrees of
freedom are treated quantum mechanically (e.g. harmonically bound
pairs of charges) while the center of mass of atoms behave in a
classical manner with pair-wise dipolar forces.  These models  capture
physically interesting effects but do not incorporate all the
effects stemming from the Coulomb potential (binding, ionization,
collective screening) in an equally consistent and fundamental
footing. Hopefully the loop formalism developed in
chapter~\ref{chapter-loops} does provide a way to treat all these
effects in a fully coherent way.

\def\({\left(}
\def\){\right)}
\def\br{{\bf r}}
\def\b\xi{\mbox{\boldmath $\xi$\unboldmath}}
\def\cf{{\cal F}}
\def\cl{{\cal L}}
\def\bX{{\bf X}}
\def\bR{{\bf R}}

\newpage \section{The gas of charged loops} \setcounter{page}{1}

\label{chapter-loops}

\subsection{The statistical mechanics of loops}
\label{sec-5.1}

\subsubsection{The magic formula}
\label{subsec-5.1.1}

We are ready to develop the general Feynman-Kac representation of the
Coulomb gas.  The first step (section \ref{sec-5.1}) is to include the
Fermi statistics of the electrons and the Bose or Fermi statistics of
the nuclei. This can be done with some combinatorial effort by a
rearrangement of the sums in the partition function (\ref{1.2.5}),
leading to the "magic formula" (\ref{5.16}), that has a marvelous
effect: as in the preceding section, the gas appears as a
classical-like assembly of extended objects, the charged loops, and
the diagrammatic techniques of classical statistical mechanics will
apply.  The magic formula has been known at least implicitly for a
long time, starting from \cite{Gin65}, see also \cite{Gin71}. A
combinatorial proof of it, following the lines of \cite{Gin65} can be
found in \cite{Cor96a}. It has been also retrieved recently in
\cite{HoSt94}.  In subsection \ref{subsec-5.1.4} we give the main
lines of a derivation that does not make use of combinatorial
arguments. From now on,
$V(\br_{1}-\br_{2})=\frac{1}{|\br_{1}-\br_{2}|}$ is the exact Coulomb
potential on all scales: a short range regularization is no more
needed since the latter is henceforth provided by quantum mechanics
that will be now fully taken into account.

The rest of the chapter is devoted to the development of a Mayer
diagrammatic for this system, following mainly \cite{Cor96a}. In
section \ref{sec-5.2} the monopole and multipole interactions are
singled out in close analogy to what was done in section~\ref{sec-4.2}, in
order to isolate the part $W(\cl_1,\cl_2)$ (\ref{5.34}) that will
ultimately be responsible for the non-exponential decay. Then,
following the classical treatment of Coulomb divergences, one sums the
Coulomb chains, providing thus the screening length $\kappa^{-1}$
(\ref{5.47}) (a comparison of this screening length with that of the
standard RPA theory is postponed to section
\ref{subsec-6.1.4}). Finally in section \ref{subsec-5.2.3} the class
of resummed diagrams is reorganized in prototype graphs involving four
types of bonds. Prototype graphs are finite, decay at least as
$r^{-3}$ at large separation and will serve as the starting point of
low density calculations. The content of this chapter as well as the
next one is mathematically formal since nothing is known about the
asymptotic nature of the infinite series of prototype diagrams.

We now come back to the partition function (\ref{1.2.5}).  We first
represent the matrix element in (\ref{1.2.5}) by a Feynman-Kac
integral.  Since $H_{\Lambda,N}$ is independent of the spins, the spin
scalar product $\prod_{i=1}^N<\sigma_{\alpha _{p(i)}}|\sigma_{\alpha
_{i}}>$ factorizes out in this matrix element, and according to
(\ref{4.6}) its positional part is represented by the functional
integral
\begin{eqnarray}
& &<\{\br_{p(i)}\}|\exp\(-\beta H_{\Lambda,N}\) |\{\br_i\}>=
\prod_{i=1}^N\(\frac{1}{2\pi\lambda_{\alpha_{i}}^2}\)^{3/2}\exp\(
-\frac{|\br_{p(i)}-\br_i|^2}{2\lambda_{\alpha_{i}}^2}\)\nonumber\\
&\times &\int\prod_{i=1}^N D(\b\xi_i)\exp\(-\frac{\beta}{2}\sum_{i\neq j}^N
e_{\alpha_{i}}e_{\alpha_{j}}\int_0^1 dsV(\br_{i,p(i)}(s)-\br_{j,p(j)}(s))\)
\label{5.1}
\end{eqnarray}
We have kept the notation (\ref{4.4}) for a path with extremities
$\br_{i,p(i)}(s=0)=\br_i$ and $\br_{i,p(i)}(s=1)=\br_{p(i)}$ 
\begin{equation}
\br_{i,p(i)}(s)=(1-s)\br_i +s\br_{p(i)}+\lambda_{\alpha_{i}}\b\xi_i(s)
\label{5.2}
\end{equation}
The same remark made after (\ref{4.9}) concerning the implementation of the
boundary conditions remains valid: all paths are constrained to stay inside
$\Lambda$.

\begin{figure}
\psfig{figure=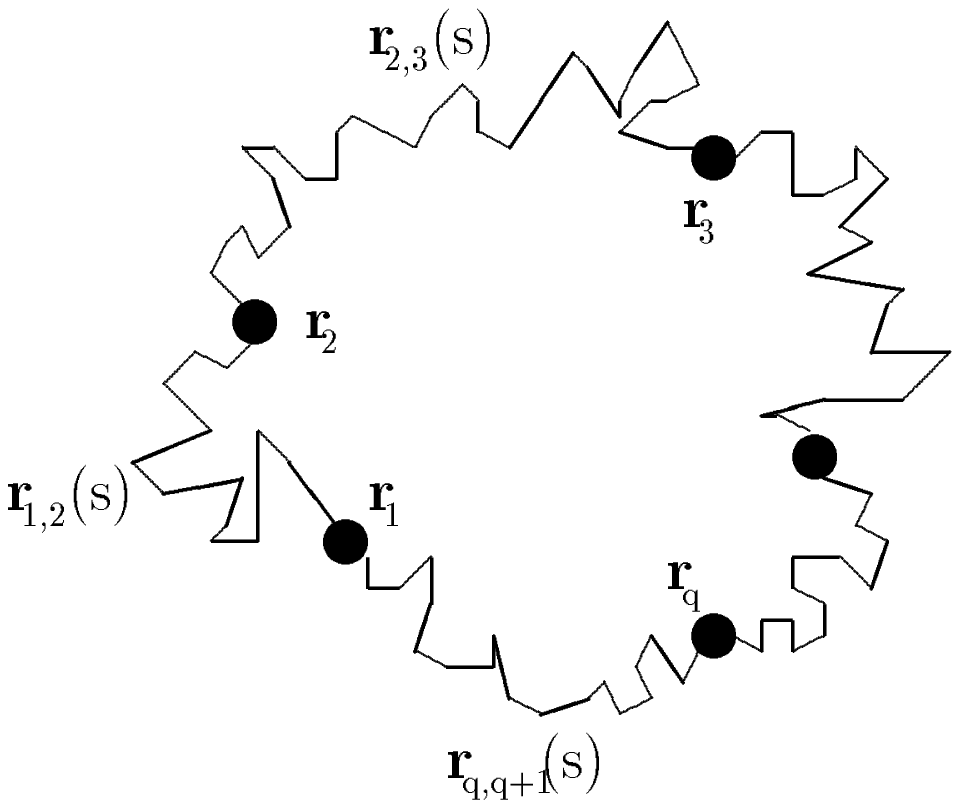}
\caption{A $q$ particle loop}
\label{fig1}
\end{figure}

The main observation to be made at this point is that by a
rearrangement of the sums, the partition function (\ref{1.2.5}) can be
formally written in the same classical form as (\ref{4.14}), provided
that one introduces a suitably enlarged phase space. This
rearrangement exploits the fact that any permutation $p_\alpha$ can
uniquely be decomposed into a product of cycles.  A cycle of length
$q$, $1\:\leq q\:\leq N_\alpha$, is a subset of the $N_\alpha$
particle indices that are permuted among themselves in a cyclic way
under $p_\alpha$, for instance \begin{equation}
(1,\,2,\ldots,q-1,\,q)\longrightarrow (2,\,3,\ldots,q,\,1) \label{5.3}
\end{equation}The objects constituting the enlarged phase space are precisely
associated with such cycles in the following way. The permutations $p$
in (\ref{5.1}) involve now open paths (except for the trivial
one). However the set of open paths belonging to a cycle of $q$
elements occurring in $p$ can be joined together to form a closed loop
$\cl$ (sometimes also called a polymer) with $q$ particle
coordinates. The loop $\cl$ corresponding to the cycle (\ref{5.3}) is
made of the collection of paths $\{\br_{k,k+1}(s)\}_{k=1,\ldots,q}$,
with $q+1$ identified to $1$ (see Figure~\ref{fig1}).

\noindent Thus a loop is specified by the variables   
\begin{equation}
\(\alpha,\;q,\;\{\br_k\}_{k=1,\ldots,q},\;\{\b\xi_k\}_{k=1,\ldots,q}\)
\label{5.4}
\end{equation}
It is convenient to unite the $q$ filaments constituting the loop into a single
path parameterized by a time parameter $s$ running from $0$ to $q$ setting
\begin{equation}
{\bf R}(s)=\br_{k,k+1}(\tilde{s}),\;\;\;k=[s]+1,\;\,\;0\leq\,s\,\leq q
\label{5.5}
\end{equation}
where $[s]$ = integer part of $s$ and $\tilde{s}=s\mbox{(mod
1)}=s-[s]$. One can then locate the loop at ${\bf R}$ by selecting the
position of one of the particles, say $\br_1={\bf R}$, and write 
\begin{equation}
{\bf R}(s)={\bf R}+\lambda_\alpha{\bf X}(s),\:\:\; {\bf X}(0)={\bf X}(q) =0
\label{5.6}
\end{equation}
Then the particles are located at the points
\begin{equation}
\br_{k}={\bf R}+\lambda_{\alpha}{\bf X}_{k},\;\;\,\;{\bf X}_{k}={\bf X}(k-1)
\label{5.10}
\end{equation}
In (\ref{5.6}), ${\bf X}(s)$ is the shape of the loop $\cl$  at 
${\bf R}$, as $\b\xi (s)$ was describing the shape of a filament at
$\br$: it is again a dimensionless Brownian bridge starting and
returning to the origin within the time $q$. Its normalized 
Gaussian measure  $D({\bf X})$ is the composition of that of the $q$
open filaments that constitute the loop (see (\ref{4.6})) 
\begin{equation}
D({\bf X})= ( 2\pi q)^{3/2}\prod_{k=1}^{q}\frac{\exp\(-\frac{1}{2}({\bf X}_{k+1}-{\bf
X}_{k})^{2}\)}{(2\pi)^{3/2}}d{\bf X}_{2}\ldots d{\bf X}_{q}
D( \b\xi_1)\ldots D(\b\xi_q )\\
\label{5.7}
\end{equation}
where one sets ${\bf X}_{1}={\bf X}_{q+1}=0$. One can calculate its covariance from the
definitions (\ref{5.5})-(\ref{5.6})  
\begin{equation}
\int D({\bf X }) X_{\mu} (s_{1})
X_{\nu}(s_{2})=\delta_{\mu\nu}q\left[\min\left(\frac{s_{1}}{q},\frac{s_{2}}{q}\right)
-\frac{s_{1}}{q}\frac{s_{2}}{q}\right] 
\label{5.8}
\end{equation}
Thus one can think of a  loop $\cl$ either as a  set of particles
coordinates (as in (\ref{5.4})) or again as a single Brownian path, setting 
\begin{equation}
\cl=(\alpha,\;\;q,\;\;\bR,\;\;\bX(s),\;\;0\leq\;s\;\leq q)
\label{5.9}
\end{equation}
It is now possible, generalizing also (\ref{4.16})-(\ref{4.18}), to define
activities and interactions of loops. The interaction between two loops
$\cl_i,i=1,2,$ is the sum of those of the particles
constituting the two loops 
$$
V(\cl_1,\cl_2)=\sum_{k_{1}=1}^{q_{1}}\sum_{k_{2}=1}^{q_{2}}\int_0^1ds
V\(\br_{1,k_{1},k_{1}+1}(s)-\br_{2,k_{2},k_{2}+1}(s)\)
$$
\begin{equation}
=\int_0^{q_{1}}ds_{1} \, \int_0^{q_{2}}ds_{2} \,\delta({\tilde
  s}_1-{\tilde s}_2)V(\bR_1+\lambda_{\alpha_1}\bX_{1}(s_1)-\bR_2-\lambda_{\alpha_2}\bX_{2}(s_2))
\label{5.11}
\end{equation}
and for $n$ loops
\begin{equation}
U(\cl_1,\ldots,\cl_n)=\sum_{1=i<j}^n e_{\alpha_{i}} e_{\alpha_{j}}
V(\cl_i,\cl_j)
\label{5.12}
\end{equation}
The activity $z(\cl)$ of a loop incorporates the effects of quantum statistics,
the spin degeneracy as well as the internal interaction $U(\cl)$ of the
particles in the same loop
\begin{equation}
z(\cl)=(2s_\alpha+1)\frac{(\eta_\alpha)^{q-1}}{q}\;
\frac{\exp(\beta\mu_\alpha q)}{(2\pi q\lambda_\alpha^2)^{3/2}} \;\exp(-\beta U(\cl))
\label{5.13} 
\end{equation}
with  
\begin{eqnarray}
U(\cl)&=&\frac{e_\alpha^2}{2}\sum_{k\neq k^\prime}^q\int_0^q ds
V\(\br_{k,k+1}(s)-\br_{k^\prime,k^\prime+1}(s)\)\nonumber\\
&=&\frac{e_{\alpha}^{2}}{2}\int_0^{q}
ds_1\int_o^{q}ds_2\(1-\delta_{[s_{1}],[s_{2}]}\)\delta({\tilde
  s}_1-{\tilde s}_2)V(\lambda_\alpha(\bX(s_1)-\bX(s_2)))\nonumber\\
\label{5.14}
\end{eqnarray}
Clearly, all the above quantities are invariant under cyclic permutations of the particles in the
same loop. Notice also that particles in the same loop have the same charge; thus their mutual
Coulomb interactions are positive and  this implies the bound
\begin{equation}
|z(\cl)|\leq \frac{2s_{\alpha}+1}{q}\;\frac{\exp(\beta\mu_\alpha  q)}{(2\pi q\lambda^2_\alpha
)^{3/2}} \label{5.15}
\end{equation}
Considering now the phase space of loops together with the above
definitions, it 
can be shown (see section~\ref{subsec-5.1.4}) that the partition
function  can be written 
in the same form as (\ref{4.14}) ({\em the magic formula})
\begin{equation}
\Xi_\Lambda=\sum_{n=0}^\infty\frac{1}{n!}\int\prod_{i=1}^nd\cl_iz(\cl_i)
\exp(-\beta U(\cl_1,\ldots,\cl_n))
\label{5.16}
\end{equation}
where the phase space integration means here
\begin{equation}
\int d\cl\cdots=\sum_{\alpha=1}^{\cal S}\sum_{q=1}^\infty
\int_\Lambda d\bR\int D(\bX)\cdots
\label{5.16a}
\end{equation}
We call the system defined by the relations (\ref{5.8})-(\ref{5.16a})
the "system of loops"\footnote{Our definitions of the weight of a loop
(\ref{5.7}), the activity (\ref{5.13}) and the self-energy
(\ref{5.14}) differ slightly from those found in \cite{Cor96a}. It is
natural to define the loop measure (\ref{5.7}) again as that of the
Brownian bridge process in time $[0,q]$.}. It reduces to the system of
filaments (i.e. charges with Maxwell-Boltzmann statistics) if all
loops with more than one particle are omitted. This view of the
quantum Coulomb system as a gas of loops is well-suited in the low
density regime, when loops remain well localized objects.

\subsubsection{Loop and particle correlations}

\label{subsec-5.1.2}
It is clear that the statistical mechanical system of loops defined by
the above relations  has a classical structure. Therefore,
introducing the "$\delta$-function" which identifies two loops,
\begin{equation}
\delta(\cl_1,\cl_2)=\delta_{\alpha_1,\alpha_2}\delta_{q_1,q_2}\delta({\bf
  R}_1-{\bf R}_2)\delta(\bX_1,\bX_2)
\label{5.17}
\end{equation}
one can define the density of loops $\rho(\cl)$ and the two loop
distribution function $\rho(\cl_1,\cl_2)$ by the formulae
(\ref{4.20})-(\ref{4.21}) with $\cl$ replacing $\cf$.
 
In order to obtain the particle distributions from those of the loops,
some care has to be exercised because loops are constituted of several
particles. For a configuration of particles distributed on a set of
loops $\cl_i$, we can write the particle density and distributions as
a summation on the loop index $i$
\begin{equation}
\hat{\rho}(\alpha,\br)=\sum_{i}\delta_{\alpha_{i},\alpha}\sum_{k=1}^{q_i}\delta(\br_{(k,i)}
-\br)
\label{5.18}
\end{equation}
where, according to (\ref{5.10}),
$\br_{(k,i)}=\bR_i+\lambda_{\alpha_i}\bX_{i}(k-1),\;\;\;k=1,\ldots,q_i$,
are the positions of the particles belonging to the loop $i$. For
$\br_{a}\neq\br_{b}$ one writes also
\begin{eqnarray}
\hat{\rho}(\alpha_a,\br_a)\hat{\rho}(\alpha_b,\br_b)&=&
\sum_{i\neq j}\delta_{\alpha_{i},\alpha_{a}}
\delta_{\alpha_{j},\alpha_{b}}\sum_{k=1}^{q_i}\sum_{k^\prime=1}^{q_j}\delta(\br_{(k,i)}-\br_a)
\delta(\br_{(k^\prime,j)}-\br_b)\nonumber\\
&&\hspace{-.75in}+\delta_{\alpha_a,\alpha_b}\sum_i\delta_{\alpha_{i},\alpha_a}\sum_{k\neq k^\prime}^{q_{i}}
\delta(\br_{(k,i)}-\br_a)\delta(\br_{(k^\prime,i)}-\br_b)
\label{5.19}
\end{eqnarray}
In (\ref{5.18}) and (\ref{5.19}), $k$ ($k^\prime$) runs on particles
belonging to the same loop.  The first term in (\ref{5.19}) refers to
particles in different loops and the second term to particles within
the same loop. Then a comparison between the density of loops
$\hat{\rho}(\cl)=\sum_i\delta(\cl_i,\cl)$ and the particle density
(\ref{5.18}) shows that
\begin{equation}
\hat{\rho}(\alpha,\br)=\sum_{q=1}^\infty\int d\bR \int D(\bX)\sum_{k=1}^q \delta(\br_k-\br)
\hat{\rho}(\cl)
\label{5.20}
\end{equation}
Taking the average of the above relation on the ensemble of loops gives the number density
\begin{equation}
\rho(\alpha,\br)=\sum_{q=1}^\infty q\int D({\bf X})\rho(\alpha,q,\br,{\bf X})
\label{5.21}
\end{equation}
where $\rho (\alpha ,q,\br ,\bX ) = \rho (\cl)$ with $\cl =
(\alpha,q,\br ,\bX )$.  We have used the fact that $\rho(\cl)$ is
invariant under cyclic permutations of the $q$ particles in the
loop. The two particle distributions are obtained by averaging the
relation (\ref{5.19}) where we distinguish again between pairs
belonging to different loops or exchanged in the same loop
($\br_{a}\neq\br_{b}$)
\begin{equation}
\rho(\alpha_a,\br_a,\alpha_b,\br_b)=\rho^{(nex)}(\alpha_a,\br_a,\alpha_b,\br_b)+
\delta_{\alpha_a,\alpha_b}\rho^{(ex)}(\alpha_a,\br_a,\br_b)
\label{5.22}
\end{equation}
with
\begin{eqnarray} \label{5.23}
&&\rho^{(nex)}(\alpha_a,\br_a,\alpha_b,\br_b)=\nn \\
&& \hspace{-1in}\sum_{q_{a}=1}^\infty\sum_{q_{b}=1}^\infty
q_a q_b\int D({\bf X}_a)\int D({\bf X}_a)\rho(\alpha_a,q_a,\br_a,{\bf X}_a;\alpha_b,q_b,\br_b,{\bf
X}_b)
\end{eqnarray}
and
\begin{equation}
\rho^{(ex)}(\alpha_a,\br_a,\br_b)=\sum_{q=2}^{\infty}\int d\bR \int D(\bX)
\sum_{k\neq k^\prime}^{q}\delta(\br_k-\br_a)\delta(\br_{k^{\prime}}-\br_b)\rho(\alpha_{a},q, \bR,\bX)
\label{5.23a}
\end{equation}
For a translation invariant state ($\rho(\alpha,q, \bR,\bX)$ independent of $\bR$), one has, using
(\ref{5.10})
\begin{eqnarray}
\rho^{(ex)}(\alpha_a,\br_a,\br_b)&=&\sum_{q=2}^\infty\int D(\bX)\sum_{k\neq k^\prime}^{q}
\delta(\lambda_{\alpha_{a}}(\bX_{k}-\bX_{k^{\prime}})+\br_{b}-\br_{a})\rho(\alpha_{a},q,\bX)
\nonumber\\
&=&\sum_{q=2}^\infty\int D(\bX)q\sum_{k=2}^{q}
\delta(\lambda_{\alpha_{a}}\bX_{k}+\br_{b}-\br_{a})\rho(\alpha_{a},q,\bX)
\label{5.23b}
\end{eqnarray}
The second line follows again from the invariance of $\rho(\alpha_{a},q,\bX)$ under cyclic
permutations of the particles in the loop.

One can also write the corresponding similar representations for the truncated distributions.

\subsubsection{Loop interactions}

The same fundamental observation made for the system of filaments remains valid here: 
the interaction
(\ref{5.11}) is not the standard electrostatic Coulomb energy of two charged loops, which would be
\begin{equation}
V_{{\rm cl}}(\cl_1,\cl_2)=\int_0^{q_{1}}ds_1\int_0^{q_{2}}ds_2
V({\bf R}_1+\lambda_{\alpha_{1}}{\bf X}_1(s_1)-{\bf R}_2-\lambda_{\alpha_{2}}{\bf X}_2(s_2))
\label{5.25}
\end{equation}
and the discussion following (\ref{4.25}) can be reproduced word by word. In particular, the Coulomb
potential has the multipolar expansion
$$
V({\bf R}_1+\lambda_{\alpha_{1}}{\bf X}_1(s_1)-{\bf R}_2-\lambda_{\alpha_{2}}{\bf
X}_2(s_2))= V({\bf R}_1-{\bf R}_2)+
$$
\begin{equation}
M_1(s_1)V({\bf R}_1-{\bf R}_2)+ M_2(s_2)V({\bf  R}_1-{\bf R}_2)+
M_1(s_1)M_2(s_2)V({\bf R}_1-{\bf R}_2)
\label{5.26}
\end{equation}
with now
\begin{equation}
M_i(s)=\sum_{k=1}^{\infty}\frac{(\lambda_{\alpha_{i}}{\bf X}_{i}(s)\cdot\nabla_{{\bf
R}_i})^k}{k!},\;\;\;\;i=1,2 \label{5.27}
\end{equation}
and  hence the loop interaction can be decomposed into its charge-charge, charge-multipole and
multipole-multipole components
\begin{equation}
V(\cl_1,\cl_2)=V_{cc}(\cl_1,\cl_2)+V_{cm}(\cl_1,\cl_2)+V_{mc}(\cl_1,\cl_2)+V_{mm}(\cl_1,\cl_2)
\label{5.28}
\end{equation}
\begin{eqnarray}
V_{cc}(\cl_1,\cl_2)&=&q_1 q_2 V({\bf R}_1-{\bf R}_2)\label{5.29}\\
V_{cm}(\cl_1,\cl_2)&=&q_1\int_0^{q_{2}}ds_2 M_2(s_2)V({\bf  R}_1-{\bf R}_2)\label{5.30}\\
V_{mc}(\cl_1,\cl_2)&=&q_2\int_0^{q_{1}}ds_1 M_1(s_1)V( {\bf R}_1- {\bf R}_2)\label{5.31}\\
V_{mm}(\cl_1,\cl_2)&=&\int_0^{q_{1}}ds_1\int_0^{q_{2}}ds_2\delta(\tilde{s}_1-\tilde{s_2})
M_1(s_1)M_2(s_2)V({\bf R}_1-{\bf R}_2)\nonumber\\
\label{5.32}
\end{eqnarray}

The relevant quantity which will again embody the long range effects is the difference between
the  quantum and classical interaction 
\begin{equation}
W(\cl_1,\cl_2)=e_{\alpha_{1}}e_{\alpha_{2}}(V(\cl_1,\cl_2)-V_{{\rm cl}}(\cl_1,\cl_2))
\label{5.33}
\end{equation}
It depends only on the multipole-multipole part of the interaction as in (\ref{4.30}) 
\begin{eqnarray}
W(\cl_1,\cl_2)&=&e_{\alpha_{1}}e_{\alpha_{2}}
\int_0^{q_{1}}ds_1\int_0^{q_{2}}ds_2(\delta(\tilde{s}_1-\tilde{s_2})-1)
\nn \\
&&\hspace{.5in}\times
M_1(s_1)M_2(s_2)V({\bf R}_1-{\bf R}_2)
\label{5.34}
\end{eqnarray}
and has the asymptotic dipolar character (\ref{4.31}) as $|{\bf R}_1-{\bf R}_2|\rightarrow\infty$\begin{eqnarray}
&&\hspace{-1in}W(\cl_1,\cl_2) \sim e_{\alpha_{1}}e_{\alpha_{2}}
\int_0^{q_{1}}ds_1\int_0^{q_{2}}ds_2\nn \\
&&\hspace{-1in}
\times (\delta(\tilde{s}_1-\tilde{s}_2)-1)
(\lambda_{\alpha_{1}}{\bf X}_1(s_1)\cdot\nabla_{{\bf {\bf R}}_{1}})(\lambda_{\alpha_{2}}
{\bf X}_{2}(s_{2})\cdot\nabla_{{\bf R}_{2}})
V({\bf R}_1-{\bf R}_2)
\label{5.35}
\end{eqnarray}

\subsubsection{Derivation of the magic formula }
\label{subsec-5.1.4}

We derive first the formula (\ref{5.16}) in the special case where there is
an external field $\phi (\br)$ and no two-body interaction. In this case the
magic formula is essentially the same as (\ref{4.6.6b}).

It is enough to consider just one species with charge $e_{\alpha }$
and spin $s_{\alpha }$ because the partition function factors.  In
terms of annihilation and creation operators $\hat{a}_{\sigma}(\br ),
\hat{a}_{\sigma}^{\ast}(\br )$ the many-body Hamiltonian for the external field
case is
\begin{equation}\label{5.1.4.0}
	H_{\alpha } = \sum _{\sigma} 
            \int \hat{a}_{\sigma}^{\ast}(\br ) h_{\alpha }
 \hat{a}_{\sigma}(\br ) \,d\br; 
\ \ \ 
	h_{\alpha } = 
	\frac{\hbar^{2}}{2m_{\alpha }}(-\Delta) + e_{\alpha } \phi 
\end{equation}
We begin with the averaged particle number 

\begin{eqnarray}\label{5.1.4.1}
\langle N\rangle&=&\frac{d}{d(\beta \mu) } \ln \Xi_\Lambda =
\frac{1}{ \Xi_\Lambda}
\sum _{\sigma} \int \,d\br \,
\mbox{Tr}  \left(\hat{a}_{\sigma}^{\ast } (\br ) \hat{a}_{\sigma}(\br ) 
e^{-\beta( H_{\alpha }-\mu N)} \right)\nonumber\\  
N&=&\sum _{\sigma} 
            \int \hat{a}_{\sigma}^{\ast}(\br )
 \hat{a}_{\sigma}(\br ) \,d\br;\;\;\;\;\Xi_\Lambda=\mbox{Tr} 
e^{-\beta( H_{\alpha }-\mu N)}
\end{eqnarray}
The canonical commutation relations for the annihilation and creation
operators $\hat{a}_{\sigma}(\br ), \hat{a}_{\sigma}^{\ast } (\br )$ imply that
\begin{equation}\label{5.1.4.2}
	\hat{a}_{\sigma}(\br ) 
	e^{-\beta H_{\alpha }} 
=
	e^{-\beta H_{\alpha }} \int 
	e^{-\beta h_{\alpha }} (\br ,\br  ')\hat{a}_{\sigma}(\br ') \, d\br '\nn \\
\end{equation}
where $\exp (- \beta h_{\alpha } )(\br ,\br ' )$ is the kernel of the
operator $\exp (-\beta h_{\alpha })$. Inserting this relation in
(\ref{5.1.4.1}), using the cyclicity of the trace and the canonical commutation
relations gives \begin{eqnarray}\label{5.1.4.5a}
	\frac{d}{d(\beta \mu) } \ln \Xi_\Lambda 
&=&
	\sum _{q=1}^{\infty } \eta^{q-1}
	\mbox{Tr}  e^{-q \beta [h_{\alpha } - \mu ]}
\end{eqnarray}
where $\mbox{Tr}$ is now the one-particle trace  and it
includes the sum over spin. In the case of zero external field this
is the standard formula for the density of an ideal Bose or
Fermi gas by summing the geometric series.

Integrating (\ref{5.1.4.5a})  from $-\infty$ to $\beta\mu$ yields
\begin{equation}\label{5.1.4.6}
	\Xi_\Lambda 
=
	\exp \left(
	\sum _{q=1}^{\infty } \frac{\eta^{q-1}}{q}
	\mbox{Tr} e^{- q \beta [h_{\alpha } - \mu ]}
	\right)
\end{equation}
If we replace $q$ in the exponent by $\int_{0}^{q}\dots ds$ and
time-order the exponential then this formula and our derivation remain
valid for a periodic time-dependent external field $\phi (\br ,s)$, and
we obtain (\ref{4.6.6b}).

The ``magic formula'' in the case of external field, no two-body
interaction and one particle species follows by substituting in the Feynman-Kac
representation \begin{eqnarray}\label{5.1.4.5b}
	\sum _{q=1}^{\infty } \frac{\eta^{q-1}}{q}
	\mbox{Tr}  e^{- \beta\int_{0}^{q} [h_{\alpha }(s) - \mu ] \,ds}
&=&
	\int d\cl \, z(\cl) e^{-\beta e_{\alpha}\phi (\cl)}
\nn \\
	\phi (\cl) = \int_{0}^{q} \phi(\bR (s),s)  \,ds
\end{eqnarray}
where we use the notation described in subsection~\ref{subsec-5.1.1}.

The ``magic formula'' (\ref{5.16}) in the case of two-body
interactions follows by integrating over imaginary time-dependent
fields $i\phi (\br ,s)$ with a Gaussian measure.  According to our
discussion of the quantum Sine-Gordon transform in
section~\ref{sec-4.6} this reconstructs the two-body interaction with
self-energies. The formulas (\ref{5.13},\ref{5.14}) follow by
absorbing the self-energies into the activities.

\subsection{The loop Mayer expansion}
\label{sec-5.2}

\subsubsection{The loop Ursell function}

Since the system of loops has a classical structure, it can be treated by the
well known methods of classical statistical mechanics. In this section we apply
the standard techniques of the virial expansion to the two-point Ursell function
$h(\cl_a,\cl_b)$ together with the partial resummations needed to deal with the
long range of the Coulomb potential. 

The Ursell function is defined in terms of
the two-loops truncated distribution $\rho_T(\cl_a,\cl_b)$ by
\begin{equation}
\rho_T(\cl_a,\cl_b)=\rho(\cl_a)\rho(\cl_b)h(\cl_a,\cl_b)
\label{5.36}
\end{equation}
Although loops are extended objects, we still call them "points" and represent
them graphically by points in the diagrammatic language. The non-integrated loop
variables $\cl_a$, $\cl_b$ are called root points. We recall that the Ursell
function has the following simple loop-density expansion \cite{HaMc76}
\begin{equation}
h(\cl_a,\cl_b)=\sum_\Gamma\frac{1}{S_\Gamma}\int\prod_{n=1}^Nd\cl_n\rho(\cl_n)
[\prod f]_\Gamma
\label{5.37}
\end{equation}
In (\ref{5.37}), the sum runs over all unlabeled topologically different connected
diagrams $\Gamma$ with two root points $\cl_a$, $\cl_b$ and $N$ internal points 
$N=0,1,2,\ldots$, without articulations (an articulation point is such that,
when removed, the diagram splits into two pieces, at least one of which is
disconnected from any root point). To each bond $(\cl_i, \cl_j)$ in the diagram
is associated the Mayer factor (called f-bond)
\begin{equation}
f(\cl_i, \cl_j)=\exp(-\beta_{i,j}V(\cl_i, \cl_j))-1,\;\,\;\;\beta_{i,j}=\beta
e_{\alpha_{i}}e_{\alpha_{j}}
\label{5.38}
\end{equation}
and $[\prod f]_\Gamma$ is the product of all f-bonds in the $\Gamma$-diagram;
the symmetry factor $S_\Gamma$ is the number of permutations of internal points
$\cl_n$ that leave this product invariant. 

It is clear from (\ref{5.28}) and (\ref{5.29}) that the loop potential
$V(\cl_i, \cl_j)$ is Coulombic at large distances, so that the bond
(\ref{5.38}) is not integrable. In the classical case it is known that
by resumming appropriate classes of graphs (the convolution chains
leading to the Debye-H\"{u}ckel potential), one can introduce new
bonds and diagrams (the prototype diagrams) that are free of
divergences. One applies here the same procedure to the quantum gas
in the loop formalism. The main guidance comes from the fact that the
pure monopole part (\ref{5.29}) would lead to Debye screening.
Therefore the subsequent operations aim at disentangling the monopole
from the multipolar effects, and then extracting as much classical
screening as possible in order to reduce the range of the
interaction. When all these steps are completed, no bond will decay
slower than $|\br|^{-3}$, thus converting the gas of Coulombic loops
into a certain gas of dipoles (together with higher order multipoles).

\subsubsection{Summing the Coulomb chains} \label{subsec-5.2.2}

\begin{figure}
\psfig{figure=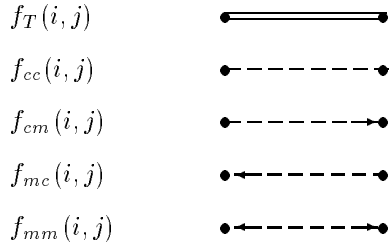}
\caption{Graphical definitions for bonds}
\label{fig2}
\end{figure}

The basic f-bond (\ref{5.38}) is split into the sum of five contributions
according to
\begin{equation}
f=f_T+f_{cc}+f_{cm}+f_{mc}+f_{mm}
\label{5.39}
\end{equation}
with the following analytical and graphical definitions (Figure~\ref{fig2}) (abbreviating\\
$f(\cl_i,\cl_j)=f(i,j)$) 
\begin{eqnarray}
f_T(i,j)&=&f(i,j)+\beta_{i,j}V(i,j)
\label{5.40}\\
f_{cc}(i,j)&=&-\beta_{i,j}V_{cc}(i,j)
\label{5.41}\\
f_{cm}(i,j)&=&-\beta_{i,j}V_{cm}(i,j)
\label{5.42}\\
f_{mc}(i,j)&=&-\beta_{i,j}V_{mc}(i,j)
\label{5.43}\\
f_{mm}(i,j)&=&-\beta_{i,j}V_{mm}(i,j)
\label{5.44}
\end{eqnarray}

\begin{figure}
\psfig{figure=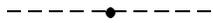}
\caption{A Coulomb point}
\label{fig3}
\end{figure}

\noindent 
Then, the Ursell function is given by the same formula as (\ref{5.37}), summing
now diagrams $\tilde{\Gamma}$ with the same topological structure as the
$\Gamma$-diagrams, but where the bonds can be any one of the quantities (\ref{5.40})-(\ref{5.44}).
Points without arrows in a  $\tilde{\Gamma}$-diagram and to which are attached only two dashed lines
are called Coulomb points (Figure~\ref{fig3}). A Coulomb point involves only interactions with the
total charge $qe_\alpha$ of the loop (considered as a point like object) without
multipolar effects.

\begin{figure}
\psfig{figure=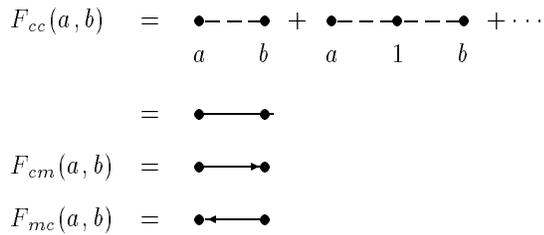}
\caption{Resummed Coulomb chains}
\label{fig4}
\end{figure}

\begin{figure}
\psfig{figure=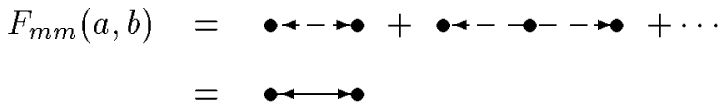}
\caption{Multipole-multipole resummed bond}
\label{fig5}
\end{figure}

A Coulomb chain is a convolution chain where all intermediate points are
Coulomb points. There are four types of Coulomb chains associated with  the
four bonds  (\ref{5.41})-(\ref{5.44}) (Coulomb chains without intermediate
points are defined as the bonds themselves). As in the classical theory, we
sum all Coulomb chains of a given type, and denote these sums $F_{cc},F_{cm},
F_{mc},F_{mm}$ (Figures~\ref{fig4} and \ref{fig5}). One has
\begin{eqnarray}
F_{cc}(\cl_a,\cl_b)&=&f_{cc}(\cl_a,\cl_b)\nonumber\\
&+&\sum_{N=1}^\infty\int\prod_{i=1}^{N}d\cl_i\rho(\cl_i)f_{cc}(\cl_a,\cl_1)f_{cc}(\cl_1,\cl_2)
\cdots f_{cc}(\cl_N,\cl_b)\nonumber\\
\label{5.44a}
\end{eqnarray}
Since by (\ref{5.29}) $f_{cc}(\cl_i,\cl_j)=-\beta_{i,j}q_iq_jV({\bf
R}_i-{\bf R}_j)$ does not depend on the configurations ${\bf X}_i,{\bf X}_j$ of the loops, the
summation on the internal variables can immediately be performed, giving a factor
$\sum_{\alpha}\sum_{q}\int D({\bf X})q^{2}e_{\alpha}^{2}\rho(\alpha,q,{\bf X})$ for each
intermediate point in (\ref{5.44a}). Because of translation invariance the
loop-density $\rho(\cl)=\rho(\alpha,q,{\bf X})$ does not depend on the location ${\bf R}$ of the
loop, therefore the multiple convolution (\ref{5.44a}) in the ${\bf
R}$ variables sum up as in the classical chains to    
\begin{equation}
F_{cc}(a,b)=-\beta_{a,b}q_{a}q_{b}\Phi({\bf R}_a-{\bf R}_b)
\label{5.45}
\end{equation}
where $\Phi({\bf R})$ is a Debye-like potential
\begin{equation}
\Phi({\bf R})=\frac{\exp(-\kappa |{\bf R}|)}{|{\bf R}|}
\label{5.46}
\end{equation}
with inverse screening length $\kappa$ given by
\begin{equation}
\kappa^2=4\pi\beta \sum_{\alpha}\sum_{q}\int D({\bf X})q^{2}e_{\alpha}^{2}\rho(\alpha,q,{\bf X})
\label{5.47}
\end{equation}
From the definitions (\ref{5.30})-(\ref{5.31}) and (\ref{5.42})-(\ref{5.43}),
the sum of Coulomb chains corresponding to the charge-multipole bonds are
obtained from (\ref{5.45}) by application of the multipolar operator (\ref{5.27}) on
$\Phi({\bf R}_1-{\bf R}_2)$ (Figure~\ref{fig4}) 
\begin{eqnarray}
F_{cm}(1,2)=-\beta_{1,2}q_1\int_0^{q_{2}}ds_2M_2(s_2)\Phi
({\bf R}_1-{\bf R}_2)\label{5.48}\\
F_{mc}(1,2)=-\beta_{1,2}q_2\int_0^{q_{1}}ds_1M_1(s_1)\Phi
({\bf R}_1-{\bf R}_2)
\label{5.48a}
\end{eqnarray}

Obviously the resummed bonds (\ref{5.48}) and (\ref{5.48a}) have a short range.
Let us have a closer look at the multipole-multipole resummed bond 
(Figure~\ref{fig5}) 
\begin{eqnarray}
F_{mm} (1,2)&=&f_{mm}(1,2)+f_{mc}\star f_{cm}(1,2)+
f_{mc}\star f_{cc}\star f_{cm}(1,2)+\cdots\nonumber\\
&&\hspace{-.5in}=-\beta_{1,2}V_{mm}(1,2)\label{5.49}\\
&&\hspace{-.5in}-\beta_{1,2}\int_0^{q_{1}}ds_1M_1(s_1)
\int_0^{q_{2}}ds_2M_2(s_2)(\Phi({\bf R}_1-{\bf R}_2)-V({\bf R}_1-{\bf R}_2))
\nonumber\\
&&\hspace{-.5in}=-\beta W(1,2)-\beta_{1,2}\int_0^{q_{1}}ds_1M_1(s_1)
\int_0^{q_{2}}ds_2M_2(s_2)\Phi({\bf R}_1-{\bf R}_2)\nonumber\\
\label{5.50}
\end{eqnarray}

The first term in (\ref{5.49}) is the multipole-multipole quantum potential
that has only "equal time" contributions, whereas the rest of the chain can be
obtained by an application of multipolar operators at both ends (omitting
here the term with no intermediate Coulomb points). The expression (\ref{5.50}) results from
(\ref{5.34}), showing the asymptotic dipolar character of $F_{mm}$ (see (\ref{5.35})).  Notice
that the sum of the resummed bonds equals
\begin{eqnarray}
F(1,2)&\equiv& F_{cc}(1,2)+ F_{cm}(1,2)+ F_{mc}(1,2)+ F_{mm}(1,2)\nonumber\\
&=&-\beta W(1,2)-\beta_{1,2}\Phi_{{\rm cl}}(1,2) 
\label{5.51}
\end{eqnarray}
where $\Phi_{{\rm cl}}(1,2)$ is a short-range interaction defined as in (\ref{5.25})
\begin{equation}
\Phi_{{\rm cl}}(\cl_1,\cl_2)=\int_0^{q_1}ds_1\int_0^{q_2}ds_2
\Phi({\bf R}_1+\lambda_{\alpha_{1}}{\bf X}_1(s_1)-{\bf R}_2-\lambda_{\alpha_{2}}{\bf X}_2(s_2))
\label{5.52}
\end{equation}
representing the screened classical interaction between loops.

The screening length $\kappa^{-1}$ in (\ref{5.47}) can be written in a
more familiar form by noting the identity coming from (\ref{5.23b}) 
\begin{equation}
\int d\br \rho^{(ex)}_{T}(\alpha,\br,{\bf 0})=\sum_{q=2}^{\infty}\int D({\bf X})q(q-1)
\rho(\alpha,q,{\bf X})
\label{5.52a}
\end{equation}
hence, using also (\ref{5.21}) 
\begin{equation}
\kappa^{2}=4\pi\beta\sum_\alpha e_\alpha^2\(\rho_\alpha+\int d\br
\rho^{(ex)}_{T}(\alpha,\br,{\bf 0})\)
\label{5.52b}
\end{equation}
An alternative expression for  $\kappa^{2}$ follows from the quantum charge sum rule
(\ref{2.2.5a})
\begin{equation}
\int d\br c_{T}(\br, {\bf 0})=\sum_{\alpha_{1}\alpha_{2}} e_{\alpha_{1}}e_{\alpha_{2}}\int d\br
\left(\rho_{T}(\alpha_{1},\br,\alpha_{2},{\bf
0})+\rho_{\alpha_{1}}\delta_{\alpha_{1}\alpha_{2}}\delta (\br)\right)=0
\label{5.52bb}
\end{equation}
which leads to the following relation, once expressed in terms of the
loop distributions by (\ref{5.22})
\begin{equation}
\sum_\alpha e_\alpha^2\(\rho_\alpha+\int d\br
\rho^{(ex)}_{T}(\alpha,\br,{\bf 0})\)=-\sum_{\alpha_{1},\alpha_{2}}e_{\alpha_{1}}
e_{\alpha_{2}}\int d\br\rho_{T}^{(nex)}(\alpha_1,\br,\alpha_2,{\bf 0})
\label{5.52c} 
\end{equation}
and thus
\begin{equation}
\kappa^{2}=-4\pi\beta \sum_{\alpha_{1},\alpha_{2}}e_{\alpha_{1}}
e_{\alpha_{2}}\int d\br\rho_{T}^{(nex)}(\alpha_1,\br,\alpha_2,{\bf 0})
\label{5.52d} 
\end{equation}
In the integrals in (\ref{5.52c}) and (\ref{5.52d}), the contribution
of coincident points is not included.  On physically reasonable
grounds, $\kappa ^{2}$ is indeed a positive quantity.  One sees from
(\ref{5.52b}) that $\kappa ^{2}$ reduces to the classical Debye value
when exchange effects can be neglected, and it will coincide with the
RPA value at high density (section~\ref{subsec-6.1.4}), both
quantities being positive.  Moreover, the screening cloud
$\sum_{\alpha_{2}} e_{\alpha_{2}}\int
d\br\rho_{T}^{(nex)}(\alpha_1,\br,\alpha_2,{\bf 0})$ of charges of
different species around the charge $e_{\alpha_{1}}$ should have the
sign opposite to $e_{\alpha_{1}}$, implying that the right hand side
of (\ref{5.52d}) is positive.

\subsubsection{The prototype diagrams} \label{subsec-5.2.3}

The $\tilde{\Gamma}$-diagrams can be divided into classes such that all diagrams in a class
lead to the same so-called prototype $\Pi$-diagram when all the Coulomb points are integrated out.
Points in a  $\tilde{\Gamma}$-diagram that survive as points in a $\Pi$-diagram are either
convolutions which do not involve Coulomb points, or points linked to three or more other points of
$\tilde{\Gamma}$. There will be four different bonds (called F-bonds) in
prototype diagrams:
\begin{description}
\item[(i)] the screened charge-charge bond (\ref{5.45}) and the two charge-multipole bonds
(\ref{5.48})-(\ref{5.48a}). Since Coulomb points cannot occur in prototype graphs, the use of these
bonds is subjected to the excluded convolution rule: in a $\Pi$-diagram, convolutions $F_{cc}\star
F_{cc},F_{mc}\star F_{cc}, F_{cc}\star F_{cm},F_{mc}\star F_{cm}$ are forbidden.
\item[(ii)] a bound $F_l$ that incorporates all the bonds that are not taken into account in
({\bf i}).  \end{description}
Let us now construct the bond  $F_l$. We distinguish two cases: 
\begin{description}
\item[(a)] $f_T$ and $F_{mm}$ occur as individual bonds in $\Pi$-diagrams since they are never
attached to Coulomb points in $\tilde{\Gamma}$-diagrams,
\item[(b)] single bonds in a $\Pi$-diagram attached to points that had more than two links in a 
$\tilde{\Gamma}$-diagram, as the result of the suppression of Coulomb points.
\end{description}

\begin{figure}
\psfig{figure=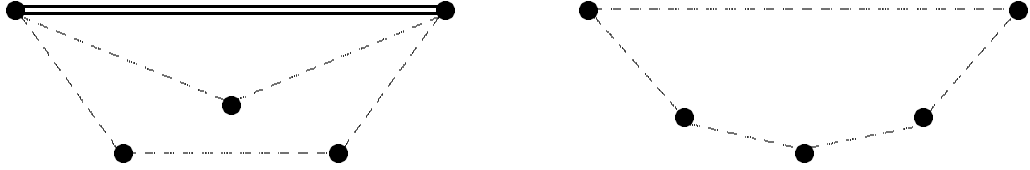}
\caption{f-bonds with Coulomb chains in parallel}
\label{fig6}
\end{figure}

\begin{figure}
\psfig{figure=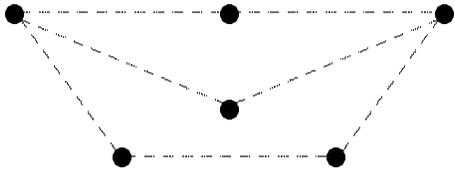}
\caption{Coulomb chains in parallel}
\label{fig7}
\end{figure}

The latter situation is obtained when two points in a $\tilde{\Gamma}$-diagram were linked  
\begin{description}
\item[(b.1)] by any one of the f-bonds together with Coulomb chains in parallel
(Figure~\ref{fig6})
\item[(b.2)] by several Coulomb chains (Figure~\ref{fig7}).
\end{description}

\begin{figure}
\psfig{figure=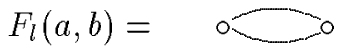}
\caption{Graphical symbol for right hand side of (\ref{5.55}) bond}
\label{fig8}
\end{figure}

The Coulomb chains occurring in ({\bf b.1}) and ({\bf b.2}) must have
intermediate points since there is at most one f-bond between two
points in a $\tilde{\Gamma}$-diagram. The sum of such Coulomb
chains,$F_{ch}$, is obtained by removing from $F$ (\ref{5.51}) the
single bonds (\ref{5.41})-(\ref{5.44}). Thus, with (\ref{5.28})
\begin{equation}
F_{ch}(i,j)=F(i,j)+\beta_{i,j} V(i,j)=-\beta W(i,j)+\beta_{i,j}(V(i,j)-\Phi_{{\rm cl}}(i,j))
\label{5.53}
\end{equation}
The contributions described in ({\bf a}) and ({\bf b}) sum up to
\begin{equation}
F_l=f_T +F_{mm}+f(\exp (F_{ch})-1)+\exp (F_{ch})-1-F_{ch}
\label{5.54}
\end{equation}
The last two parts of (\ref{5.54}) correspond to the contributions
({\bf b.1}) and ({\bf b.2}) (for the combinatorial aspects, see
\cite{Cor96a}. Combining the definitions (\ref{5.38}), (\ref{5.40}),
(\ref{5.51}) and (\ref{5.53}) in (\ref{5.54}) leads to the final
result (Figure~\ref{fig8})
\begin{equation}
F_l(i,j)=\exp(-\beta W(i,j)-\beta_{i,j}\Phi_{{\rm cl}}(i,j) )-1 -F_{cc}(i,j)-F_{cm}(i,j)-F_{mc}(i,j)
\label{5.55}
\end{equation}

Eventually the loop-Ursell function is given by a sum of prototype graphs $\Pi$
\begin{equation}
h(\cl_1,\cl_2)=\sum_\Pi\frac{1}{S_{\Pi}}\int\prod_{n=1}^Nd\cl_n\rho(\cl_n)[\prod F]_{\Pi}
\label{5.56}
\end{equation}
with F-bonds $F_{cc}, F_{mc},F_{cm}$ and $F_{l}$. The $\Pi$-diagrams have the same structure as the
$\Gamma$-diagrams, with the additional excluded convolution rule formulated in ({\bf i}).

All F-bonds are rapidly decreasing except $F_l$. Obviously the
asymptotic behavior of $F_l$ is dominated by that of $W(\cl_1,\cl_2)$
\begin{equation}F_l(\cl_1,\cl_2) \sim\exp(-\beta W(\cl_1,\cl_2))-1\sim -\beta
W(\cl_1,\cl_2),\;\;\;|{\bf R}_1-{\bf R}_2| \rightarrow\infty
\label{5.57} \end{equation}which is of dipole type as shown by (\ref{5.35}). It
is interesting to note that each $F_l$ bond has the same form (up to
short range contributions) as that found in the model of
section ~\ref{sec-4.3} when only two quantum mechanical
charges are present. The long range part $W$ depends solely on the
intrinsic quantum fluctuations of the two particles (or the two
loops), and not on the temperature and density of the state.

There are of course several ways of reorganizing the diagrams with
bare Coulomb bonds (\ref{5.38}) into prototype graphs by partial
resummations. Other resummations may be more adequate depending on the
density and temperature regime and the physical quantity to be
computed. The present reorganization appears to be optimal in the
following sense. The remaining long range part has been isolated in
the single bond $F_{l}$.  The decay cannot be improved any more by
chain resummations. By (\ref{5.57}) and (\ref{5.35}), the longest
range part of such chains would arise from convolutions of $W$-bonds,
but such convolutions keep their dipolar character \cite{Cor96b}; the
same is true for classical dipoles, see for instance
\cite{HoSt74}. Moreover, our prototype diagrams are the natural
generalization of those used in the classical case. Indeed, if we set
the de Broglie lengths equal to zero, the multipolar operator
(\ref{5.27}) vanishes as well as $W$, $F_{cm}$ and $F_{mc}$. The
effects of the quantum statistics disappear, and both $\Phi$
(\ref{5.46}) and $\Phi_{{\rm cl}}$ (\ref{5.52}) reduce to the usual
Debye potential $\Phi_{D}$. Hence one finds that the only surviving
bonds are
$$
\lim_{\hbar\rightarrow 0}F_{cc}(1,2)=-\beta_{1,2}\Phi_{D}(1,2) 
$$
and from (\ref{5.55})
$$
\lim_{\hbar\rightarrow 0}F_{l}=e^{-\beta_{1,2}\Phi_{D}(1,2) }-1+\beta_{1,2}\Phi_{D}(1,2) 
$$
They are precisely the two bonds occurring in the Meeron theory of the classical plasma
\cite{Mee58,Mee61}.

\subsubsection{Integrability of prototype diagrams} 
\label{subsec-5.2.4}

Let us present some crude arguments for the integrability of
$\Pi$-diagrams; they can be substantiated by the more thorough
analysis of \cite{Cor96a}, but a proof is still missing. Each internal
point involves a density factor $\rho(\cl)=\rho(\alpha,q,\bX)$ which
depends only on the internal variables of the loop and not on its
location in space because of translation invariance.  Integration on
an internal point $\cl$ will be of the form \begin{equation}\int d{\bf
R}\sum_\alpha\sum_{q=1}^\infty\int D(\bX) \rho(\alpha,q,\bX)
G(\alpha,q,{\bf R},\bX\;;\ldots) \label{5.58} \end{equation}where
$G(\cl\;;\ldots)$ denotes the $\cl$-dependence of the bonds attached
to this point. It can be argued that the loop-density $\rho(\cl)$ is
bounded in the shape variables $\bX$ and inherits a prefactor
$\exp(\beta\mu_\alpha q)$ from the activity of the loop (see
(\ref{5.13}), (\ref{5.15}) and (\ref{5.60}) below).  Thus for a
sufficiently negative chemical potential this factor should insure
convergence for the $q$ summation on the size of the loops. If these
conditions are met, the integrability of $\Pi$-diagrams, when the
distance between the loops becomes large, is determined by the the
decay of the F-bonds. Because of the absence of articulation points,
any cluster of points is linked by at least two bonds to the remaining
points of the diagram. Hence the product $[\prod F]_{\Pi}$, once
integrated on the internal variables, decays at least as $|{\bf
R}|^{-6}$ in the loop distances, showing the integrability for the
whole diagram. One should add that the splitting (\ref{5.39}) of the
original f-bond may introduce spurious non-integrable singularities at
the origin in some diagrams. These singularities will cancel out when
suitably collecting together such dangerous diagrams (the f-bond
(\ref{5.38}) gives a finite contribution at the origin because of the
smoothening provided by the functional integration on the filaments).

To summarize the situation, once the topological reduction to
prototype diagrams has been performed, the two-particle correlation is
reduced, in the sense of the formal diagrammatic cluster expansion
(\ref{5.56}), to a system that resembles a classical gas of multipoles
described in the loop formalism by the four types of bonds
$F_{cc},F_{mc},F_{cm}$ and $F_{l}$.

\subsubsection{The density expansion}
\label{subsec-5.2.5}

An important remark has to be made when one comes down to computations, in particular when
one seeks to obtain explicit low density expansions of physical quantities. The diagrammatic
representation (\ref{5.56}) of the loop Ursell function is a straightforward expansion in the
density of loops, but not in the particle densities $\rho_\alpha$ themselves, for two reasons:
\begin{itemize}
\item 
the resummed F-bonds depend on the particle densities through the density dependent screened
potential $\Phi$ (\ref{5.46}),
\item
the loop densities entering as weights of the internal points in $\Pi$-diagrams are not simply
proportional to the particle densities. 
\end{itemize}
The first point is not new and occurs of course already in the
diagrammatic treatment of the classical gas. The second point is more
delicate: the loop and particle densities are related by the formula
(\ref{5.21}); hence the particle-particle distributions (\ref{5.22})
calculated from (\ref{5.56}) have a very implicit dependence on the
$\rho_\alpha$. In principle this dependence can be extracted as
follows. According to the standard rules of the Mayer diagrammatics,
the loop density is represented by the activity expansion
\begin{equation}
\rho(\cl)=z(\cl)\sum_G\frac{1}{S_G}\int\prod_{n=1}^{N}d\cl_nz(\cl_n)[\prod f]_G
\label{5.60}
\end{equation}
The $G$-diagrams are defined as the $\Gamma$-diagrams (with one root
point) except that articulation points are allowed and weights are
given by activities. The integrals diverge in (\ref{5.60}) because of
the long range of the $f$-bonds, but after a suitable topological
reduction (resumming Coulomb chains), the representation of
$\rho(\cl)$ becomes a series of prototype graphs, each of them being
finite.  The procedure is similar to that given above for the Ursell
function and will not be described here in detail, see
\cite{Cor96a}. Once this is done, formula (\ref{5.21}) yields in
principle the particle densities as functions of the activities
\begin{equation}
\rho_\alpha=\rho_\alpha(z_1,\ldots,z_{\cal S}),\;\;\;\,\alpha=1,\ldots,{\cal S}
\label{5.61}
\end{equation}
Because of the overall neutrality $\sum_{\alpha=1}^{{\cal S}}e_\alpha\rho_\alpha=0$
the system  (\ref{5.61}) consists only of ${\cal S}-1$ independent relations, and so does not
determine the $z_\alpha$ uniquely. A unique determination of the $z_a$ is usually made by imposing
the convenient additional relation (already discussed in section~\ref{sec- neutrality})
\begin{equation}
\sum_{\alpha=1}^{{\cal S}}e_\alpha z_\alpha=0 
\label{5.62}
\end{equation}
Thus by inverting the system (\ref{5.61})-(\ref{5.62}) one can express
first $\rho(\cl)$ and then the particle-particle correlations in terms
of the $\rho_\alpha$. Explicit calculations in this formalism can of
course only be performed in the low density regime. Applications to
the determination of the asymptotic behavior of the correlations and
to the virial equation of state will be presented in the next chapter.

\def\({\left(}
\def\){\right)}
\def\br{{\bf r}}
\def\bx{\mbox{\boldmath $\xi$\unboldmath}}
\def\cf{{\cal F}}
\def\cl{{\cal L}}
\def\bk{{\bf k}}
\def\bX{{\bf X}}

\newpage \section{Correlations and the equation of state} \setcounter{page}{1}

\label{chapter-6}

\subsection{Asymptotic behavior of particle and charge correlations}
\label{sec-6.1}

The first application of the general formalism concerns the
determination of the large distance behavior of the correlations
without recourse to $\hbar$- expansions, including the quantum
statistics and using the exact Coulomb potential on all scales. The
calculation remains perturbative in the sense that we examine the
large distance behavior of each $\Pi$-diagram in the series
(\ref{5.56}) representing the Ursell function, but, as before, without
control on the sum of these series.

The semi-classical analysis of section~\ref{sec-4.3} has revealed that
the two particle correlation, at order $\hbar^{4}$, decay as $r^{-6}$
at large distance. We show that the same result holds for each of the
$\Pi$-diagrams contributing to the particle- particle correlation
$\rho_{T}(\alpha_{a},\br_{a},\alpha_{b},\br_{b})$, i.e.
$$
\rho_{T}(\alpha_{a},\br_{a},\alpha_{b},\br_{b})\sim 
\frac{A_{ab}(\beta,\{\rho_{\alpha}\})}{|\br_{a}-\br_{b}|^{6}}
$$
where the coefficient $A_{ab}(\beta,\{\rho_{\alpha}\})$ is the formal
sum of the asymptotic contributions of the $\Pi$-diagrams as
$|\br_{a}-\br_{b}|\to\infty$ at densities $\rho_{\alpha}$ and inverse
temperature $\beta$.  It is then possible to calculate the lowest
order contributions in the densities $\rho_{\alpha}$ to
$A_{ab}(\beta,\{\rho_{\alpha}\})$ leading to the explicit formula
(\ref{6.23}) below.  We will describe the general strategy and
illustrate some key points by explicit calculations, details can be
found in \cite{Cor96b}. A short review can be found in \cite{AlCo97}.

\subsubsection{Large distance behavior of the loop Ursell function}
\label{sec-6.1.1} 

The first task is to study the behavior of the Ursell function
(\ref{5.56}) in the space of loops as the loops are separated.  This
large distance behavior will be determined by that of the four bonds
$F_{cc},F_{cm},F_{mc}$ and $F_{l}$, once integration on the internal
loop variables has been performed in prototype graphs. Among these
bonds, only $F_{l}$ has the long range multipolar character
(\ref{5.57}) that may induce a slow decay of the graph as the two
loops are separated. Therefore one has to keep track of the effects of
$F_{l}$-bonds (and more specifically of the basic quantum interaction
$W$ (\ref{5.33})) in the convolutions occurring in prototype diagrams.
For this it is appropriate to recall some facts on the asymptotic form
of convolutions in three dimensional space.

The forthcoming analysis depends on the assumption (made throughout
the whole section \ref{sec-6.1}) that the leading asymptotic term has
a monotone decay, whose behavior is governed by that of the small wave
numbers ${\bf k}\sim 0$ of the Fourier transform.  Possible decay with
oscillation, due to singularities at non-zero values of ${\bf k}$, is
not considered. Such oscillating decays are not expected to occur at
positive temperature and low density.

Consider a (non-oscillating) function $g(\br)$ which decays as
$r=|\br|\rightarrow\infty$ as (up to a multiplicative constant)  
\begin{equation}
g(\br)\sim \frac{(\ln r)^{m}}{r^{n}},\;\;\;n,m\;\mbox{integers},\;\,\,n\geq 3,\;\;\;m\geq 0
\label{6.1b}
\end{equation}
This type of decay is reflected by the existence of a non-analytic part $\tilde{g}_{sing}({\bf k})$
in its Fourier transform $\tilde{g}({\bf k})$ as ${\bf k}\rightarrow 0$. One can split the low
${\bf k}$-expansion of $\tilde{g}({\bf k})$ into $\tilde{g}({\bf k})=\tilde{g}_{reg}({\bf
k})+\tilde{g}_{sing}({\bf k})$ where  $\tilde{g}_{reg}({\bf k})
=a_{0}+\sum_{\mu=1}^{3}a_{\mu}k_{\mu} +\sum_{\mu,\nu=1}^{3}a_{\mu\nu}k_{\mu}k_{\nu} +\cdots$ has a
Taylor expansion around ${\bf k}={\bf 0}$ and $\tilde{g}_{sing}({\bf k})$ behaves (up to
multiplicative constants) as \cite{GeSh64} 
\begin{equation}
\tilde{g}_{sing}({\bf k})\sim \left\{ \begin{array}{ll} |{\bf k}|^{n-3}(\ln|{\bf k}|)^{m+1},
& n\;\mbox{odd}\\
|{\bf k}|^{n-3}(\ln|{\bf k}|)^{m},  & n\;\mbox{even}\end{array}\right.,\;\;\,\,{\bf
k}\rightarrow 0  
\label{6.1a}
\end{equation}
The singular part of Fourier transforms of derivatives 
$\partial_{\mu_{1}}\cdots \partial_{\mu_{j}}g(\br) $ is\\ 
${ k}_{\mu_{1}}\cdots { k}_{\mu_{j}}\tilde{g}_{sing}({\bf k})$.
Since the Fourier transform of the convolution $(g_{1}\star g_{2})(\br)$ is the product 
$\tilde{g}_{1}({\bf k})\tilde{g}_{2}({\bf k})$, one deduces from the above correspondence rules
that for  $n_{1},n_{2}>3$ 
\begin{equation}
(g_{1}\star g_{2})(\br) \sim g_{1}(\br)\int d\br^{\prime}g_{2}(\br^{\prime})+ 
g_{2}(\br)\int d\br^{\prime} g_{1}(\br^{\prime}),\,\;\,r\rightarrow\infty
\label{6.3}
\end{equation}
In particular, if $n_{2}>n_{1}>3$, the convolution decays as the slowest of the two functions
\begin{equation}
(g_{1}\star g_{2})(\br) \sim g_{1}(\br)\int d\br^{\prime}
g_{2}({\br^{\prime}),\,\;\,r\rightarrow\infty }
\label{6.2}
\end{equation}
At the border $n_{1}=n_{2}=3$ of the integrability domain one has
\begin{equation}
(g_{1}\star g_{2})(\br)\sim \frac{(\ln r)^{m_{1}+m_{2}+1}}{r^{3}}
\label{6.4}
\end{equation}
One sees from (\ref{6.4}) that convolutions of $r^{-3}$ functions can generate in principle 
logarithmic factors in the power law decays.
To simplify the discussion we abbreviate (\ref{6.1b}) by $g(\br)\approx r^{-n}$, omitting from the
notation the possible logarithmic dependence. Since $F_{l}$ decays as $r^{-3}$, it is clear
from the above convolution rules that no $\Pi $-diagram (or subdiagram) decays slower than $\approx
r^{-3}$. The slowest possible decay comes then from diagrams where the two root points can be
disconnected by cutting a single $F_{l}$-bond: otherwise chains containing a $F_{l}$-bond will be
multiplied by another chain (or subdiagram) decaying at least as $\approx r^{-3}$ so the decay of
the whole diagram will be at least as $\approx r^{-6}$. To put this idea at work, one splits the
$F_{l}$-bond into two pieces
\begin{equation}
F_{l}=-\beta W+\tilde{F}_{l},\,\;\;\,\,\,\tilde{F}_{l}\equiv F_{l}+\beta W   
\label{6.6}                      
\end{equation}
and enlarge accordingly the set of $\Pi$-diagrams to $\tilde{\Pi}$-diagrams equipped with the new
bonds $-\beta W$ and $\tilde{F}_{l}$. One has from (\ref{5.57}) 
\begin{eqnarray}
\tilde{F}_{l}(1,2)&\sim& \exp(-\beta W(1,2))-1+\beta W(1,2)\sim\frac{1}{2}(\beta W(1,2))^{2}\nonumber\\
\mbox{and}\;\;\;\;\, (W(1,2))^{2}&\sim&\frac{C}{|{\bf R}_{1}-{\bf R}_{2}|^{6}},\;\;\;\;|{\bf
R}_{1}-{\bf R}_{2}|\rightarrow\infty 
\label{6.7}
\end{eqnarray}
thus $\tilde{F}_{l}$ is integrable and $W$ is the slowest bond in the new set of
$\tilde{\Pi}$-diagrams. Introduce now the subclass of $\tilde{\Pi}_{C}$-diagrams that remain connected
when one removes a $W$-bond, and denote $H(\cl_{a},\cl_{b})$ the value of the sum of these diagrams.
With this definition, the loop truncated distribution function (\ref{5.36}) can be expressed as an
exact Dyson equation in terms of convolution chains involving $H$ and $W$
\begin{eqnarray}
&&\rho_{T}(\cl_{a},\cl_{b}) = \rho(\cl_{a})\rho(\cl_{b})H(\cl_{a},\cl_{b})
-\beta \int d\cl_{1}d\cl_{1}^{\prime}G(\cl_{a},\cl_{1})
W(\cl_{1},\cl_{1}^{\prime})G(\cl_{1}^{\prime},\cl_{b})\label{6.9}\\
&&\hspace*{.25in}+\beta^{2}\int d\cl_{1}d\cl_{1}^{\prime}\int d\cl_{2}d\cl_{2}^{\prime}G(\cl_{a},\cl_{1})
W(\cl_{1},\cl_{1}^{\prime})G(\cl_{1}^{\prime},\cl_{2})
W(\cl_{2},\cl_{2}^{\prime})G(\cl_{2}^{\prime},\cl_{b}) + \cdots \label{6.10}
\end{eqnarray}
In (\ref{6.9})-(\ref{6.10}) we have set
\begin{equation}
G(\cl_{1},\cl_{2})=\rho(\cl_{1})\rho(\cl_{2})H(\cl_{1},\cl_{2})+\delta(\cl_{1},\cl_{2})
\rho(\cl_{1})
\label{6.11}
\end{equation}
where the term $\delta(\cl_{1},\cl_{2})\rho(\cl_{1})$ takes into account the case when
$W$-bonds are in direct convolution with themselves. By construction, no diagram in
$H(\cl_{a},\cl_{b})$ decays slower than $\approx |{\bf R}_{a}-{\bf R}_{b}|^{-6}$. Indeed
if a  $\tilde{\Pi}_{C}$-diagram contains no $W$-bonds, the slowest possible decay is determined by
that of $\tilde{F}_{l}$ according to (\ref{6.7}). If it contains a $W$-bond, this $W$ bond (or the
chain where it belongs to) is necessarily multiplied by another chain that decays not slower than
$\approx r^{-3}$, hence the overall decay of the diagram is at least $\approx r^{-6}$. In this way
we have isolated the origin of
 decays that are slower than $\approx r^{-6}$ in the chains of $W$-bonds (\ref{6.9})-(\ref{6.10}).
Since  (\ref{6.9}) contains in particular the term $\rho(\cl_{a})\rho(\cl_{b})W(\cl_{a},\cl_{b})$
the loop correlation decays indeed as $\approx r^{-3}$.

\subsubsection{Large distance behavior of the two-particle
correlation} \label{sec-6.1.2} In the loop formalism, the two-particle
distribution consists of the two terms displayed in
(\ref{5.22})-(\ref{5.23a}). The dominant part at large distances will
come from $\rho^{(nex)}$ involving particles belonging to different
loops. The contribution $\rho^{(ex)}$ of particles exchanged in the
same loop will be discussed at the end of this subsection. To obtain
$\rho^{(nex)}$ we have to integrate
$\rho_{T}(\cl_{a},\cl_{b})=\rho(\cl_{a})\rho(\cl_{b})h(\cl_{a},\cl_{b})$
over the internal variables of the root loops $\cl_{a},\cl_{b}$
according to (\ref{5.23}):
\begin{equation}
\rho^{(nex)}(\alpha_{a},\br_{a},\alpha_{b},\br_{b})=\sum_{q_{a}q_{b}}q_{a}q_{b}\int D({\bf X_{a}})
D({\bf X_{b}}) \rho(\cl_{a})\rho(\cl_{b})h(\cl_{a},\cl_{b})
\label{6.11a} 
\end{equation}
The point to be made
below is that, {\em once these integrations have been performed}, the asymptotic contribution of
the chains of $W$-bonds (\ref{6.9})-(\ref{6.10}) is strictly faster than $\approx |{\bf R}_{a}-{\bf
R}_{b}|^{-6}$. Hence the decay of $\rho^{(nex)}$ is entirely determined by that of the
$\tilde{\Pi}_{C}$-diagrams constituting the function $H$, i.e. $\approx \br^{-6}$ according to the
analysis of the preceding subsection.     

Consider for simplicity the diagram consisting of a single $W$-bond in  (\ref{6.9}): its
contribution to $\rho^{(nex)}(\br_{a},\br_{b})$ is 
\begin{equation}
-\beta\sum_{q_{a}=1}^{\infty}\sum_{q_{b}=1}^{\infty}q_{a}q_{b}\int D({\bf X}_{a})\int D({\bf
X}_{b}) \rho(\alpha_{a},q_{a},{\bf X}_{a})\rho(\alpha_{b},q_{b},{\bf X}_{b})W(\cl_{a},\cl_{b})
\label{6.12}
\end{equation}
The same mechanism that was put forward in the model of section~\ref{sec-4.3} to show that the
term linear in $W$ was in fact short-ranged works also in the present case. Since in a homogeneous
phase of the Coulomb gas the loop-density $\rho(\alpha,q,{\bf X})$ as well as the measure
$D({\bf X})$ are invariant under global rotations of the shape ${\bf X}$ of the loop, the expressions
\begin{equation}
\int D({\bf X})\rho(\alpha,q,{\bf X})\frac{({\bf X}(s)\cdot{\bf \nabla_{\bf
R}})^{k}}{k!},\;\,\,k=1,2,\ldots
\label{6.13}
\end{equation}
occurring in (\ref{6.12}) according to the multipole expansion (\ref{5.34}) of $W$ have to vanish for 
$k$ odd and are necessarily proportional to $(\nabla^{2}_{\bf R})^{k/2}$ for $k$ even. Then the
contribution (\ref{6.13}) to (\ref{6.12}) is strictly local since $\nabla^{2}_{\bf R}\(\frac{1}{|{\bf
 R}|}\)=0,{\bf R} \neq{\bf 0}$: thus  (\ref{6.12}) decays faster than any inverse power. More
generally, it can be shown that as a consequence of the rotational symmetry and of the harmonicity
of the Coulomb potential, all chains  (\ref{6.10}) involving more than one $W$-bond decay also
strictly faster than $\approx |\br_{a}-\br_{b}|^{-6}$.

As an example, we give the argument for the term (\ref{6.10}) involving a two $W$-bond
convolution and keeping only the slowest part of these $W$-bonds, i.e. the dipole-dipole
interaction (\ref{5.35}). Once (\ref{5.35}) has been inserted in (\ref{6.10}) one obtains
convolutions relative to the loop positional variables ${\bf R}$ of the form
\begin{equation}
(G_{\nu_{1}}\star \partial_{\nu_{1}}\partial_{\mu_{1}}V\star G_{\mu_{1}\nu_{2}}\star
\partial_{\nu_{2}}\partial_{\mu_{2}}V\star G_{\mu_{2}})({\bf R}_{a}-{\bf R}_{b})   
\label{6.14}
\end{equation}
with
\begin{eqnarray}
G_{\nu}({\bf R})&=&\int D({\bf X}_{a})\int D({\bf X}){ X}_{\nu}(s)
G({\bf R},{\bf X}_{a},{\bf X})
\label{6.15}\\
G_{\mu\nu}({\bf R})&=&\int D({\bf X}^{\prime})\int D({\bf X}){ X}^{\prime}_{\mu}(s^{\prime})
{ X}_{\nu}(s)G({\bf R},{\bf X}^{\prime},{\bf X})
\label{6.16}
\end{eqnarray}
where we write 
\begin{equation}
G(\cl_{1},\cl_{2})=G({\bf R}_{1}-{\bf R}_{2},{\bf X}_{1},{\bf X}_{2})
\label{6.17}
\end{equation}
because of translation invariance. In  (\ref{6.14})-(\ref{6.17}) we have omitted the loop indices
$\alpha,q$ and the times $s$ as well as the corresponding summations. The loop internal variables have
been integrated  and sums on repeated Cartesian indices are understood. To analyze the long
distance behavior of  (\ref{6.14}) it is convenient to first consider its Fourier transform
\begin{equation}
\tilde{G}_{\nu_{1}}({\bf k})\frac{{ k}_{\nu_{1}}{ k}_{\mu_{1}}}{|{\bf
k}|^{2}}\tilde{G}_{\mu_{1}\nu_{2}}({\bf k}) \frac{{ k}_{\nu_{2}}{ k}_{\mu_{2}}}{|{\bf
k}|^{2}}\tilde{G}_{\mu_{2}}({\bf k}) 
\label{6.18}
\end{equation}
Since the measure $D({\bf X})$ as well as $G({\bf R},{\bf X}_{1},{\bf X}_{2})$ are invariant under
spatial rotations , $\tilde{G}_{\nu}({\bf k})$ and $\tilde{G}_{\mu\nu}({\bf k})$ transform as
tensors: hence they are necessarily of the form 
\begin{eqnarray}
\tilde{G}_{\nu}({\bf k})&=&{ k}_{\nu}a(|{\bf k}|)\nonumber\\
\tilde{G}_{\mu\nu}({\bf k})&=&\delta_{\mu\nu}b(|{\bf k}|)+{ k}_{\mu}{ k}_{\nu}c(|{\bf k}|)
\label{6.19}
\end{eqnarray}
for some functions $a, b$ and $c$ of $|{\bf k}|$. We know that $G({\bf
R},{\bf X}_{1},{\bf X}_{2})$ and hence $G_{\nu}({\bf R}),\;\;
G_{\mu\nu}({\bf R})$ do not decay slower than $\approx |{\bf
R}|^{-6}$. Taking the correspondence (\ref{6.1b})-(\ref{6.1a}) into
account, this implies that the small $|{\bf k}|$ expansion of
$\tilde{G}_{\nu}({\bf k})$ and $\tilde{G}_{\mu\nu}({\bf k})$ must be
of the form 
\begin{eqnarray}
\tilde{G}_{\nu}({\bf k})&=&{ k}_{\nu}(a+O\(|{\bf k}|^{2}(\ln|{\bf k}|)^{m_{1}+1}\)\nonumber\\
\tilde{G}_{\mu\nu}({\bf k})&=&\delta_{\mu\nu}(b_{1}+b_{2}|{\bf k}|^{2}+O(|{\bf
k}|^{3}(\ln |{\bf k}|)^{m_{2}})\nonumber\\
&+&{ k}_{\mu}{ k}_{\nu} \(c+O(|{\bf k}|(\ln|{\bf k}|)^{m_{3}}\) 
\label{6.20}
\end{eqnarray}
for some $m_{1},m_{2},m_{3}\geq 0$. Upon inserting (\ref{6.20}) into
(\ref{6.18}) one sees that the Coulomb singularities $|{\bf k}|^{-2}$
are compensated and (\ref{6.18}) behaves as $C|{\bf k}|^{2}+O(|{\bf
k}|^{4}(\ln|{\bf k}|)^{m+1})$ as $|{\bf k}|\rightarrow 0$, hence
(\ref{6.14}) decays strictly faster than $\approx |{\bf R}|^{-6}$
(possibly as $\approx |{\bf R}|^{-7}$).  The same argument extends to
all higher order multipoles and chains in (\ref{6.10}) with arbitrary
numbers of $W$-bonds, showing that
$\rho^{(nex)}(\alpha_{a},\br_{a},\alpha_{b},\br_{b})$ decays at least
as $\approx |\br_{a}-\br_{b}|^{-6}$.

We briefly comment about the exchange part
$\rho^{(ex)}(\alpha_{a},\br_{a},\br_{b})$ of the particle-particle
distribution. One sees from (\ref{5.23a}) that this exchange part
involves the loop density $\rho(\cl)$ with the position of two
particles fixed at $\br_{a}$ and $\br_{b}$. The Mayer expansion of the
loop density (\ref{5.60}) has an activity factor $z(\cl)$ that has the
bound (\ref{5.15}). One can show that after the needed resummation of
Coulomb chains, all prototype graphs representing $\rho(\cl)$ inherit
the same activity factor $z(\cl)$. Furthermore prototype graphs are
bounded functions of the loop variables
$\br_{1},\ldots,\br_{q}$. Taking these facts into account in
(\ref{5.23b}), the contribution of a prototype graph to
$\rho^{(ex)}(\alpha_{a},\br_{a},\br_{b})$ will be majorized by
$$
\sum_{q=2}^{\infty}q\int
D(\bX)|z(\cl)|\sum_{k=2}^{q}\delta(\lambda_{\alpha_{a}}\bX_{k}+\br_{a}-\br_{b}) \leq
\frac{2s_{\alpha_{a}}+1}{\lambda_{\alpha_{a}}^{3}} $$
$$
\sum_{q=2}^{\infty}e^{\beta\mu_{\alpha_{a}} q}
\int d\bX_{2}\ldots d\bX_{q}\prod_{k=1}^{q}
\frac{\exp(-\frac{1}{2}|\bX_{k+1}-\bX_{k}|^{2})}{(2\pi)^{3/2}}
\sum_{k=2}^{q}\delta(\lambda_{\alpha_{a}}\bX_{k}+\br_{a}-\br_{b})
$$
\begin{equation}
=(2s_{\alpha_{a}}+1)\left[\frac{1}{(2\pi)^{3}}\int d{\bf k}\exp(-i{\bf k}\cdot(\br_{a}-\br_{b}))
\frac{1}{\exp\(\beta\(\frac{\hbar^{2}}{2m}|{\bf k}|^{2}-\mu_{\alpha_{a}}\)\)-1}\right]^{2}
\label{6.22}
\end{equation}
The integral in (\ref{6.22}) is nothing else than the off-diagonal
part of the one-body density matrix of a free Bose gas (the Bose
distribution occurs because we have replaced the fermionic sign
$\eta_{\alpha_{a}}$ by $1$ in the bound). At sufficiently low density
($\exp(\beta\mu_{\alpha_{a}}<1$), this function, the Fourier
transform of a infinitely differentiable function, decays faster than
any inverse power of $|\br_{a}-\br_{b}|$. Under the assumption that
the same property remains true for the sum of all prototype graphs, we
conclude that the asymptotic behavior of the particle-particle
distribution is dominated by its non-exchange part previously
discussed.

The diagrams contributing to the coefficient of the $r^{-6}$ tail can
also be analyzed in the low density regime \cite{Cor97a}. At the
lowest order in the densities, one finds that the tail has the same
form (\ref{4.54}) as in the semi-classical gas
\begin{equation}
\rho_{T}(\alpha_{a},\br_{a},\alpha_{b},\br_{b})\sim \frac{1}{|\br_{a}-\br_{b}|^{6}}
\frac{\hbar^{4}\beta^{4}}{240}\int d\br G^{D.H.}(\br,\alpha_{a})
\int d\br G^{D.H.}(\br,\alpha_{b})
\label{6.23} 
\end{equation}
where $G^{D.H.}(\br,\alpha)$ is defined as in (\ref{4.53}), but
calculated in the Debye-H\"{u}ckel approximation 
\begin{equation}
\int d\br G^{D.H.}(\br,\alpha)=\rho_{\alpha}\(\frac{e_{\alpha}^{2}}{m_{\alpha}}
-\frac{4\pi\beta
e_{\alpha}}{\kappa^{2}}\sum_{\gamma}\frac{e_{\gamma}^{3}\rho_{\gamma}}{m_{\gamma}}\)
\label{6.23a}
\end{equation}
This is to be expected, since at lowest order in the densities, the
particles surrounding the two fixed charges at $\br _{a}$ and $\br
_{b}$ are fully ionized and will behave as a classical plasma.
It is interesting to specialize (\ref{6.23}) to the electron-proton
system. In this case one finds from  
(\ref{6.23}) and (\ref{6.23a}) with
$\alpha$=electron or proton, $e_{{\rm el}}=-e_{{\rm pr}}=e,\;\rho_{{\rm el}}
=\rho_{{\rm pr}}=\rho$ 
\begin{equation}
\rho_{T}(\alpha_{a},\br_{a},\alpha_{b},\br_{b})\sim \frac{\rho^{2}}{|\br_{a}-\br_{b}|^{6}}
\frac{\hbar^{4}\beta^{4}}{240}\frac{e^{4}}{4}\(\frac{1}{m_{{\rm el}}}+\frac{1}{m_{{\rm pr}}}\)^{2}
\end{equation}
showing that in this system all tails have exactly the same decay rate at low density.

The tail of the particle correlation in presence of a uniform magnetic
field ${\bf B}$ has also been calculated 
\cite{Cor97,Cor98a}. Because of breaking of the full rotation
invariance, it has the slower decay
$\sim\frac{P_{4}(\cos\theta)}{|\br_{a}-\br_{b}|^{5}}$ where
$P_{4}(\cos\theta)$ is a Legendre polynomial and $\theta$ is the angle
between ${\bf B}$ and $\br_{a}-\br_{b}$.

\subsubsection{ Correlations of the total charge density} 
\label{subsec-6.1.3}

As in the semi-classical case, the total charge density enjoys better
screening properties than individual particles. This is reflected by
the fact that correlations involving the total charge density
$\hat{c}(\br)=\sum_{\alpha}e_{\alpha}\hat{\rho}(\alpha,\br)$ such as
the charge-particle correlation
\begin{equation}
<\hat{c}(\br_{a})\hat{\rho}(\alpha_{b},\br_{b}))>_{T}=\sum_{\alpha_{a}}e_{\alpha_{a}}
\rho_{T}(\alpha_{a},\br_{a},\alpha_{b},\br_{b})
\label{6.24}
\end{equation}
will have a faster decay than the particle-particle correlation.

To illustrate the origin of this additional screening mechanism, it is convenient to construct a
certain "dressing" of the root point $\cl_{a}$ in the Ursell function. For this we call $\cl_{a}$ a
Coulomb root point in a $\Pi$-diagram if $\cl_{a}$ is linked to the rest of the diagram by exactly one
$F_{cc}$ bond or one $F_{cm}$ bond, and split the Ursell function accordingly
\begin{equation}
h(\cl_{a},\cl_{b})=h^{c}(\cl_{a},\cl_{b})+h^{nc}(\cl_{a},\cl_{b})
\label{6.25}
\end{equation}
$h^{c}(\cl_{a},\cl_{b})$ $(h^{nc}(\cl_{a},\cl_{b}))$ is the sum of $\Pi$-diagrams where $\cl_{a}$
is  (is not) a Coulomb point. This definition implies immediately the following integral equation
\begin{equation}
h^{c}=F_{cc} + F_{cc}\rho\star h^{nc}+F_{cm}+F_{cm}\rho\star h
\label{6.26}
\end{equation}
and thus
\begin{equation}
h=F_{cc}+F_{cm}+(F_{cc}\rho+1)\star h^{nc} +F_{cm}\rho\star h
\label{6.27}
\end{equation}
where here the symbol $\star$ means both integration on the internal variables and convolution with
respect to the positions of the loops; $\rho=\rho(\cl)$ is the weight of intermediate points. Note 
that $h^{nc}$ occurs in the second term of (\ref{6.26}) because convolutions of $F_{cc}$ bonds are
forbidden in $\Pi$-diagrams. As we know the two first terms of (\ref{6.27}) are rapidly
decreasing; the third term contributes to $\rho_{T}(\cl_{a},\cl_{b})$ as   
$$
\rho(\cl_{a})\int d\cl \(F_{cc}(\cl_{a},\cl)\rho(\cl)
+\delta(\cl_{a},\cl)\)h^{nc}(\cl,\cl_{b})\rho(\cl_{b})
$$
\begin{equation}
=\int d\cl S(\cl_{a},\cl)\rho(\cl)h^{nc}(\cl,\cl_{b})\rho(\cl_{b})
\label{6.28}
\end{equation}
with "dressing" factor of the point $\cl_{a}$ defined by
\begin{equation}
S(\cl_{a},\cl)=\rho(\cl_{a})F_{cc}(\cl_{a},\cl)+\delta(\cl_{a},\cl)
\label{6.29}
\end{equation}
When summed on the internal variables of the loop $\cl_{a}$ and the charges,
it is immediately seen, using (\ref{5.45}) and (\ref{5.47}) that (\ref{6.29}) gives
\begin{equation}
\sum_{\alpha_{a}}e_{\alpha_{a}}\sum_{q_{a}}q_{a}\int D({\bf X}_{a})S(\cl_{a},\cl)=
e_{\alpha}q_{\alpha}\(\delta({\bf r})-\kappa^{2}\frac{\exp(-\kappa r)}{4\pi r}\),\;\;\;{\bf r}={\bf
R}_{a}-{\bf R} 
\label{6.30}
\end{equation}
with Fourier transform
\begin{equation}
e_{\alpha}q_{\alpha}\(1-\frac{\kappa^{2}}{\kappa^{2}+|{\bf k}|^{2}}\)=
e_{\alpha}q_{\alpha}\frac{|{\bf k}|^{2}}{\kappa^{2}+|{\bf k}|^{2}}
\label{6.31}
\end{equation}
Hence the convolution (\ref{6.28}) will contribute to the Fourier
transform of the charge-particle correlation as
\begin{equation}
\frac{|{\bf k}|^{2}}{\kappa^{2}+|{\bf
k}|^{2}}\times\left\{\begin{array}{ll}\mbox{Fourier transform of
$\Pi$-diagrams constituting $h^{nc}$}\\
\mbox{integrated on the internal variables of their root
points}\end{array}\right\} \label{6.32}
\end{equation}
Since the singular term of these diagrams is $|{\bf k}|^{3}(\ln|{\bf
k}|)^{m}$ (corresponding to a $\approx r^{-6}$ decay as seen in
section~\ref{sec-6.1.2}), the singular term in (\ref{6.32}) is $|{\bf
k}|^{5}\ln|{\bf k}|^{m}$ implying a $\approx r^{-8}$ decay. The same
conclusion can be drawn for the contribution of the last term of
(\ref{6.27}) on the grounds of rotational invariance and of the short
range of $F_{cm}$. The final result is that the charge-particle has
the following large distance behavior in the low density regime
\cite{Cor97a}
\begin{equation}
\langle\hat{\rho}(\alpha_{a},\br_{a})\hat{c}(\br_{b})\rangle_{T}
\sim\frac{1}{|\br_{a}-\br_{b}|^{8}}
\frac{\hbar^{4}\beta^{4}}{16}\int d\br G^{D.H.}(\br, \alpha_{a})
\(\frac{1}{3}\int d\br |\br|^{2}K^{D.H.}(\br)\) \label{6.33}
\end{equation}
where the last factor is the expression (\ref{4.57}) calculated in the
Debye-H\"{u}ckel approximation
\begin{equation}
\frac{1}{3}\int d\br
|\br|^{2}K^{D.H.}(\br)=-\frac{2}{\kappa^{2}}\sum_{\alpha}\frac{e_{\alpha}^{3}\rho_{\alpha}}{m_{\alpha}}
\label{633a}
\end{equation}

For the charge-charge correlation, both root points $\cl_{a}$ and
$\cl_{b}$ can be dressed as before, leading to a $\approx r^{-10}$
decay with
\begin{eqnarray}
\langle\hat{c}(\br_{a})\hat{c}(\br_{b})\rangle_{T}&\sim&\frac{1}{|\br_{a}-\br_{b}|^{10}}
\frac{7\hbar^{4}\beta^{4}}{4}
\(\frac{1}{3}\int d\br |\br|^{2}K^{D.H.}(\br)\)^{2}\nonumber\\
&\sim&\frac{1}{|\br_{a}-\br_{b}|^{10}}\frac{7\hbar^{4}\beta^{4}}{\kappa^{4}}
\(\sum_{\alpha}\frac{e_{\alpha}^{3}\rho_{\alpha}}{m_{\alpha}}\)^{2}
\label{6.34}
\end{eqnarray}
The fact that the space integral of (\ref{6.30}) vanishes (or
equivalently (\ref{6.31}) vanishes at ${\bf k}=0$) is the elementary
form of the classical charge sum rule in the Debye-H\"{u}ckel theory
(see (\ref{2.1.13})). In the quantum gas, this vestige of classical
screening can still operate to improve the decay of the charge-charge
correlation from $\approx r^{-6}$ to $\approx r^{-10}$, but not
beyond.

Finally the susceptibility (\ref{1.2.7d}) (i.e. the response to an external
classical charge)  decays as $r^{-8}$. Its tail at low density is given by
\begin{equation}
\chi(r)\sim\frac{4\pi}{r^{8}}\frac{\hbar^{4}\beta^{5}}{8\kappa^{4}}\(\sum_{\alpha}
\frac{e_{\alpha}^{3}\rho_{\alpha}}{m_{\alpha}}\)^{2}
\label{6.34a}
\end{equation}
We emphasize that (\ref{6.23}), (\ref{6.33}), (\ref{6.34}) and (\ref{6.34a}) are exact asymptotic
results for the multicomponent quantum gas.

\subsubsection{A comparison with the standard many-body perturbation theory}
\label{subsec-6.1.4}

\begin{figure}
\psfig{figure=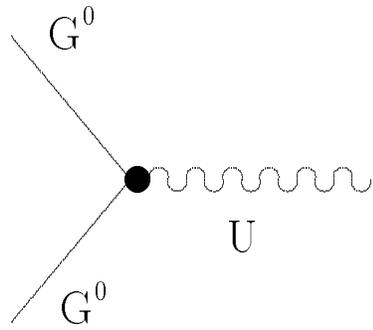}
\caption{Vertex with two propagator lines and one interaction line}
\label{fig9}
\end{figure}

The standard finite temperature many-body perturbation theory is an
expansion in powers of the coupling constant (the charge
$e^{2}$). Here we adopt the definitions and normalizations of \cite{FeWa71} and 
\cite{CoMa91}\footnote{There is a minus sign missing in front of the
right hand side of \cite[formula (2.4)]{CoMa91}.}.  We recall a few
facts. The terms of the expansion are represented by Feynman graphs in
which bonds $(\alpha_{1},\br_{1},s_{1} ;\alpha_{2},\br_{2},s_{2})$ are
joined either by free propagators
$\delta_{\alpha_{1},\alpha_{2}}G^{0}_{\alpha_{1}}(\br_{2}-\br_{1},s_{1}-s_{2})$
or by interaction lines
$U_{\alpha_{1},\alpha_{2}}(\br_{2}-\br_{1},s_{1}-s_{2})$
(Figure~\ref{fig9}). At each vertex $(\alpha,\br,s)$ are attached two
propagator lines and one interaction line.

A basic link between physical quantities and perturbation theory can
be established through the imaginary time displaced charge correlation
(\ref{1.2.7c}) by
\begin{equation}c_{T}(\br, s, {\bf 0})
=-\sum_{\alpha\gamma}e_{\alpha}e_{\gamma} {\cal P}_{\alpha\gamma}(\br,
s) \label{6a1} 
\end{equation}
The total polarization ${\cal P}_{\alpha\gamma}(\br, \tau)$ consists
of all connected Feynman graphs in which the points $(\alpha,\br,s)$
and $(\gamma,{\bf 0},0)$ are joined by internal lines. ${\cal
P}_{\alpha\gamma}(\br, \tau)$ is a time periodic function of period
$\beta$. Introducing its space and time Fourier transform ${\cal
P}(\bk,n)$, the structure factor (\ref{1.2.7cc}) and the
susceptibility (\ref{1.2.7d}) are given by ($n$ indexes even Matsubara
frequencies $\omega_{n}=\frac{\pi n}{\beta}$, $n$ even)\footnote {In
general, $n$ is odd for fermionic propagators and even for bosonic
propagators, but the charge correlation (\ref{1.2.7c}) involves only
quadratic expressions of the Fermi or Bose fields and only even
frequencies occur in its time Fourier series.}  \begin{equation}
S(\bk)=-\sum_{n=-\infty}^{\infty}\sum_{\alpha\gamma}e_{\alpha}e_{\gamma}
{\cal P}_{\alpha\gamma}(\bk,n) \label{6a2}
\end{equation}\begin{equation}\chi(\bk)=\beta
V(\bk)\sum_{\alpha\gamma}e_{\alpha}e_{\gamma} {\cal
P}_{\alpha\gamma}(\bk, n=0), \;\;\;\;\;\;
V(\bk)=\frac{4\pi}{|\bk|^{2}} \label{6a3} \end{equation}It is
convenient to consider the proper polarization ${\cal
P}_{\alpha\gamma}^{\ast}(\bk,n)$, the set of all polarization parts
that cannot be separated into two polarization parts by cutting a
single interaction line. ${\cal P}_{\alpha\gamma}(\bk,n)$ and ${\cal
P}_{\alpha\gamma}^{\ast}(\bk,n)$ are related by a Dyson equation
\begin{equation} {\cal P}_{\alpha\gamma}(\bk,n)=\frac{{\cal
P}_{\alpha\gamma}^{\ast}(\bk,n)} {1-\beta
V(\bk)\sum_{\alpha\gamma}e_{\alpha}e_{\gamma}{\cal
P}_{\alpha\gamma}^{\ast}(\bk,n)} \label{6a4} \end{equation}An
effective frequency dependent interaction can be defined by
\begin{equation} U^{eff}(\bk,n)=\frac{V(\bk)}{1-\beta
V(\bk)\sum_{\alpha\gamma}e_{\alpha}e_{\gamma}{\cal
P}_{\alpha\gamma}^{\ast}(\bk,n)} \label{6a5} \end{equation}This
definition is justified by the fact that its zero frequency component
coincides with the static effective potential (\ref{1.1.10f}) (with
$\varepsilon (\bk )$ given by (\ref{1.2.7d}))
\begin{equation}
U^{eff}(\bk,n=0)=\varepsilon^{-1}(\bk)V(\bk) \label{6a6} \end{equation}One has
also the relation \begin{equation}{\cal P}_{\alpha\gamma}(\bk, n)={\cal
P}_{\alpha\gamma}^{\ast}(\bk,n)+\beta\sum_{\delta\eta} {\cal
P}_{\alpha\delta}^{\ast}(\bk,n)e_{\delta}e_{\eta}U^{eff}(\bk,n) {\cal
P}_{\eta\gamma}^{\ast}(\bk,n) \label{6a6a} \end{equation}

\begin{figure}
\psfig{figure=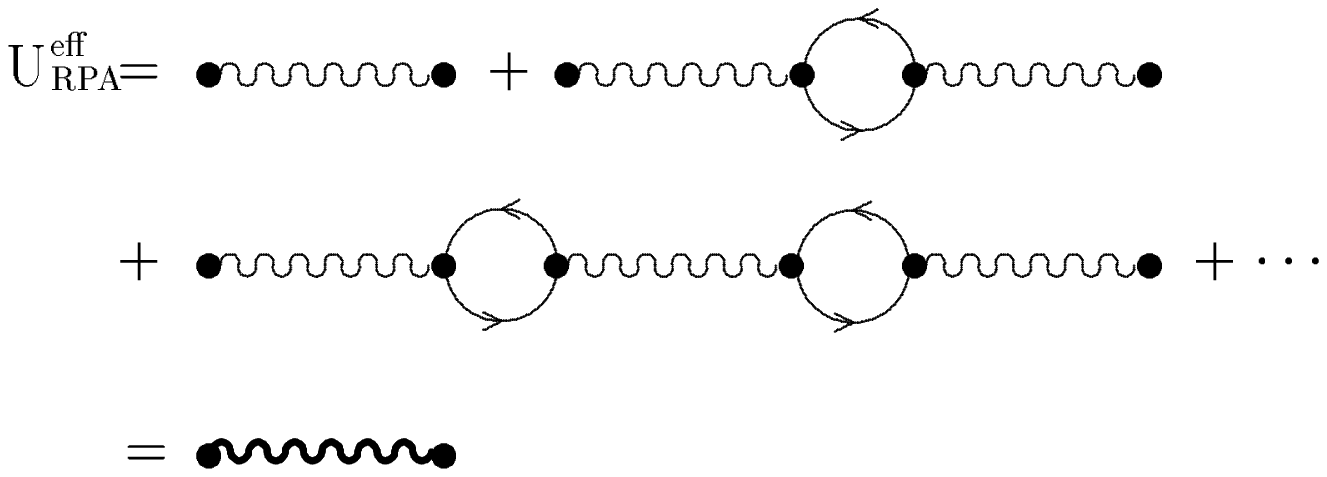}
\caption{The random phase approximation}
\label{fig10}
\end{figure}

The well known random phase approximation (RPA) amounts to replacing
${\cal P}_{\alpha\gamma}^{\ast}(\br,\tau)$ by the lowest order proper polarization 
term, i.e. the product of two free propagators
\begin{equation}
{\cal P}_{RPA,\alpha\gamma}^{\ast}(\br,s)=
-\eta_{\alpha}\delta_{\alpha\gamma}(2s_{\alpha}+1)
G^{0}_{\alpha}(\br,s)G^{0}_{\alpha}(-\br,-s)=\delta_{\alpha\gamma}
{\cal P}_{RPA,\alpha}^{\ast}(\br,s)
\label{6a7}
\end{equation}
Hence from (\ref{6a4}) and (\ref{6a5}) (see Figure~\ref{fig10}) 
\begin{equation}
{\cal P}_{RPA,\alpha\gamma}(\bk,n)=\frac{\delta_{\alpha\gamma}
{\cal P}_{RPA,\alpha}^{\ast}(\bk,n)}
{1-\beta
V(\bk)\sum_{\alpha}e_{\alpha}^{2}{\cal P}_{RPA,\alpha}^{\ast}(\bk,n)}
\label{6a8}
\end{equation}
\begin{equation}
U^{eff}_{RPA}(\bk,n)=\frac{V(\bk)}{1-\beta
V(\bk)\sum_{\alpha}e_{\alpha}^{2}{\cal P}_{RPA,\alpha}^{\ast}(\bk,n)}
\label{6a9} 
\end{equation}
and with (\ref{6a2}) and (\ref{6a3}), this defines also $S_{RPA}(\bk)$ 
and $\chi_{RPA}(\bk)$.

Let us show that the RPA approximation fails to capture the algebraic
tails in the correlations\footnote{The present considerations apply
only if the temperature is different from zero. At zero temperature,
the effective potential shows the long range Friedel oscillations
$\cos(2k_{F}r)/r^{3}$ due to the sharpness of the Fermi surface.}.  The
free propagator has the form
\begin{equation}G^{0}_{\alpha}(\bk,s)=\exp\(-\beta
s\(\frac{\hbar^{2}|\bk|^{2}}{2m_{\alpha}}- \mu_{\alpha}\)\)
[n^{0}_{\alpha}(\bk)-\theta(s)] \label{6a10}
\end{equation}
where $n^{0}_{\alpha}(\bk)=\(\exp\(\frac{\hbar^{2}\beta |\bk|^{2}}{2m_{\alpha}}-
\beta\mu_{\alpha}\)\pm 1\)^{-1}$ is the free Fermi or Bose distribution and $\theta(s)$ 
is the Heaviside function. It is clearly an infinitely differentiable function of $\bk$
(if $\mu_{\alpha}<0$ for Bose particles). One can deduce from (\ref{6a7}) that the same is true for
${\cal P}_{RPA,\alpha}^{\ast}(\br,n)$ and one finds the small $\bk$ behaviour
\begin{eqnarray}
{\cal P}_{RPA,\alpha}^{\ast}(\bk,n=0)&=&
-\frac{\partial \rho^{0}_{\alpha}}{\partial(\beta\mu_{\alpha})} +O(|\bk|^{2})\neq 0\nonumber\\
{\cal P}_{RPA,\alpha}^{\ast}(\bk,n\neq 0)&=&O(|\bk|^{2}),\;\,\,\bk\rightarrow 0
\label{6a11}
\end{eqnarray}
where $\rho^{0}_{\alpha}=(2s_{a}+1)\int d\bk n_{\alpha}^{0}(\bk) $ is the density of the free
gas.  With (\ref{6a8}) this implies that the ${\cal P}_{RPA,\alpha}^{\ast}(\bk,n)$ are infinitely
differentiable at $\bk =0$ and by (\ref{6a2}) and (\ref{6a3}) $S_{RPA}(\bk)$ and $\chi_{RPA}(\bk)$
inherit the same property. Thus, in space, these functions decay faster than any inverse power of
the distance. From (\ref{6a8}) and (\ref{6a11}) one sees that $\chi_{RPA}(\bk)$ can be approximated
for small $\bk$ by the Debye-like form 
\begin{equation}
\chi_{RPA}(\bk)\simeq -\frac{\kappa_{RPA}^{2}}{|\bk|^{2}+\kappa_{RPA}^{2}}
\label{6a12}
\end{equation}
with
\begin{equation}
\kappa_{RPA}^{2}=4\pi \beta\sum_{\alpha}e_{\alpha}^{2}
\frac{\partial \rho^{0}_{\alpha}}{\partial(\beta\mu_{\alpha})}
\label{6a13}
\end{equation}
so defining the RPA screening length $\kappa_{RPA}^{-1}$.
It is instructive to compare it with the loop screening length
$\kappa^{-1}$ (\ref{5.52b}). For this, we introduce the general relation
\begin{eqnarray}
\frac{\partial}{\partial(\beta\mu_{\alpha})}\rho_{\alpha}&=&\rho_{\alpha}+\int d\br\rho_{T}
(\alpha,\br,\alpha,{\bf 0})\nonumber\\
&=&\rho_{\alpha}+\int d\br\rho_{T}^{(ex)}(\alpha,\br,{\bf 0})+
\int d\br\rho_{T}^{(nex)}(\alpha,\br,\alpha,{\bf 0})
\label{6a14}
\end{eqnarray}
in (\ref{5.52b}), so that $\kappa^{2} $ can also be written as  
\begin{equation}
\kappa^{2}=4\pi\beta\(\sum_{\alpha}e_{\alpha}^{2}
\frac{\partial \rho_{\alpha}}{\partial(\beta\mu_{\alpha})}-\sum_{\alpha}e_{\alpha}^{2}
\int d\br \rho_{T}^{(nex)}(\alpha,\br,\alpha,{\bf 0})\)
\label{6a15}
\end{equation}
In the integrals in (\ref{6a14}) and (\ref{6a15}), there are no
contributions of coincident points. The loop screening length
interpolates between the classical Debye length at low density (as
seen from (\ref{5.52b}) when exchange effects can be neglected) and
the RPA length at high density.  Indeed, in the high density regime,
the kinetic energy dominates the Coulomb interaction energy: the gas
becomes free so the non-exchange part of the correlation can now be
neglected in (\ref{6a15}) and $\kappa^{2}$ tends to the RPA value.

As far as the effective potential (\ref{6a5}) is concerned, one finds
in the RPA approximation that its zero frequency component
$U^{eff}_{RPA}(\bk,n=0)$ is also Debye-like for small $\bk$ with a
screening length slightly different from $\kappa_{RPA}^{-1}$ (for
explicit formulae see \cite{CoMa91,Cor96b}). However its non-zero
frequency components $U^{eff}_{RPA}(\bk,n\neq 0)$ are not shielded and
keep a pure Coulombic singularity $|\bk|^{-2}$, but within the RPA
theory these singularities do not induce any algebraic tails in the
particle and charge correlations. For instance, if ${\cal
P}_{\alpha\gamma}^{\ast}(\bk,n)$ is approximated by ${\cal
P}_{RPA,\alpha\gamma}^{\ast}(\bk,n)$ in (\ref{6a6a}), the vanishing of
the non-zero frequency components as $|\bk|\rightarrow 0$ kills the
$|\bk|^{-2}$ singularity coming from $U^{eff}_{RPA}(\bk,n\neq 0)$.

\begin{figure}
\psfig{figure=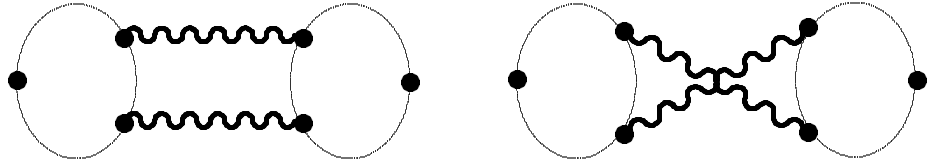}
\caption{A first correction to RPA}
\label{fig11}
\end{figure}

One may however wonder how these tails emerge in the framework of the
many-body perturbation theory. The role played by $U^{eff}_{RPA}$ in
Feynman graphs is analogous to that of the resummed bond $F_{l}$
(\ref{5.55}) in loop prototype diagrams. $F_{l}$ has a long range part
$W$ (see (\ref{6.6})), but this long range part generates tails in the
correlations only if it occurs at least quadratically in prototype
graphs (see section \ref{sec-6.1.2}). In the same way, $U^{eff}_{RPA}$
must enter quadratically in Feynman graphs to generate tails in the
correlations. This leads to including as a first correction to the RPA
approximation the proper polarization insertion ${\cal P}^{\ast}_{1}$
made of two propagator loops linked by two RPA lines of
Figure~\ref{fig11}. The effect of this correction has been studied in
the one component system \cite{CoMa91}. The Coulombic singularities of
$U^{eff}_{RPA}(\bk,n\neq 0)$ come now non-trivially into the game and
one finds that ${\cal P}_{1}^{\ast}(\bk, n=0)$ and ${\cal
P}_{1}^{\ast}(\bk, n\neq0)$ have respectively a $|\bk|^{3}$ and a
$|\bk|^{7}$ singular term in their small $\bk$ expansion. One can then
infer through the Dyson equation (\ref{6a4}) that these singular terms
induce $r^{-10}$ and $r^{-8}$ tails in $S(\br)$ and $\chi(\br)$
respectively, in accordance with the findings of
section~\ref{subsec-6.1.3}.  The graphs of Figure~\ref{fig11} have
been considered previously in \cite{MaAs87,LaVo87} in connection with
calculating the cohesive energy of metals and inter-atomic potentials
at zero temperature.

It is not possible to establish in general strict correspondence rules
between the many-body perturbation algorithm and that of the loop
formalism. Since the reference system of the many-body perturbation
scheme is the noninteracting gas, any approximation there gives
expressions in terms of free quantities. In the loop expansion,weights
and bonds in Mayer diagrams contain already the coupling constant
$e^{2}$ and the statistics in a non-perturbative way, and they involve
the exact particle densities.  This difference is clearly seen in the
two screening lengths $\kappa^{-1}_{RPA}$ and $\kappa^{-1}$ that
depend respectively on the free densities $\rho^{0}_{\alpha}$ and
exact densities $\rho_{\alpha}$. One can however establish a precise
relation between the loop chain potential and the RPA effective
potential in the Boltzmann limit when statistics are neglected (see
\cite[section IV]{CoMa91}).

The two formalisms play complementary roles. The many-body formalism
is suitable for performing low temperature and ground state
calculations. By analytic continuation in time it can also be used to
compute real time displaced correlations. Information on ground state
quantities in the loop formalism would necessitate applying the
Feynman-Kac formula backwards at some point before taking the
$\beta\rightarrow\infty$ limit. However the loop formalism provides a
better insight into Coulombic correlations and screening. For instance
the reduction to a gas of multipoles is not visible in the many-body
framework where the effective potential $U_{RPA}^{eff}$ keeps an
unscreened bare Coulomb part. Low density behaviors are much more
conveniently obtained from the loop expansion: it would not be
straightforward to recover the exact results of section \ref{sec-6.1}
from an analysis of Feynman graphs.  The next section offers another
example of the usefulness of the loop formalism.

\subsection{The virial equation of state}
\label{6.1}

The derivation of the equation of state of a quantum mechanical plasma
is an old problem of practical and theoretical interest. In
particular, a precise knowledge of this equation plays an important
role in astrophysical conditions, and it is not possible to give
credit here to the large literature devoted to this question. We quote
the thorough study of the quantum Coulomb gas undertaken by Ebeling
and collaborators by the usual methods of the many-body problem. This
work is reported in \cite{KKER86} and some comments will be given at
the end of this section. In the sequel, we put forward that the loop
representation is also a convenient tool for calculating the equation
of state at low density and fixed temperature. In particular the
formalism enables simultaneously keeping track, in a coherent and
systematic way, of the quantum effects and of the long range of the
Coulomb potential, a notoriously difficult problem.  A detailed
presentation of all calculations can be found in
\cite{AlPe92,AlCoPe94,AlCoPe95,AlPe96} and a review on this subject is
presented in \cite{Ala94}. Here we outline the general computational
scheme and illustrate some salient points.

To evaluate the density expansion of the thermodynamic functions, it
is appropriate to start from a standard identity relating the free
energy $f(\beta, \{\rho_\alpha\})$ per unit volume to the
two--particle correlation with a variable coupling parameter
\begin{equation}
\beta f =\beta f_{{\rm ideal}} +\frac{\beta}{2}
\sum_{\alpha_{a}\alpha_{b}}e_{\alpha_a}
e_{\alpha_b}\int_0^1 dg\int
d\br\rho_{T,g}(\alpha_a,\br,\alpha_b,0)\frac{1}{|\br|}
\label{6.35}
\end{equation}
Here $\rho_{T,g}(\alpha_a,\br,\alpha_b,{\bf 0})$ is the truncated
two-particle correlation (without the contribution of coincident
points) for a system of charges interacting by the Coulomb potential
$g/|\br|$ with a dimensionless coupling parameter
$g,\;\;0\leq\;g\;\leq 1$. For each $g$,
$\rho_{T,g}(\alpha_a,\br,\alpha_b,0)$ must be evaluated at the same
values of the densities and temperature that occur as arguments of
$f(\beta, \{\rho_\alpha\})$.  The particle correlation
$\rho_{T,g}(\alpha_a,\br_a,\alpha_b,\br_b)$ is in turn related to the
loop correlations through (\ref{5.22}), (\ref{5.23}) and (\ref{5.23a})
and the loop correlations have the diagrammatic expansion discussed in
section \ref{sec-5.2}. Then one has to organize the contributions of
this expansion in terms that are of increasing order in the particle
densities and insert them in (\ref{6.35}). This will yield the desired
expansion for the free energy. Finally the pressure $P(\beta,
\{\rho_\alpha\})$ follows from the usual formula
\begin{equation}
P=\sum_\alpha \rho_\alpha \frac{\partial}{\partial\rho_\alpha }f\;-\;f
\label{6.36}
\end{equation}
The calculation has been carried out up to order $\rho^{5/2}.$

\subsubsection{The Maxwell-Boltzmann pressure} \label{subsec-6.2.1)}
At this point one can make the following comment about the treatment
of exchange effects. It is clear from (\ref{5.13}) that $q$-particle
exchanges are of decreasing order $\exp(\beta\mu_\alpha q)$ as $q$
grows ($\mu_\alpha$ negative); hence they can also be treated
perturbatively at low density. This is the view point adopted in
\cite{AlCoPe95} and that is reported here. One considers first the
Maxwell-Boltzmann (MB) Coulomb gas, and one includes the exchange
effects in a second perturbative stage. The MB system is simpler since
loops reduce to filaments. Also the exchange part (\ref{5.23a}) of the
correlation can be ignored in the first stage.  Dropping here the
Fermi statistics should not alarm the reader. All terms of the low
density expansion remain well defined, since these terms involve only
finitely many body Hamiltonians, and thus the stability of matter
plays no role at any finite order. As already noted, the local
singularity of the Coulomb potential is smoothed by the usual laws of
quantum mechanics (the uncertainty principle) as in atomic physics.

In the works \cite{AlCoPe94,AlCoPe95}, prototype diagrams are
constructed with a slightly different definition of resummed bonds
than in section \ref{subsec-5.2.3}. One has here, keeping similar
notations as in section \ref{subsec-5.2.2} and \ref{subsec-5.2.3}
\begin{eqnarray}\label{6.36a}
&&\text{the charge-charge bond } F_{cc}, \text{ the charge-dipole bonds
} F_{cd},\;F_{dc}\nn \\
&&\text{the dipole-dipole bond } F_{dd}, \text{ the residual bond } F_{r}
\end{eqnarray}
The bond $F_{cc}$ is identical to (\ref{5.45}) (with $q_{a}=q_{b}=1$); $F_{cd}$ and $F_{dc}$ are
defined as in (\ref{5.48}) and (\ref{5.48a}) but retaining only the dipole part 
\begin{equation}
F_{cd}(1,2)=-\beta e_{1}e_{2}\int_{0}^{1}ds_{2}(\lambda_{2}\bx_{2}(s_{2})\cdot{\bf \nabla}_{2})
\frac{\exp(-\kappa |{\bf R}_{1}-{\bf R}_{2}|)}{|{\bf R}_{1}-{\bf R}_{2}|}
\label{6.37}
\end{equation}
The bond $F_{dd}$ is equal to the dipole part of the second term in  (\ref{5.50}), i.e 
\begin{eqnarray}
&&
F_{dd}(1,2)=-\beta e_{1}e_{2} \nn \\
&&\hspace{-1in}\times
\int_{0}^{1}ds_{1}\int_{0}^{1}ds_{2}
(\lambda_{1}\bx_{1}(s_{1})\cdot{\bf \nabla}_{1})(\lambda_{2}\bx_{2}(s_{2})\cdot{\bf \nabla}_{2})
\bigg (\frac{\exp(-\kappa |{\bf R}_{1}-{\bf R}_{2}|)-1}{|{\bf R}_{1}-{\bf R}_{2}|}\bigg)
\label{6.38}
\end{eqnarray}
while $F_{r}$ incorporates all the other contributions (higher order multipoles and quantum
effects). Note that in this decomposition we have the two types of long range $ r^{-3}$-bonds
$F_{dd}$ and $F_{r}$ (in contrast to the prototype graphs of subsection \ref{subsec-5.2.3} where only 
$F_{l}$ is dipolar).

As already explained in subsection \ref{subsec-5.2.5}, there are two sources for the density
dependence: the weight $\rho(\alpha, \bx)$, that has itself an expansion in the particle densities
$\rho_{\alpha}$,  and the screening length $\kappa$ occurring in the resummed bonds. One performs
first the activity expansion of $\rho_{\alpha}$ with the help of appropriate prototype graphs
(here the bonds are as in (\ref{6.36a}) but weights are activities), and one eliminates
activities in favor of densities with the help of the relation
$\rho_{\alpha}=\int D(\bx) \rho(\alpha,\bx)$. The analysis shows that  $\rho(\alpha,\bx)$ is given
as double power series in $\rho^{1/2}$ and $\ln\rho$ with first terms  
\begin{eqnarray}
\rho(\alpha,\bx)&=&\rho_{\alpha}+\sum_{\gamma}\rho_{\alpha}\rho_{\gamma}\int d\br\int
D(\bx_{1})\label{6.39}\\ &\times &\left[\exp\(-\beta
e_{\alpha}e_{\gamma}\int_{0}^{1}dsV(|\br+\lambda_{\gamma}\bx_{1}(s)-
\lambda_{\alpha}\bx(s)|)\)\right.\nonumber\\ &-&\left.\int D(\bx_{2})\exp\(-\beta
e_{\alpha}e_{\gamma}\int_{0}^{1}dsV(|\br+\lambda_{\gamma}\bx_{1}(s)-
\lambda_{\alpha}\bx_{2}(s)|)\)\right]\nn \\
&+& O(\rho^{5/2})\nonumber
\end{eqnarray}
These series are introduced in the statistical weights of the
expansion of the Ursell function.  Finally one obtains also the free
energy (\ref{6.35}) and the pressure (\ref{6.36}) as double power
series in $\rho^{1/2}$ and $\ln\rho$. Half powers of $\rho$ come from
the Debye length $\kappa^{-1}\sim\rho^{-1/2}$ in the screened
potential, while $\ln\rho$ terms occur because the $r^{-3}$-bonds
$F_{dd}$ and $F_{r}$ are at the border line of integrability in three
dimensions.

\begin{figure}
\psfig{figure=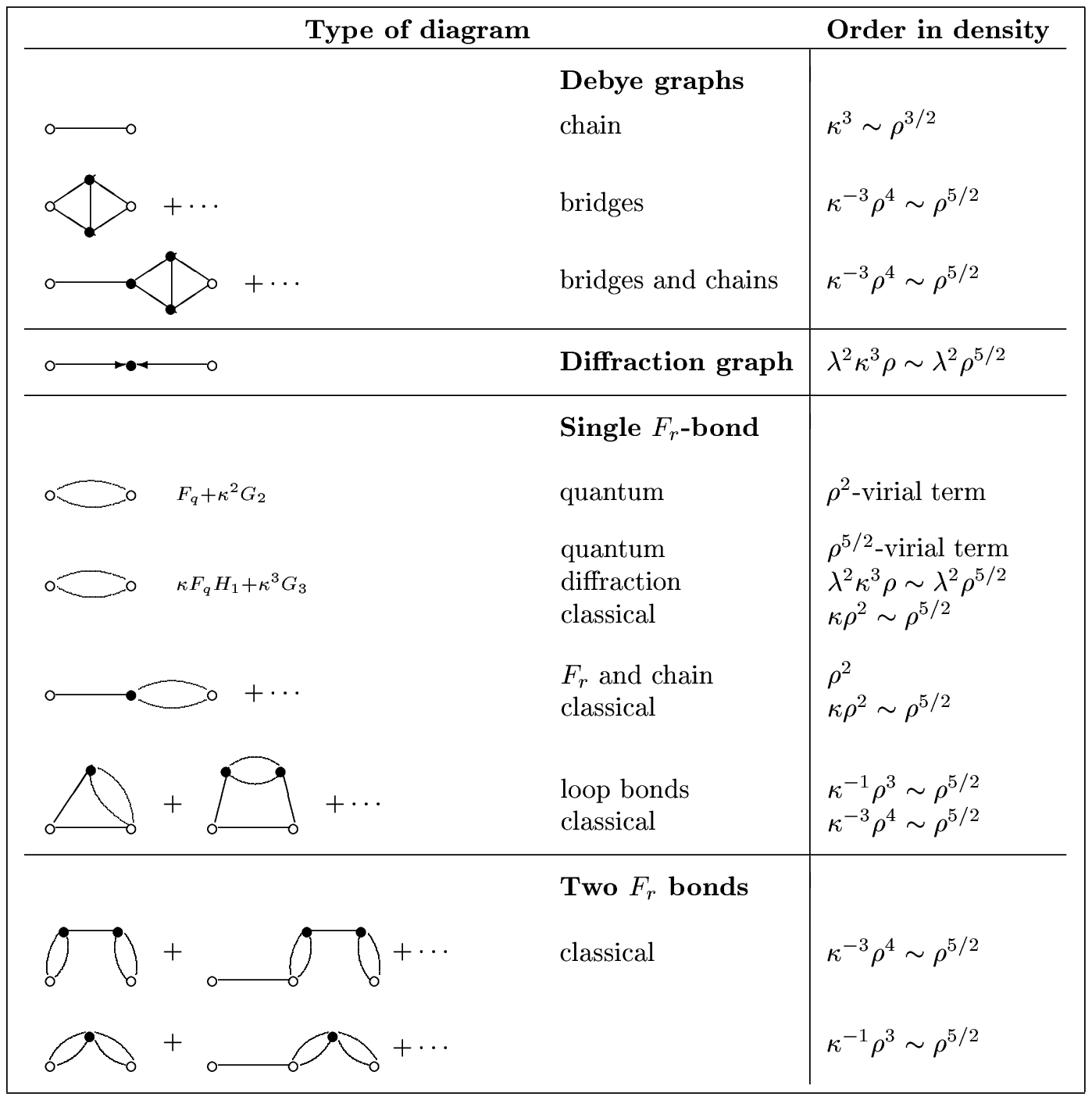,height=8in,width=6.5in}
\caption{Examples of contributions to Maxwell-Boltzmann free energy}
\label{fig12}
\end{figure}

The density dependence contained in the Debye length can be extracted by a scaling analysis.
Introducing the dimensionless variable $x=\kappa|{\bf R}_{1}-{\bf R}_{2}|$ one sees that $F_{cc}$
scales as $\kappa$, $F_{cd}\;(F_{dc})$ as $\kappa^{2}$, and $F_{dd}$ as $\kappa^{3}$. More
generally a higher order multipole term of the form
$$
(\bx_{1}(s_{1})\cdot{\bf \nabla}_{1})^{k_{1}}(\bx_{2}(s_{2})\cdot{\bf \nabla}_{2})^{k_{2}}
\frac{\exp(-\kappa |{\bf R}_{1}-{\bf R}_{2}|)}{|{\bf R}_{1}-{\bf R}_{2}|}
$$
scales as $\kappa^{k_{1}+k_{2}+1}$. In $F_{r}$ the quantum effects (associated with the de
Broglie length $\lambda_{\alpha}$) and classical screening (associated with the Debye length
$\kappa^{-1}$) can be disentangled by a decomposition of the form
\begin{equation}
F_{r}=F_{q}(1+\kappa H_{1}+\kappa^{2}H_{2}+\cdots)+\kappa^{2}G_{2}+\kappa^{3}G_{3}+\cdots
\label{6.40}
\end{equation}
where $F_{q}$ is the quantum (truncated) bond
\begin{eqnarray}
F_{q}(1,2)&=&\exp\left(-\beta_{12}\int_{0}^{1}dsV(|\br+\lambda_{\alpha_{1}}\bx_{1}(s)-
\lambda_{\alpha_{2}}\bx_{2}(s)|)\right)\nonumber\\
&-&1+\frac{\beta_{12}}{r}-\frac{\beta_{12}^{2}}{2r^{2}}+\beta_{12}\int_{0}^{1}ds
[\lambda_{\alpha_{1}}\bx_{1}(s)\cdot{\bf \nabla}-
\lambda_{\alpha_{2}}\bx_{2}(s)\cdot{\bf \nabla}]\(\frac{1}{r}\)\nonumber\\
\label{6.41}
\end{eqnarray}
and the $H_{i}$ and $G_{i}$ are scaled functions involving contributions of higher order
multipoles. We distinguish three different types of effects in this game:
\begin{itemize}
\item classical screening manifested by the occurrence of the Debye length in the scaled bonds
\item quantum diffraction effects due to the coupling of quantum fluctuations $\bx$ to the 
classical Debye potential as in $F_{cc}, \;F_{cd}$ and $F_{dd}$
\item quantum bound and scattering states appearing through $F_{r}$.
\end{itemize}
As an illustration the table in Figure~\ref{fig12} gives a (non
exhaustive) list of examples of graphs contributing to different
effects in the MB free energy.  The calculations are tractable up to
order $\rho^{5/2}.$ There are two simplifications up to this order:
internal weights $\rho(\alpha,\bx)$ can always be replaced by their
lowest order $\rho_{\alpha}$ and only low order multipoles contribute
in view of the above mentioned scaling arguments. The MB pressure is
given by the terms (\ref{6.48a}) to (\ref{6.51}) in the final result
stated in the next subsection.

\subsubsection{The exchange contributions and the final result}
\label{subsec-6.2.1}

To include the exchange contributions perturbatively, one has to come
back to the expression (\ref{1.2.5}) of the grand-partition function
with statistics and collect together all the terms involving the
exchange of exactly n particles
\begin{equation}
\Xi_{\Lambda}=\Xi_{\Lambda}^{{\rm
MB}}+\sum_{n=2}^{\infty}\Xi_{\Lambda}^{(n)}=\Xi_{\Lambda}^{{\rm MB}}
\(1+\sum_{n=2}^{\infty}\frac{\Xi_{\Lambda}^{(n)}}{\Xi_{\Lambda}^{{\rm
MB}}}\) \label{6.44}
\end{equation}
where $\Xi_{\Lambda}^{{\rm MB}}$ is the Maxwell-Boltzmann partition function
and the contribution with $N_{\alpha}$ exchanged particles has
necessarily an activity factor
$\exp(\beta\mu_{\alpha}N_{\alpha})$. The expansion (\ref{6.44})
generates in turn an expansion of the grand-canonical pressure
\begin{equation}
P=\beta^{-1}\lim_{|\Lambda|\rightarrow\infty}\frac{1}{
|\Lambda|}\ln\Xi_{\Lambda}=P^{{\rm MB}}+\sum_{n=2}^{\infty}P^{(n)}
\label{6.45}
\end{equation}
with $P^{{\rm MB}}$ the Maxwell-Boltzmann pressure and $P^{(n)}$ the
contribution of $n$ exchanged particles. The matrix element in
(\ref{1.2.5}) corresponding to a two particle exchange, say
$\langle\br_{2},\br_{1},\br_{3},\ldots,\br_{N}|\exp(-\beta
H_{\Lambda,N})|\br_{1},\br_{2},\br_{3},\ldots,\br_{N}\rangle$, will
involve $N-2$ closed filaments $\br_{i,i}(s),\;i=3,\ldots,N$ together
with two open filaments $\br_{1,2}(s)$ and $\br_{2,1}(s)$. Keeping
these two open filaments fixed, but performing the configuration
integrals and grand-canonical sums on all the remaining closed
filaments leads to the consideration of the inhomogeneous
Maxwell-Boltzmann partition function $\Xi_{\Lambda}^{(2)(inhom)}$. It
is the grand-partition function of a system of (closed) filaments in
presence of the potential due to these two open filaments. Similarly,
higher order exchange terms can be viewed as partition functions of
closed filaments in presence of an external potential created by the
open ones. The density of filaments $\rho^{(inhom)}({\cal F})$ in the
inhomogeneous system can be expanded in terms of the correlations
$\rho({\cal F}), \;\rho_{T}({\cal F}_{1},{\cal F}_{2})$ of the
homogeneous gas by standard methods. Finally, one can apply to the
latter correlations all the diagrammatic techniques previously
described to obtain the density expansion of the MB gas. There is
however one caveat. The particle densities are given in principle by
the usual relation
\begin{equation}
\rho_{\alpha}=\frac{\partial}{\partial \mu_{\alpha}}P(\{\mu_{\gamma}\})
\label{6.46}
\end{equation}
where $P$ is the total grand-canonical pressure (\ref{6.45}) including
the statistics. But $P^{{\rm MB}}$ has been computed directly as function of
densities that we call now $\rho^{{\rm MB}}_{\alpha}$ within the MB
approximation. Then the $\rho^{{\rm MB}}_{\alpha}$ can be eliminated in
favor of the true densities $\rho_{\alpha}$ by noting that both $P$
and $P^{{\rm MB}}$ are evaluated at the same values of the chemical
potentials $\mu_{\alpha}$. Hence we have the thermodynamic relation
\begin{equation}
\mu_{\alpha}=\frac{\partial}{\partial
\rho_{\alpha}^{{\rm MB}}}f^{{\rm MB}}(\{\rho_{\gamma}^{{\rm MB}}\}) \label{6.47}
\end{equation}
with $f^{{\rm MB}}$ the free energy of the MB gas. Using (\ref{6.46})
together with (\ref{6.47}) enables us to express the $\mu_{\alpha}$ and
$\rho^{{\rm MB}}_{\alpha}$ in terms of the desired particle densities
$\rho_{\alpha}$.  The final result is \cite{AlPe96}
\begin{eqnarray}
\beta P&=&\sum_{\alpha}\rho_{\alpha}-\frac{\kappa^{3}}{24\pi}\label{6.48a}\\
&+&\pi\beta^{3}\sum_{\alpha\gamma}\rho_{\alpha}\rho_{\gamma}e_{\alpha}^{3}
e_{\gamma}^{3}\left[    \frac{(\ln 2-1)}{6}+\(\frac{1}{3}-\frac{3}{4}\ln 2+\frac{1}{2}\ln 3\)\beta
\kappa e_{\alpha}e_{\gamma}\right]\nonumber\\
&+&C_{1}\beta^{5}\kappa^{-1}\sum_{\alpha\gamma\delta}
\rho_{\alpha}\rho_{\gamma}\rho_{\delta}e_{\alpha}^{3}e_{\gamma}^{4}e_{\delta}^{3}
+C_{2}\beta^{6}\kappa^{-3}\sum_{\alpha\gamma\delta\eta}
\rho_{\alpha}\rho_{\gamma}\rho_{\delta}\rho_{\eta}e_{\alpha}^{3}e_{\gamma}^{3}
e_{\delta}^{3}e_{\eta}^{3}\nonumber\\
& &\label{6.49}\\
&+&\frac{1}{16}\beta^{2}\hbar^{2}\kappa^{3}\sum_{\alpha}\rho_{\alpha}
\frac{e_{\alpha}^{2}}{m_{\alpha}}\label{6.50}\\
&-&\frac{\pi}{\sqrt{2}}\sum_{\alpha\gamma}\rho_{\alpha}\rho_{\gamma}(1+\frac{3}{2}\beta\kappa
e_{\alpha }e_{\gamma})\left[\lambda_{\alpha\gamma}^{3}
Q(x_{\alpha\gamma})+\frac{\sqrt{2}}{3}\beta^{3}e_{\alpha}^{3}e_{\gamma}^{3}
\ln(\kappa\lambda_{\alpha\gamma})\right]\nonumber\\
& &\label{6.51}\\
&+&\frac{\pi}{\sqrt{2}}\sum_{\alpha}\frac{(-)^{2\sigma_{\alpha}+1}}{2\sigma_{\alpha}+1}
\rho_{\alpha}^{2}\lambda_{\alpha\alpha}^{3}\(1+\frac{3}{2}\beta\kappa
e_{\alpha}^{2}\)E(x_{\alpha\alpha})\label{6.51a}\\
&+&O(\rho^{3}\ln \rho)\nonumber
\end{eqnarray}
with 
\begin{eqnarray*}
&&
\kappa= (4\pi\beta\sum_{\alpha}e_{\alpha}^{2}\rho_{\alpha})^{1/2}, \;\lambda_{\alpha\gamma}
=(\beta\hbar^{2}/m_{\alpha\gamma})^{1/2},\;\\
&&
m_{\alpha\gamma}=
m_{\alpha}m_{\gamma}/(m_{\alpha}+m_{\gamma}),\;
C_{1}=15.205\pm.001,\;C_{2}=-14.733\pm.001
\end{eqnarray*}
In (\ref{6.51}), $x_{\alpha\gamma}=-\sqrt{2}\beta
e_{\alpha}e_{\gamma}/\lambda_{\alpha\gamma}$ and $Q(x_{\alpha\gamma})$
is the quantum second order virial coefficient first introduced in
\cite{Ebe67,Ebe68}
\begin{eqnarray}
Q(x_{\alpha\gamma})&=&\frac{1}{\sqrt{2}\pi\lambda_{\alpha\gamma}^{3}}\lim_{R\rightarrow\infty}
\left\{\int_{r<R}d\br\left[(2\pi\lambda_{\alpha\gamma}^{2})^{3/2}\langle\br|\exp(-\beta
H_{\alpha\gamma})|\br \rangle
\right.\right.\nonumber\\
&-&\left.\left.1+\frac{\beta e_{\alpha}e_{\gamma}}{r}-\frac{\beta^{2}
e_{\alpha}^{2}e_{\gamma}^{2}}{2r^{2}}\right]
+\frac{2\pi}{3}\beta^{3}e_{\alpha}^{3}e_{\gamma}^{3}
\(\ln\(\frac{3\sqrt{2}R}{\lambda_{\alpha\gamma}}\)+C\)\right\}\nonumber\\ 
\label{6.52}
\end{eqnarray}
while $E(x_{\alpha\alpha})$ is the exchange integral
\begin{equation}
E(x_{\alpha\alpha})=2\sqrt{\pi}\int d\br \langle-\br|\exp(-\beta
H_{\alpha\alpha})|\br\rangle \label{6.53}
\end{equation}
In (\ref{6.52}) and (\ref{6.53}) $H_{\alpha\gamma}$ is the relative one-body Coulomb Hamiltonian 
for a particle of mass $m_{\alpha\gamma}$ submitted to the potential 
$\frac{e_{\alpha}e_{\gamma}}{r}$,
and $C=.577216\ldots$ is the Euler-Mascheroni's constant. 

The formula (\ref{6.48a})-(\ref{6.51a}) incorporates and generalizes
several earlier results. The perfect gas contribution
$\sum_{\alpha}\rho_{\alpha}$ (the dominant one) has first been
established rigorously in \cite{LePe73}, completed by \cite{Hug85}
\footnote{Hughes studies the continuity of the free energy at
vanishing density in relation to the techniques of Fefferman
discussed in the next chapter.}. The next term is the lowest order
correction to the free gas: one recognizes the familiar
Debye-H\"{u}ckel contribution (\ref{2.1.2b}). The quantum diffraction
effects appear in the term (\ref{6.50}) at the order $\rho^{5/2}$. The
term (\ref{6.51}) incorporates the total contribution of bound and
scattering states of the two-body Coulomb Hamiltonian\footnote{The
$\ln (\kappa\lambda_{\alpha\gamma})$ term can be recombined with
$Q(x_{\alpha\gamma})$. The splitting is introduced to have
$Q(x_{\alpha\gamma})$ defined as in \cite{KKER86}. Note also that our
de Broglie length differs from that in \cite{KKER86} by a factor
$\sqrt{2}$.}.  The regularization of the integral in (\ref{6.52}),
which insures that $Q(x_{\alpha\gamma})$ is finite, is not arbitrary
but comes from the structure of the bond $F_{q}$ (\ref{6.41}). One has
here a $\rho^{2}$ contribution, the proper second order virial term,
and a $\rho^{5/2}$ multiplicative correction that arises as a
many-body mean field effect. Finally, the exchange term (\ref{6.51a})
has a similar structure.

The terms up to order $\rho ^{5/2}$, including the exact $\rho
^{2}$-virial contribution (\ref{6.51}) are found in the work of
Ebeling \cite{Ebe67,Ebe68} and related references \cite{KKER86}. The
full formula (\ref{6.48a}-\ref{6.51a}), up to the diffraction term
(\ref{6.50}) appears in \cite{KKER86}.  These works use the method of
the effective potential (or method of Morita).  In this method, one
associates to the quantum Coulomb gas an equivalent classical system
of point particles interacting with an effective many-body
interaction. In the case of the MB gas, this effective many-body
potential $\Psi_{N}$ is defined by interpreting the diagonal element
$\langle\{\br_{i}\}|\exp(-\beta H_{N})|\{\br_{i}\}\rangle$ as a
classical Boltzmann factor
\begin{equation}\langle\{\br_{i}\}|\exp(-\beta
H_{N})|\{\br_{i}\}\rangle=\prod_{i=1}^{N}\(\frac{1}{2\pi\lambda_{\alpha_{i}}^{2}}\)^{3/2}
\exp(-\beta\Psi_{N}(\br_{1},\ldots,\br_{N})) \label{6.54}
\end{equation}One can decompose $\Psi_{N}$ into a sum of two-body,
three-body and higher order interactions, the two-body potential being
\begin{equation} \Psi_{2}(\br_{1},\br_{2})=-\beta^{-1} \ln
\left[\prod_{i=1}^{2}\(2\pi\lambda_{\alpha_{i}}^{2}\)^{3/2}
\langle\br_{1},\br_{2}|\exp(-\beta
H_{2})|\br_{1},\br_{2}\rangle\right] \label{6.55} \end{equation}
Calculations up to order $\rho^{5/2}$ have been performed
by keeping only the two-body potential (\ref{6.55}), see references in
\cite{KKER86}. The long range Coulombic part of (\ref{6.55}) can again
be eliminated by summing the convolution chains and one can use the
Abe-Meeron diagrammatics. This procedure gives the above formula for
the pressure except (\ref{6.50}). It turns out that one has to take
the three-body interaction into account to recover the diffraction
term.

One can specialize the equation of state to the one component plasma
and take the classical limit\footnote{This can be achieved by
considering an asymmetrical two component system and letting
$m_{2}\rightarrow\infty,\; e_{2}\rightarrow
0,\rho_{2}\rightarrow\infty $ with $\rho_{2}e_{2}$ finite and
$\rho_{2}e_{2}+\rho_{1}e_{1}=0$.  The one component plasma remains
well behaved in the classical limit because of the spreading of the
positive charge.}: these classical terms coincide with those
calculated in \cite{CoMu69}. One retrieves also the $\hbar$-correction
$\pi\beta^{2}e^{2}\rho^{2}\hbar^{2}/6m$ to the classical equation of
state \cite{PoHa73}. This correction has to be of order $\rho^{2}$:
indeed one checks that the $\hbar^{2}\rho^{5/2}$ contributions coming
from (\ref{6.50}) and (\ref{6.51}) compensate exactly.  Finally high
temperature series are recovered by expanding $Q(x)$ and $E(x)$ in
powers of $\sqrt{\beta}$ (such expansions are derived in
\cite{KKER86}, and these series agree with the expressions obtained by
the methods of \cite{Dew62,Dew66,DSSK95}. Explicit formulae for
various thermodynamic functions as well as their specialization to
these particular cases can be found in \cite{AlPe96}.

The generalization of the low density equation of state to a uniformly
magnetized plasma is given in \cite{Cor97,Cor98b}.

\def\({\left(}
\def\){\right)}

\def\br{{\bf r}}
\def\bx{{\bf \xi}}
\def\cf{{\cal F}}
\def\cl{{\cal L}}
\def\um{\underline{\bf m}}
\def\ue{\underline{\bf e}}
\def\uN{\underline{\bf N}}
\def\umu{\underline{\mbox{\boldmath $\mu$\unboldmath}}}
\def\bmu{\mbox{\boldmath $\mu$\unboldmath}}
\def\blambda{\mbox{\boldmath $\lambda$\unboldmath}}

\newpage \section{The atomic and molecular limit} \setcounter{page}{1}

\label{chapter-7}

\subsection{The electron-proton gas in the Saha regime}
\label{sec-7.1}

The diagrammatic technique developed in chapter~\ref{chapter-loops}
has provided a systematic low density expansion of the equation of
state at fixed temperature. In particular, the lowest order term
$\beta P=\sum_{\alpha=1}^{\cal S}\rho_{\alpha}$ in (\ref{6.48a})
represents a mixture of perfect Maxwell-Boltzmann gases constituted by
the ${\cal S}$ species of fully dissociated charges: it is a free
plasma state.  Possible two-body bound states appear only as
corrections in the second order virial coefficient.  In this chapter,
we will treat another regime where the basic constituents are now
chemically bound entities behaving as ideal substances. This regime (
called the Saha regime) is characterized by a joint limit where both
the density and the temperature go to zero in a coupled way. The
(exponentially fast) rate at which the density is reduced as $T
\rightarrow 0$ determines a certain energy-entropy balance, selecting
in turn the formation of some specific chemical species. Indeed low
temperature favors binding over ionization, whereas low density, by
increasing the available phase space per particle, favors
dissociation. This limit is called the molecular (or atomic) limit.
It was first formulated in precise terms by \cite{Fef85} and further
studied by \cite{CLY89,GrSc95a}. Equilibrium ionization phases in this
context are discussed in \cite{MaMa90}. The Saha regime is also
described in the usual language of many-body perturbation theory in
\cite{EKK76,KKER86}.

In order to understand the issues involved in a simple setting, we
consider the electron-proton (e-p) system and the possible formation
of hydrogen atoms (see also \cite{Fef86,Mar93} for general
discussion and background). Let us adopt for a moment the
thermodynamic view point that a chemist would take in presence of
three preformed species, the electrons (e), the protons (p) and the
hydrogen atoms (a) in their ground state. As a first approximation he
takes into account the binding energy of each chemical species, but
otherwise treats them as perfect gases of point particles. In this
approximation, the grand-canonical densities are \footnote{This
density $\rho_{j}=\rho_{{\rm ideal},j}$ as well as the densities
(\ref{7.32}) are those of ideal gases to which the interacting system
will be eventually compared. For brevity, we drop the mention ideal
here and in (\ref{7.32}).}  \begin{equation}\rho_{j}=\(\frac{m_j}{2\pi \beta
\hbar^2}\)^{3/2}\exp({-\beta(E_j-\mu_j))},\;\;\,\;\;\;j=e,\:p,\:a
\label{7.1} \end{equation}where $\mu_j$ are the respective chemical potentials,
$m_a=m_e+m_p,\;\; E_e=E_p=0$ and $E_a<0$ is the ground state energy of
the hydrogen atom. The law of chemical equilibrium for the
dissociation reaction $e+p\leftrightarrow a$ requires \begin{equation}
\mu_a=\mu_e+\mu_p \label{7.2} \end{equation}and one also must have the
neutrality $\rho_e=\rho_p $. Taking (\ref{7.2}) into
account and introducing the combinations \begin{equation}
\mu=\frac{\mu_e+\mu_p}{2},\;\;\,\nu=\frac{\mu_e-\mu_p}{2} \label{7.4}
\end{equation}it is easily seen that the neutrality condition imposes the choice
\begin{equation}\nu=\nu_{0}(\beta)=\frac{3}{4\beta}\ln\frac{m_p}{m_e} \label{7.5}
\end{equation}and hence (\ref{7.1}) becomes
\begin{equation}
\rho_e =
\rho_p=\(\frac{\sqrt{m_em_p}}{2\pi\beta\hbar^2}\)^{3/2}\exp(\beta\mu),
\ \ \
\rho_a = \(\frac{m_e+m_p}{2\pi\beta\hbar^2}\)^{3/2}\exp(-\beta(E_a-2\mu))
\label{7.6}
\end{equation}
Finally, since the species are treated as non-interacting, the
pressure obeys the law of perfect gases
\begin{eqnarray}
\beta P &=& \rho_e+\rho_p+\rho_a =\(\frac{1+\gamma}{2}\)\rho\label{7.7}
\end{eqnarray}
In (\ref{7.7}), the pressure is expressed in terms of the total number
density $\rho $ of protons and electrons $\rho=\rho_e+\rho_p+2\rho_a$
and we have introduced the Saha coefficient for the degree of
ionization
\begin{equation}
\gamma=\frac{\rho_e}{\rho_e+\rho_a},\;\;\;\;\;\;\,\;\,0\:\leq\:\gamma\:\leq\:1
\label{7.10}
\end{equation}
A low density-low temperature regime can clearly be obtained by choosing $\mu$ negative and letting
$\beta\rightarrow \infty$. Let us examine various cases noting from (\ref{7.6}) that
$\rho_a/\rho_e\sim \exp(-\beta(E_a-\mu)$.
\begin{description}
\item[(i)]
Fix $\mu<E_a$. The atomic density is exponentially small compared to that of the electrons and
protons; there is full dissociation ($\gamma\rightarrow 1$) and the equation of state reduces to
$\beta P=\rho$ as in the low density expansion (\ref{6.48a}).
\item[(ii)]
Fix $\mu>E_a$ (but $\mu<E_a/2$). One obtains the opposite situation where the pure atomic phase
dominates; now $\gamma\rightarrow 0$ and $\beta P=\frac{1}{2}\rho$ as $\beta\rightarrow \infty$.
\item[(iii)]
At the borderline $\mu=E_a$, $\rho_e$ and $\rho_a$ are of the same order. More precisely, replace
$\mu$ by 
\begin{equation}
\tilde{\mu}(\beta)=E_a +\lambda\beta^{-1}+o(\beta^{-1}) 
\label{7.11}
\end{equation}
for some $\lambda,-\infty<\lambda<\infty$, i.e.  approach the point $(E_a,0)$ in the $(\mu,T)$
plane along a direction having a finite slope $\lambda$. Then one obtains ionization equilibrium
phases that interpolate between the fully ionized and the atomic ones. Their degree of ionization
found from (\ref{7.6}), (\ref{7.10}) and (\ref{7.11}) is
\begin{equation}
\gamma=\(\(\frac{m_e+m_p}{\sqrt{m_em_p}}\)^{3/2}\exp (\lambda) +1\)^{-1}
\label{7.12}
\end{equation}
and they obey the equation of state (\ref{7.7}). Written in terms of the pressure, the Saha
coefficient reads after a short computation
\begin{equation}
\gamma=\(1+\beta P \(\frac{2\pi \beta \hbar^2 (m_e+m_p)}{m_em_p}\)^{3/2}\exp(-\beta E_a)\)^{-1/2}
\label{7.13}
\end{equation}
\end{description}
This discussion makes clear that to an increase of $\mu$ corresponds an increase of the densities
that favors the electron-proton binding when $\mu$ crosses the value $E_a$. How can we justify
this picture from statistical mechanics? Consider now the exact infinite volume pressure
$P(\beta,\mu_e,\mu_p)$ of the e-p system in the grand-canonical
ensemble, and recall from section~\ref{sec- neutrality} that it
depends only on $\mu=\frac{\mu_e+\mu_p}{2}$ 
\begin{equation}
P(\beta,\mu_e,\mu_p)=P(\beta,\mu)
\label{7.14}
\end{equation}
One may expect that the equation of state associated with $P(\beta,\mu)$  becomes close to
(\ref{7.7}) as $\beta\rightarrow \infty$ for appropriate values of $\mu$ as in cases (i)-(iii).
The theorem on the atomic and molecular limit will gives a precise formulation of this assertion.

In the elementary considerations just presented, we have not taken
into account, among other things, that electrons and protons can form
other complexes as, for instance, hydrogen molecules.  Let us examine
this issue from the view point of statistical mechanics. If typical
configurations of charges are indeed those of a dilute gas of hydrogen
atoms (case (ii)), these configurations will be mainly formed of e-p
pairs, the extension of a pair being of the order of the Bohr radius
$a_B$, whereas two different pairs are at distance $\rho_a^{-1/3}
\;(a_B\ll \rho_a^{-1/3}) $.  If we choose at random in this
configuration a region $D$ of linear extension $R$ with \begin{equation}a_B\:\ll
\:R\:\ll\:\rho_a^{-1/3} \label{7.15} \end{equation}we will observe that this
region is empty most of the time, but if it contains something it is
exactly one hydrogen atom, except on rare occasions. In the
grand-canonical formalism, it means that one must be able to find
chemical potentials $\mu_e,\mu_p$ such that for $\beta\:\gg\:1$ and
all $(N_e,N_p)\neq (0,0),(1,1)$ \begin{equation}p_{00}(\beta,\mu_e,\mu_p,D)\gg
p_{11}(\beta,\mu_e,\mu_p,D)\gg p_{N_eN_p}(\beta,\mu_e,\mu_p,D)
\label{7.16} \end{equation}where \begin{equation}
p_{N_eN_p}(\beta,\mu_e,\mu_p,D)=\frac{1}{\Xi_{D}}\exp(\beta(\mu_eN_e+\mu_pN_p))
\mbox{Tr}\exp(-\beta H_{D,N_eN_p}) \label{7.17} \end{equation}is the probability
to find exactly $N_e$ electrons and $N_p$ protons in $D$. Thus the
condition (\ref{7.16}) means indeed that if $D$ is not void, it is
more probable to find an e-p pair than anything else. The dominant
term in $\mbox{Tr}\exp(-\beta H_{D,N_eN_p})$ as $\beta\to\infty$ will
be $\exp(-\beta E_{D,N_{e}N_{p}})$ where $E_{D,N_{e}N_{p}}$ is the
ground state energy of $H_{D, N_{e}N_{p}}$. Hence to obtain the
inequalities (\ref{7.16}), one must be able to find values of
$\mu_{e}$ and $\mu_{p}$ such that \begin{equation}
E_{D,N_{e}N_{p}}-(\mu_{e}N_{e}+\mu_{p}N_{p})>E_{D,11}-(\mu_{e}+\mu_{p})>0
\;\;\mbox{for all}\;(N_e,N_p)\neq (0,0),\:(1,1) \label{7.18} \end{equation}Since
$\rho_{a}$ tends to zero exponentially fast as $\beta\to\infty$ we can
also let the region $D$ grow with $\beta$ (but maintaining the
inequalities (\ref{7.15})) so that $E_{D,N_{e}N_{p}}$ will differ from
the bottom of the spectrum of $H_{N_{e}N_{p}}$ in infinite space \begin{equation}
E_{N_eN_p}=\mbox{inf spectrum}(H_{N_eN_p}) \label{7.19} \end{equation}by a
vanishingly small error. Moreover, introducing $\mu,\nu$ as in
(\ref{7.4}) \begin{equation}\mu_eN_e+\mu_pN_p=\mu(N_e+N_p)+\nu(N_e-N_p)
\label{7.20} \end{equation}we remark that the part $\mu$ of the chemical
potential controls the total number density and $\nu$ the total charge
density. We can take advantage of the fact that the infinite volume
limit of the pressure will not depend on $\nu$ to make a convenient
choice. As it was done in (\ref{7.5}), we fix $\nu$ to ensure
neutrality in the ideal gas when the lowest energy state corresponds
to charged complexes. Then $\nu$ will be of order $0(\beta^{-1})$ as
in the dissociated e-p system and can be dropped from (\ref{7.20}) as
$\beta$ gets large.  Therefore, in view of these remarks (that will be
made mathematically more precise in section~\ref{sec-7.3}) the
relevant inequalities to be satisfied to produce the situation
(\ref{7.16}) as $\beta\to\infty$ are (with $E_{11}=E_{a}$) \begin{equation}
E_{N_eN_p}-\mu(N_e+N_p)>E_{a}-2\mu>0\;\;\mbox{for all}\;(N_e,N_p)\neq
(0,0),\:(1,1) \label{7.21} \end{equation}for some values of $\mu$.

\begin{figure}
\psfig{figure=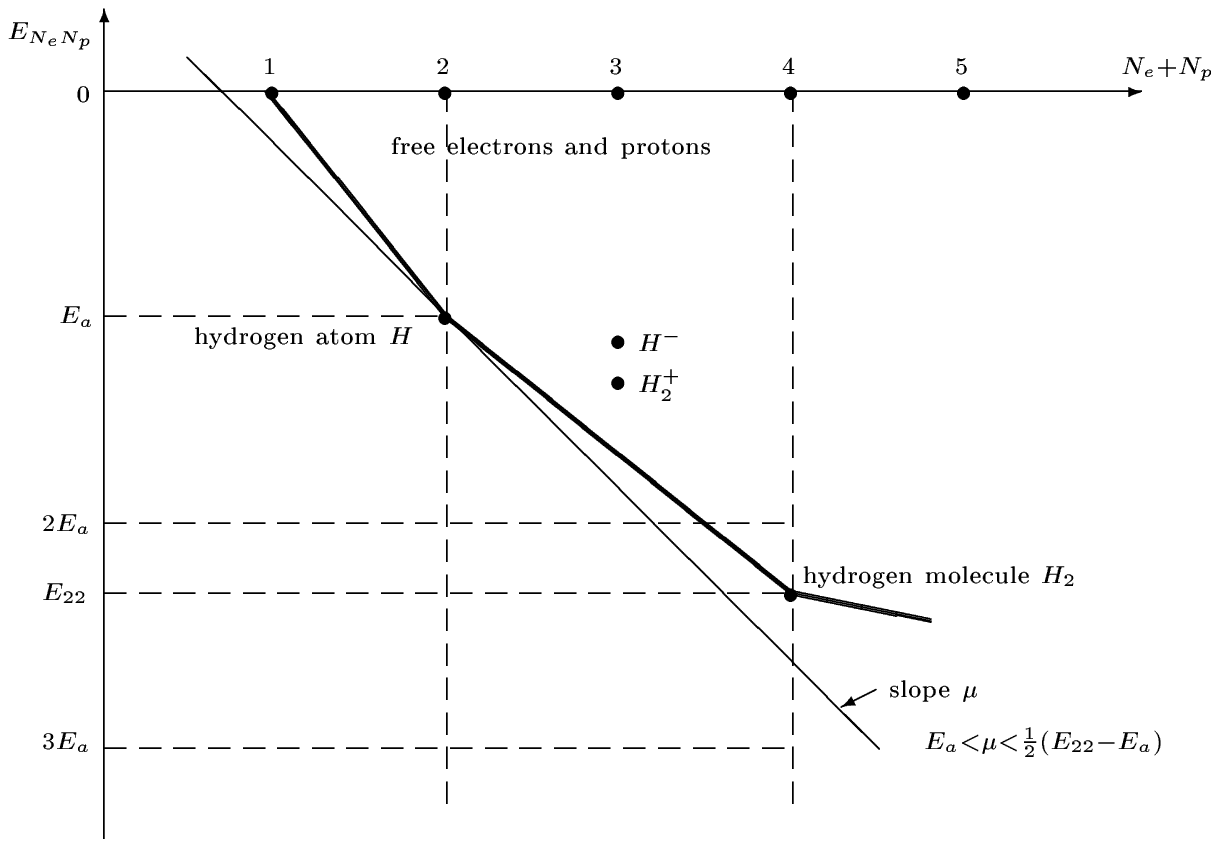,height=8in,width=6.5in}
\caption{The inequalities (\ref{7.21}) }
\label{fig13}
\end{figure}

Let us examine some implications of (\ref{7.21}). If $(N_e,N_p)=(1,0)$ (single electron) $E_{10}=0$
implies in  (\ref{7.21})
\begin{equation}
E_{a}<\mu<0
\label{7.22}
\end{equation}
If $(N_e,N_p)=(2,2)$,  (\ref{7.21}) gives
\begin{equation}
\mu<\frac{1}{2}(E_{22}-E_{a})
\label{7.23}
\end{equation}
where $E_{22}$ is the ground state energy of the hydrogen
molecule. One can find $\mu$ satisfying (\ref{7.22}) and (\ref{7.23})
only if $E_{a}<\frac{1}{2}(E_{22}-E_{a})$ or equivalently if
$|E_{22}-2E_{a}|<|E_{a}|$. The latter inequality means that the
binding energy gained by the formation of an hydrogen molecule must be
less than the binding energy of the atom itself, a well known fact. It
allows the entropy to win over molecular binding at low
density. However the numerical values of $(E_{22}-E_{a})/2 \sim - 9
\text{ev}$ and of $\mu \sim -8 \text{ev}$, $k_{\rm B}T \sim 0.026
\text{ev}$ under prevailing Earth conditions give an exceedingly small
value of the ratio $\rho _{a}/\rho _{{\rm H}_{2}} \sim \exp (\beta
(E_{22}-E_{a} - 2\mu ))$. Only molecular hydrogen appears on Earth,
while the formation of the atomic gas discussed here can only occur in
very dilute extraterrestrial medium. These considerations on molecular
hydrogen do not suffice to guarantee the validity of the full set of
inequalities (\ref{7.21}): this necessitates a simultaneous
examination of all ground state energies $E_{N_{e}N_{p}}$, a highly
nontrivial problem.  We shall discuss it further at the end of the
next section. A geometric illustration of the inequalities
(\ref{7.21}) is given in Figure~\ref{fig13}. Notice that the binding
energies of the ion $H^{-}$ and the ionized molecule $H_{2}^{+}$ are
too weak: they do not meet the inequalities analogous to (\ref{7.21}),
appropriate to these cases, for any value of $\mu$.

Coming back to the e-p pressure, we may conjecture that if $\mu$ is
increased just beyond $\frac{1}{2}(E_{22}-E_{a})$, the chances for
binding also increase, and the pressure $P(\beta,\mu)$ will become
close to that a free gas of hydrogen molecules as $\beta\rightarrow
\infty$. More generally, if one fixes $\mu <0$, one expects that
$P(\beta,\mu)$ approaches the pressure of free gases of some complexes
that are selected at this particular value of the chemical potential.

\subsection{The main theorem}
\label{sec-7.2}

With these preliminaries in mind, we are ready to formulate the main
theorem about the Saha regime for electrons and an arbitrary number of
nuclei with masses $m_{\alpha}$, charges $e_{\alpha}$, particle
numbers $N_{\alpha}$ and chemical potentials $\mu_{\alpha},\:
\alpha=1,\ldots,{\cal S}$ ($\alpha=1$ refers to electrons). As in
section~\ref{sec- neutrality} we adopt a vector notation for
these. Set
\begin{eqnarray}
\um &=& (m_{1},\ldots,m_{{\cal S}}),
\ \ \
\ue = (e_{1},\ldots,e_{{\cal S}})\nonumber\\
\uN &=& (N_{1},\ldots,N_{{\cal S}}), \ \ \ N=\sum_{\alpha}^{{\cal S}}N_{\alpha},
\ \ \ 
\umu = (\mu_{1},\ldots,\mu_{{\cal S}})
\label{7.24}
\end{eqnarray}
As we know (see (\ref{2.1.1ee})), the infinite volume pressure
depends only on the component $\bmu$ of the chemical potential orthogonal to the charge vector.
As a consequence, the kind of complexes that will occur for a certain value of the chemical
potentials $\umu$ depends only on the value of $\bmu$.
 We will use  this freedom to make a convenient choice of $\nu$ along the proof.

The study of the molecular limit of $P(\beta, \bmu)$ (i.e. $\beta\rightarrow \infty,\;\bmu$ fixed)
will be carried out under the basic assumption on $\bmu$: 
\begin{description}
\item[(A)]
there exists $\kappa>0$ ($\kappa$ independent of $\uN$) such that
\end{description}
\begin{equation}
H_{\uN}-\bmu\cdot\uN \:\geq\: \kappa N\;\;\;\;\;\;\;\;\;\; \mbox{for all} \;\uN
\label{7.27}
\end{equation}
The complexes (atoms, ions, molecules) that will be formed in this limit are defined as follows.
Consider the lowest energy state of the $N$-body Hamiltonian $H_{\uN}$
\begin{equation}
E_{\uN}=\mbox{inf spectrum}(H_{\uN})=\inf\{(\Phi,H_{\uN}\Phi):\:\Phi\in {\cal H}_{N},\:||\Phi||=1\}
\label{7.28}
\end{equation}
and set
\begin{equation}
E(\bmu)=\inf_{\uN\neq \underline{{\bf 0}}}(E_{\uN}-\bmu\cdot\uN)
\label{7.29}
\end{equation}
A value $\uN_{j}$ for which this infimum is taken, i.e.
\begin{equation}
E_{\uN_{j}}-\bmu\cdot \uN_{j}=E(\bmu)
\label{7.30}
\end{equation}
is called the composition of the complex (j) having ground state energy $E_{j}=E_{\uN_{j}}$.
Notice that (\ref{7.27}) imposes the bound $N_{j,\alpha}\:\leq\:\kappa^{-1}E(\bmu)$ for all
$\alpha$, so complexes are made out of a finite number of constituents, hence their number g is
also finite. The complex (j), made of $N_{j,\alpha},\;\alpha=1,\ldots,{\cal S}$, elementary
constituents, has total particle number, mass and charge 
\begin{equation}
N_{j}=\sum_{\alpha}^{\cal S}N_{j,\alpha},\,\;\;\;\;\,\,m_{j}=\um\cdot\uN_{j},
\;\;\,\,\;\;\;e_{j}=\ue\cdot\uN_{j} 
\label{7.31}
\end{equation}

Strictly speaking, $H_{\uN}$ has no ground state because of
translation invariance. However for $N \geq 2$ one can decompose
$H_{\uN}=K_{\uN}^{{\rm cm}}+H_{\uN}^{{\rm rel}}$ into the sum of the
kinetic energy of the center of mass and the relative Hamiltonian,
expressed in the coordinates relative to the center of mass. Since the
spectrum of $K_{\uN}^{{\rm cm}}$ starts from zero, one has also
$E_{\uN}=\mbox{inf spectrum}(H_{\uN}^{{\rm rel}})$. Consider now the
threshold $\sigma_{c}(H_{\uN}^{{\rm rel}})$ of the continuous spectrum
of $H_{\uN}^{{\rm rel}}$. We know from the HVZ theorem \cite{ReSi79}
that
\begin{equation}
\sigma_{c}(H_{\uN}^{{\rm rel}})=\inf _{a,b;N_{a}+N_{b}=N}(E_{\uN_{a}}+E_{\uN_{b}}),\;\;\;N\geq 2 
\label{7.31a}  
\end{equation}
where the infimum is taken over all partitions of the $N$ particles into two clusters $a$ and $b$ 
having $N_{a}$ and $N_{b}$ particles, and $N_{a}=1,\ldots,N-1$. From the very definition of
$E(\bmu)$, one concludes from (\ref{7.31a}) that\linebreak[3] $\sigma_{c}(H_{\uN}^{{\rm
rel}}-\bmu\cdot \uN)\geq 2E(\bmu), N\geq 2$. This means that the part of the spectrum of
$H_{\uN}^{{\rm rel}}-\bmu\cdot \uN$ located  in $[E(\bmu), \;2E(\bmu))$ consists of isolated
eigenvalues ($E(\bmu)$ is strictly positive). In particular $E_{\uN_{j}}$ has to be an eigenvalue of
$H_{\uN_{j}}^{\rm rel}$ and the corresponding eigenstate is the the wave function of the bound
complex $(j)$. When $N=1$, complexes reduce to the individual electrons or  nuclei themselves with
pure kinetic energy and we take then $E_{\uN}=0$. 

When there are charged complexes, we add to (A) the assumption     
\begin{description}
\item[(B)]
if a complex has a charge $e_{j}\neq 0$, there is another complex having a charge of
sign opposite to $e_{j}$ (but not necessarily equal to $-e_{j}$) .
\end {description}
insuring that an overall neutral gas can be formed.

We describe now the mixture of free gases made of these complexes.  To
each complex (j) one associates the density of a perfect gas of point
particles \begin{equation}
\rho_{j}(\beta,\umu)=\(\frac{m_{j}}{2\pi\beta\hbar^{2}}\)^{3/2}
\exp[-\beta(E_{j}-\umu\cdot\uN_{j})] \label{7.32} \end{equation}and define the
pressure of an ideal mixture of such gases by \begin{equation}\beta P_{{\rm
ideal}}(\beta,\umu)=\sum_{j=1}^{g}\rho_{j}(\beta,\umu) \label{7.33}
\end{equation}The corresponding number density of electrons and nuclei is \begin{equation}
\rho_{{\rm ideal}}(\beta,\umu)=\sum_{\alpha=1}^{{\cal
S}}\frac{\partial}{\partial\mu_{\alpha}} P_{{\rm
ideal}}(\beta,\umu)=\sum_{j=1}^{g}N_{j}\rho_{j}(\beta,\umu)
\label{7.34} \end{equation}If some complexes are charged, $ P_{{\rm
ideal}}(\beta,\umu)$ for a general $\umu$ may belong to a non-neutral
ensemble. Then, under assumption (B), the neutrality
$\sum_{j}e_{j}\rho_{j}=0$ is implemented by setting
\begin{equation}P_{{\rm ideal, neutral}}(\beta,\umu)=\inf_{\nu}
P_{{\rm ideal}}(\beta,\umu+\nu\ue) \label{7.35} \end{equation}Then
$P_{{\rm ideal, neutral}}(\beta,\umu)=P_{{\rm ideal,
neutral}}(\beta,\bmu)$ verifies (\ref{2.1.1ee}), and one can formulate
the theorem on the molecular limit as

\begin{description}
\item[Theorem]

{\em Let $\bmu$ be a chemical potential satisfying the condition (A) and determining
complexes (j), $j=1,\ldots,g$, as described above, fulfilling the condition (B).
Then there exists $\varepsilon>0$ such that 
\begin{equation}
P(\beta,\tilde{\bmu})=P_{{\rm ideal,
neutral}}(\beta,\tilde{\bmu})(1+O(\exp(-\beta\varepsilon))) \label{7.36}
\end{equation}
as $\beta\rightarrow\infty$ and $\tilde{\bmu}\rightarrow\bmu$.}
\end{description}

The theorem deserves several comments that will be again illustrated
in the e-p system. One should first appreciate that all the residual
effects due to thermal excitations and interactions between complexes
when the temperature and the densities are not strictly equal to zero
are controlled by the $O(\exp(-\beta\varepsilon))$ correction. The
issue of the theorem is precisely about obtaining this control.

Concerning the assumption (A), it is a difficult task to find what are
the complexes associated with a chemical potential $\bmu$ and to
verify (\ref{7.27}). Conversely, if we ask for the occurrence of a
certain chemical species, it is still beyond the present possibilities
to rigorously determine the range of $\bmu$ (if any) allowing for the
formation of this molecule.  Suppose that we ask for the formation of
hydrogen atoms in the e-p system. This means that there should be an
interval $\mu\in (\mu_{1},\mu_{2})$ such that (A) holds and
$\inf_{(N_e,N_p)\neq (0,0)}(E_{N_eN_p}-\mu(N_e+N_p))$ is taken at
$(N_e,N_p)=(1,1)$ with $E(\mu)=E_{a}-2\mu$. This is equivalent to the
fact that the inequalities (\ref{7.21}) hold when $\mu\in
(\mu_{1},\mu_{2})$ (if (\ref{7.21}) is true for such a $\mu$, one can
pick $\kappa>0$ small enough so that the same inequalities are true
for $\mu+\kappa$; then $\kappa$ provides the constant in the lower
bound (\ref{7.27})). A little bit of thinking reveals that the
validity of the inequalities (\ref{7.21}) for a range of $\mu$ ($\mu$
slightly above $E_{a}$) is equivalent in turn with the following
statement about the stability of matter: there exists a constant $B$
with $0<B<|E_{a}|$ such that 
\begin{equation}H_{N_eN_p}\geq -B(N_e+N_p-1)
\;\;\,\mbox{for all }(N_e,N_p)\neq (0,0),(1,1) \label{7.37} 
\end{equation}
The point in (\ref{7.37}) is that the constant $B$ can be chosen
strictly less than $|E_{a}|$ for all cases except of course for the
hydrogen atom itself. Therefore a complete proof of the existence of
the atomic phase relies on exhibiting a sufficiently small stability
constant. Although (\ref{7.37}) must hold on experimental and
numerical grounds, a rigorous proof has not yet been provided (see
\cite{Fef86} for a more detailed discussion).

Taking from now on (\ref{7.37}) as a working hypothesis, we recover
easily the previous cases (i)-(iii) when applying the theorem to the
e-p system. If $\mu$ is slightly above $E_{a}$, there is a single
complex, the hydrogen atom (a) with $E(\mu)=E_{a}-2\mu$ (case
(ii)). If $\mu=E_{a}$, we have three complexes (e,p,a) and
$E(\mu)=|E_{a}|$ (case (iii)). If $\mu<E_{a}$, the complexes are (e,p)
and $E(\mu)=|\mu|$ (case (i)). When there are charged complexes, the
minimizer $\nu$ of $P_{ideal}(\beta,\umu+\nu\ue)$ has the value
(\ref{7.5}) so that the result of the theorem is precisely the
equation of state (\ref{7.7}) up to an exponentially small correction.

For arbitrary nuclei, the theorem describes generically similar situations involving more
complicated atoms, ions and molecules. If $\bmu$ selects a single complex, we have a free gas
of that chemical species. If $\bmu$ selects several complexes, we can have coexisting free
gases as $T\rightarrow 0$, generalizing the situation (iii) of the section \ref{sec-7.1}. As a
corollary of the theorem, we can specify the relative proportion of these gases
by taking $\tilde{\bmu}$  a function of the temperature that approaches $\bmu$
linearly as $T\rightarrow 0$ (as in (\ref{7.11}))   
\begin{equation}
\tilde{\bmu}(\beta)=\bmu+{\blambda}\beta^{-1}+o(\beta^{-1})
\label{7.38}
\end{equation}
The densities of the gases will have weights
proportional to $\exp({\blambda}\cdot\uN_{j})$, leading then to general Saha formulae.

Although the determination of the kind of complexes occurring for a
given value of $\bmu$ remains unsolved, it is pleasing to see how the
thermodynamics of ideal substances emerges in principle from the basic
statistical mechanics of electrons and nuclei in the molecular limit.

\subsection{Elements of proof}
\label{sec-7.3}

The theorem was first proven \cite{Fef85} for the e-p system under the
assumption (\ref{7.37}) (with a weaker control ${\rm O}(\beta^{-1})$
of the error term). All the main ideas and subsequent developments
summarized here have their roots in this work.

The general strategy of the proof does not rely on
low density expansions, but on an analysis of typical equilibrium configurations of dilute gases of
atoms and molecules. It proceeds along the following qualitative ideas. Let $\bmu$
be fixed and satisfy the condition (\ref{7.27}), and take $\beta$ very large. 
Then, in such configurations,  if we look into a region of linear extension $R$
in the range (\ref{7.15}), according to the same discussion as that following
(\ref{7.15}), this region is empty with high probability, but if it contains
something, this will essentially be one of the complexes determined by $\bmu$.
The idea is now to decompose the total domain $\Lambda=\bigcup_{r}D_{r}$
into subdomains $D_{r}$ of diameter $R$ and to let $R$ grow with $\beta$ in a suitable way.
One has to solve the two following problems. 
\begin{description}
\item[Problem 1]
Determine the size of $D_{r}$ such that the partition function $\Xi_{D_{r}}(\beta,\umu)$
of a single subdomain is that of a system having
at most complexes of the type $(j),\; j=1, \ldots, g$, in $D_{r}$.
\item[Problem 2]   
Show that the residual interaction between the different regions $D_{r}$ is negligible as 
$\beta\rightarrow\infty$ so that
the total partition function $\Xi_{\Lambda}(\beta,\umu)\simeq
\prod_{r}\Xi_{D_{r}}(\beta,\umu)$ is close to that of independent subdomains $D_{r}$.

\end{description}

We make the preliminary decision (valid throughout the whole proof) to
choose $\nu=\nu_{0}(\beta)$ as the minimizer of (\ref{7.35}) insuring
the neutrality of the ideal gas. This is equivalent to the
pseudo-neutrality condition met earlier on several occasions. If some
complex is not neutral and the condition (B) holds, $\nu_{0}(\beta)$
exists, is unique and $\nu_{0}(\beta)={\rm O} (\beta^{-1})$ as can be
checked from the explicit formulae (\ref{7.33}) and (\ref{7.34}) (see
(\ref{7.5}) in the e-p system). If all complexes are neutral one takes
$\nu=0$. Thus in the sequel we keep in mind that
\begin{equation}
\umu=\bmu +{\rm O}(\beta^{-1})\ue 
\label{7.38a}
\end{equation}

\subsubsection{Towards a solution of problem 1}

We envisage now problem 1, following mainly \cite{CLY89}.  We write
the grand-canonical partition function for a finite region
$D\subset{\Bbb R}^3$ of volume $|D|$ as a sum on all particle numbers
\begin{equation}\Xi_{D}(\beta,\umu)=1+\sum_{j=1}^{g}\Xi_{D}(\beta,\umu,\uN_{j})
+\sum_{\uN \neq \uN_{j}}\Xi_{D}(\beta,\umu,\uN) \label{7.39} \end{equation}where
we have singled out the contribution of the $g$ complexes that
correspond to the chosen value of $\bmu$ and
\begin{eqnarray}
\Xi_{D}(\beta,\umu,\uN)
&=&\mbox{Tr}_{D}\exp\(-\beta\( H_{D,\uN}-\umu\cdot\uN\)\)\nonumber\\
&=&\mbox{Tr}_{D}\exp\(-\beta K_{D,\uN}^{{\rm cm}}\)
\mbox{Tr}_{D}\exp\(-\beta\( H_{D,\uN}^{{\rm rel}}-\umu\cdot\uN\)\)\nonumber\\
\label{7.40}
\end{eqnarray}
In (\ref{7.40}), $\mbox{Tr}_{D}$ means that the trace is taken on states having the Dirichlet
conditions appropriate to the finite volume Hamiltonians $H_{D}$ as well as the appropriate particle
statistics.
For $R$ large enough, one expects that the center of mass contribution is close to that of a free
particle in space, and that the finite volume ground state energy $E_{D,j}$ of the complex $j$ is
close to its value $E_{j}$ in infinite space. Indeed one can establish 
\begin{equation}
\mbox{Tr}_{D}\exp\(-\beta K_{D,\uN_{j}}^{{\rm cm}}\)=\(\frac{m_{j}}{2\pi\beta\hbar^{2}}\)^{3/2}
|D|\(1+{\rm O}\(\frac{1}{R}\)\)
\label{7.41}
\end{equation}
and
\begin{equation}
E_{D,j}=E_{j}+{\rm O}\(\frac{1}{R^{2}}\)
\label{7.42}
\end{equation}
Now let
\begin{equation}
|D|=|D_{\beta}|\simeq e^{c\beta},\;\;\;\,c>0
\label{7.42a}
\end{equation}
grow exponentially fast as the temperature tends to zero, and consider the contribution of the
ground state of $ H_{D,\uN_{j}}^{{\rm rel}}$ to the partition function (\ref{7.40}) for 
$\uN=\uN_{j}$. Taking (\ref{7.41}), (\ref{7.42}) and (\ref{7.42a}) into account, this contribution
is   
\begin{eqnarray}
& &\(\frac{m_{j}}{2\pi\beta\hbar^{2}}\)^{3/2}
e^{-\beta(E_{j}-\umu\cdot\uN_{j})}|D_{\beta}|\(1+{\rm O}\(e^{-\frac{c}{3}\beta}\)\)
\nonumber\\&=&\rho_{j}(\beta,\umu)|D_{\beta}|\(1+{\rm
O}\(e^{-\frac{c}{3}\beta}\)\)
\label{7.43}
\end{eqnarray}
By (\ref{7.30}) and (\ref{7.38a}), the density of the ideal gas $\rho_{j}(\beta,\umu)$ (\ref{7.32}) 
decreases as  $\beta^{-3/2}e^{-\beta E(\bmu)},\;E(\bmu)>0$; thus one can choose $c>0$ small enough so
that (\ref{7.43}) is still exponentially small for $\beta$ large. 

We like to show that, under the condition (\ref{7.27}), all the other terms in the sum (\ref{7.39})
are negligible compared to (\ref{7.43}), provided that the rate of growth of $|D_{\beta}|$ is
suitably chosen. As $|D|\to\infty$, part of the spectrum of $H_{D,\uN}^{{\rm rel}}$ becomes
continuous, corresponding to the scattering states of all the sub-complexes belonging to 
$H_{D,\uN}^{{\rm rel}}$.  The basic idea is to ensure the convergence of the traces by
the pure kinetic energy: one will dispense with the full Hamiltonian with the help of (\ref{7.27})
and the ensuing gap property. Then the strict positivity of the constant $\kappa$ in (\ref{7.27}) 
and of the gap will be used to control the volume dependence of these traces.
For this one borrows a small fraction of the kinetic energy $K_{\uN}$ writing 
\begin{equation}
H_{\uN}=\eta K_{\uN}+(1-\eta)K_{\uN}+V_{\uN},\;\;\;\,\eta>0
\label{7.45}
\end{equation}
where $ K_{\uN}=-\sum_{i=1}^{N}\frac{\hbar^{2}\Delta_{i}}{2m_{i}}$ and
$V_{\uN}$ is the total Coulomb interaction. Clearly the Hamiltonians
$(1-\eta)K_{\uN}+V_{\uN}$ and
$(1-\eta)^{-1}(K_{\uN}+V_{\uN})=(1-\eta)^{-1}H_{\uN}$ are related by
the scaling transformation $x_{i}\longrightarrow (1-\eta)x_{i}$. Hence
they are unitarily equivalent. This implies, using (\ref{7.27})
and(\ref{7.38a})
\begin{eqnarray}
(1-\eta)K_{\uN} +V_{\uN}-\umu\cdot\uN&\geq& 
(1-\eta)^{-1}\mbox{inf spectrum}(H_{\uN}-\bmu\cdot\uN)\phantom{-------}\nonumber\\ 
 &+&(1-\eta)^{-1}\eta \bmu\cdot\uN+
{\rm O}(\beta^{-1})N\nonumber\\ &\geq&(1-\eta)^{-1}(\kappa-\eta\max_{i}|\mu_{i}|)N+
{\rm O}(\beta^{-1})N\equiv c_{1}N\nonumber\\
\label{7.46}
\end{eqnarray}
Thus
\begin{equation}
H_{\uN}-\umu\cdot\uN\geq \eta K_{\uN}+c_{1}N
\label{7.46a}
\end{equation}
where $c_{1}>0$ provided that $\eta$ is small enough and $\beta$ is large enough.
The same inequality holds when
$H_{\uN}$ is replaced by $H_{D,\uN}$ since Dirichlet boundary conditions increase the kinetic energy;
hence one concludes that 
\begin{equation}
\Xi_{D}(\beta,\umu,\uN)\leq \exp(-\beta c_{1}N)
\mbox{Tr}_{D}e^{-\beta\eta K_{\uN}}\leq |D|^{N}\prod_{i=1}^{N}
\(\frac{m_{i}}{2\pi\eta\beta\hbar^{2}}\)^{3/2}e^{-\beta c_{1}N}
\label{7.47}
\end{equation}
If now $|D|=|D_{\beta}|$ grows at the rate $e^{c\beta}$ with
$0<c<c_{1}$, $\Xi_{D_{\beta}}(\beta,\umu,\uN)$ is bounded by
$e^{-c_{2}\beta N}$ for some $c_{2}>0$.  Thus by choosing $N_{0}$
large, one can make the exponential decay of the whole sum of terms
with $N\geq N_{0}$ in (\ref{7.39}) fast enough to have
\begin{equation}
\sum_{\uN,N\geq
N_{0}}\Xi_{D_{\beta}}(\beta,\umu,\uN)=\rho_{j}(\beta,\umu)|D_{\beta}|{\rm
O}\(e^{-\beta\varepsilon}\),\;\,\;\varepsilon>0 
\label{7.48}
\end{equation}

For the remaining terms, i.e. the terms with $\uN\neq\uN_{j},\;N<N_{0}$ and the contribution of the
excited states to $\Xi_{D}(\beta,\umu,\uN_{j})$, one uses essentially the fact that 
$E(\bmu)$ is separated from the rest of the spectrum by a gap,
according to the discussion following (\ref{7.31a}). On the subspace orthogonal to the ground states
of  $H_{\uN_{j}}^{{\rm rel}}-\bmu\cdot\uN_{j}$ one has, arguing as in (\ref{7.46}) 
\begin{eqnarray}
(1-\eta)K_{\uN} +V_{\uN}-\umu\cdot\uN&\geq& 
(1-\eta)^{-1}(E(\bmu) + 2\delta)\phantom{-------}\nonumber\\ 
 &+&(1-\eta)^{-1}\eta \bmu\cdot\uN+
{\rm O}(\beta^{-1})N\nonumber\\
&\geq&(1-\eta)^{-1}(E(\bmu)+2\delta-\eta\max_{i}|\mu_{i}|N)+{\rm O}(\beta^{-1})N\nonumber\\
\label{7.48a}
\end{eqnarray}
for some $\delta>0$. Since now $N<N_{0}$, one can choose $\eta$ sufficiently small and $\beta$
sufficiently large to have on this subspace
\begin{equation}
H_{\uN}-\umu\cdot\uN\geq \eta K_{\uN}+E(\bmu)+\delta,\;\,\,\delta>0,\;\;N<N_{0}
\label{7.48b}
\end{equation}
Hence for  $\uN\neq\uN_{j},\;N<N_{0}$, one finds as in (\ref{7.47}) 
\begin{equation}
\Xi_{D}(\beta,\umu,\uN)\leq e^{-\beta (E(\bmu)+\delta)}
\mbox{Tr}_{D}e^{-\beta\eta K_{\uN}}\leq |D|^{N}\prod_{i=1}^{N}
\(\frac{m_{i}}{2\pi\eta\beta\hbar^{2}}\)^{3/2}e^{-\beta E(\bmu)}e^{-\beta\delta }
\label{7.49}
\end{equation}
Since $|D_{\beta}|^{N}$ grows at most as $e^{c\beta N_{0}}$ we can again have
\begin{equation}
\Xi_{D_{\beta}}(\beta,\umu,\uN)=
\rho_{j}(\beta,\umu)|D_{\beta}|{\rm O}(e^{-\beta\varepsilon}),\;\;\uN\neq\uN_{j},\;N<N_{0},
\;\;\varepsilon>0
\label{7.50}
\end{equation}
by taking $c$ small.
Notice that a volume factor $|D_{\beta}|$ corresponding to the free motion of the overall center of
mass has been explicitly maintained in the estimates (\ref{7.48}) and (\ref{7.50}) (at the
expense of a possibly smaller constant $\varepsilon$).  
Finally one deals with the excited states of $H_{\uN_{j}}^{{\rm rel}}-\bmu\cdot\uN_{j}$
in the same way. 
Combining this with (\ref{7.43}), (\ref{7.48})
and (\ref{7.50}) in the grand canonical sum (\ref{7.39}), one obtains 
\begin{equation}
\Xi_{D_{\beta}}(\beta,\umu)=1+\sum_{j=1}^{g}\rho_{j}(\beta,\umu)|D_{\beta}|  \(1+{\rm
O}\(e^{-\varepsilon\beta}\)\) 
\label{7.50a}
\end{equation}
Since the second term in the right hand side is exponentially small as $\beta\rightarrow \infty$
(see the comment after (\ref{7.43})) one has also 
\begin{equation}
\ln\Xi_{D_{\beta}}(\beta,\umu)=\sum_{j=1}^{g}\rho_{j}(\beta,\umu)|D_{\beta}|  
\(1+{\rm O}\(e^{-\varepsilon\beta}\)\) 
\label{7.50b}
\end{equation}
Taking into account the definition (\ref{7.33}) together with our choice of $\nu$, 
this says that the pressure $P_{D_{\beta}}(\beta,\umu)$ in $D_{\beta}$ is given
by  
\begin{equation}
\beta P_{D_{\beta}}(\beta,\umu)=\frac{1}{|D_{\beta}|}\ln \Xi_{D_{\beta}}(\beta,\umu)
=\beta P_{{\rm ideal}}(\beta,\umu )\(1+{\rm O}\(e^{-\varepsilon\beta}\)\) 
\label{7.51}
\end{equation}
This solves problem 1: in a volume $|D_{\beta}|\simeq e^{c\beta}$ with $c>0$ sufficiently small
(depending only on $\bmu$), the pressure is exclusively due
to the non-interacting  complexes $(j),\; j=1,\ldots,g$. However this does not  prove the
theorem yet. In (\ref{7.51}), $D_{\beta}$ is constrained by the growth condition (\ref{7.42a}) and not
an independent thermodynamic variable as it should be in the grand canonical ensemble. One needs to
solve problem 2.

\subsubsection{Towards a solution of problem 2}

Problem 2 is technically considerably more sophisticated, so let us
give first some very heuristic ideas. Decompose ${\Bbb
R}^3=\bigcup_{r}D_{r}$ into a disjoint union of domains $D_{r}$ with
characteristic functions $\chi_{D_{r}}$, $\chi_{D_{r}}(x) =1$ if $x$
is in $D_{r}$ and $\chi_{D_{r}}(x) =0$ otherwise.  One wants to show
that the interactions between the different domains $D_{r}$ become
negligible as $\beta\rightarrow\infty$. To this end, define an
uncorrelated Hamiltonian by 
\begin{equation}H^{{\rm
uncor}}_{\Lambda}=H_{\Lambda}-U^{{ \rm cor}} \label{7.52} 
\end{equation}
where
$U^{{ \rm cor}}$ is the potential energy between the different domains
\begin{equation}
U^{{ \rm cor}}=\sum_{i<j}e_{\alpha_{i}}e_{\alpha_{j}}V^{{ \rm
cor}}(x_{i},x_{j}) \label{7.52a} 
\end{equation}
\begin{equation}V^{{ \rm
cor}}(x_{1},x_{2})=\frac{1-\sum_{r}\chi_{D_{r}(x_{1})}
\chi_{D_{r}}(x_{2})}{|x_{1}-x_{2}|}
\label{7.53} 
\end{equation}$U^{{ \rm cor}}$ (as well as $H^{{\rm un\rm
cor}}_{\Lambda}$) will be an effective temperature dependent energy
via the condition $ |D_{r}|\simeq e^{c\beta}$ determined in problem 1.
Note that if $|x_{1}-x_{2}|<R$ and both $x_{1}$ and $x_{2}$ belong to
the same subdomain, then $V^{{\rm \rm cor}}(x_{1},x_{2})=0$; also
$V^{{\rm \rm cor}}(x_{1},x_{2})$ is Coulombic at large distances. This
resembles the properties of a regularized Coulomb interaction. Suppose
that we are able to define the uncorrelated Hamiltonian in a less
crude way than in (\ref{7.52}) and (\ref{7.53}) with the effect of
replacing (\ref{7.53}) by a smooth positive definite regularized
potential, finite everywhere for $|x_{1}-x_{2}|<R$ and asymptotically
Coulombic . A possible candidate is \begin{equation}V^{{ \rm cor}}_{{\rm
reg}}(x_{1},x_{2})=\int dy_{1}\int
dy_{2}\frac{\phi_{R}(y_{1})\phi_{R}(y_{2})}
{|x_{1}-y_{1}-x_{2}+y_{2}|} \label{7.54} \end{equation}where
$\phi_{R}(y)=R^{-3}\phi(\frac{y}{R})$ with $\phi(y)$ a smooth
function, ($\phi(y)=0,\;|y|\geq 1,\;\int dy\phi(y)=1$) representing a
charge density supported in a region of extension $R$ with total
charge equal to $1$.  Other candidates closer to the form (\ref{7.53})
are \begin{equation}V^{{\rm \rm cor}}_{{\rm
reg}}(x_{1},x_{2})=\frac{1-h\(\frac{x_{1}-x_{2}}{R}\)}{|x_{1}-x_{2}|}
\label{7.55} \end{equation}with $h(x)$ a smooth short range function, $h(0)=1$,
and $\frac{1-h(x)}{|x|}$ positive definite.  Then we can use the basic
positivity argument already introduced in section~\ref{sec- classical}
\begin{eqnarray}
U^{{\rm \rm cor}}_{{\rm reg}}&=&\frac{1}{2}\sum_{ij}^{N}e_{\alpha_{i}}e_{\alpha_{j}}V^{{\rm
\rm cor}}_{{\rm reg}}(x_{i},x_{j}) -\frac{1}{2}V^{{\rm
\rm cor}}_{{\rm reg}}(0)\sum_{i}^{N}e_{\alpha_{i}}^{2}\nonumber\\ &\geq& -\frac{1}{2}V^{{\rm
\rm cor}}_{{\rm reg}}(0)\sum_{i=1}^{N}e_{\alpha_{i}}^{2} \label{7.56}
\end{eqnarray}
because of the positive definiteness of $V^{{\rm \rm cor}}_{{\rm
reg}}(x_{1},x_{2})$.  Note that by scaling, in both forms (\ref{7.54})
and (\ref{7.55}), one has $V^{{\rm \rm cor}}_{{\rm reg}}(0) = {\rm
O}\(\frac{1}{R}\)$. Therefore if one can find such a regularized
correlation energy, one will have (being still very sketchy) that $
H^{{\rm un\rm cor}}_{\Lambda}$ provides a lower bound to the full
Hamiltonian
\begin{eqnarray}
H_{\Lambda}-\umu\cdot \uN &=& H^{{\rm uncor}}_{\Lambda}+U^{{\rm cor}}_{{\rm reg}}-\umu\cdot
\uN \geq H^{{\rm uncor}}_{\Lambda}-(1+{\rm O}(R^{-1}))\umu\cdot\uN
\label{7.57}
\end{eqnarray}
Thus, up to a small correction to the chemical potential, one finds
that the partition function is dominated by that of the uncorrelated
domains
\begin{equation}
\Xi_{\Lambda}(\beta,\umu)\leq \prod_{D_{r}\cap \Lambda \neq \emptyset }
\Xi_{D_{r} }(\beta,\umu(1+{\rm O}(R^{-1})))
\label{7.58}
\end{equation}
We set now $|D_{r}|=|D_{r,\beta}|\simeq e^{c\beta}$, $\beta$ sufficiently large, and apply to each of
the $D_{r}$ the result (\ref{7.50b}) of problem 1. Taking the logarithm of (\ref{7.58}) and absorbing
the error in the chemical potential in the ${\rm O}(e^{-\varepsilon \beta)})$ correction, one obtains
in the thermodynamic limit 
\begin{eqnarray} 
\beta P(\beta, \umu)&\leq&
\lim_{|\Lambda|\rightarrow\infty}\left[\frac{1}{|\Lambda|}\sum_{D_{r,\beta}\cap
\Lambda\neq\emptyset}|D_{r,\beta}|\right]
\beta P_{{\rm ideal}}(\beta,\umu )\(1+{\rm
O}\(e^{-\varepsilon\beta}\)\)\nonumber\\  &=&\beta P_{{\rm ideal}}(\beta,\umu )\(1+{\rm
O}\(e^{-\varepsilon\beta}\)\)   \label{7.59}
\end{eqnarray} 

The technically most elaborate parts of the works
\cite{Fef85,CLY89,GrSc95a} consists in constructing an uncorrelated
Hamiltonian having the requested positivity and smoothness properties,
and comparing it with the true Hamiltonian $H_{\Lambda}$. In its main
lines the construction of \cite{CLY89,GrSc95a} goes as follows.

The decomposition ${\Bbb R}^3=\bigcup_{r}D_{r}$ is realized with
regions $D_{r}$ that are translates and dilations (with the scale
parameter $R$) of a fundamental domain $D_{0}$.  Let ${\cal
F}_{D_{r}}$ be the Fock space for a Coulomb system localized in
$D_{r}$, having Hamiltonian $H_{D_{r}}$ with Dirichlet boundary
conditions on $\partial D_{r}$. The most obvious way to define states
of independent Coulomb systems, each of them confined in a region
$D_{r}$, is to introduce the product space ${\cal F}_{\Lambda}^{\rm
uncor}=\prod_{D_{r}\cap \Lambda \neq \emptyset}^{\bigotimes} {\cal
F}_{D_{r}}$ and the Hamiltonian $\sum_{D_{r}\cap \Lambda \neq
\emptyset}H_{D_{r}}$ on it. One wants to compare this obviously
strictly uncorrelated Hamiltonian with the true $H_{\Lambda}$ acting
on ${\cal F}_{\Lambda}$ when the scale parameter $R\to \infty$.  For
this, one introduces a partition of the unity on ${\Bbb R}^3$ of the
form $\sum_{r}\chi^{2}_{r,R}(x)=1$ for all $x\in {\Bbb R}^3$ where
$\chi_{r,R}$ are suitably chosen smooth functions essentially
supported in the region $D_{r}$ of linear size $R$. With the help of
this partition, one defines an isometry $J : {\cal F}_{\Lambda}\to
{\cal F}_{\Lambda}^{\rm uncor}$ which maps the states of the system
onto the uncorrelated ones. Then one compares the image
$J^{-1}(\sum_{D_{r}\cap \Lambda \neq \emptyset}H_{D_{r}})J$ of the
strictly uncorrelated Hamiltonian with $H_{\Lambda}$ on ${\cal
F}_{\Lambda}$. The calculation gives
\begin{equation}
H_{\Lambda}-J^{-1}(\sum_{D_{r}\cap \Lambda \neq \emptyset}H_{D_{r}})J=U^{\rm
cor}_{{\rm reg}}+O\left(\frac{1}{R^{2}} \right)
\label{7.60}
\end{equation}
where $U^{{\rm cor}}_{{\rm reg}}$ has the form (\ref{7.52a}) and (\ref{7.55}), and 
\begin{equation}
h(x)=\int dy \chi^{2}(x+y)\chi^{2}(y) 
\label{7.60a}
\end{equation}
is a smooth function localized in the neighborhood of $D_{0}$:
(\ref{7.60}) is a more precise version of (\ref{7.52}). The ${\rm
O}\left(\frac{1}{R^{2}}\right)$ term is the price of the increase in
kinetic energy paid for the additional Dirichlet conditions imposed at
the boundary of each $D_{r}$ in the strictly uncorrelated Hamiltonian
$\sum_{r}H_{D_{r}}$. This price can be made small as $R\to\infty$
because of the differentiability of the localization functions
$\chi_{r,R}$.

The main issue is now the lower bound (\ref{7.56}), i.e. to have
$V^{{\rm \rm cor}}_{{\rm reg}}(x_{1},x_{2})$ positive definite.  In
\cite{CLY89}, the regions $D_{r}$ are cubes, and this positivity is
obtained by replacing the Coulomb potential in $\sum_{D_{r}\cap
\Lambda \neq \emptyset}H_{D_{r}}$ by a Debye potential
$\frac{e^{-\varepsilon|x|}}{|x|}$ with small $\varepsilon$
($\varepsilon\simeq e^{-c\beta}$). Then it turns out that
$$
V^{{\rm \rm cor}}_{{\rm reg}}(x_{1},x_{2})=
\frac{1}{|x_{1}-x_{2}|}\(1-\frac{h((x_{1}-x_{2})/R)}{h(0)}
e^{-\varepsilon |x_{1}-x_{2}|}\)
$$
is similar to $ V_{\infty ,\lambda }$ (\ref{1.1.3bc}).  Its Fourier
transform can be shown to be positive for $\varepsilon$ and $R$ in
appropriate ranges by using properties of $h(x)$ (\ref{7.60a}), and the
arguments (\ref{7.56})-(\ref{7.59}) can eventually be cast in a
rigorous form.  However, to complete the proof, one has to show that
the assumption (A) in the main theorem as well as the results obtained
in problem 1 are stable when the Coulomb potential is approximated by
the Debye potential. In particular, if (A) holds with a constant
$\kappa$, then one has also for the Debye Hamiltonian
$H^{\varepsilon}_{\uN}$ \begin{equation}
H^{\varepsilon}_{\uN}-\bmu\cdot\uN\geq\kappa(\varepsilon)N\;\;\,{\rm
with}\;\,\kappa(\varepsilon)\to\kappa, \;\;\varepsilon\to 0
\label{7.61} \end{equation}This theorem is proven in \cite{CLY89} with
the same localization techniques

In \cite{GrSc95a}, the decomposition ${\Bbb R}^3=\bigcup_{r}D_{r}$ is
realized with simplices. Then the stability (\ref{7.56}) follows
remarkably from the geometrical properties of a simplex.  Namely for
the spherical average $\bar{h}(|x|)$ of $h(x)$ (\ref{7.60a}) (here
$\chi=\chi_{D_{0}}$ is the characteristic function of a simplex), the
function $\frac{\bar{h}(0)-\bar{h}(|x|)}{|x|}$ has a positive Fourier
transform.

The original proof \cite{Fef85} uses certain coverings of ${\Bbb
R}^{3}$ (called Swiss cheese) made of small cubes and large balls of
different radii, and a positive definite potential of the form
(\ref{7.54}). The delicate point is to estimate the difference
$V^{{\rm error}}= V^{{\rm cor}}- V^{{\rm cor}}_{{\rm reg}}$ between
(\ref{7.53}) and (\ref{7.54}). $V^{{\rm error}}$ still contains a
local singularity as well as non-positive definite Coulomb energy
terms. By averaging on the different radii of the balls, Fefferman
shows that these dangerous contributions in $V^{{\rm error}}$ are
majorized by a Coulomb potential $\frac{{\rm O}(\beta^{-1})}{|x|}$
with a small amplitude, and that these dangerous terms become
eventually negligible. Then the stability (\ref{7.56}) follows from
the positive definiteness of $ V^{{\rm cor}}_{{\rm reg}}$ (\ref{7.54}).

In all cases the proofs involve fixing a number of important details not mentioned
here.

Finally, (\ref{7.59}) must be completed by the converse inequality; the job is fortunately easier
and relies on an application of the variational principle. We refer to the original papers for this
final step.

\newpage \section{Convergent expansions} \setcounter{page}{1}

\label{chapter-convergent-expansions}

In previous chapters we have alluded to the Mayer expansion and
results on the convergence of Mayer expansions. This is an old subject
reviewed in numerous places including \cite{Bry84} but since that
review some beautiful new combinatoric formulas have been discovered.
The main objective of this summary is to update \cite{Bry84} to bring
these to the attention of our readers, together with some applications
and ramifications. This is done in
sections~\ref{sec-mayer1},~\ref{sec-mayer2}. 

Related ideas convert the considerations of
section~\ref{sec-tunneling} into convergent expansions and thereby
prove there is screening. We outline these arguments in
sections~\ref{sec-truncated-expectations}, \ref{sec-polymer-rep2},
\ref{sec-polymer-rep1}. This updates an older review
\cite{BrFe81}.

We are accustomed to expansions in theoretical physics where the terms
are labeled by connected graphs.  It is usually the case that the
number of connected graphs at order $N$ grows more rapidly than $N!$
so that such expansions appear to be divergent at least until
cancellations between graphs have been taken into account.  The older
review \cite{Bry84} was centered on rigorous bounds that say that the
sum over all connected graphs at order $N$ is actually smaller than the
sum over just the least connected graphs, i.e. tree graphs. The
new part of this review is theorem~\ref{thm-forest} which makes the
relation between all graphs and tree graphs much clearer.

\subsection{Tree graph formulas and the Mayer expansion} 
\label{sec-mayer1}

We begin with a little review of the theory of activity expansions
\cite{Rue69,HaMc76,MaMa77}. Suppose $N$ particles labeled $1, 2,
\dots , N$ have two-body potentials $V_{ij}$. The total potential of a
subset $X$ of these particles is
\begin{equation}\label{C.Gibbs1}
U_X =  \sum_{ij \in X} V_{ij}
\end{equation}
where $ij = ji$ is an unordered pair of distinct particles and $ij \in
X$ means $i,j \in X$.  
Consider the formulas
\begin{eqnarray}\label{C.truncated2}
e^{-U_X} &=& \sum _{G} \prod _{ij \in G}
\left(e^{-V_{ij}}-1\right)\nn \\
\left(e^{-U_X}\right)_{c} &=& \sum _{G \ {\rm connected}} \prod _{ij \in G}
\left(e^{-V_{ij}}-1\right)
\end{eqnarray}
In the first the sum is over all graphs $G$, connected or
disconnected, on vertices in $X$. It is an immediate consequence of
expanding the product over $ij$ of $\exp \left(-V_{ij} \right) - 1 +
1$, remembering that a graph on $X$ is, by definition, an arbitrary
subset of $\{ij:ij\in X \}$.  The second formula is the usual way to
define the {\it connected part \/} $\exp \left(-U_X \right)_{c}$, but
there is another equivalent definition, namely the connected part is
the (unique) recursive solution of
\begin{eqnarray}\label{C.truncated1}
&& e^{-U_X} 
= \sum_{M \geq 1} \frac{1}{M!}
\sum _{\stackrel{X_{1},\dots ,X_{M}}{\cup X_{j} = X}}
\prod_{j=1 }^{M} \left(e^{-U_{X_{j}}}\right)_{c}
\mbox{ if } |X| > 1\nn \\
&& \left(e^{-U_X}\right)_{c} =  1 \mbox{ if } |X| = 1
\end{eqnarray}
where $X_{1},\dots ,X_{M}$ are disjoint. The equivalence of the two
definitions is not a hard exercise and is the main step in obtaining
the Mayer expansion, which is

\begin{theorem}\label{thm-mayer1} Let $U_{N} = U_{\{1,\dots ,N \}}$
where $V_{ij} = V(\br _{i},\br _{j})$ is the pair interaction
between particles in a grand canonical ensemble $\Xi$. Then 
\begin{equation}\label{C.2}
\ln \Xi = \sum _{N \geq 1} \frac{1}{N!} \int \prod _{j=1}^{N}
d\cE_{j}\, z(\cE_{j} ) \left(e^{-\beta U_{N}}\right)_{c}
\end{equation}
\end{theorem}

The proof that this formal expansion results from (\ref{C.truncated1})
is in most textbooks in statistical mechanics.  The classical proofs
that this expansion is convergent for small activity were given in
\cite{Gro62,Pen63,Rue63}. Our line of development begins with
\cite{Pen66} who showed that when the interactions are repulsive the
sum over {\it all} connected graphs at order $N$ is dominated by the
sum over tree graphs at the same order. By Cayley's theorem the number
of tree graphs on $N$ vertices is $N^{N-2}$ which is comparable to
$N!$. The convergence of the Mayer expansion at small activity will be
seen to be an immediate consequence.

{\it Notation:\/} We consider the class $\cF$ of graphs called
\underline{f}orests: A graph $G$ is a forest if it has no closed loops
or equivalently if each connected component of $G$ is a tree graph. A
single isolated vertex which is not in any bond is considered to be an
empty tree graph. The empty graph which has no lines, so that every
vertex is isolated, is therefore a forest. A forest $G$ has the
property that for any pair of vertices $ij$, either there is a unique
path that joins $i$ to $j$ consisting of bonds in $G$ or there is no
path at all.  To every bond $ij \in \{1,2,\dots ,N \}$ is associated a
non-negative parameter $s_{ij}$. The vector whose components are all
these parameters is denoted by $\bs $ and $\1 $ is the vector where
each $s_{ij} = 1$.  Let $G$ be a forest. For each bond $ij$ we set
\begin{equation} \label{C.1} 
\sigma_{ij}(G, \bs ) = \left\{
\begin{array}{l} \inf\{s_{b}:b \in \mbox{ path in G joining } i
\mbox{ and } j \}\\
0 \mbox{ if no path}
\end{array}
\right.
\end{equation}
Thus if $ij$ belongs to the forest $G$ then $\sigma_{ij} = s_{ij}$. Also,
$\sigma_{ij}(G,\bs )$ depends only on the $s_{ij}$ parameters assigned to
bonds in $G$. We let $\bsigma (G,\bs )$ denote the vector whose components
are $\sigma_{ij}(G,\bs )$. We set $\partial ^{G} = \prod _{b \in G}
\partial /\partial s_{b}$. The symbol $\int ^{t} d^{G}s$ means that
for each $b \in G$ $s_{b}$ is integrated over the interval $[0,t]$. We
write $f \circ \bsigma (G,\bs ) = f(\bsigma (G,\bs ))$ so that the
notation $(\partial ^{G} F) \circ \bsigma(G,\bs )$ will mean: first do
the derivatives with respect to $s_{b}$ with $b \in G$ and then
evaluate at $\bsigma (G,\bs)$.

The following theorem is a type of fundamental theorem of calculus.
It reached this simple form through the series of papers
\cite{GJS74,BrFe78,BaFe84,BrKe87,AbRi95}.

\begin{theorem}\label{thm-forest}
Let $F(\bs )$ be any continuously differentiable function of
parameters $s_{ij}$ where $ij=ji$ and $i \not = j \in \{1,2,\dots
,N\}$ and $N \geq 2$. Let $t \geq 0$. Then
\begin{equation}\label{C.4}
F({t \1} )
=
\sum _{G \in \cF} \int^{t} d^{G}s \, (\partial ^{G} F) \circ \bsigma(G,\bs )
\end{equation}
where
\[
\int^{t} d^{G}s \, (\partial ^{G} F) \circ \bsigma(G,\bs )
= F (0\1 )
\]
when $G$ is the empty forest.
\end{theorem}

Note that the theorem is valid when $t=0$ because all terms in the sum
over $G$ vanish except when $G$ is the empty forest and in this case
the right hand side is $F({0 \1} )$, by definition. Also, if $N=2$,
the theorem reduces to $F(t) = F(0) + \int^{t} \partial_{12} F(s_{12})
ds_{12}$, which is the fundamental theorem of calculus.  A proof of
this theorem is given in section~\ref{sec-proofs}.

To study the Gibbs factor $\exp \left(- \beta \sum V_{ij} \right)$ by
this theorem we introduce parameters $s_{ij}$ in such a way that the
Gibbs factor is the value of
\begin{equation}\label{C.3}
F(\bs ) = e^{-\beta U_{N}(\bs )} = \exp \left( -\beta \sum _{ij}
V_{ij}(s_{ij} )\right) \end{equation} when $\bs = \1$. We choose
$V_{ij}(s_{ij})$ to vanish at $s_{ij}=0$ so that $\bs = 0$ corresponds
to all interactions being switched off. We will call such a choice of
dependence on $s_{ij}$ an interpolation.

Assume we have such an interpolation.  Given a forest $G$ on vertices
$\{1,2,\dots ,N \}$, we can decompose it into connected
components. These connected components are subgraphs, trees, on
subsets of vertices $ X_{1},\dots ,X_{M}$. Thus $G$ determines a
partition of the set of vertices into ``clusters'' $X_{1},\dots
,X_{M}$. Vertices label particles and from the definition of $\bsigma
\equiv \bsigma(G,\bs )$ there are no interactions in $ U_{N}(\bsigma)$
between particles in different trees so that $(\partial ^{G} F) \circ
\bsigma$ factors across the partition $X_{1},\dots ,X_{M}$. By
comparing the definition of connected part (\ref{C.truncated1}) with
the result of the theorem one obtains

\begin{theorem}\label{thm-mayer2} For $N \geq 2$
\begin{equation}\label{C.2b}
\left(e^{-\beta U_{N}}\right)_{c} = \sum _{T} \int^{1} d^{T}s \,
\left( \partial ^{T} e^{-\beta U_{N}} \right) \circ \bsigma(T,\bs )
\end{equation} where $T$ is summed over all connected trees $T$ on $N$
vertices. 
\end{theorem}

\noindent {\em Tree Graph domination:\/} Suppose that the potential is
repulsive, $V_{ij} \geq 0$. Then let $V_{ij}(s_{ij}) = s_{ij}V_{ij}$
or indeed any other interpolation with $V_{ij}'(s) \geq 0$. By
\begin{equation}\label{C.tree-lower}
U_{N}(\bsigma(T,\bs )) 
\geq \sum_{ij \in T} V_{ij}(\sigma_{ij}(T,\bs ))
= \sum_{ij \in T} V_{ij}(s_{ij})
\end{equation}
and theorem~\ref{thm-mayer2} 
\begin{equation} \label{C.2c}
|\left(e^{-\beta U_{N}}\right)_{c}| \leq \sum _{T} \int^{1} d^{T}s \,
\prod _{ij \in T} \beta V_{ij}'(s_{ij})e^{-\beta V_{ij}(s_{ij})} =
\sum _{T} \prod _{ij \in T} \left( 1 - e^{-\beta V_{ij}} \right)
\end{equation}
which says that the connected part --- the sum over all connected
graphs --- is dominated by the sum just over tree graphs, which is the
remarkable result \cite{Pen66}. 

However there are also tree graph domination results when the
interaction is not purely repulsive. $\bsigma (G,\bs )$ has the
following unobvious property to be proved in section~\ref{sec-proofs}.

\begin{theorem}\label{thm-C.stability.a}
If
\begin{equation}\label{C.stability1}
\sum _{ij} V_{ij} \geq -B N
\end{equation}
then
\begin{equation}\label{C.stability2}
\sum _{ij} \sigma_{ij}(G,\bs )V_{ij} \geq -B N
\end{equation}
The theorem remains valid when self-energies $1/2 \sum _{ii} V_{ii}$
are added to both left hand sides.
\end{theorem}

\noindent 
{\it Tree Graph Domination} (\ref{C.2c}) now becomes
\begin{equation} \label{C.2da}
|\left(e^{-\beta U_{N}}\right)_{c}| 
\leq e^{\beta BN} \sum _{T} \prod _{ij \in T} \beta |V_{ij}|  
\end{equation}

\noindent {\em Tree Graph limit:\/} Consider the limit
\begin{equation}\label{C.9a2}
z \rightarrow \infty, \ \ \ \beta \rightarrow 0 \mbox{ with }
z\beta \mbox{ fixed}
\end{equation}
By theorems~\ref{thm-mayer1} and \ref{thm-mayer2} the leading terms 
are
\begin{equation}\label{C.9a}
\ln \Xi \sim 
\sum _{N} \frac{1}{N!}
\sum _{T}
\int \prod_{k=1}^N z(\cE_{k}) \, d\cE_{k} \,
\prod _{ij \in T} \left(
-\beta V(\cE _{i} , \cE _{j})
\right)
\end{equation}
where $T$ is summed over all connected trees $T$ on $N$
vertices. There are $N-1$ lines in a connected tree graph on $N$
vertices so every term in the sum is $O (z^{N}\beta ^{N-1}) = O (z)$.
The estimates in section~\ref{sec-mayer2} show that the Mayer
expansion converges uniformly [after dividing out an overall factor of
$z \times \mbox{volume}$] as this tree graph limit is taken so the
limit can be taken under the sum over $N$.  This series can even be
summed exactly.  It is identical with the series in powers of $z$ for
\begin{equation}\label{C.9b} \frac{1}{2}\int h (\cE ) V^{-1} h (\cE
) \, d\cE + \int z(\cE) \, d\cE \, e^{-\beta h(\cE )}
\end{equation}
evaluated at the $h$ that makes it stationary.  $V^{-1}$ is the
operator inverse of of the operator whose kernel is $V$. This is the
well known folk theorem that Feynman tree graphs sum to the
classical action.

More general interpolations besides $s_{ij}V_{ij}$ are actually
useful because sometimes one has interactions which are really the sum
of several interactions at different scales and then it is useful to
use more complicated interpolations that turn them off one after the
other. Since there will be examples of this in the next section we
also record a more general stability result

\begin{theorem}\label{thm-C.stability.b}
If the interpolation satisfies 
\begin{equation}\label{C.stability3}
\sum _{ij} V_{ij}'(t) \geq -B(t)N
\end{equation}
then
\begin{equation}\label{C.12}
U_{N}(\bsigma (G,\bs ))
\geq
-2\sum _{ij\in T}\int _{0}^{s_{ij}}B(s)ds 
\end{equation}
\end{theorem}
\noindent 
{\it Tree Graph Domination} (\ref{C.2c}) now becomes
\begin{equation} \label{C.2db}
|\left(e^{-\beta U_{N}}\right)_{c}| 
\leq \sum _{T} \int^{1} d^{T}s \,
\prod _{ij \in T} \beta |V_{ij}'(s_{ij})| e^{2\beta \int
_{0}^{s_{ij}}B(s)ds } 
\end{equation}

\subsection{Convergence of Mayer expansions}\label{sec-mayer2}

We demonstrate some uses of tree graph domination by giving
estimates on the radius of convergence for the Mayer expansion for a
variety of systems including ones discussed earlier in this review.

We will use the following two principles: A tree graph on $N$
vertices whose lines represent factors $f(\br - \br ')$ contributes
exactly
\begin{equation}\label{E.tree1}
\left( \int f(\br ) d\br \, \right)^{{N-1}} \times {\rm Volume}
\end{equation}
because tree graphs, being connected, have $N-1$ lines.  The integrals
over vertices can be evaluated in order starting with ones on lines
that are outermost branches (Trim the tree!).  Furthermore the number
of tree graphs at order $N$ is $N^{N-2}$ (Cayley's theorem).

\subsubsection{Repulsive potentials}\label{subsubsec-repulsive
potentials}

As an instructive exercise we recover the result in
\cite{Pen66}.  Consider for simplicity one species. By tree graph
domination (\ref{C.2c}) the Mayer expansion is bounded by
\begin{eqnarray} \label{E.tree1a}
&&
\sum _{N} \frac{1}{N!}  
\int |\left(e^{-\beta U_{N}}\right)_{c} z^{N}| d^{N}\br\nn \\
&& 
\leq \sum _{N} \frac{1}{N!}\sum _{T} \left(|z|\int (1 - e^{-\beta V(\br )})
d\br \, \right)^{N-1} |z| |\Lambda |\nn \\ 
&& 
\leq
\sum _{N} \frac{N^{N-2}}{N!} \left(|z|\int (1 - e^{-\beta V(\br )}) d\br \,
\right)^{N-1} |z| |\Lambda |
=\sum _{N} \frac{N^{N-2}}{N!} Q^{N-1} |z| |\Lambda |
\end{eqnarray}
which is convergent provided
\begin{equation} \label{E.tree1b}
Q := |z|\int d\br \, (1- e^{-\beta V(\br ) }) <
e^{-1} 
\end{equation}
This estimate is also used in the next two examples

\subsubsection{Debye spheres}\label{subsubsec-debye spheres}

In section~\ref{sec-debyesphere} we encountered
conditions (\ref{2.4.1.5}). The system consisted of charged particles
interacting either by two body potentials
\begin{equation}\label{E.9-1}
V_{L,\lambda }(\br ) = |\br |^{-1} \left( e^{-|\br |/L} -
e^{-|\br |/\lambda } \right) 
\end{equation}
or by the Yukawa potential $V_{L}(\br )$ and hard cores of radius
$\lambda $. First we consider two species interacting by
$(\ref{E.9-1})$. Suppose we can find an interpolation as in
theorem~\ref{thm-C.stability.b}, then we can repeat the argument given
above for repulsive potentials, but now using the tree graph
domination (\ref{C.2db}) instead and find convergence when
\begin{equation}\label{C.8}
Q: = 2|z| \int _{0}^{1} ds\, \int d\br \, |\beta V'(\br,s )| \exp
\left( 2\beta\int_{0}^{s}B(s^{\prime}) ds^{\prime} \right) < e^{-1}
\end{equation}
The factor $2$ is there because there are two species.

The interpolation is chosen to
be $V(\br ,t) = V_{L,\ell(t)}(\br)$ where $\ell(t)$ is a length scale
which decreases from $L$ when $t=0$ to $\lambda $ when $t=1$. Then we
show below that
\begin{equation}\label{E.10}
V'(\br,t) = - \frac{\partial }{\partial t} 
\frac{e^{- |\br|/\ell(t)}}{|\br |}
\end{equation}
obeys a stability estimate (\ref{C.stability3}) with
\begin{equation} \label{E.9-2}
B(t) = \frac{\partial }{\partial t}
\frac{1}{2\ell(t)} 
\end{equation}
Also
\begin{equation}
\int |V'(\br,s )|d\br
=
-\frac{\partial }{\partial s} 
\int \frac{e^{- |\br|/\ell(s)}}{|\br |}d\br =
-\frac{\partial }{\partial s} 4\pi \ell(s)^{2} 
\end{equation}
because $|V'(\br,s )| = V'(\br,s )$. Therefore (\ref{C.8}) reads
\begin{eqnarray}
Q &=& - 2 |z|\beta \int_{0}^{1} ds \, \frac{\partial }{\partial s} 4\pi
\ell(s)^{2} \exp \left(\beta \int _{0}^{s} \frac{\partial }{\partial
s^{\prime}} \frac{1}{\ell(s^{\prime})}ds^{\prime}\, \right) \nn \\
&=&
16\pi |z| \beta \int_{\lambda }^{L} d\ell \, \ell
\exp \left(
\frac{\beta }{\ell} - \frac{\beta }{L}
\right) < e^{-1}
\end{eqnarray}
When  $\lambda < \beta < L$ we break up the range of the
$d\ell$ integration into $[\lambda ,\beta ]$ and $[\beta ,L]$ and find
that this criterion is satisfied if
\begin{equation}
|z|\beta L ^{2} \ll 1, \ \ \ |z|\beta^{3} e^{\beta /\lambda } \ll 1
\end{equation} 
which is the condition (\ref{2.4.1.5}). The Ruelle estimate
(\ref{2.4.ruelle}) is not as good because it does not distinguish the
length scales and essentially associates the bad stability factor
$\exp(\beta /\lambda )$ to all scales. In the Debye-H\"uckel limit $Q
= O (L^{2}/l_{D}^{-2})$ is determined by the first criterion because
$|z|\beta^{3} e^{\beta /\lambda } \rightarrow 0$. By this remark and
(\ref{E.tree1a}) estimate (\ref{2.4.1.5c}) follows. Results related to
(\ref{2.4.1.5}) were given in \cite{GoMa81,Imb83a,BrKe87}.

To obtain the stability estimate (\ref{E.9-2}), note that by the
argument given near (\ref{1.1.3bb}) the stability constant $B (\ell)$ for
$V_{L,\ell }$ is half the self-energy, which can be calculated
using the Fourier transform to be
\begin{equation}\label{E.9}
B = \left(\frac{1}{2\ell } - \frac{1}{2L}\right) 
\end{equation} 
Now take $L$ in (\ref{E.9}) equal to $\ell + d\ell$ .  

It is not a hard exercise to rework this calculation for two
dimensions.  One recovers a result of \cite{Ben85} that if $\beta < 2$
[ $4\pi$ using the units in \cite{Ben85}] that the Mayer expansion is
convergent for $|z|$ small uniformly in the short distance cutoff
$\lambda $.  In other words for $\beta $ smaller than the first
threshold the Mayer expansion converges for the two dimensional Yukawa
gas without any cutoff. This is notable because this interaction is
singular at $\br =0$ and is not stable in the Ruelle sense.  The
reader will find this exercise done in \cite{BrKe87}.

The Yukawa gas with hard core is convergent under the conditions
\cite{BrKe87}, see also \cite{Imb83a}, \begin{equation} |z|\lambda
^{3} \ll 1; \ \ \ |z|\beta L ^{2} \ll 1; \ \ \ |z|\beta^{3} e^{\beta
/\lambda } \ll 1
\end{equation}
Briefly, this is derived by combining the hard core example with the
Yukawa by letting $s$ vary over $[0,2]$ so that $V(\br ,s) $ has no
interactions at $\bs = 0$ and has the complete interaction when $s =
2\1 $. For $s \in [0,1]$ we choose $V'$ as in the repulsive example
and for $s \in [1,2]$ we have $V'$ given by (\ref{E.10}) with $\ell(s)
= L$ at $s=1$ and $\ell(s) = 0$ at $s=2$. The main point is that the
hard core is already there when the Yukawa is turned on so that
$2B(s)$ for $s$ corresponding to $\ell(s) \approx 0$ is the stability
constant including the hard cores, which prevents a divergence.

Tree graph formulas were applied iteratively by \cite{GoMa81} to
expand successive scales in the interaction in a renormalization group
analysis. They showed how to adapt stability estimates to scales and
particular clusters of particles.  They were motivated to look for
better estimates on the convergence of Mayer expansions in order to
prove confinement at all parameters for three dimensional Euclidean
lattice quantum electromagnetism by exploiting a connection with
screening. The iterated tree graph formulas of G\"opfert and Mack were
then simplified \cite{BrKe87} by passing to a limit in which the
scales are replaced by a continuous parameter (the $t$ in
Theorem~\ref{thm-forest}). This was partly motivated by the study of
renormalization \cite{Pol84}. \cite{AbRi95} then found
Theorem~\ref{thm-forest} as a corollary of the results in
\cite{BrKe87}.

The ideas in \cite{GoMa81} were independently developed by Gallavotti
and Nicolo and applied by Benfatto to the two dimensional Yukawa gas
\cite{Gal85,GaNi85,Ben85,GaNi86}.  Note that the trees in these papers
express the hierarchical structure of the clusters on different scales
and should not be confused with our trees which are Feynman or Mayer
graphs.

\subsubsection{The Mayer expansion for a polymer
gas}\label{subsubsec-polymer-gas}

This application is technical and should be omitted on first reading.
It will be used later in this chapter. Consider the grand canonical
partition function 
\begin{equation}\label{E.poly1}
Z(\Lambda) = \sum_{N} \frac{1}{N!}
\sum _{\stackrel{X_{1},\dots ,X_{N} \subset \Lambda}{\cup X_{j} =
\Lambda } }
\prod _{j} A(X_{j})
\end{equation}
where $\Lambda $ is a finite set. In applications to lattice models
$\Lambda $ is all lattice points in a big box. A typical element of
$\Lambda $ is denoted by $\eta $.  Swimming around in this big box are
the ``polymers'' $X_{1}, \dots ,X_{N}$, which are disjoint subsets of
$\Lambda $.  For lattice models they are often nearest neighbor
connected but instead of making this assumption we assume that the
activity $A (X)$ of polymer $X$ obeys a bound of the form
\begin{equation}\label{E.conv1}
|A(X)| < a^{|X|} \sum _{G}\prod _{\eta \eta ' \in G}
w(\eta, \eta ')
\end{equation}
where $|X|$ is the number of elements $\eta $ in $X$, $G$ is summed
over all tree graphs on $X$. For example when $\Lambda $ is a box in a
lattice, we could take $w(\eta, \eta ')$ to vanish whenever $\eta, \eta '$
are not nearest neighbors and to be one otherwise.  In this case $A (
X)= 0$ unless $X$ is connected.

Now we show that the Mayer expansion is convergent when
\begin{equation}\label{E.poly2}
Q:= a \sum _{\eta '} \left(
\delta_{ \eta ,\eta '} + w(\eta,\eta ')
\right) < e^{-1}
\end{equation}
where $\delta_{ \eta ,\eta '}$ is the Kronecker delta function on the
elements of the set $\Lambda $.  Suppose there is a notion of distance
between elements in $\Lambda $ and $w(\eta,\eta ')$ decays
exponentially as ${\rm dist} (\eta, \eta ' )\rightarrow \infty $. If
$eQ <1$ then correlations obtained by differentiating $\ln Z(\Lambda)
$ with respect to external fields $\psi $ with $A (X) = A (X,\psi )$
decay exponentially.

For example $\eta$ could be a point in a lattice in ${\Bbb R}^{3}$
which actually represents a unit box centered on $\eta$ so that $X$
represents a union of these boxes.  If $A(X,\psi )$ is independent of
variations of $\psi (\br)$ at points $\br$  outside  $X$ then
\begin{equation}\label{E.poly3}
\frac{\delta ^{2}}{\delta \psi (\br ) \delta \psi (\br ')}
\ln Z (\Lambda )
\end{equation}
and higher variational derivatives will decay exponentially , uniformly
in the size of $\Lambda $ as the points $\br , \br '$ are separated.

We drop the constraint that polymers are disjoint from the sum in
(\ref{E.poly1}) and impose it instead by a hard core Gibbs factor
$\exp (-U_{N} )$. Let
\begin{equation}\label{E.log4}
J(X_{1},\dots ,X_{N}) = A(X_{1})\cdots A(X_{N})
\left( e^{- U_{N}(X_{1},\dots ,X_{N})} \right )_{c}
\end{equation}
In the Mayer expansion
\begin{equation}\label{E.log1}
\sum_{N \geq 1} \frac{1}{N!}
\sum _{X_{1},\dots ,X_{N} \subset \Lambda } J(X_{1},\dots ,X_{N})
\end{equation}
we write
\begin{equation}\label{E.log1c}
\sum _{X_{i}} = \sum \frac{1}{n_{i}!} 
\sum _{\eta _{1},\dots ,\eta _{n_{i}}} \dots 
\end{equation}
Then we set $M = \sum_{i} n_{i}$ and note that $M!/(\prod n_{i}!)$ is
the number of ways to partition $\{1,\dots ,M \}$ into subsets $\gamma
_{1},\dots ,\gamma _{N}$. Therefore (\ref{E.log1}) is the same as
\begin{eqnarray}\label{E.log1b}
&&
\sum _{M\geq
1}\frac{1}{M!}
\sum_{N\geq 1} 
\frac{1}{N!} \sum _{\gamma
_{1},\dots ,\gamma _{N}}
\sum _{\eta _{1},\dots ,\eta _{M}} 
J(X_{1},\dots ,X_{N}) \nn \\
&&
X_{i} = \{\eta _{j}: j \in \gamma _{i}\}
\end{eqnarray}
with the constraint on the sum over $\eta _{1},\dots ,\eta _{M}$ that
$\eta _{j}$ with $j$ in the same partition are distinct. We
substitute in the tree graph domination bound (\ref{C.2c})
\begin{eqnarray}\label{E.log2}
|\left( e^{- U_{N}(X_{1},\dots ,X_{N})} \right )_{c}|
&\leq&
\sum _{T}\prod _{ij \in T} 
\bigg | e^{-V(X_{i}, X_{j})} - 1 \bigg | \nn \\
&\leq&
\sum _{T}\prod _{ij \in T} 
\sum _{\eta \in X_{i}, \eta ' \in X _{j}}
\bigg |e^{-V(\eta, \eta')} - 1 \bigg |
\end{eqnarray}
and (\ref{E.conv1}). 
Note that the hard core interaction $V(\eta, \eta')$ means that
$\exp \left(-V(\eta, \eta') \right) - 1$ is $-\delta_{ \eta
,\eta '}$.  The trees with $w$ bonds and the tree with $\delta_{
\eta ,\eta '}$ bonds link into one connected tree $G$ on $\{1,\dots M
\}$ and the respective sums are equivalent to a sum over all possible
connected trees on $\{1,\dots M \}$.  The partition $\{\gamma
_{1},\dots ,\gamma _{N} \}$ determines uniquely which bonds in this
tree are $V$ bonds and which are $w$ bonds.  Indeed the sum over
partitions, including the $1/N!$ is equivalent to summing over the
choice $w$ bond or $V$ bond for each bond in the tree.  Therefore the
sum over partitions is the same as assigning to each bond $ij$
\begin{equation}\label{E.log3}
\delta_{\eta_{i},\eta_{j}} + w(\eta_{i} , \eta_{j})
\end{equation}
Therefore the expansion (\ref{E.log1b}) is bounded by
\begin{equation}\label{E.log5}
\leq \sum _{M\geq
1}\frac{a^{M}}{M!}
\sum _{\eta _{1},\dots ,\eta _{M}}  
\sum _{G}\prod _{ij \in G} 
(\delta_{\eta_{i},\eta_{j}} + w(\eta_{i}, \eta_{j}))
\end{equation}
dropping the constraints on the distinctness of $\eta _{1},\dots
,\eta _{M}$.  As above, (\ref{E.tree1}--\ref{E.tree1b}), this is
convergent if (\ref{E.poly2}) holds.

The additional claim we made concerning exponential decay of correlations
is an immediate consequence: suppose we perform variational
derivatives such as 
\[
	\frac{\delta }{\delta \psi (\eta _{a})}
	\frac{\delta }{\delta \psi (\eta _{b})} 
\]
on $\ln Z$ where $\psi $ is an external field that enters through 
dependence $A(X) = A (X,\psi )$. Then all terms in the Mayer expansion
(\ref{E.log1}) vanish except those where the union of the sets $X_{j}$
contains both points $\eta _{a},\eta _{b}$, because $\delta A (X,\psi )/
(\delta \psi (\eta ) ) = 0$ when $\eta \not \in X$.  When $\eta _{a}, \eta
_{b}$ are far apart the surviving terms are exponentially small since
they are connected graphs with exponentially decaying
propagators. This argument that is being made term by term in the
Mayer expansion is valid when all the terms are summed because we 
established that the expansion is convergent, provided the convergence
has some uniformity in $\psi $.

\subsection{Gaussian integrals and truncated
expectations}\label{sec-truncated-expectations}

In this section, following \cite{Bry84,BrKe87} we derive tree graph
formulas for truncated Gaussian expectations. The point of the
formulas is that all the loops in the associated Feynman diagrams are
resummed back into a positive Gaussian measure, leaving only tree
diagrams, which carry the connectedness information.

Truncated expectations for arbitrary observables will be defined later
in (\ref{B.14}) but for products of polynomials they can equivalently
be defined by applying Wick's theorem and then discarding all but the
connected graphs.

Suppose $\av{ \ }$ is a Gaussian average over variables $\phi _{i},
\ i = 1,\dots N$, 
\begin{eqnarray}\label{D.4}
&& \av{P} = Z^{-1} \int P 
            e^{-1/2 \sum _{i,j} \phi _{i} A_{ij} \phi _{j}}
            \prod _{i}d\phi _{i} \nn \\
&& \av{ \phi_{i} \phi_{j}} = A^{-1}_{ij} \equiv C_{ij} 
\end{eqnarray} 
$A$ is a matrix with positive eigenvalues. 
Suppose  $P_{i} = P_{i}(\phi _{i})$ are polynomials each of which
depends only on one of the variables, then by Wick's theorem 
\begin{equation}\label{D.laplacian1}
\av{\prod _{l}P_{l}} = e^{\frac{1}{2}\sum _{i,j}\DDelta_{ij}} 
\prod _{l}P_{l}\arrowvert_{\phi=0} \nn \\
\end{equation}
where the exponential is defined by power series and
\begin{equation}\label{D.laplacian2}
\DDelta _{ij} = C_{ij}
                \frac{\partial }{\partial \phi _{i}}
                \frac{\partial }{\partial \phi _{j}}
\end{equation}
We apply theorem~\ref{thm-forest} with
\begin{equation}\label{D.laplacian3}
F(\bs ) = \exp \left( \frac{1}{2} \sum _{i,j} s_{ij}\DDelta_{ij} \right) P
\end{equation}
with $P = \prod _{l}P_{l}$ and the conventions $s_{ij} = s_{ji}$ and
$s_{ii}=1$.  The right hand side of theorem~\ref{thm-forest} will then
contain
\begin{equation}\label{D.laplacian4}
 \exp \left(  \frac{1}{2}\sum _{i,j} \sigma_{ij}(G,\bs )\DDelta_{ij} \right) 
 \DDelta ^{G} P \arrowvert_{\phi=0}
\end{equation}
where $\DDelta ^{G} = \prod _{ij \in G} \DDelta _{ij}$.  This is a new
Gaussian expectation $\av{ \ }^{G,\bs }$ characterized by
\begin{equation}\label{D.4b}
\av{ \phi_{i} \phi_{j}}^{G,\bs } = \sigma_{ij}(G,\bs ) C_{i,j} 
\end{equation} 
The key point is that it truly is a Gaussian expectation: we shall
show below that $C_{ij}\sigma_{ij}$ has positive eigenvalues, so that it
has an inverse which defines the expectation $\av{ \phi_{i}
\phi_{j}}^{G,\bs }$ in parallel to (\ref{D.4}). We define
\begin{equation}\label{D.expectation}
\av{P}^{G} := \int^{1} d^{G}s \, \av{ P }^{G,\bs }
\end{equation}
and conclude from theorem~\ref{thm-forest} that
\begin{equation}\label{D.truncated1}
\av{\prod _{i}P_{i}} = \sum _{G} \av{\DDelta ^{G} \prod _{i}P_{i}}^{G}
\end{equation}
where $G$ is summed over all forests. 

Notice that the previous equations contained exponentials of $\DDelta$
which only make immediate sense on polynomials, but now that we have
got rid of these we can use this formula when the $P_{i}$ are
arbitrary smooth functions.

By the discussion above theorem~\ref{thm-mayer2} there is a
factorization $\av{ \ }^{G} = \prod_{T\subset G} \av{ \ }^{T}$ where
$T$ runs over trees in the forest $G$.  Therefore by definition of the
truncated expectation \footnote{A superscript $T$ refers to a tree
graph and a subscript $T$ means the expectation is truncated. We
apologize for the notational collision.} $\av{ \ }_{T}$ given in
(\ref{B.14}) we obtain 
\begin{equation}\label{D.truncated2}
\av{\prod _{i=1}^{N}P_{i}}_{T} = \sum _{G} \av{\DDelta ^{G}\prod
_{i}P_{i}}^{G}
\end{equation}
where $G$ runs only over connected trees on $N$ vertices.  This
argument easily generalizes to allow each $P_{i}$ to depend on more
variables $\phi _{x}, \ x \in X_{i}$. This formula expresses a
truncated Gaussian expectation as a sum only over tree graphs. The
loops that would appear in a standard application of Wick's theorem
are resummed into the expectation $\av{ - }^{G}$ defined in terms of the
Gaussian expectation with the altered propagator $C_{ij}\sigma _{ij}$.

To see that the matrix $C_{ij}\sigma _{ij}$ has positive eigenvalues: since
$A_{ij}$ has positive eigenvalues, $C_{ij}$ has positive eigenvalues
which is equivalent to 
\begin{equation}\label{D.4a}
\sum _{i,j} \eta _{i}C_{ij}\eta _{j} 
\geq \lambda _{\rm min} \sum _{j}\eta _{j}^{2}
\end{equation}
By theorem~\ref{thm-C.stability.a} with $V_{ij} = \eta _{i}[C_{i,j} -
\lambda _{\rm min}\delta _{ij}] \eta _{j}$, $B=0$ and self energies
included, 
\begin{equation}\label{D.4c}
\sum _{i,j} \eta _{i}C_{ij}\sigma_{ij}\eta _{j} \geq 
\lambda _{\rm min} \sum _{j}\eta _{j}^{2}
\end{equation}
which implies that the matrix $A (\bsigma )^{-1} = C_{ij}\sigma_{ij}$
has positive eigenvalues.  We define the expectation $\av{ \ }^{G,\bs
}$ by replacing the matrix $A$ by $A (\bsigma )$ in (\ref{D.4}).

\subsection{Polymer representations and exponential
decay}\label{sec-polymer-rep2}

In this section we describe how to obtain a convergent expansion for
the logarithm of functional integrals of the form
\begin{equation}\label{8.pol-rep-1}
\av{Z(\Lambda )e^{i\int \phi f}} = \int d \mu (\phi ) Z(\Lambda,\phi)
e^{i\int \phi f}
\end{equation}
where $d\mu $ is a Gaussian measure with a propagator [covariance] $C
(\br ,\br ')$ that has both infra-red and ultraviolet cutoffs so that
it is smooth on the diagonal and has exponential decay as $\br - \br '
\rightarrow \infty $.  The $\Lambda $ in the functional $Z(\Lambda
,\phi )$ means that variational derivatives at points outside $\Lambda
$ vanish:
\begin{equation}\label{8.locality1}
Z_{1}(\Lambda ,\phi ; f) \equiv 
\frac{\partial }{\partial \alpha }
Z(X,\phi+\alpha f)_{|_{\alpha =0}} = 0
\end{equation}
if $f = 0$ in $\Lambda $ and the same is true for all higher
derivatives.

Functionals such as 
\begin{equation}\label{8.ex1}
Z(\Lambda,\phi) = \exp ( z \int_{\Lambda } \cos \phi (\br ) \, d\br )
\end{equation}
factorize
\begin{equation}\label{8.locality2}
Z(X_{1}\cup X_{2},\phi) = Z(X_{1},\phi) Z( X_{2},\phi)
\end{equation}
but we will work with a less restrictive property, which is that
$Z(\Lambda ,\phi )$ has a {\it polymer representation}
\begin{equation}\label{8.locality3}
Z(\Lambda ,\phi ) = \sum_{N} \frac{1}{N!}
\sum _{\stackrel{X_{1},\dots ,X_{N} \subset \Lambda}{\cup X_{j} =
\Lambda } }
\prod _{j} K(X_{j},\phi)
\end{equation}
where the interiors of the sets $X_{1},\dots ,X_{N}$ are disjoint. In
other words $Z(\Lambda ,\phi )$ is at least a sum over contributions
that factorize.  This type of representation is not useful if the sets
$X_{j}$ are very complex so we insist that all sets are finite
unions of unit cubes in ${\Bbb R} ^{3}$, centered on points with integral
coordinates.

Recall the well known combinatoric miracle that when the partition
function has a graphical expansion, the logarithm has the same
expansion except that only the connected graphs appear.  This comes
about as a result of the following easily verified relation (cumulant
expansion) between expectations and exponentials
\begin{equation}\label{B.connected}
\av{e^{F}} = \exp \left(\sum_{N\geq 1} \frac{1}{N!}\av{F^{N}}_{T}
\right)
\end{equation}
where the \underline{t}runcated expectation
$\av{F_{1} \cdots F_{N}}_{T}$ is recursively defined by solving
\begin{eqnarray}\label{B.14}
&&
\av{F_{1} \cdots F_{N}}
=
\sum _{\pi }\prod _{\gamma \in \pi }
\av{F^{\gamma } }_{T} \mbox{ if } N > 1 \nn \\
&&
\av{F_{1}}_{T} = \av{F_{1}}
\end{eqnarray}
where $\pi $ is summed over all partitions of $\{1,2,\dots ,N \}$ and
$F^{\gamma } = \prod _{j\in \gamma }F_{j}$. For the case (\ref{8.ex1})
the expansion for the logarithm of (\ref{8.pol-rep-1}) based on these
relations is convergent at small $z$, essentially because the cosine
has the special property that it and all of its derivatives are
bounded uniformly in $\phi $ \footnote{By reversing the Sine-Gordon
transformation the expansion becomes the Mayer expansion for $\pm 1$
charges interacting by  two body potential $C (\br ,\br ')$.}.  The
next steps can be viewed as a reduction of more general functionals
such as $\exp (-\lambda \int \phi ^{4} )$ to this bounded case.

Relations similar to (\ref{B.connected},\ref{B.14}) hold for the
polymer representation (\ref{8.locality3}). This is plausible once one
realizes that the polymer representation is an exponential in a
different commutative product $\circ$ defined by
\begin{equation}\label{B.17}
A \circ B(X) = \sum_{Y\subset X} A(Y)B(X \setminus Y). 
\end{equation} 
As always $Y, X$ are finite unions unit cubes. $Y, X$ are permitted to
be the empty set.  We do not distinguish between open and closed
cubes, that is we identify two cubes if they have the same interior
and regard them as disjoint if their interiors are disjoint. Define
\begin{equation}\label{B.16b}
\cE xp(K)
=
{\cal I} + K + \frac{1}{2!}K \circ K + \cdots 
\end{equation}
${\cal I}(X) = 0$ unless $X$ is the empty set, in which case it is
one.  ${\cal I}$ is the identity for this product.  We only define
$\cE xp(K)$ for functionals $K$ that vanish when applied to the empty
set. It is easy to verify that
\begin{equation}\label{B.16}
\cE xp(K)(X ,\phi ) = \left\{ \begin{array}{l} \sum_{N \geq 0}
\frac{1}{N!}  \sum _{\stackrel{X_{1},\dots ,X_{N} \subset X}{\cup
X_{j} = X} }
\prod _{j} K(X_{j},\phi)\\
1       \mbox{ if } X = \emptyset
\end{array}
\right.
\end{equation}
Thus $Z (\Lambda ,\phi ) = \cE xp (K) (\Lambda ,\phi )$.  Furthermore
$\cE xp(A + B) = \cE xp(A) \circ \cE xp(B)$.  As an example, consider
the function $\Box (X)$ that equals one if $X$ is a cube and vanishes
in all other cases.  Then by (\ref{B.16}), for any set $X$,
\begin{equation}\label{B.expbox}
\cE xp(\Box )(X) =1
\end{equation}

Returning to our functional integral (\ref{8.pol-rep-1}), let us
define
\begin{equation}\label{B.19}
K_{T}(X) = \sum_{N\geq 1} \frac{1}{N!}
\av{ K^{\circ N}(X)}_{T},
\ \ \
= 0 \mbox{ if } X = \emptyset
\end{equation}
More explicitly
\begin{equation}\label{B.ncirc}
	\av{ K^{\circ N}(X)}_{T} 
=
	\av{K\circ \dots \circ K(X) }_{T}
=
	\sum _{\stackrel{X_{1},\dots ,X_{N} \subset X}
	{\cup X_{j} = X } } 
	\av{\prod _{j} K(X_{j}) }_{T}
\end{equation}
where the truncated expectation $\av{ \ }_{T}$ is defined as above.
Then $K_{T}$ is the logarithm in the sense
\begin{equation}\label{B.18}
\av{\cE xp(K)(X )}
=
\cE xp(K_{T})(X )
\end{equation}
which is the $\circ$ equivalent of (\ref{B.connected}). However these
series are terminating after a finite number of terms that depends on
the set $X$ because there are at most a finite number of partitions of
any given $X$.  As an extreme example, if $X = \Delta $ is a single
cube, then there are no partitions of $X$ into proper subsets so only
the $N= 1$ term in (\ref{B.19}) survives and accordingly $K_{T}(\Delta
) = \av{K(\Delta )} $.

This analysis is unaffected by the external field term $\exp (\int
\phi f )$ in the partition function (\ref{E.poly1}) because it can be
absorbed into the polymer activities $K$. From the definition of $\cE
xp$
\begin{equation}\label{8.pol-rep-2}
	\cE xp (K (\phi)) (\Lambda ) e^{\int_{\Lambda }\phi f}
=
	\cE xp (K (\phi ,f)) (\Lambda ), 
\ \ \
	K (X,\phi ,f) := K (X,\phi ) e^{\int_{X}\phi f}
\end{equation}
We want an expansion for the standard logarithm of the partition
function. Comparing the definition of $\cE xp$ in (\ref{B.16}) with
the polymer gas partition function (\ref{E.poly1}), we see that the
standard logarithm is given by the Mayer expansion described in
section~\ref{subsubsec-polymer-gas}, provided we can weaken the
constraint $\cup X_{j} = X$ that occurs in $\cE xp(K_{T})(X )$ to
$\cup X_{j} \subset X$. There is a simple way to achieve this: for any
function of sets $\tilde{A} (X)$ declare that $\tilde{A}$ is
normalized if $\tilde{A} (X) = 1$ whenever $X$ is a unit box, then
\begin{eqnarray}
\cE xp(\tilde{A})(X)
&=&	
\cE xp(\Box) \circ \cE xp(\tilde{A} - \Box)  (X) \nn \\
&=&
\sum_{N} \frac{1}{N!}
\sum _{X_{1},\dots ,X_{N} \subset X}
\cE xp(\Box) (X \setminus \cup X_{j})
\prod _{j} A(X_{j}) \nn \\
&=&
\sum_{N} \frac{1}{N!}
\sum _{X_{1},\dots ,X_{N} \subset X}
\prod _{j} A(X_{j}), \ \ \
A  := \tilde{A} - \Box
\end{eqnarray}
The constraint $\cup X_{j} = X$ has been lifted because by
(\ref{B.expbox}) $\cE xp(\Box) = 1$ on $X \setminus \cup X_{j}$. If
$\tilde{A}$ is
\begin{equation}\label{B.hat}
	\tilde{A}(X) 
=
	K_{T}(X)/\prod _{\Delta \subset X}K_{T}(\Delta ),
\end{equation}
where $/$ has its usual division meaning, then it is normalized. Also,
by the same principle as was used in (\ref{8.pol-rep-2})
\begin{equation}\label{B.norm1}
	\cE xp(K_{T})(X) 
=
	\left(
	\prod _{\Delta \subset X}K_{T}(\Delta )
	\right)
	\cE xp(\tilde{A})(X)
\end{equation}
so 
\begin{equation}\label{B.poly}
	\cE xp(K_{T})(X) 
=
	\left(
	\prod _{\Delta \subset X}K_{T}(\Delta )
	\right)
	\sum_{N} \frac{1}{N!}
	\sum _{X_{1},\dots ,X_{N} \subset X}
	\prod _{j} A(X_{j})
\end{equation}

In section~\ref{subsubsec-polymer-gas} we gave a criterion
(\ref{E.conv1},\ref{E.poly2}) for how small the activity $A(X)$ must
be for convergence of the Mayer expansion and exponential decay of
correlations.  $A$ is built out of $K_{T}$ so we need to see that
$K_{T}$ obeys a bound similar to (\ref{E.conv1},\ref{E.poly2}). This
is where section~\ref{sec-truncated-expectations} comes into play. By
the tree graph formula (\ref{D.truncated2}) $K_{T}$ is small provided
variational derivatives with respect to $\phi $ of the initial polymer
activity $K (X,\phi)$ in the partition function (\ref{8.pol-rep-1})
are small. We will give a partial explanation of this through the
following example.

Example: suppose 
\begin{equation}\label{B.poly2}
	Z (\Lambda) = e^{-V (\Lambda )},
\ \ \
	V (\Lambda,\phi  ) = \lambda \int_{\Lambda }\phi ^{4} 
\end{equation}
Because this factors into a product over boxes in $\Lambda $, by the
same argument that shows that $\cE xp (\Box) = 1$ this
equals $\cE xp (K)$ with 
\[
	K (X,\phi ) = \Box (X) e^{-V (X,\phi )}
\]
When $\lambda $ is small variational derivatives of $K (X,\phi )$ are
small, uniformly in $\phi $, because
\[
	\frac{d}{d\alpha } e^{-V (\Delta,\phi +\alpha g)} 
=
	-4\lambda \int_{\Delta } \phi ^{3} g \, d\br \,  
	e^{-\lambda \int_{\Delta } \phi ^{4} } 
\sim
	O (\lambda ^{1/4})
\]
because $\phi ^{4} $ controls $\int \phi ^{3}g$ when $\phi $ is
large. This is somewhat of an oversimplification because the
dependence of the estimate on the test function $g$ is not uniform.
Unfortunately a gentle reference to cover these specific missing
details is not available, but \cite{Lem95} is detailed and close in
spirit to the discussion here.  More general background references are
\cite{GlJa87,Riv91,BrFe92}.  The ability to work with the bounded
$\exp (-\lambda \int \phi ^{4} )$ in place of $\int \phi^{4}$ is half
the reason why we obtain a convergent expansion when $\lambda $ is
small.  The other half is that all formulas involve at most tree
graphs whose numbers are compensated by $N!$ factors, see the
discussion at the beginning of section~\ref{sec-mayer2}. Exponential
decay of correlations for functional integrals of the form
(\ref{8.pol-rep-1}) follows as in the discussion of the polymer gas in
section~\ref{sec-mayer2}.

It is possible that polymer representations could be useful in
numerical renormalization group calculations. This has been considered
\cite{MaPo85,MaPo89}. Their program can be viewed as a natural step
beyond methods that do work well for hierarchical models
\cite{KoWi91,KoWi94}.

\subsection{The polymer representation}\label{sec-polymer-rep1}

In this section we indicate how exponential decay of correlations for
the Coulomb system is deduced from the considerations of
section~\ref{sec-polymer-rep2}. The missing step, outlined in this
section, is to show that the Sine-Gordon representation is an integral
of the form (\ref{8.pol-rep-1}) with $Z (\Lambda ,\phi ) = \cE xp
(K)(\Lambda ,\phi )$ for polymer activity $K (X,\phi )$ which becomes
very small as the Debye-H\"uckel limit is approached.

As introduced in section~\ref{sec-debye-huckel2} we use units where
$l_{D} = 1$.  We also simplify the presentation by discussing a
Coulomb system with two species with equal but opposite charges and
equal activities $z$ interacting by a smoothed Coulomb potential of the
form $V_{\infty ,L}$ with $L = l_{D}$.  Then there is no need for the
Mayer expansion of section~\ref{sec-debyesphere}, which complicates
the original proofs.

With these simplifications the left hand side $\exp \left(F \right)$
in the ``Villain'' approximation (\ref{2.7.villain}) becomes the ideal
gas partition function
\begin{equation}\label{A.local1}
\exp \left(\int_{\Lambda } d\cE \, z^{(L)}(\cE )
        e^{-\beta ^{1/2} e_{\alpha } i\phi} \right)
\end{equation}
We make the dependence on $\Lambda $ explicit by writing $\exp
\left(F(\Lambda ,i\phi ) \right)$. The left hand side of the
Villain approximation (\ref{2.7.villain}) factors 
\begin{equation}
e^{F(X \cup Y,i\phi )} = e^{F(X ,i\phi )} e^{F(Y,i\phi )}
\end{equation}
whenever $X$ and $Y$ are disjoint. The sum over $h$ in the right hand
side of the Villain approximation also factors, if $X,Y$ are unions of
unions of cubes. From this there follows the same property 
\begin{equation}\label{A.local2}
e^{R_{3}(X \cup Y,i\phi )} = e^{R_{3}(X ,i\phi )} e^{R_{3}(Y,i\phi )}
\end{equation}
for $R_{3}$. 

From (\ref{2.7.6})
 \begin{eqnarray}\label{A.1}
&&\hspace{-.5in} \Xi_{\Lambda } /(\Xi_{{\rm ideal},\Lambda }\Xi_{{\rm
DH},\Lambda }) = \int d\mu _{{\rm DH}}(\phi) \, \sum _{h} e^{- S_{{\rm
DH}}(- ih, ig)} \nn \\ && \times e^{-{S'}_{{\rm DH}}(-ih , ig; i\phi)
+ R_{3}(ig + i\phi)} 
\end{eqnarray}
For simplicity we set the external field $\psi = 0$.  In the formula
(\ref{2.7.Sdef}) for $S_{{\rm DH}}(- ih, ig)$, $R_{2}, R_{0}$ vanish
for this simplified model.  The propagator for $d\mu _{{\rm DH}}(\phi)
$ decays exponentially which is the property we need in order to use
the expansion from section~\ref{sec-polymer-rep2}.

Now we write the $d\mu $ integrand in the form
\begin{equation}\label{A.2}
\cE xp (K) (\Lambda ,\phi ) = 
\sum_{N} \frac{1}{N!}
\sum _{\stackrel{X_{1},\dots ,X_{N} \subset \Lambda}{\cup X_{j} =
\Lambda } }
\prod _{j} K(X_{j},i\phi)
\end{equation}
The rough idea will be that each $X_{j}$ is a contour $\Gamma _{j}$ or
several conglomerated contours together with a collar of width $M$
which is thick enough so that the effect of the contour is locked
inside $X_{j}$. To take care of regions without contours there will
also be the dominant possibility that $X$ is a single cube.

The outcome of our manipulations will be a polymer activity $K (X,\phi )$
whose $p-$th variational derivative with respect to $\phi $ is of size
\begin{eqnarray}\label{A.K1} 
	\frac{p!}{a^{p}} 
	\exp \left(
	\kappa \|\phi \|^{2}_{X} - O(a)|X| 
	\right ),
\ \ \
	a = [z l_{D}^{3}]^{1/2},
\ \ \
	|X| > 1
\end{eqnarray}
$a^{2}$ is the number of particles in a Debye sphere and is very large
near the Debye-H\"uckel limit. The $\exp (- O(a)|X|)$ is the tunneling
factor promised in section~\ref{sec-tunneling}. $|X|$ is the number of
unit boxes in $X$. $\| \ \|_{X}$ is a norm involving $\int _{X}|\phi
|^{2}$ and integrals of derivatives of $\phi $. The constant $\kappa $
is small enough so that $\exp \left(\kappa \|\phi \|^{2}_{X} \right)$
can be integrated with respect to $d\mu _{{\rm DH}}$. Since polymers
never overlap this is enough for a proof that the expansion of
section~\ref{sec-polymer-rep2} is convergent. The non-overlapping of
polymers is the reason why they are useful for obtaining convergent
expansions.

To understand why this crucial limitation on $\kappa $ cannot be
achieved unless the tunneling effects represented by contours are
taken into consideration consider the one dimensional version
of the Villain approximation.
\begin{equation}\label{A.villain}
e^{a ^{2} \cos (\phi/a )}
=
e^{R_{3}(\phi)}
\sum _{n} e^{-\frac{1}{2}(\phi-2\pi a n)^{2}+a ^{2}} 
\end{equation}
If we restrict $R_{3}(\phi)$ to the period $[-\pi a ,\pi a ]$ and note
that it is largest at $\phi = a \pi $ we find, by throwing away all
but the $n = 0$ term, an upper bound
\begin{equation}\label{A.Rsub3-1}
e^{R_{3}(\phi)} \leq e^{\frac{1}{2} \kappa \phi^{2}}; \ \ \ 
\kappa = 1 - \frac{4}{\pi ^{2}} < 1
\end{equation}
Since $R_{3}(\phi)$ is a periodic function the bound holds for all $\phi$.
$R_{3}$ is analytic with more or less the same Gaussian bound in the
strip $\phi + iy$ with $|y| / a \ll 1$ so 
\begin{equation}\label{A.Rsub3-3}
\frac{d^{p}}{d\phi ^{p}} e^{R_{3}(\phi )} 
\sim \frac{p!}{a ^{p}} e^{\frac{1}{2} \kappa \phi^{2}}
\end{equation}
$\kappa  < 1$ means that $\exp \left(R_{3} \right)$ and all derivatives
can be integrated against the one dimensional analogue of $d\mu _{{\rm
DH}}$, namely the Gaussian $\exp \left(-\phi^{2}/2 \right)$
\begin{equation}\label{A.Rsub3-2}
\frac{1}{\sqrt{2\pi }} \int e^{-\phi^{2}/2}
\frac{d^{p}}{dx^{p}}e^{R_{3}(\phi)}dx =
\left\{ \begin{array}{ll}
O(a ^{-p})&	\mbox{if } p \not = 0\\
\rightarrow 1&	\mbox{if } p = 0
\end{array} \right.
\end{equation}
as $a \rightarrow \infty $. This fails if $R_{3}$ is instead
defined as the error in the single well approximation $a ^{2}\cos (\phi/a)
\approx 1 - \phi^{2}/2 + R_{3}$.

The tunneling phenomena contained in the Villain approximation are
invisible in perturbation theory but 
they are the reason why perturbation theory will not converge. The
appearance of $\exp (-a |X| )$ in (\ref{A.K1}) quantifies the physical
idea that the tunneling effects are very rare in the Debye-H\"uckel
limit when $a$ is large.  This bound (\ref{A.K1}) together with the
considerations of section~\ref{sec-polymer-rep2} implies the
exponential decay of correlations.

The first task on the way to (\ref{A.2}) is to obtain an exact version
of (\ref{2.7.Sdecomp}) so 
that the sum over $h$ will factor as in (\ref{2.7.contour}).  Let $M$
be a (large) integer.  Given $h$ we choose the translation $g$ to
minimize $S_{{\rm DH}}(- ih, ig)$ subject to the constraint that
$g(\br ) = h(\br )$ at every point that is distant by more than $M-1$
from any contour in $h$. The price of imposing this constraint is that
in (\ref{A.1}) we no longer have ${S'}_{{\rm DH}}(-ih , ig; i\phi) =
0$ as was the case in section~\ref{sec-tunneling}.  Now ${S'}_{{\rm
DH}}(-ih , g; i\phi)$ is an integral of $\phi$ and derivatives of
$\phi$ times derivatives of $g$ over the surface of the region where
the constraint holds.  These derivatives are of order $O(\exp(-M))$
because the constraint is almost irrelevant since the minimizer
without the constraint satisfies $g(\br ) = h(\br ) + O(\exp
\left(-\dist(\br ,\Gamma \right))$.  We choose $M$ large so that
${S'}_{{\rm DH}}(-ih , g; i\phi)$ is as small as the error term
$R_{3}(ig + i\phi)$.

With this choice of $g$ we have an exact version of the additivity
property of $S_{{\rm DH}}(- ih, ig)$ in (\ref{2.7.Sdecomp}):
whenever $h = h_{1} + h_{2}$,
\begin{equation}\label{A.local3}
S_{{\rm DH}}(- ih, ig) = \sum _{j=1,2} S_{{\rm DH}}(- ih_{j},
ig_{j}) 
\end{equation}
provided every contour in $h_{1}$ is separated by a distance of $2M$
or more from every contour in $h_{2}$. See the discussion below
(\ref{2.7.ising}) and note that our constraint forces the cross
terms to vanish. Under the same conditions we also have 
\begin{equation}\label{A.local4}
{S'}_{{\rm DH}}(-ih , g; i\phi) = \sum _{j} {S'}_{{\rm
DH}}(-ih_{j} , g_{j}; i\phi)
\end{equation}

Define $K(X,i\phi )$ to vanish if $X$ is the empty set or not a connected
union of cubes, otherwise
\begin{equation}\label{A.5}
K(X,i\phi) = e^{R_{3}(X,ig + i\phi)}
\left\{
\begin{array}{l}
\sum_{h: X(h) = X}
e^{- S_{{\rm DH}}(- ih, ig)}
e^{-{S'}_{{\rm DH}}(-ih , ig; i\phi) } \\[.1in]
1
\end{array}
\right.
\end{equation}
The second line applies when $X$ is a single cube. The constraint
$X(h) = X$ is defined below.  The constraint forces $X$ to hold
contours when it is not a single cube and contours are responsible for
the $\exp (-O (a)|X| )$ in (\ref{A.K1}). This definition leads to the
desired representation 
\begin{equation}\label{A.6} \Xi_{\Lambda } /(\Xi_{{\rm ideal},\Lambda
}\Xi_{{\rm DH},\Lambda }) = \int d\mu _{{\rm DH}}(\phi) \, \sum_{N}
\frac{1}{N!}\sum  _{\stackrel{X_{1},\dots ,X_{N} \subset \Lambda}{\cup X_{j} =
\Lambda } } \prod _{j}
K(X_{j},i\phi)
\end{equation}

It is convenient to measure distance using
\begin{equation}\label{A.dist}
\dist (\br ,\br ') = \max _{i} |r_{i} - r_{i}'|
\end{equation}
so that the unit cube centered on the origin is $\{\br : \dist (0,\br
) < 1/2\}$.  $X(h)$ is defined to be the open set consisting of all
points closer than $M$ to some contour in $h$. To derive (\ref{A.6})
note that any $h$ occurring in the sum (\ref{A.1}) determines 
$X(h)$ which decomposes uniquely into disjoint open connected subsets
$X_{j}$. Each $X_{j}$ determines a unique $h_{j}$ such that 
\begin{eqnarray}\label{A.hdecomp}
&&X(h_{j}) = X_{j}; \ \ \ 
h_{j} = 0 \mbox{ outside }\Lambda \nn \\
&&h_{j} = h + \mbox{ a constant when restricted to } X_{j}\nn \\
&&h = \sum _{j}h_{j}
\end{eqnarray}
The first two conditions determine $h_{j}$ and the last is implied
by them.  Therefore the sum over $h$ in (\ref{A.1}) is equivalent
to summing over $N$, $X_{1},\dots ,X_{N}$ followed by $h_{1},\dots
,h_{N}$. By (\ref{A.local3}) and the symmetry
\begin{equation}\label{A.symmetry}
e^{R_{3}(X_{k},i\sum _{j}g_{j} + i\phi)}
=
e^{R_{3}(X_{k},ig_{k} + i\phi)} 
\end{equation}
we can factor the summands. This symmetry is the one we remarked in
(\ref{2.7.1}) and it holds in this form because for $j \not = 
k$ $g_{j} = h_{j}$ is constant on $X_{k}$.  Incidentally a fractional
charge immersed in the system destroys this symmetry.  See the remarks
on fractional charges in section~\ref{sec-tunneling}.

\subsection{Proofs for theorems in
chapter VIII  } 
\label{sec-proofs}

\bigskip \noindent {\em Proof of theorems~\ref{thm-C.stability.a} and
\ref{thm-C.stability.b}:\/} 
Following \cite{BrKe87}.  Let 
$\chi (a \geq \tau )$ denote a function of $\tau $ which is $1$ when
$a \geq \tau$ and zero otherwise. Then
\begin{eqnarray}
U_{N}\circ \bsigma (T,\bs ) &=&  \sum _{ij} 
\int _{0}^{\sigma_{ij}(T,\bs )} d\tau \,
V_{ij}'(\tau ) \nn \\
&=& \int _{0}^{1} d\tau \, \sum _{ij}
V_{ij}'(\tau ) \chi (\sigma_{ij}(T,\bs ) \geq \tau ) 
\end{eqnarray}
A value of $\tau $ and a tree $T$ determine a partition of the
particles $=$ vertices of $T$ into clusters: simply erase all bonds
$ij \in T$ for which $s_{ij} < \tau $. The rest of the tree falls into
connected subtrees. Each cluster consists of the vertices in each
subtree. The bonds $ij \in T$ we erased are in fact eliminated by the
$\chi$ in the above equation. The key point is that if $lk$ is another
bond not in $T$ with $k$ and $l$ is separate clusters that used to be
connected by $ij$ then it too is eliminated by $\chi$ because $s_{ij}
< \tau $ implies $\sigma_{kl}(T,\bs) < \tau $. Therefore at each value
of $\tau $ we have a $\tau $ dependent partition of the vertices into
clusters $C_{\alpha }$ such that
\begin{equation} 
\sum _{ij} V_{ij}'(\tau ) \chi (\sigma_{ij}(T,\bs ) \geq
\tau ) = \sum _{\alpha }\sum_{ij \in C_{\alpha }} V_{ij}'(\tau )
\end{equation} 
Thus a stability estimate $\sum V_{ij}'(\tau ) \geq
-B(\tau )N$ is inherited by $U_{N}\circ \bsigma (T,\bs )$, because we can
apply it to each cluster in the last equation and then integrate over
$\tau $. Theorem~\ref{thm-C.stability.a} follows easily.

{\it Theorem~\ref{thm-C.stability.b}:\/} Following \cite{GoMa81} 
there is a systematic way to improve stability
estimates by using numerically calculated constants for small
clusters. The simplest instance is to note that for clusters of single
particles we have no interaction to bound so that
\begin{equation} \sum _{ij} V_{ij}'(\tau ) \chi (\sigma_{ij}(T,\bs )
\geq \tau ) \geq -B(\tau ) N_{1}
\end{equation}
where $N_{1}$ is the number of particles not in single clusters,
\begin{eqnarray} N_{1} &=&
|\{i: i \in \mbox{ some } C_{j} \mbox{ with } |C_{j}| > 1 \}| \nn \\
&\leq &
|\{i: s_{ij} \geq \tau \mbox{ some } ij \mbox{ in } T \}| \nn \\ &\leq &
\sum _{i,j} \chi (s_{ij} \geq \tau ) =
2 \sum _{ij} \chi (s_{ij} \geq \tau )
\end{eqnarray}
Therefore
\begin{equation}
U_{N}\circ \bsigma (T,\bs )
\geq
-\int _{0}^{1} d\tau \, 2 B(\tau )
\sum _{ij} \chi (s_{ij} \geq \tau )
\end{equation}
which is (\ref{C.12}). \qed

{\em Proof of Theorem ~\ref{thm-forest}:\/} We know the case $N =
2$. We assume $N > 2$ and make the inductive hypothesis that the
theorem holds for all cases with less than $N$ vertices. It is
sufficient to prove that both sides have the same $t$ derivative and
agree at $t=0$. The $t$ derivative of the right hand side is
\begin{equation} \sum _{G \in \cF}\sum _{b \in G} \int^{t}
d^{G\setminus b}s \, \partial ^{G} F \circ \bsigma(G,\bs )_{s_{b}=t}
\end{equation} We interchange the sums and obtain $\sum X_{b}$ where
\begin{equation}\label{F.5} X_{b} = \sum _{G \in \cF, G \ni b}
\int^{t} d^{G\setminus b}s \, \partial ^{G} F \circ \bsigma(G,\bs
)_{s_{b}=t} \end{equation}

Now we fix $b$. To use the inductive hypothesis, we reduce $N$ by one
by rewriting $X_{b}$ as a sum over all forests $\bar{G}$ on $N-1$
vertices. The $N-1$ vertices are obtained from the original vertices
$\{1,2,\dots ,N \}$ by identifying the vertices at either end of
$b$. The possible bonds $\bar{c}$ are viewed as sets of identified
bonds on $\{1,2,\dots ,N \}$. For example, if $b = 12$, then $1i$ and
$2i$ become the same bond which we would denote by $\bar{1i}$ or
$\bar{2i}$.

$\bar{\bs }$ denotes a vector of parameters $\bar{s}_{\bar{c}}$
assigned to bonds $\bar{c}$. Given such parameters, we determine
parameters $\bs $ by assigning $\bar{s}_{\bar{c}}$ to all bonds $c \in
\bar{c}$ and $t$ to bond $b$. Thus define $\bs = \bj(\bar{\bs})$ where
the $c$ component is given by \begin{equation} j_{c}(\bar{\bs}) =
\left \{ \begin{array}{ll} \bar{s}_{\bar{c}}
&       \mbox{ if } c \not = b\\
t
&       \mbox{ if } c = b
\end{array}
\right.
\end{equation}
To each forest $G \ni b$ on $N$ vertices we associate a unique forest
$G/b$ on $N-1$ vertices by identifying the vertices in the pair
$b$. Thus
\begin{equation}
G/b = \{\bar{c}: c \in G \}
\end{equation}

\bigskip
\noindent
For any function $f$ of parameters $\bs $, 
\begin{equation}\label{F.6}
\sum _{G: G/b = \bar{G}}
(\partial ^{G \setminus b} f) \circ \bj(\bar{\bs} ) =
\partial ^{\bar{G}} (f \circ \bj)(\bar{\bs }) 
\end{equation}
To understand this, consider
$$
\left( \frac{\partial f}{\partial u} \right)_{u=s} + \left(
\frac{\partial f}{\partial v} \right)_{v=s} = \frac{\partial
}{\partial s} \left(f(u,v)_{u=v=s}\right).
$$

\bigskip \noindent Also, for any function $f$ of parameters $\bs $ and
$G \ni b$,
\begin{equation}\label{F.7} 
\int ^{t} d^{G\setminus b}s \, f \circ \bsigma
(G,\bs )_{s_{b}=t} = \int ^{t} d^{G/b}\bar{s} \, f \circ \bj \circ
\bar{\bsigma}(G/b,\bar{\bs }) 
\end{equation}
The key point here is that $\sigma _{c} (G,\bs )_{s_{b}=t}$ depends
only on $\bar{c}$.

\bigskip \noindent We return to (\ref{F.5}). By summing over $G$ with
$G/b$ held fixed 
\begin{equation} X_{b} = \sum_{\bar{G}}\sum
_{G:G/b=\bar{G}} \int^{t} d^{G\setminus b}s \, \partial ^{G} F \circ
\bsigma(G,\bs )_{s_{b}=t}. 
\end{equation} By (\ref{F.7}) this equals
\begin{equation}
\sum_{\bar{G}}\sum _{G:G/b=\bar{G}} \int^{t} d^{\bar{G}} \bar{s} \,
\partial ^{G} F \circ \bj \circ \bar{\bsigma}(\bar{G},\bar{\bs } )
\end{equation} which by (\ref{F.6}), with $f = \partial _{b}F$ is
\begin{equation} \sum_{\bar{G}} \int^{t} d^{\bar{G}} \bar{s} \,
(\partial ^{\bar{G}} \partial _{b} F \circ \bj) \circ
\bar{\bsigma}(\bar{G},\bar{\bs } )
\end{equation}
which by the inductive hypothesis is
\begin{equation}
(\partial _{b}F )\circ \bj(t\bar{\1}) = \partial _{b}F(t\1)
\end{equation} Therefore the $t$ derivative of the right hand side of
(\ref{C.4}) is $\sum _{b}\partial _{b}F(t\1)$ which is the $t$
derivative of the left hand side of (\ref{C.4}). \qed

\section*{Acknowledgments}

D. C. Brydges thanks the National Science Foundation for partial
support under grants DMS-9401028 and DMS-9706166 and P. Federbush,
T. Kennedy and G. Keller for collaborations on which some of this
review is based.  Ph. Martin acknowledges several informative
discussions with A. Alastuey, F. Cornu, and B. Jancovici.

\bibliographystyle{agsm} 
 

\end{document}